\documentclass[twoside,12pt]{article}

\usepackage{graphicx,color}
\usepackage{equations}

\makeatletter
\def\@cite#1#2{{[{#1}]\if@tempswa\typeout
{IJCGA warning: optional citation argument
ignored: `#2'} \fi}}

\newcount\@tempcntc
\def\@citex[#1]#2{\if@filesw\immediate\write\@auxout{\string\citation{#2}}\fi
  \@tempcnta\z@\@tempcntb\m@ne\def\@citea{}\@cite{\@for\@citeb:=#2\do
    {\@ifundefined
       {b@\@citeb}{\@citeo\@tempcntb\m@ne\@citea\def\@citea{,}{\bf ?}\@warning
       {Citation `\@citeb' on page \thepage \space undefined}}%
    {\setbox\z@\hbox{\global\@tempcntc0\csname b@\@citeb\endcsname\relax}%
     \ifnum\@tempcntc=\z@ \@citeo\@tempcntb\m@ne
       \@citea\def\@citea{,}\hbox{\csname b@\@citeb\endcsname}%
     \else
      \advance\@tempcntb\@ne
      \ifnum\@tempcntb=\@tempcntc
      \else\advance\@tempcntb\m@ne\@citeo
      \@tempcnta\@tempcntc\@tempcntb\@tempcntc\fi\fi}}\@citeo}{#1}}
\def\@citeo{\ifnum\@tempcnta>\@tempcntb\else\@citea\def\@citea{,}%
  \ifnum\@tempcnta=\@tempcntb\the\@tempcnta\else
   {\advance\@tempcnta\@ne\ifnum\@tempcnta=\@tempcntb \else \def\@citea{--}\fi
    \advance\@tempcnta\m@ne\the\@tempcnta\@citea\the\@tempcntb}\fi\fi}
\makeatother

\newlength{\captsize}           \let\captsize=\footnotesize
\newlength{\captwidth}          
\newlength{\beforetableskip}    
\newcommand{\capt}[1]{\begin{minipage}{\captwidth}
              \let\normalsize=\captsize
              \caption[0]{#1}
              \end{minipage}\\ \vspace{\beforetableskip}}

\makeatletter
      \long\def\@makecaption#1#2{\vskip 10 \p@
      \setbox\@tempboxa\hbox{\textbf{#1:} #2}
      \ifdim \wd\@tempboxa >\hsize
            \textbf{#1:} #2\par                 
      \else
         \hbox to \hsize{\box\@tempboxa\hfil}
      \fi}
\makeatother
         
\newcommand{\fnref}[1]{~\ref{#1}}

\newcommand{\pp}{p\bar{p}}

\newcommand{\bb}{b\bar{b}}

\newcommand{\lpm}{\ell^+\ell^-}
\newcommand{\nn}{\nu\bar{\nu}}

\newcommand{\trilep}{\ell^\pm\ell^{\prime\pm}\ell^\mp}
\newcommand{\lsdilep}{\ell^\pm\ell^\pm jj}

\newcommand{\lsim}{\stackrel{<}{\sim}}
\newcommand{\gsim}{\stackrel{>}{\sim}}

\newcommand\longdash{\hbox{\rm{\phantom{a}---\phantom{a}}}}

\def\nicefrac#1#2{\hbox{${#1\over #2}$}}

\def\mpl{M_{\rm PL}}
\def\beq{\begin{equation}}
\def\eeq{\end{equation}}
\def\crr{\crcr\noalign{\vskip .1in}}
\newenvironment{Eqnarray}%
     {\arraycolsep 0.14em\begin{eqnarray}}{\end{eqnarray}}
\newcommand{\mathbold}[1]{\mbox{\boldmath $\bf#1$}}
\def\beqa{\begin{Eqnarray}}
\def\eeqa{\end{Eqnarray}}
\def\beqno{\begin{eqalignno}}
\def\eeqno{\end{eqalignno}}
\def\ifmath#1{\relax\ifmmode #1\else $#1$\fi}
\def\half{\ifmath{{\textstyle{1 \over 2}}}}
\def\lsim{\mathrel{\raise.3ex\hbox{$<$\kern-.75em\lower1ex\hbox{$\sim$}}}}
\def\gsim{\mathrel{\raise.3ex\hbox{$>$\kern-.75em\lower1ex\hbox{$\sim$}}}}
\def\eq#1{eq.~(\ref{#1})}

\def\fig#1{fig.~\ref{#1}}
\def\Fig#1{Fig.~\ref{#1}}
\def\figns#1#2{figs.~\ref{#1} and~\ref{#2}}
\def\figs#1#2{figs.~\ref{#1}--\ref{#2}}

\def\Ref#1{ref.~\cite{#1}}
\def\Refs#1#2{refs.~\cite{#1} and \cite{#2}}
\def\Rref#1{Ref.~\cite{#1}}

\def\Sec#1{Section~\ref{#1}}
\def\Secs#1#2{Sections~\ref{#1} and \ref{#2}}
\def\eqs#1#2{eqs.~(\ref{#1})--(\ref{#2})}
\def\Eq#1{Eq.~(\ref{#1})}
\def\Eqs#1#2{Eqs.~(\ref{#1})--(\ref{#2})}
\def\eqns#1#2{eqs.~(\ref{#1}) and (\ref{#2})}
\def\tanb{\tan\beta}
\def\sinb{\sin\beta}
\def\cosb{\cos\beta}

\def\cosa{\cos\alpha}
\def\sinbma{\sin(\beta-\alpha)}
\def\cosbma{\cos(\beta-\alpha)}
\def\sinbmaii{\sin^2(\beta-\alpha)}
\def\cosbmaii{\cos^2(\beta-\alpha)}
\def\hsm{h_{\rm SM}}
\def\mhsm{m_{h_{\rm SM}}}
\def\hl{h}
\def\ha{A}
\def\hh{H}
\def\hp{H^+}
\def\hpm{H^\pm}
\def\mha{m_{\ha}}
\def\mhl{m_{\hl}}
\def\mhh{m_{\hh}}
\def\mhpm{m_{\hpm}}
\def\mhmax{m_h^{\rm max}}
\def\mz{m_Z}
\def\mw{m_W}

\def\mt{m_t}
\def\mb{m_b}
\def\mstopa{M_{\widetilde t_1}}
\def\mstopb{M_{\widetilde t_2}}
\def\msusy{M_{\rm S}}
\def\msusyy{M_{\rm S}^2}
\def\MSUSY{M_{\rm SUSY}}

\def\SM{Standard Model}
\def\phm{\phantom{-}}
\def\ls#1{\ifmath{_{\lower1.5pt\hbox{$\scriptstyle #1$}}}}
\def\app#1#2#3{{\sl Act. Phys. Pol. }{\bf B#1} (#2) #3}

\def\npb#1#2#3{{\sl Nucl. Phys. }{\bf B#1} (#2) #3}
\def\jpg#1#2#3{{\sl J. Phys. }{\bf G#1} (#2) #3}
\def\plb#1#2#3{{\sl Phys. Lett. }{\bf B#1} (#2) #3}
\def\prd#1#2#3{{\sl Phys. Rev. }{\bf D#1} (#2) #3}
\def\pR#1#2#3{{\sl Phys. Rev. }{\bf #1} (#2) #3}
\def\prl#1#2#3{{\sl Phys. Rev. Lett. }{\bf #1} (#2) #3}
\def\prc#1#2#3{{\sl Phys. Reports }{\bf #1} (#2)~#3}
\def\cpc#1#2#3{{\sl Comp. Phys. Commun. }{\bf #1} (#2) #3}
\def\nim#1#2#3{{\sl Nucl. Inst. Meth. }{\bf #1} (#2) #3}
\def\pr#1#2#3{{\sl Phys. Reports }{\bf #1} (#2) #3}
\def\sovnp#1#2#3{{\sl Sov. J. Nucl. Phys. }{\bf #1} (#2) #3}
\def\jetpl#1#2#3{{\sl JETP Lett. }{\bf #1} (#2) #3}
\def\zpc#1#2#3{{\sl Z. Phys. }{\bf C#1} (#2) #3}
\def\ptp#1#2#3{{\sl Prog.~Theor.~Phys.~}{\bf #1} (#2) #3}

\def\aop#1#2#3{{\sl Ann.~of~Phys.~}{\bf #1} (#2) #3}
\def\epjc#1#2#3{{\sl Eur.~Phys.~J.~}{\bf C#1} (#2) #3}
\def\ijmpa#1#2#3{{\sl Int.~J.~Mod.~Phys.~}{\bf A#1} (#2) #3}
\def\fP#1#2#3{{\sl Fortschr.~Phys.~}{\bf #1} (#2) #3}
\def\9{\phantom 0}     

\begin{document}

\begin{flushright}
FERMILAB-Pub-02/114-T \\
SCIPP 02/07     \\
hep--ph/0208209 \\
\end{flushright}
\begin{center}
{\LARGE \bf  Higgs Boson Theory and Phenomenology}\\[1cm]
{\Large Marcela Carena$^*$ and Howard E. Haber$^\dagger$}\\[5pt]
{\large \it $^*$  Fermi National Accelerator Laboratory \\
 P.O. Box 500, Batavia, IL 60510, USA}\\[.3cm]
{\large \it $^\dagger$ Santa Cruz Institute for Particle Physics  \\
   University of California, Santa Cruz, CA 95064, USA} 
\end{center}

\begin{abstract}
Precision electroweak data presently favors a weakly-coupled Higgs
sector as the mechanism responsible for electroweak symmetry
breaking.  Low-energy supersymmetry provides a natural framework for
weakly-coupled elementary scalars.  In this review, we summarize the
theoretical properties of the Standard Model (SM) Higgs boson and the Higgs
sector of the minimal supersymmetric extension of the Standard Model (MSSM).
We then survey the phenomenology of the SM and MSSM Higgs bosons at
the Tevatron, LHC and a future $e^+e^-$ linear collider.  We focus on
the Higgs discovery potential of present and future colliders and stress the
importance of precision measurements of Higgs boson properties.
\end{abstract}

\section{Introduction---Origin of Electroweak
Symmetry Breaking}
\label{sec:intro}

Deciphering the mechanism that breaks the electroweak symmetry
and generates the masses of the known fundamental particles
is one of the central challenges of particle physics.
The Higgs mechanism~\cite{Higgs-orig} in its most general form
can be used to explain the observed masses of the $W^\pm$
and $Z$ bosons as a consequence
of three Goldstone bosons ($G^\pm$ and $G^0$) that
end up as the longitudinal components of the gauge bosons.
These Goldstone bosons are generated by
the underlying dynamics responsible for electroweak symmetry breaking.
However, the fundamental nature of this dynamics is still unknown.
Two broad classes of electroweak
symmetry breaking mechanisms have been pursued theoretically.  In one
class of theories, electroweak symmetry breaking dynamics is
weakly-coupled, while in the second class of theories the dynamics is
strongly-coupled.

The electroweak symmetry
breaking dynamics that is employed by the Standard Model posits a
self-interacting complex doublet of scalar fields, which consists of four
real degrees of freedom~\cite{hhg}.  Renormalizable interactions are
arranged in such a way that the neutral component of the scalar doublet
acquires a vacuum expectation value, $v=246$~GeV, which sets the scale
of electroweak symmetry breaking.
Consequently, three massless Goldstone bosons are generated, while
the fourth scalar degree of freedom that remains in the physical spectrum
is the CP-even neutral Higgs boson ($\hsm$)  of the Standard Model.
It is further assumed in the Standard Model that the scalar doublet also
couples to fermions through
Yukawa interactions.  After electroweak symmetry breaking, these interactions
are responsible for the generation of quark and charged
lepton masses.  This approach is an example of weak electroweak
symmetry breaking.  Assuming that $\mhsm\lsim 200$~GeV, all
fields remain weakly interacting at energies up to the Planck scale.
In the weakly-coupled approach to electroweak symmetry breaking, the
Standard Model is very likely embedded in a supersymmetric
theory~\cite{susyreview} in
order to stabilize the large gap between the electroweak and the Planck
scales in a natural way~\cite{natural,susynatural}.
These theories predict a spectrum
of Higgs scalars~\cite{susyhiggs},
with the properties of the lightest Higgs scalar often resembling that
of the Standard Model (SM) Higgs boson.

Alternatively, strong breaking of electroweak symmetry is accomplished
by new strong interactions near the TeV scale~\cite{weinberg}.
More recently, so-called ``little Higgs models'' have been proposed in
which the scale of the new strong interactions is pushed up above
10~TeV~\cite{littlehiggs}, and the lightest Higgs scalar resembles
the weakly-coupled SM Higgs boson.  In a more speculative direction,
a new approach to electroweak symmetry breaking has
been explored in which extra space dimensions beyond
the usual $3+1$ dimensional spacetime are
introduced~\cite{extradim} with
characteristic sizes of order $(\rm TeV)^{-1}$.
In such scenarios, the mechanisms for
electroweak symmetry breaking are inherently
extra-dimensional, and the resulting phenomenology may be significantly
different from the usual approaches mentioned above.
All these alternative approaches lie outside the scope of this review.

\begin{figure}[t!]
\begin{center}
\unitlength1cm
\begin{picture}(15,8.3)
\put(-2.0,-1.0){\includegraphics[height=10.4cm,width=8.85cm]{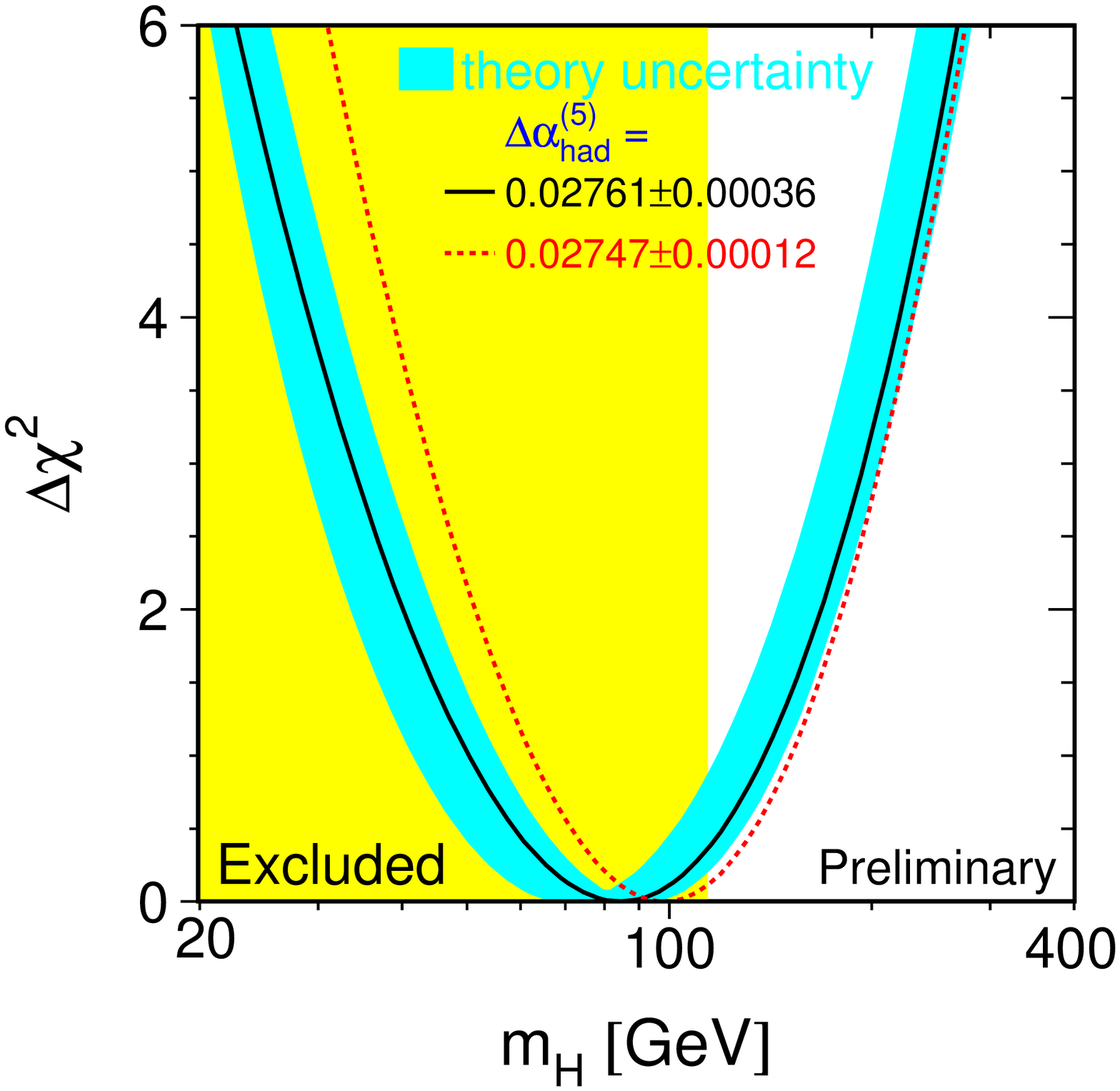}}
\put(8.05,-0.15){\includegraphics[height=8.47cm,width=7.975cm]{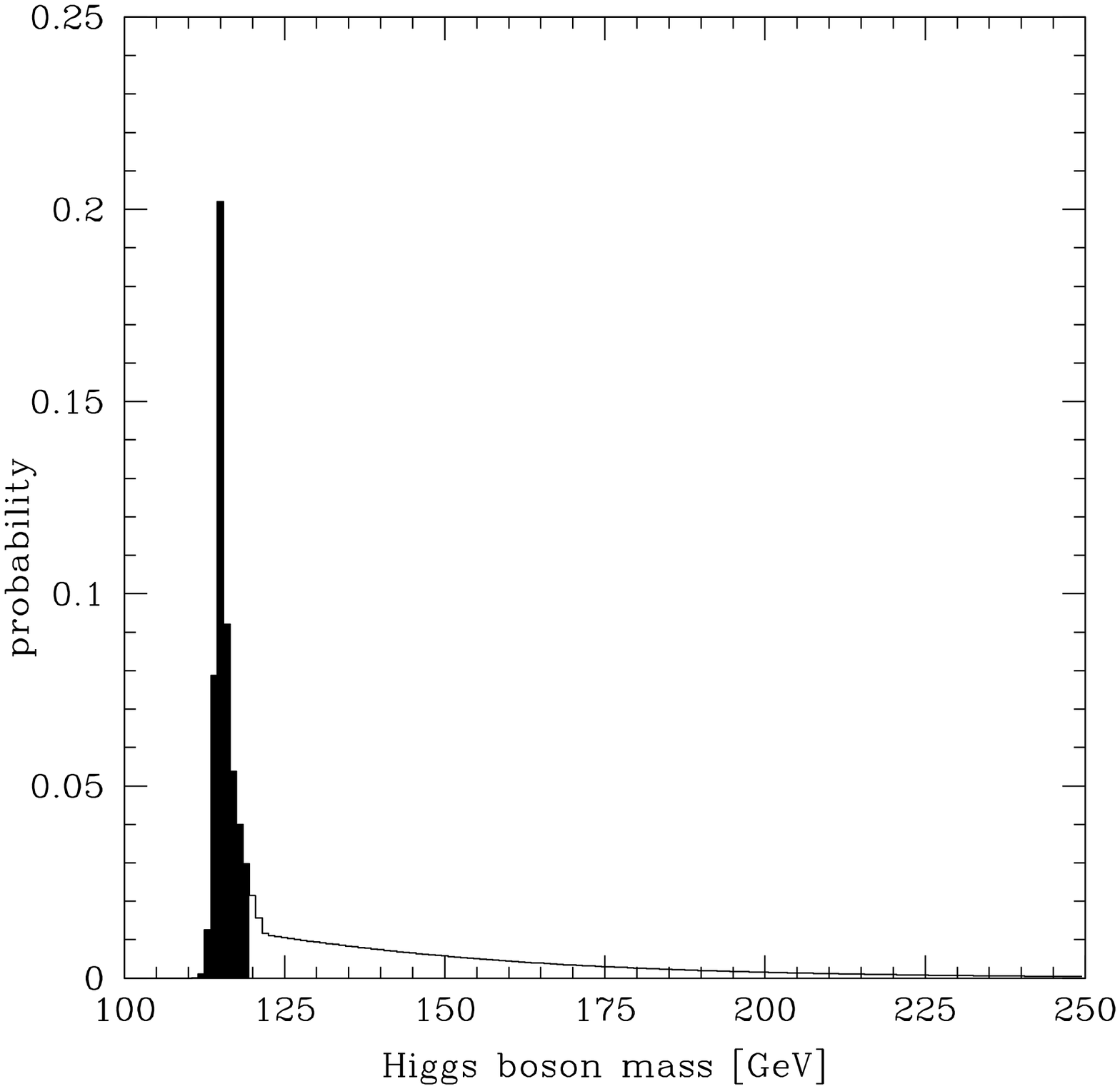}}
\end{picture}
\end{center}
\capt{\label{fig:blueband}
(a)~The ``blueband plot'' shows $\Delta
\chi^2\equiv \chi^2-\chi^2_{\min}$ as a function of the Standard Model
Higgs mass~\protect\cite{lepewwg}.
The solid line is a result of a global fit
using all data; the band represents the theoretical
error due to missing higher order corrections.  The rectangular shaded region
shows the 95\% CL exclusion limit on the Higgs mass from direct
searches at LEP~\protect\cite{LEPHiggs}.
(b)~Probability distribution function for the Higgs boson
mass, including all available direct and indirect data~\protect\cite{erler}.
The probability is shown for 1~GeV bins.  The shaded and unshaded
regions each correspond to an integrated probability of 50\%}
\end{figure}

Although there is as yet no direct evidence for the nature of
electroweak symmetry breaking dynamics, present data
can be used to discriminate among the different approaches.  For example,
precision electroweak data, accumulated in the past decade at LEP, SLC,
the Tevatron and elsewhere,
strongly support the Standard Model with a weakly-coupled
Higgs boson~\cite{lepewwg}.
Moreover, the contribution of new physics, which
can enter through $W^\pm$ and $Z$ boson vacuum polarization
corrections, is severely constrained.  This fact has already
served to rule out several models of strongly-coupled electroweak symmetry
breaking dynamics.  The Higgs boson contributes
to the $W^\pm$ and $Z$ boson vacuum polarization through loop effects, and
so a global Standard Model fit to the electroweak data yields information
about the Higgs mass.  The results of the LEP
Electroweak Working Group analysis shown in \fig{fig:blueband}(a)
yield \cite{lepewwg}:
$\mhsm=81^{+52}_{-33}~{\rm GeV}$, and provides a 95\% CL upper limit of
$\mhsm<193$~GeV.
These results reflect the logarithmic sensitivity to the Higgs mass via
the virtual Higgs loop contributions to the various electroweak
observables.  The 95\% CL upper limit is consistent with the
direct searches at LEP~\cite{LEPHiggs}
that show no conclusive evidence for the Higgs
boson, and imply that $\mhsm> 114.4$~GeV at 95\%~CL.
\Fig{fig:blueband}(b) exhibits the most probable range of values for
the SM Higgs mass~\cite{erler}.  This mass range is consistent
with a weakly-coupled Higgs scalar that is expected to emerge from the
Standard Model scalar dynamics.

There are some loopholes that can be exploited to circumvent
this conclusion.  It is possible to construct models of new physics
where the goodness of the global Standard
Model fit to precision electroweak data is not compromised
while the strong upper limit on the Higgs mass is
relaxed.  In particular, one can construct effective
operators~\cite{newoperators,murayama}
or specific models~\cite{Peskin-Wells}
of new physics where the Higgs mass is significantly larger, but the new
physics contributions to the
$W^\pm$ and $Z$ vacuum polarizations, parameterized by the
Peskin-Takeuchi~\cite{peskin} parameters $S$ and $T$,
are still consistent with the experimental data.
In addition, some have argued that the
global Standard Model fit exhibits
possible internal inconsistencies~\cite{chanowitz}, which would suggest that
systematic uncertainties have been underestimated and/or new physics
beyond the Standard Model is required.
Thus, although weakly-coupled electroweak
symmetry breaking seems to be favored
by the strong upper limit on the Higgs mass, one cannot definitively
rule out all other approaches.

Nevertheless, one additional piece of data is very suggestive.  Within
the supersymmetric extension of the Standard Model, grand unification
of the electromagnetic, the weak and the strong gauge interactions can
be achieved in a consistent way, strongly supported by the prediction
of the electroweak mixing angle at low energy scales with an accuracy at the
percent level~\cite{IbanezRoss,susygut}.
The significance of this prediction is not easily matched by other
approaches.  For example, in strongly-coupled electroweak symmetry breaking
models, unification of couplings is not addressed {\it per se}, whereas in
extra-dimensional models it is often achieved by introducing new
structures at intermediate energy scales.
Unless one is willing to regard
the apparent gauge coupling unification as a coincidence, it is
tempting to conclude that weak electroweak symmetry breaking
with low-energy supersymmetry is the
preferred mechanism, leading to an expected mass of the lightest Higgs
boson below 200~GeV (less than 135~GeV in the simplest supersymmetric
models), and a possible
spectrum of additional neutral and charged Higgs bosons
with masses up to of order 1~TeV.

Henceforth, we shall assume that the dynamics of electroweak
symmetry breaking is a result of a weakly-coupled scalar sector.
The Standard Model is an effective field theory and provides a very good
description of the physics of elementary particles and their
interactions at an energy scale of ${\cal O}(100)$~GeV and below.
However, there must exist some energy scale, $\Lambda$, at which the
Standard Model breaks down.  That is, the Standard Model is no longer
adequate for describing the theory above $\Lambda$, and degrees of
freedom associated with new physics become relevant.
In particular, we know that $\Lambda\leq\mpl$, since
at an energy scale above the Planck scale, $\mpl\simeq 10^{19}$~GeV, quantum
gravitational effects become significant and the Standard Model must
be replaced by a more fundamental theory that incorporates
gravity.  (Similar conclusions
also apply to recently proposed extra-dimensional theories in which
quantum gravitational effects can become
significant at energies scales as low as ${\cal{O}}$(1 TeV)~\cite{extradim}.)
Of course, it is possible that new physics beyond the Standard Model
exists at an energy scale between the electroweak and Planck scale, in
which case the value of $\Lambda$ might lie significantly below
$\mpl$.\footnote{For example, the recent experimental evidence for neutrino
masses of order $10^{-2}$~eV or below cannot be strictly explained in
the Standard Model.  Yet, one can easily write down a dimension-5
operator responsible for neutrino masses that is suppressed by $v/\Lambda$.
If $m_\nu\sim 10^{-2}$~eV, then one obtains as a rough estimate
$\Lambda\lsim 10^{15}$~GeV.}

\begin{figure}[t!]
\begin{center}
\includegraphics*[width=9.1cm]{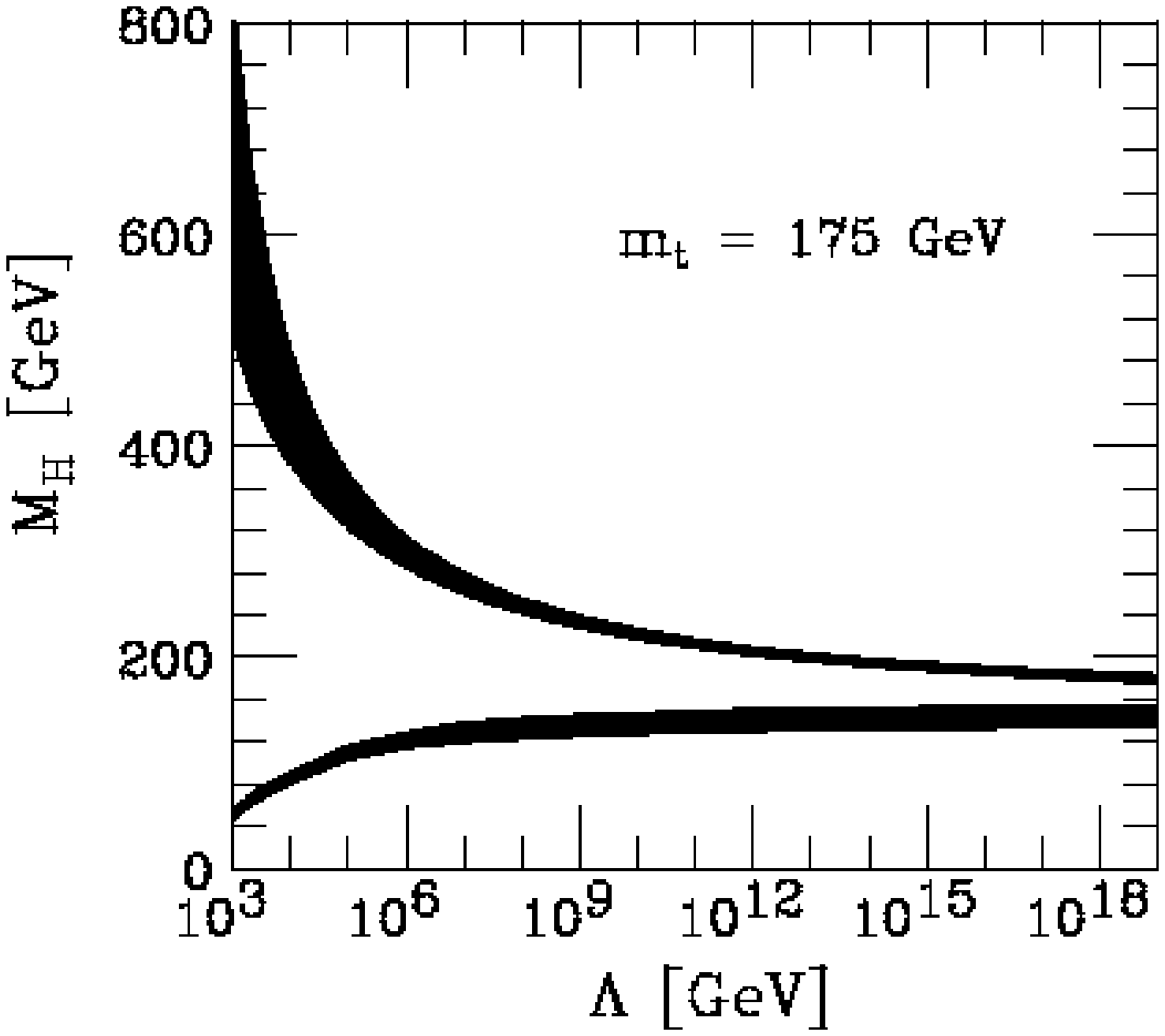}
\hspace*{5mm}
\includegraphics*[width=8.5cm]{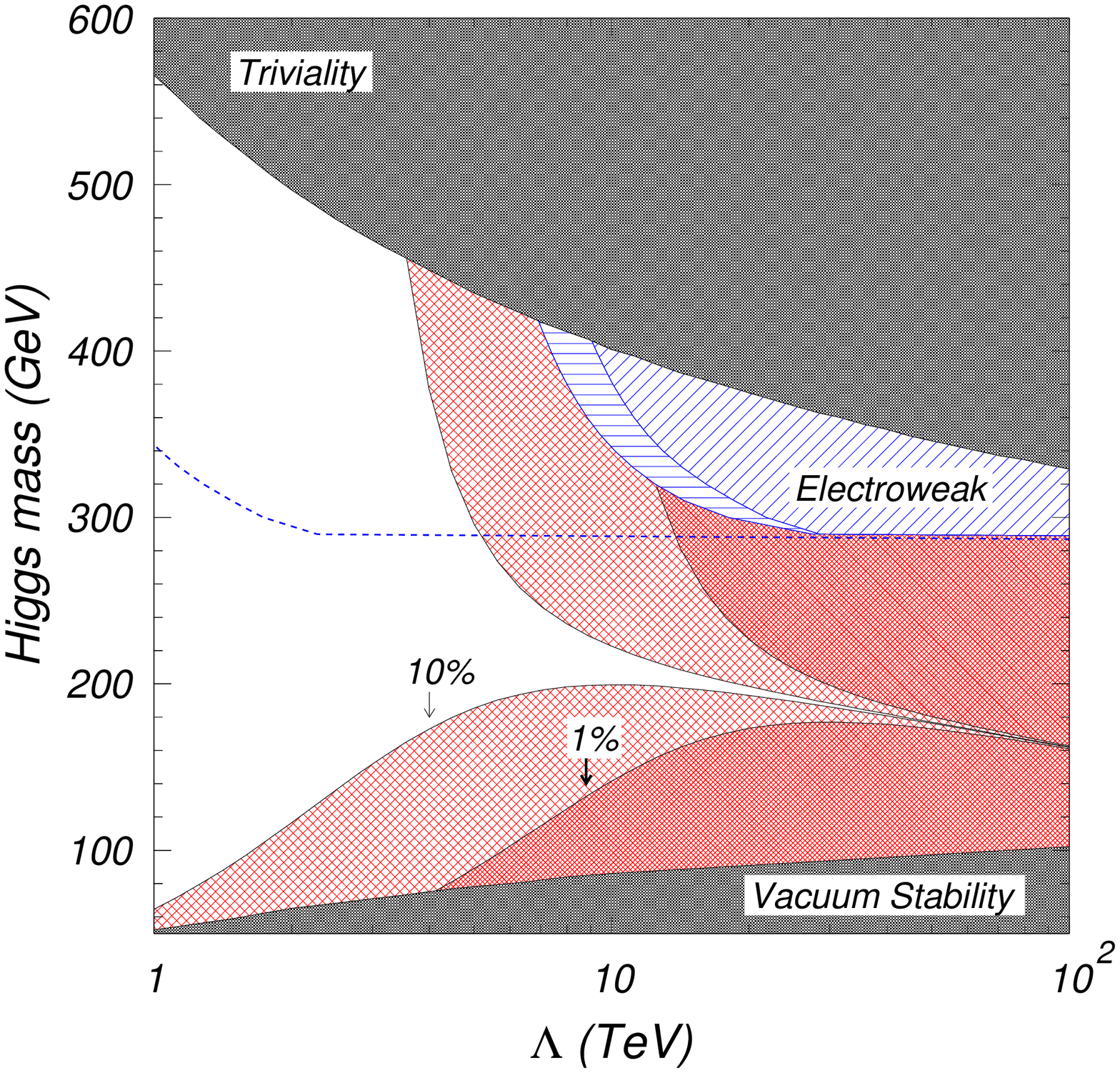}
\end{center}
\capt{\label{trivial} (a)~The upper \protect\cite{hambye}
and the lower \protect\cite{quiros}
Higgs mass bounds as a function of the
energy scale $\Lambda$ at which the Standard Model breaks down,
assuming $M_t=175$~GeV and $\alpha_s(m_Z)=0.118$, taken from
\protect\Ref{Riesselmann}.  The shaded
areas above reflect the theoretical uncertainties in the
calculations of the Higgs mass bounds.  (b)~Following \protect\Ref{murayama},
a reconsideration of the $\Lambda$ {\it vs.} Higgs mass plot with a
focus on $\Lambda<100$~TeV.  Precision electroweak measurements
restrict the parameter space to lie below the dashed line, based on a
95\% CL
fit that allows for nonzero values of $S$ and $T$ and the existence of
higher dimensional operators suppressed by $v^2/\Lambda^2$.  The
unshaded area has less than one part in ten fine-tuning.}
\end{figure}

The value of the Higgs mass itself can provide an important constraint
on the value of $\Lambda$.  If $\mhsm$ is too large, then the
Higgs self-coupling blows up at some scale $\Lambda$ below the
Planck scale \cite{hambye}.  If $\mhsm$ is too small, then the Higgs
potential develops
a second (global) minimum at a large value of the scalar field of order
$\Lambda$ \cite{quiros}.  Thus new physics must enter at a scale
$\Lambda$
or below in order that the global minimum of the theory correspond to the
observed SU(2)$\times$U(1) broken vacuum with
$v=246$~GeV.  Thus, given a value of
$\Lambda$, one can compute the minimum and maximum Higgs mass allowed.
The results of this computation (with shaded bands indicating the
theoretical uncertainty of the result) are illustrated in
\fig{trivial}(a)~\cite{Riesselmann}.
Consequently, a
Higgs mass range 130~GeV~$\lsim\mhsm\lsim 180$~GeV
is consistent with an effective Standard Model
that survives all the way to the Planck scale.\footnote{The
constraint on $\Lambda$ due to vacuum stability in \fig{trivial}
is less stringent if one allows for the electroweak vacuum
to be metastable, with a lifetime greater than the age of the universe.
An analysis of \Ref{Isidori:2001bm} finds that for a
sufficiently long-lived electroweak vacuum, the Higgs mass
lower limit of 130~GeV just quoted is reduced to about 115~GeV.}

However, the survival of the Standard Model as an effective theory all
the way up to the Planck scale is unlikely based on
the following ``naturalness''~\cite{natural} argument.  In
an effective field theory, particle masses and dimensionless couplings
of the low-energy theory are calculable in terms of parameters
of a more fundamental theory that
describes physics at the energy scale $\Lambda$.  All
low-energy couplings and fermion masses are logarithmically sensitive to
$\Lambda$.  In contrast, scalar squared-masses are {\it quadratically}
sensitive to $\Lambda$.  Thus, in this framework, the observed Higgs
mass (at one-loop) has the following form:
\beq \label{natural}
\mhsm^2= (m_h^2)_0+{kg^2\Lambda^2\over 16\pi^2}\,,
\eeq
where $(m_h)_0$ is a parameter of the fundamental theory, $g$ is an
electroweak coupling and  $k$ is a
constant, presumably of ${\cal O}(1)$, that is calculable within the
low-energy effective theory.  Because these two contributions arise
from independent sources,
it is very unlikely that the magnitude of $\mhsm^2$ is
significantly smaller than either of the two terms.   That is,
the ``natural'' value for the physical
scalar squared-mass is at least of order $g^2\Lambda^2/16\pi^2$.
In order for this
value to be consistent with the requirement that the Higgs mass is of
order the electroweak symmetry breaking scale (as required from
unitarity constraints~\cite{unitarity,thacker}), the value of $\Lambda$
must satisfy
\beq \label{tevscale}
\Lambda\simeq {4\pi \mhsm\over g}\sim {\cal O}(1~{\rm TeV})\,.
\eeq
If $\Lambda$ is significantly larger than 1~TeV (often called the
hierarchy problem in the literature), then the only way
to generate a Higgs mass of $\mathcal{O}(\mz)$
is to have an ``unnatural'' cancellation between the two terms
of \eq{natural}.
This seems highly unlikely given that the two terms of
\eq{natural} have completely different origins.
The requirement of $\Lambda\sim\mathcal{O}(1~{\rm TeV}$) as a condition for
the absence of fine-tuning of the Higgs mass parameter
is nicely illustrated in \fig{trivial}(b), taken from
\Ref{murayama}.

A viable theoretical framework that incorporates weakly-coupled Higgs
bosons and satisfies the constraint of \eq{tevscale} is that of
``low-energy'' or ``weak-scale'' supersymmetry~\cite{susyreview}.  If
supersymmetry (which relates fermion and boson masses and
interactions) is exact, then boson masses must exhibit the same
logarithmically sensitivity to $\Lambda$ as do the fermion masses.
Since no supersymmetric partners of the Standard Model particles have
been found, it follows that supersymmetry is not an exact symmetry of
the fundamental particle interactions.  Hence, in the framework of
low-energy supersymmetry, $\Lambda$ should be identified with the
energy scale of supersymmetry-breaking.  The naturalness constraint of
\eq{tevscale} is still relevant, which implies that the scale of
supersymmetry breaking should not be much larger than a few TeV in
order that the naturalness of scalar masses be preserved.  Moreover,
low-energy supersymmetry with a supersymmetry-breaking scale of
$\mathcal O(1~{\rm TeV})$ is precisely what is needed to explain the
observed gauge coupling unification as previously noted.  We conclude
that a suitable replacement for the Standard Model is a supersymmetric
extension of the Standard Model as the effective field theory of the
TeV scale.  One good feature of the supersymmetric approach is that
the effective low-energy supersymmetric theory {\it can} be valid all
the way up to the Planck scale, while still being natural!

The physics of the Higgs bosons will
be explored by experiments now underway at the upgraded
proton-antiproton Tevatron collider at Fermilab and
in the near future at the Large Hadron Collider (LHC) at CERN.
Once evidence for electroweak symmetry breaking dynamics is
obtained, a more complete understanding of the mechanism involved
will require experimentation at a future $e^+e^-$ linear
collider (LC) now under development.
In this review we focus primarily on the theory and phenomenology of
the Standard Model Higgs boson and the Higgs bosons of low-energy
supersymmetry.  In \Sec{sec:2}, we review the theoretical properties of
the Standard Model Higgs boson, and exhibit its main branching ratio
and production rates at hadron colliders and at the LC.
The main Higgs boson search techniques at the
Tevatron, LHC and the LC are described.  In \Sec{sec:3}, we examine the
Higgs bosons of the minimal supersymmetric Standard Model (MSSM).  We
summarize the tree-level properties of the MSSM Higgs sector and
describe the most significant effects of the radiative corrections
to the computation of the Higgs masses and couplings.  We then
exhibit the main branching ratios and production rates of the MSSM
Higgs bosons and survey the phenomenology of the MSSM Higgs sector at
the Tevatron, LHC and LC.  A brief summary concludes this review in
\Sec{sec:4}.

\section{The Standard Model Higgs Boson}
\label{sec:2}

In the Standard Model, the Higgs mass is given by: $\mhsm^2=\half\lambda
v^2$, where $\lambda$ is the Higgs self-coupling parameter.  Since
$\lambda$ is unknown at present, the value of the Standard Model Higgs
mass is not predicted.  However, other theoretical considerations,
discussed in \Sec{sec:intro}, place constraints on the Higgs mass
as exhibited in \fig{trivial}.  In contrast, the Higgs couplings to
fermions [bosons] are predicted by the theory to be
proportional to the corresponding particle masses [squared-masses].
In particular, the SM Higgs boson is a CP-even scalar, and its
couplings to gauge bosons, Higgs bosons and
fermions are given by:\footnote{The corresponding Feynman rules are
obtained by multiplying the Higgs boson-$VV$ couplings by $ig^{\mu\nu}$
and the other Higgs couplings of \eqns{hsmcouplings1}{hsmcouplings2}
by a factor of $-i$.
The appropriate combinatorial factors have been included.}
\beqa 
&& g_{h f\bar f}= {m_f\over v}\,,\qquad\qquad
\qquad\qquad
g_{hVV} = {2m_V^2\over v}\,, \qquad\qquad
g_{hhVV} = {2m_V^2\over v^2}
\,,\label{hsmcouplings1}\\[5pt]
&& g_{hhh} = \nicefrac{3}{2}\lambda v ={3\mhsm^2\over v}\,,\,\,\,
\quad\qquad
g_{hhhh} = \nicefrac{3}{2}\lambda= {3\mhsm^2\over
v^2}\,,\label{hsmcouplings2}
\eeqa
where $h\equiv\hsm$, $V=W$ or $Z$ and $v=2m_W/g=246$~GeV.
In Higgs production and decay processes, the dominant mechanisms involve
the coupling of the Higgs boson to the $W^\pm$, $Z$ and/or
the third generation quarks and leptons.
Note that a $\hsm gg$ coupling ($g$=gluon)
is induced by virtue of a one-loop graph
in which the Higgs boson couples to a virtual $t\bar t$ pair.
Likewise, a $\hsm\gamma\gamma$ coupling is generated, although in this
case the one-loop graph in which the Higgs boson couples to
a virtual $W^+W^-$ pair is the dominant contribution.   Further details
of the SM Higgs boson properties are given in \Ref{hhg}.  
A review of the SM Higgs properties and its phenomenology, with an emphasis on
the impact of loop corrections to the Higgs decay rates and
cross-sections can be found in \Ref{kniehlreview}. 

\subsection{Standard Model Higgs Boson Decay Modes}
\label{sec:21}

\begin{figure}[t!]
\begin{center}
\resizebox{0.65\textwidth}{!}{
\includegraphics*{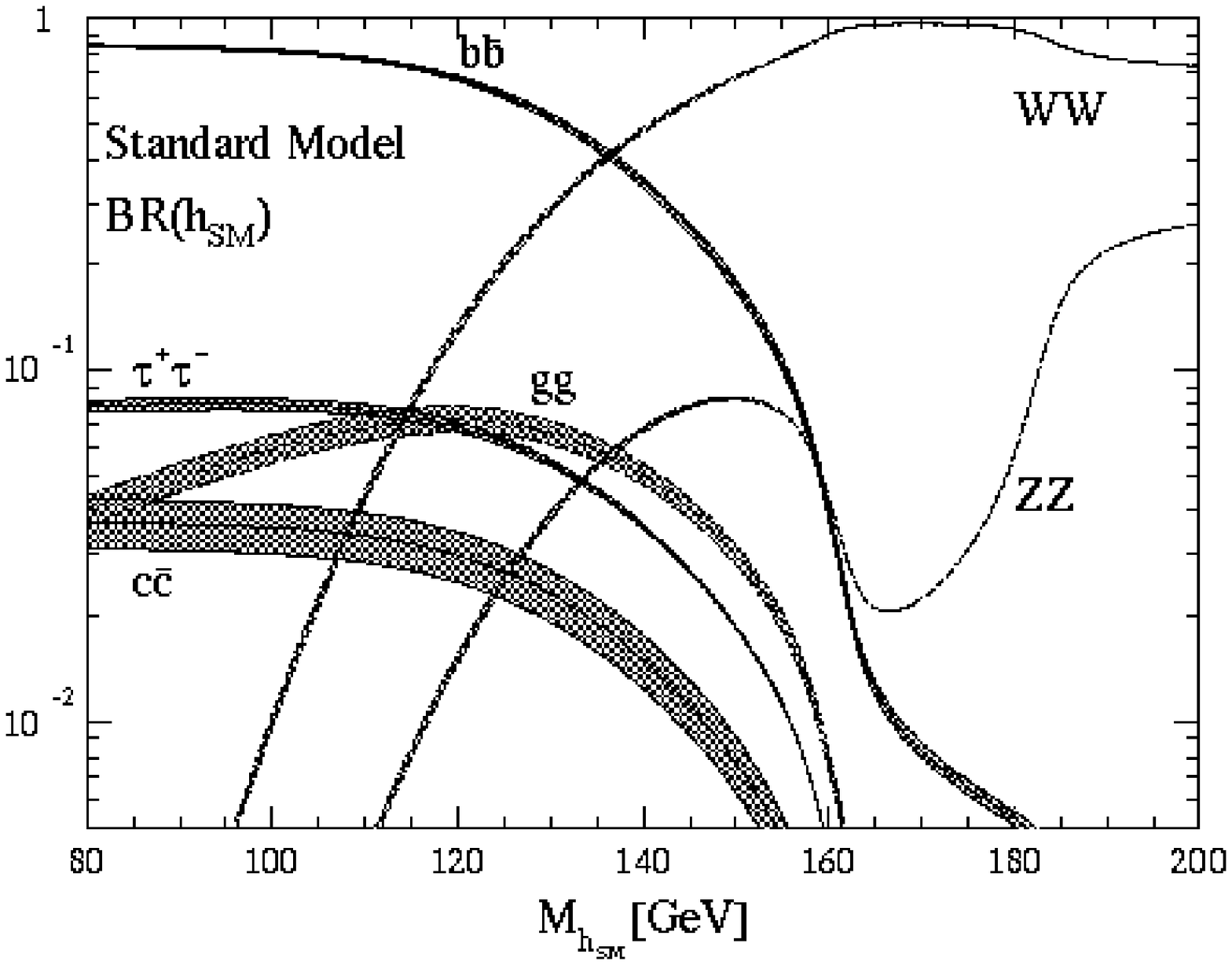}
}
\end{center}
\capt{\label{fg:1} Branching ratios of the dominant decay
modes of the \SM\ Higgs boson as a function of Higgs mass
for $\mhsm\leq 200$~GeV, taken from \protect\Ref{9}.  These results
have been obtained with the program 
\texttt{HDECAY} \protect\cite{hdecay}, and include
QCD corrections beyond the leading order~\protect\cite{DSZ}.
The shaded bands represent the
variations due to the uncertainties in the input parameters:
$\alpha_s(M_Z^2) = 0.120 \pm 0.003$, $\overline{m}_b(M_b) = 4.22
\pm 0.05$~GeV, $\overline{m}_c(M_c) = 1.22 \pm 0.06$~GeV, and $M_t =
174 \pm 5$~GeV.}
\end{figure}

\begin{figure}[t!]
\begin{center}
\resizebox{\textwidth}{!}{
\includegraphics*{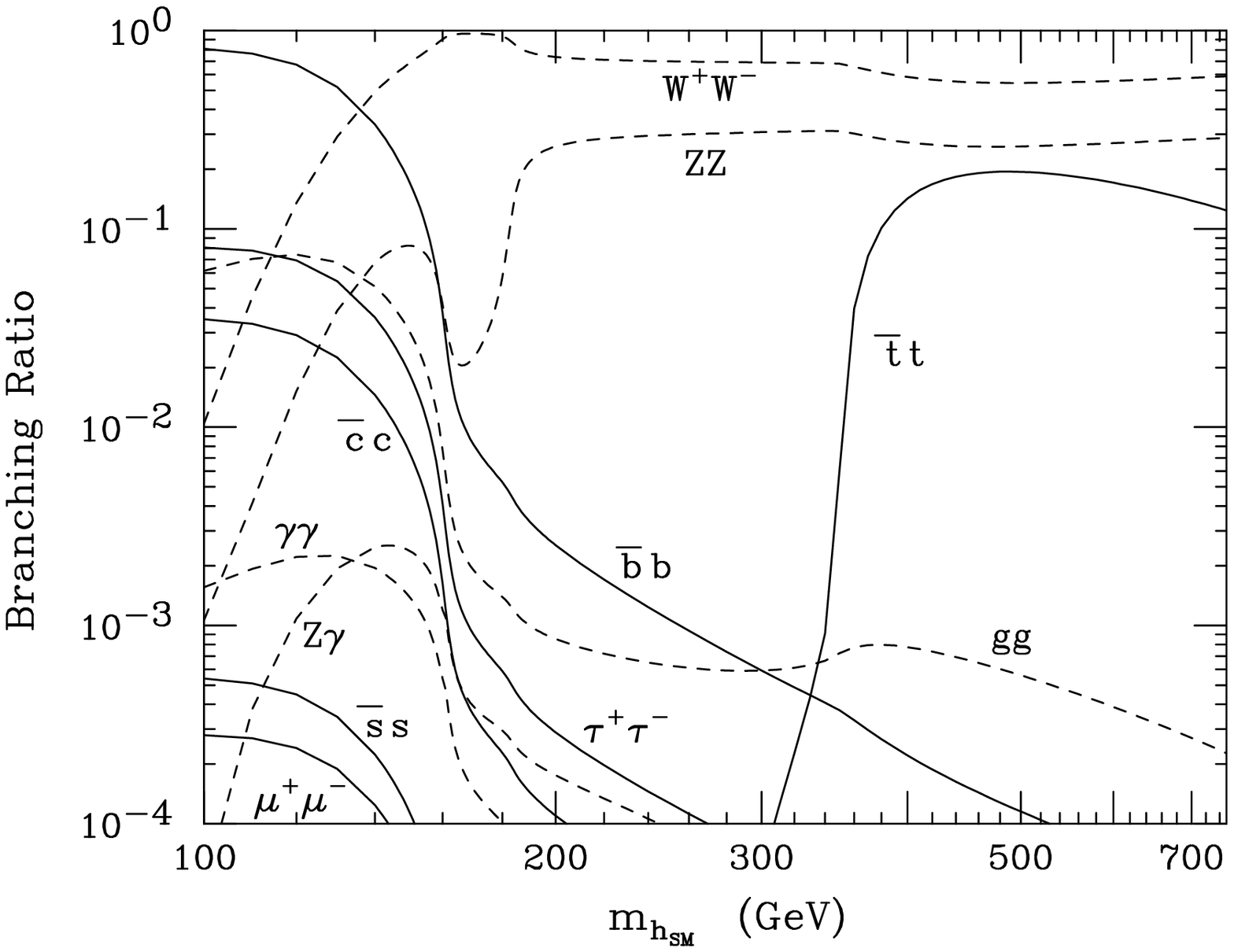}
\hspace*{3mm}
\includegraphics*{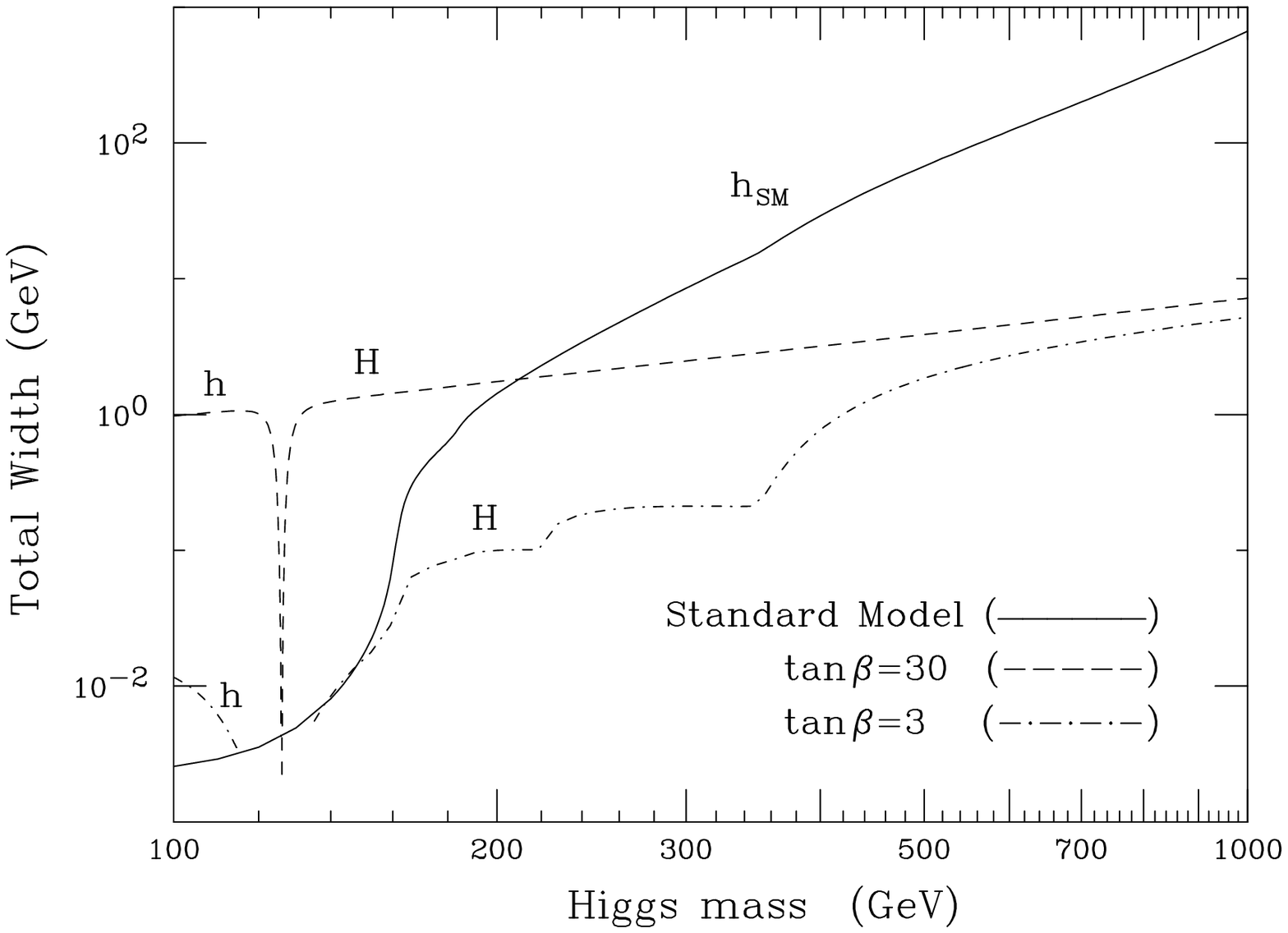}
}
\end{center}
\capt{\label{fg:hwidth} (a) Branching ratios of the \SM\ Higgs
boson as a function of Higgs mass.  Two-boson [fermion-antifermion]
final states are exhibited by solid [dashed] lines.
As compared with \fig{fg:1},
a larger range of Higgs masses and branching ratios are shown.
(b) The total width of the Standard Model
Higgs boson is shown as a function of its mass.  For comparison, we
exhibit the widths of the two CP-even scalars, $\hl$ and $\hh$ of the
MSSM for two different choices of MSSM parameters ($\tan\beta=3$ and
30 in the maximal mixing scenario; the onset of the $\hh\to\hl\hl$ and
$\hh\to t\bar t$ thresholds in the $\tan\beta=3$ curve are clearly evident).
The central values of $\alpha_s$, $\overline{m}_b(M_b)$ and
$\overline{m}_c(M_c)$ quoted in the caption of \protect\fig{fg:1}
are employed in both (a) and (b).}
\end{figure}

The branching ratios for the main decay modes of a SM
Higgs boson are shown as a function of Higgs boson mass in \fig{fg:1}
and \ref{fg:hwidth}(a), based on the results obtained using
the \texttt{HDECAY} program~\cite{hdecay}.
For Higgs boson masses below 135 GeV, the
decay $\hsm\to b\bar b$
dominates, whereas above 135 GeV, the dominant decay mode is
$\hsm\to WW^{(*)}$ (below $W^+W^-$ threshold, one of the
$W$ bosons is virtual as indicated by the star).
Above $t\bar t$ threshold, the branching ratio into top-quark pairs
increases rapidly as a function of Higgs mass, reaching a maximum of
about 20\% at $\mhsm\sim 450$~GeV.
The total Higgs width is obtained by summing all the Higgs partial
widths and is displayed as a function of Higgs mass in \fig{fg:hwidth}(b).

The leading effects of the QCD corrections to the Higgs decay to
quark pairs~\cite{hffsmqcd}\footnote{The formulae for the
leading order QCD-corrections to $\Gamma(\hsm\to q\bar q)$ are nicely
summarized in \Ref{DSZ}.  The leading electroweak radiative
corrections have been obtained in \Ref{hbbewloop}.  A useful summary
of results, which includes new two-loop contributions to the
radiatively corrected $\hsm\to b\bar b$ partial width, can be found
in \Ref{hbbradcor}.}
can be taken into account by using
the tree-level formula for the Higgs partial width (which depends on
the quark mass), and identifying the quark mass with the {\it running}
quark mass evaluated at the Higgs mass, $\overline{m}_Q(\mhsm)$.
The running quark mass, $\overline{m}_Q(\mhsm)$ is obtained from the
$\overline{\rm MS}$ mass, $\overline{m}_Q(M_Q)$ [where $M_Q$ is the
corresponding quark pole mass], by renormalization group evolution.
The $\overline{\rm MS}$ quark masses
are obtained from fits to experimental data~\cite{quarkmasses}.  Note that the
large decrease in the charm quark mass due to QCD running is responsible
for suppressing ${\rm BR}(c\bar c)$ relative to ${\rm BR}(\tau^+\tau^-)$,
in spite of the color enhancement of the former, thereby reversing the
naively expected hierarchy.
Below the corresponding two-body thresholds, the $WW^{(*)}$,
$ZZ^{(*)}$ and $t^{(*)}\bar t$ decay modes (where the asterisk indicates an
off-shell particle)
are still relevant as shown in \fig{fg:hwidth}.

The $\hsm gg$, $\hsm\gamma\gamma$ and $\hsm Z\gamma$ vertices are
generated at one-loop.
The partial width for $\hsm\to gg$ is
primarily of interest because it determines the $gg\to\hsm$ production
cross-section.  The $\hsm\gamma\gamma$ vertex is especially relevant
both for the $\hsm\to\gamma\gamma$
discovery mode at the LHC and for the $\gamma\gamma\to\hsm$ 
production mode at the LC operating as a $\gamma\gamma$ collider.

\subsection{Standard Model Higgs Boson Production at Hadron Colliders}
\label{sec:22}

\subsubsection{Cross-sections at hadron colliders}
\label{sec:221}

\begin{figure}[t!]
\begin{center}
\includegraphics[width=9cm,angle=-90]{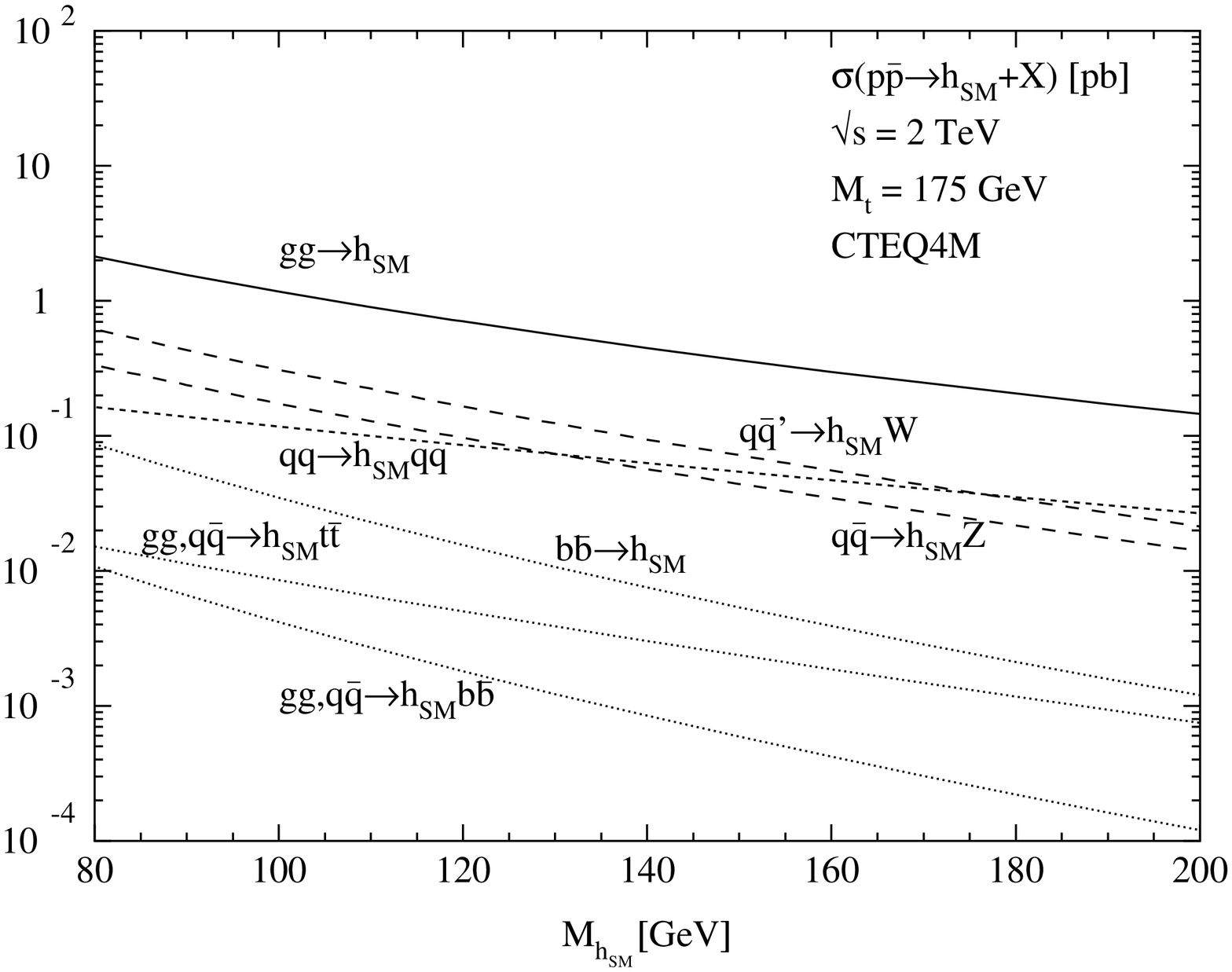}
\end{center}
\capt{\label{fg:4} Higgs production cross-sections
(in units of pb) at the Tevatron [$\sqrt{s}=2$ TeV],
for the various production mechanisms as a function of the
Higgs mass, taken primarily from \protect{\Refs{9}{private}}.
The full NLO QCD-corrected results are employed for the gluon
fusion $gg \to \hsm$, vector boson fusion $qq\to V^*V^*qq \to \hsm qq$
(here, $qq$ refers to both $ud$ and $q\bar q$ scattering),
Higgs-strahlung processes $q\bar q \to V^* \to V\hsm$ (where
$V=W^\pm$, $Z$), $b\bar b\to\hsm$ (taken from \protect\Ref{DSSW}),
and $gg,q\bar q \to \hsm t\bar t$.
Tree-level cross-sections are exhibited for
$gg,q\bar q \to \hsm b\bar b$.  In the latter case,
the cross-section has been computed with a running $b$-quark mass,
$\alpha_s$ and the parton distribution functions all
evaluated at the corresponding Higgs mass.}
\end{figure}

This section describes the most important Higgs production processes at
the Tevatron ($\sqrt{s}=2$~TeV) and the LHC ($\sqrt{s}=14$~TeV).
The relevant cross-sections are exhibited
in \figns{fg:4}{hsmlhc}~\cite{9,hxsec,private,DSSW}.
Combining these Higgs production mechanisms with the decays discussed
in \Sec{sec:21}, one obtains the most promising signatures.

Due to the large luminosity of gluons at high energy hadron colliders,
$gg\to \hsm$ is the Higgs
production mechanism with the largest cross-section at the Tevatron
and the LHC~\cite{29a,sally}.
The two-loop, next-to-leading order (NLO)
QCD corrections enhance the gluon fusion cross-section by about a
factor of two~\cite{sally,balazs}.  The corresponding NLO differential
cross-section (as a function of the Higgs boson $p_T$ and rapidity) 
has also been obtained~\cite{smith}.
Recently, the next-to-NLO (NNLO) QCD corrections have
been evaluated~\cite{kilgore},
and show a further enhancement of about 10\% to 30\%
depending on the Higgs mass and center-of-mass energy of the collider.
The remaining scale dependence and the effects of higher order terms
not yet computed are estimated to give a  theoretical
uncertainty of 10--$20\%$. The dependence of the gluon fusion
cross-section on different parton densities yields roughly an additional
uncertainty of order 10\%.

The cross-section for $q\bar q\to W^\pm\hsm$ (summed over both $W$
charge states) is the second largest Higgs cross-section at the
Tevatron for $\mhsm\lsim 175$~GeV.  At the LHC, the $W^\pm\hsm$
cross-section is not as prominent over the Higgs mass range of interest.
The corresponding $q\bar q\to Z\hsm$ cross-section is roughly a
factor of two lower than the corresponding $W^\pm\hsm$ cross-section.
The QCD corrections
to $\sigma(q\bar q\to V\hsm)$ [$V=W$ or $Z$] coincide with those
of the Drell-Yan process and increase the cross-sections by about 30\%
\cite{23,23a0,xseca}. The
theoretical uncertainty is estimated to be about $15\%$ from the
remaining scale dependence. The dependence on different sets of parton
densities is rather weak and also leads to a variation of the
production cross-sections by about 15\%.

Vector boson fusion is a shorthand notation for the full
$q\bar q\to q\bar q\hsm$ process, where the quark and anti-quark both radiate
virtual vector bosons ($V^*$)
which then annihilate to produce the Higgs boson.  Vector
boson fusion via $ud\to du\hsm$ (and its charge-conjugate process)
is also possible. In \figns{fg:4}{hsmlhc}, all contributing
processes are included, and the sum of all such contributions is
labeled $qq\to qq\hsm$ for simplicity.
The QCD corrections enhance the cross-section by about 10\% \cite{xseca,32}.
The vector boson fusion process is the second largest Higgs
cross-section at the LHC; its cross-section approaches the $gg\to\hsm$
cross-section for $\mhsm\sim 1$~TeV.

\begin{figure}[t!]
\begin{center}
\includegraphics[width=12cm]{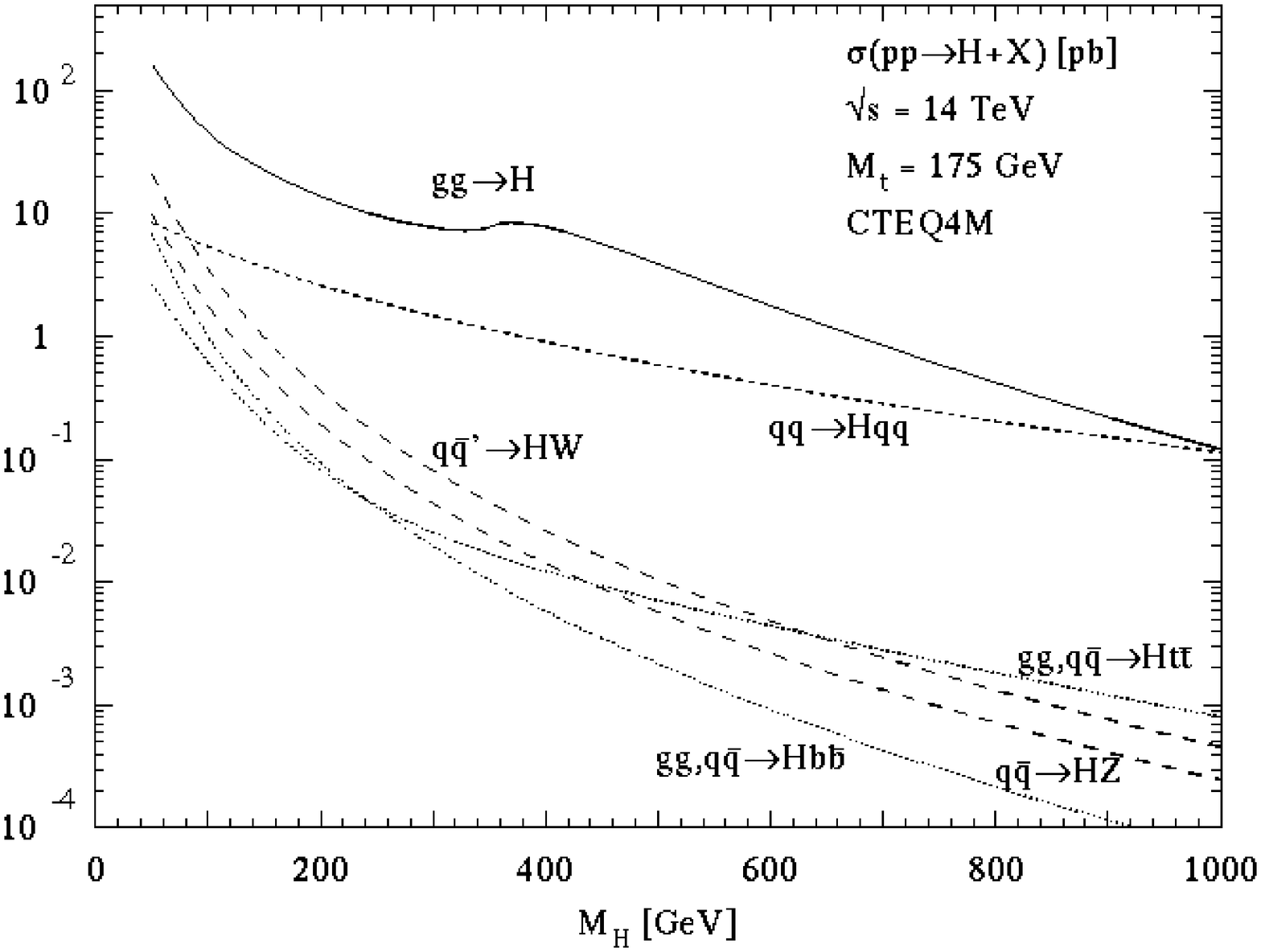}
\end{center}
\capt{\label{hsmlhc} Higgs production cross-sections
(in units of pb) at the LHC
[$\sqrt{s}=14$~TeV], for the various production mechanisms
as a function of the Higgs mass, taken from \protect\Ref{xseca}.
The cross-section curves for
$gg$, $q\bar q \to \hsm t\bar t$
(which has not been updated to include the
NLO calculation of \protect\Ref{Beenakker}) and $gg$, $q\bar q\to \hsm
b\bar b$ are based on a tree-level calculation with $t$-quark
and $b$-quark pole
masses and $\alpha_s$
and the parton distribution functions evaluated at the
partonic center-of-mass energy.}
\end{figure}

The cross-sections for $gg$, $q\bar q\to t\bar t\hsm$ at the Tevatron
and the LHC are displayed in \figns{fg:4}{hsmlhc}.
The NLO QCD corrections to $q\bar q\to t\bar t\hsm$ have recently been
computed in \Refs{Reina}{Beenakker}, and the corrections to $gg\to
t\bar t\hsm$ have been obtained in \Ref{Beenakker}.
\Fig{fg:4} includes the complete NLO QCD corrections from
\Ref{Beenakker} (at Tevatron energies, $t\bar t\hsm$
production is dominated by the $q\bar q\to t\bar t\hsm$ subprocess).
The size of the QCD corrections depends
sensitively on the choice of scale, $\mu$, employed in the running
coupling constant and parton distribution functions.   Changes in
$\mu$ can significantly modify the tree-level cross-sections, whereas
the NLO-corrected cross-sections are rather insensitive to
reasonable changes in $\mu$.  In \fig{hsmlhc}, the tree-level $t\bar t\hsm$
cross-section is shown for $\mu^2=\hat s$ (the square of the
partonic center-of-mass
energy), and is roughly a factor of two smaller than the corresponding
cross-section with $\mu=m_t$.  With respect to the latter choice,
the NLO-corrections of \Ref{Beenakker}
are rather small, typically of order 10--20\%
depending on the precise choice of $\mu$.

The tree-level $gg$, $q\bar q\to b\bar b\hsm$
cross-section (as a function of $\mhsm$) shown in \fig{fg:4}
has been computed by fixing
the scales of the parton distribution functions,
the running coupling $\alpha_s$ and the running Higgs--bottom-quark
Yukawa coupling (or equivalently, the running $b$-quark mass) at the
value of the corresponding $\hsm$ mass.  In \fig{hsmlhc}, the $b$-quark
pole mass is employed,\footnote{The effect of using the $b$-quark pole
mass [$M_b\simeq 5$~GeV] as opposed to the running $b$-quark mass
[$\overline{m}_b(\mhsm)$] is to {\it increase} the cross-section by
roughly a factor of two.}
while the scales for $\alpha_s$ and the parton
distribution functions were set equal to the partonic center-of-mass energy.

By using the running $b$ quark mass, one implicitly resums
large logarithms associated with the QCD-corrected Yukawa coupling.
Thus, the tree-level $b\bar b\hsm$ cross-section displayed in
\fig{fg:4} implicitly includes a part of the QCD corrections to the
full inclusive cross-section.  However, the most significant effect of
the QCD corrections to $b\bar b\hsm$ production arises from the
kinematical region where the $b$ quarks are emitted near the forward
direction.  In fact, large logarithms arising in this region spoil
the convergence of the QCD perturbation series since
$\alpha_s\ln(\mhsm^2/m_b^2)\sim {\cal O}(1)$.  These large logarithms
(already present at lowest order) must be resummed to all orders, and
this resummation is accomplished by the generation of the $b$-quark
distribution function~\cite{soper,dicuswill}.  Thus, the QCD-corrected
{\it fully} inclusive $b\bar b\hsm$ cross-section can be approximated
by $b\bar b\to\hsm$ and its QCD corrections.\footnote{This result,
although correct in the far asymptotic regime (where $\sqrt{s} \gg
\mhsm\gg m_b$), may still not be reliable for Higgs production
at the Tevatron and LHC.  In \Ref{Rainwater:2002hm},
it is argued that even at the
LHC for $\mhsm=500$ GeV, sizable $m_b$-effects still remain and
$\sigma(b\bar b\to \hsm)$ is an overestimate of the true
QCD-corrected $b\bar b \hsm$ cross-section.}
The latter is also exhibited in \fig{fg:4}
and is seen to be roughly an order of magnitude larger than
the tree-level $b\bar b\hsm$ cross-section~\cite{DSSW}.
Of course, this result is not very relevant for the searches at hadron
colliders in which transverse momentum cuts on the
$b$-jets are employed.  Ultimately, one needs the QCD-corrected
{\it differential} cross-section for $b\bar b\hsm$ (as a function of the final
state $b$-quark transverse momentum) in order to do realistic simulations of
the Higgs signal in this channel.  However, if only one $b$-quark
jet is tagged,
it may be sufficient to consider the process $bg\to b\hsm$.  The
cross-section for $bg\to b\hsm$ at lowest order can be found in
\Ref{ddr}; the NLO QCD-corrected cross-section has been recently
obtained in \Ref{maltoni}.  For example, assuming that $p_T>15$~GeV,
and the pseudorapidity $|\eta|<2$ for the observed $b$-quark jet,
the NLO cross-section at the Tevatron ranges from about 6~fb to $0.25$~fb for
$100~{\rm GeV}\leq\mhsm\leq 200$~GeV.  Increasing the
cuts to $p_T>30$~GeV and $|\eta|<2.5$ at the LHC yields a range of
cross-sections from about 200~fb to $1.2$~fb for
$100~{\rm GeV}\leq\mhsm\leq 500$~GeV.

Not shown in \figns{fg:4}{hsmlhc} is the cross-section for
inclusive double Higgs production ($\hsm\hsm+X$).  
Double Higgs production is not
observable at the Tevatron, but may be possible to detect at the LHC
given sufficient luminosity.  The main contributions to double Higgs
production in order of importance
are: (i) $gg\to\hsm\hsm$; (ii) $VV\to\hsm\hsm$; and
(iii) $q\bar q\to V\hsm\hsm$, where $V=W$ or $Z$.  The gluon-gluon
fusion cross-section dominates by at least an order of magnitude, so
we focus on this
subprocess~\cite{glover,ggdoublehiggs,doublehiggs,doublehiggstwo}.
Including NLO QCD corrections,  typical cross-sections for
$pp\to\hsm\hsm+X$ at $\sqrt{s}=14$~TeV range from about 40~fb to 10~fb
for $100<\mhsm<200$~GeV~\cite{doublehiggs,doublehiggstwo}.
There are two classes of diagrams that contribute to
$\hsm\hsm$ production via gluon fusion: $gg\to \hsm^*\to\hsm\hsm$ (via the
top-quark triangle diagram) which is sensitive to the
triple-Higgs vertex, and $gg\to t^*t^*\to\hsm\hsm$
(via the top-quark box diagram) which
is independent of the triple-Higgs vertex.
Due to the relatively low cross-sections,
it will be very challenging to extract information on
the Higgs self-coupling parameter from LHC data.

Finally, assuming that very forward protons can be tagged,
diffractive production of Higgs bosons may provide a viable signal at
the LHC
(the corresponding cross-sections at the Tevatron are probably too
small)~\cite{diffraction,diffraction2}.\footnote{A critical comparison of
various theoretical approaches to diffractive Higgs production can be
found in \Ref{diffraction2}.} 
Such events are characterized by rapidity gaps, {\it i.e.}, the
absence of particle production between the forward protons and the
centrally produced Higgs boson. In particular, for $\mhsm\sim 130$~GeV,
the exclusive process $pp\to pp\hsm$ (with $\hsm\to b\bar b$) can be used 
at the LHC to obtain
the Higgs mass with an accuracy of about 1\% by measuring the
invariant mass recoiling against the final state protons~\cite{diffraction3}.

\subsubsection{Standard Model Higgs Boson Searches at the Tevatron}
\label{sec:222}

In the mass region of interest to the Tevatron Higgs search
(100~GeV$\lsim\mhsm\lsim 200$~GeV),
the SM Higgs boson is produced most copiously via $gg$ fusion, with a
cross-section from about 1.0---0.1~pb.  For $\mhsm\lsim 135$~GeV,
the Higgs boson decays
dominantly to $\bb$.  Since the cross-section for the QCD
production of $\bb$ dijet events is orders of magnitude larger than
the Higgs production cross-section, the $gg \to h_{SM} \to b \bar b$
channel is not a promising channel.
For $\mhsm\gsim 135$~ GeV, the Higgs boson decays
dominantly to $WW^{(*)}$ (where $W^{*}$ is a virtual $W$),
and the channel $gg\to \hsm\to WW^{(*)}$ is accessible to the
Tevatron Higgs search~\cite{turcot}.

Given sufficient luminosity, the most promising
SM Higgs discovery mechanism at the Tevatron for $\mhsm\lsim 135$~GeV
consists of $q\bar q$ annihilation into a virtual $V^*$ ($V=W$ or $Z$),
where $V^*\to V\hsm$ followed by a leptonic decay of the $V$ 
and $\hsm\to b\bar b$~\cite{Stange:1994ya}.  The sum of the $W\hsm$
and $Z\hsm$ cross-sections is about 0.2---0.5~pb in the mass region of
interest ($100\lsim\mhsm\lsim 135$~GeV), in which the dominant Higgs
decay is $\hsm\to b\bar b$.  These processes lead to three
main final states, $\ell\nu\bb$, $\nn\bb$ and $\lpm\bb$,
that exhibit distinctive signatures on which
the experiments can trigger (high $p_T$ leptons and/or missing $E_T$).
The backgrounds are manageable and are typically dominated by
vector-boson pair production, $t\bar t$ production and QCD dijet
production.  The
signal efficiencies and backgrounds have all been estimated with both
the CDF Run~1 detector simulation and with the simple SHW simulation
\cite{tevreport}.  Monte Carlo estimates have been used for the
backgrounds everywhere except in the $\nn\bb$ channel, where there is
a significant contribution from QCD $\bb$ dijet production.  To be
conservative, in \Ref{tevreport} the unknown QCD $b\bar b$ dijet
background to the $\nn\bb$ channel has been taken to be equal in size
to the sum of all other contributing background processes.  In
addition, the separation of signal from background was optimized using
neural network techniques~\cite{neuralnet}, resulting in a
demonstrable gain in the significance of the Higgs signal for the
$\ell\nu\bb$ and $\nn\bb$ channels \cite{tevreport}.  The $b$-tagging
efficiencies and the $\bb$ mass resolution play a key role in
determining the ultimate efficiency and background rejection.  Much
work remains, using real data studies, to optimize the performance in
both these areas.

For larger Higgs masses ($\mhsm\gsim 135$~GeV) it is possible to
exploit the distinct signatures present when the Higgs boson decay
branching ratio to $WW^{(*)}$ becomes appreciable.  In this case,
there are final states with $WW$ (from the gluon-fusion production of
a single Higgs boson), and $WWW$ and $ZWW$ arising from associated
vector boson--Higgs boson production.
Three search channels were identified in \Ref{tevreport} as
potentially sensitive at these high Higgs masses: like-sign dilepton
plus jets ($\lsdilep$) events, high-$p_T$ lepton pairs plus missing
$E_T$ ($\lpm\nn$), and trilepton ($\trilep$) events.  Of these, the
first two were found to be most sensitive.  The strong
angular correlations of the final state leptons resulting from
$WW^*$ is one of the crucial ingredients for these
discovery channels \cite{nelson,dreiner,turcot}.

\begin{figure}[t!]
\begin{center}
\resizebox{\textwidth}{3.5in}{
\scalebox{1.3}[1.99]{
\includegraphics*{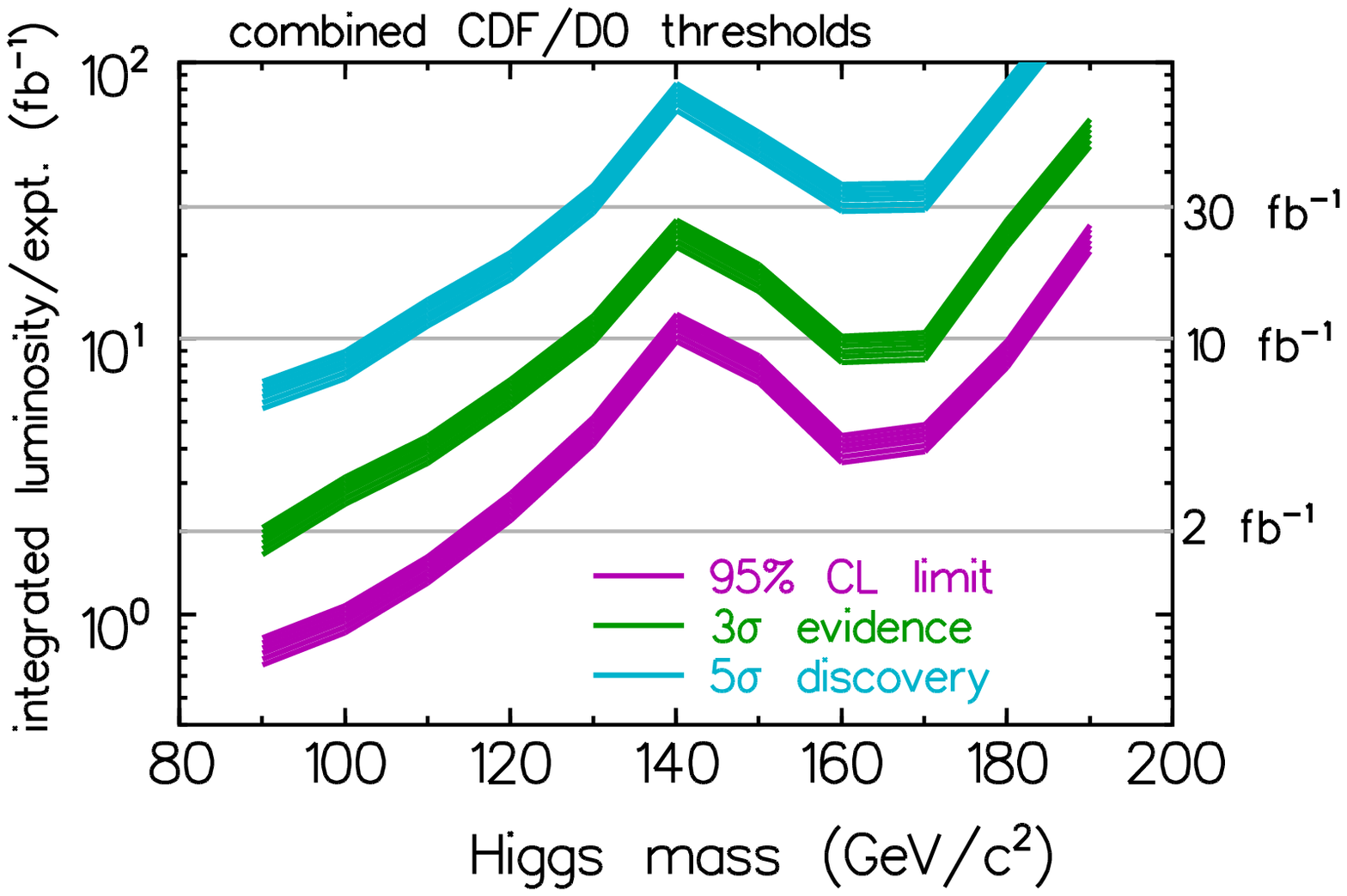}}
\scalebox{1.12}[1.12]{
\includegraphics*{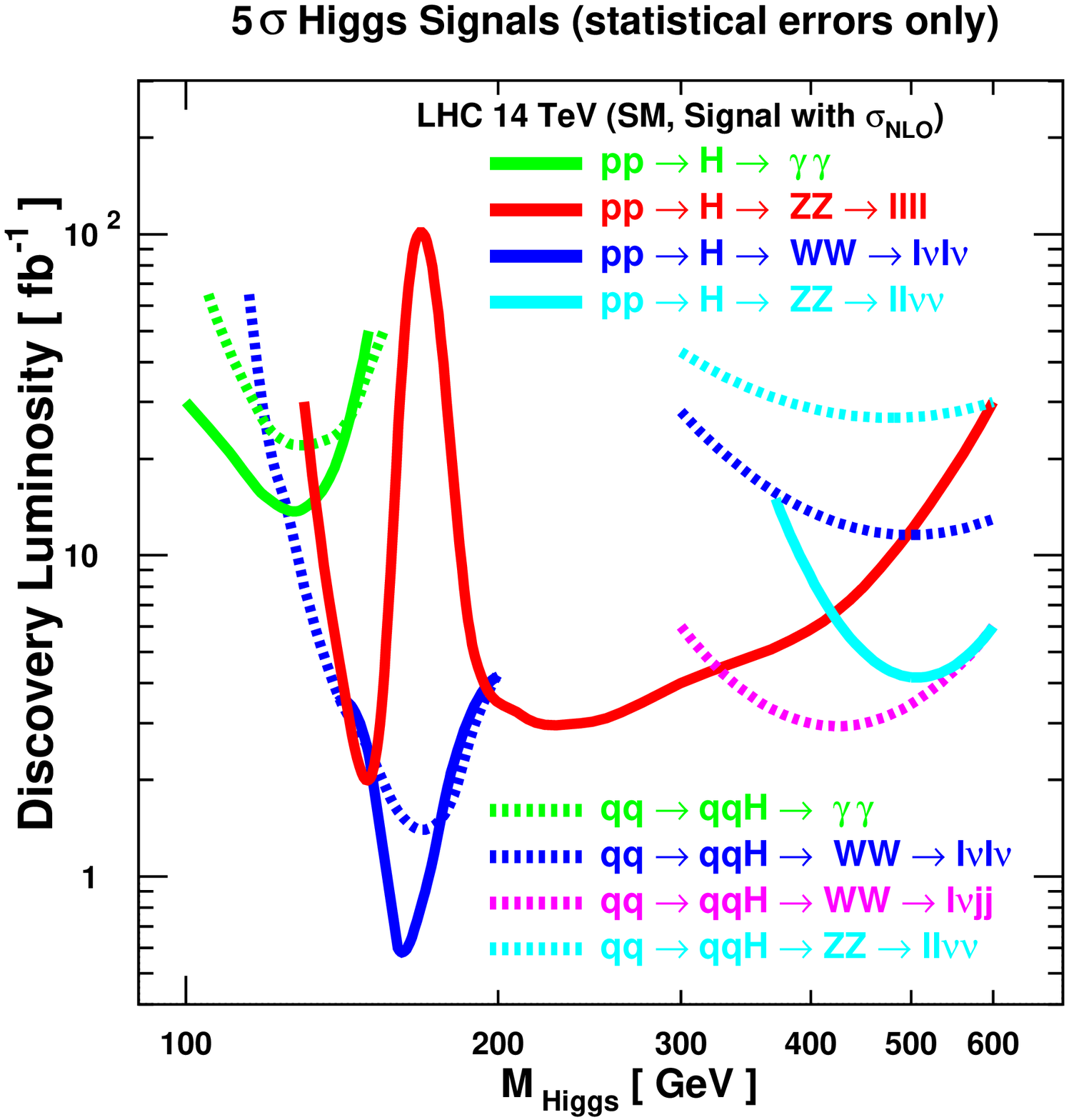}}
}
\end{center}
\capt{\label{fig:smhiggsathadron} (a)~The
integrated luminosity required per Tevatron experiment, to
either exclude a SM Higgs boson at 95\% CL or observe it at the
$3\sigma$ or $5\sigma$ level, as a function of the Higgs
mass~\protect\cite{tevreport}.  (b)~Expected $5\sigma$ discovery 
luminosity requirements for the SM Higgs boson 
at the LHC for one experiment, based on
a study performed with CMS fast detector simulation, assuming
statistical errors only~\protect\cite{Higgs_lumi}.  The $gg$ and
$W^+W^-$ fusion processes are indicated respectively 
by the solid and dotted lines.}
\end{figure}

The integrated luminosity required per Tevatron experiment
as a function of Higgs mass
to either exclude the SM Higgs boson at 95\% CL or discover it at the
$3\sigma$ or $5\sigma$ level of significance, 
for the SHW analyses with neural net selection (see \Ref{tevreport}
for details),
is shown in Fig.~\ref{fig:smhiggsathadron}(a).
These results are based on the combined statistical power of {\it both}
the CDF and D\O\ experiments.
The bands extend from the neural net result
on the low side upward in required integrated luminosity by 30\% to
the high side, as an indication of the range of uncertainty in the
$b$-tagging efficiency, $b\bar b$ mass resolution and background
uncertainties.
As the plots show, the required integrated luminosity increases
rapidly with Higgs mass to 140 GeV, beyond which the high-mass
channels play the dominant role.
If $\mhsm=115$~GeV, which lies just above
the 95\% CL exclusion limit achieved by LEP \cite{LEPHiggs}, then
5~fb$^{-1}$ of integrated luminosity per experiment would provide
sufficient data to see a 3$\sigma$ excess above
background.  With 15~fb$^{-1}$
of integrated luminosity per experiment, a $5\sigma$ discovery of the
Higgs boson would be possible.

The final result shows that for an integrated luminosity of
10~fb$^{-1}$, if the SM Higgs boson mass lies beyond the discovery
reach of the Tevatron, then one can attain a 95\% CL exclusion
for masses up to about 180~GeV.  Moreover, if the SM Higgs happens to
be sufficiently light ($\mhsm\lsim 125$~GeV), then a tantalizing
$3\sigma$ effect will be visible with the same integrated luminosity.
With about 25~fb$^{-1}$ of data, $3\sigma$ evidence for the Higgs
boson can be obtained for the entire Higgs mass range up to 180~GeV.
However, the discovery reach is considerably more limited for a
5$\sigma$ Higgs boson signal.
With 30~fb$^{-1}$ integrated luminosity delivered per
detector, a $5\sigma$ Higgs boson discovery may be possible for Higgs
masses up to about 130 GeV, a significant extension of the LEP
Standard Model Higgs search.  The latter figure of merit is
particularly significant when applied to the search for the lightest
Higgs bosons of the MSSM.  We address this
case in \Sec{sec:353}.

Other Higgs signatures could help improve the sensitivity of the Higgs
search at the Tevatron.  In \Ref{tevreport},
channels containing the $\hsm\to\tau^+\tau^-$ decay mode
have not been studied, as the small branching ratio
(less than $8\%$) makes the corresponding signal rates small.
Still, a significant improvement of $\tau$-lepton identification
could lead to a viable Higgs signal in the Higgs mass region
120~GeV$\lsim\mhsm\lsim 140$~GeV~\cite{bhr}.
Another possibility which has been explored
is the detection of the Higgs boson via
$t\bar t\hsm$ production (the Higgs boson is radiated off the
top-quark), followed by $\hsm\to b\bar b$.
Initial studies
\cite{rainwater} suggested that this channel could be observable at the
upgraded Tevatron for $\mhsm\lsim 140$~GeV, with a statistical significance
comparable to the Higgs signals in the $W\hsm$ and
$Z\hsm$ channels.

If a Higgs boson is discovered at the Tevatron, one can begin to
measure some of its properties.  The Higgs mass can be measured
with an accuracy of about 2~GeV~\cite{precisiongroup}.  However, the
determination of the Higgs couplings to $W$ and $Z$ bosons and to
$b\bar b$ will be model-dependent and rather crude.   To improve and
expand the possible Higgs measurements and determine its phenomenological
profile will require Higgs studies at the LHC.

\subsubsection{Standard Model Higgs Boson Searches at the LHC}
\label{sec:223}

Production rates for the Higgs boson in the Standard Model are
significantly larger at the LHC~\cite{xseca,xsecsnow}. The dominant Higgs
production process, gluon fusion, can be exploited in conjunction with
a variety of other channels, {\it e.g.}, $WW/ZZ$ fusion of the Higgs
boson and Higgs radiation off top
quarks~\cite{LHCreps,perini,Higgs_lumi,trefzger_summary,branson,leshouches}.
Integrated luminosities between 30 and 100
fb$^{-1}$, achievable within the first few years of LHC
operation, will be
sufficient to cover the entire canonical Higgs mass range of the
Standard Model up to values close to 1 TeV with a significance greater
than $5\sigma$.  The required LHC luminosities for a Higgs discovery
in various channels are
shown in \fig{fig:smhiggsathadron}(b)~\cite{Higgs_lumi}.  
Thus, there is no escape route for the SM Higgs boson at the LHC.

The properties of the SM Higgs boson
can be determined with some precision at the LHC.  The $\hsm\to ZZ^{(*)}\to
\ell^+\ell^-\ell^+\ell^-$ channel allows for an accurate Higgs mass
determination of about $0.1\%$ for $120~{\rm GeV} \lsim\mhsm\lsim
400$~GeV, assuming an integrated luminosity of
300~fb$^{-1}$~\cite{trefzger}.
For larger Higgs masses, the precision in the Higgs
mass measurement deteriorates due to the effect of the increasing
Higgs width; nevertheless a $1\%$ Higgs mass measurement is possible
for $\mhsm\simeq 700$~GeV.  The Higgs width can be extracted with a
precision of 5 to $6\%$ over the mass range $300$---700~GeV from the
Breit-Wigner shape of the Higgs resonance~\cite{trefzger}.  Below
300~GeV, the instrumental resolution becomes larger than the Higgs width,
and the accuracy of the Higgs width measurement degrades.  For
example, the four-lepton invariant mass spectrum from $\hsm\to ZZ$ yields a
precision of about $25\%$ at $\mhsm=240$~GeV~\cite{precisiongroup}.
For lower Higgs masses, indirect methods
must be employed to
measure the Higgs width.

For Higgs masses below 200 GeV, a number of different Higgs decay
channels can be studied at the LHC.
The most relevant processes are
\beqa
&& gg\to\hsm\to\gamma\gamma\,, \nonumber \\
&& gg\to\hsm\to VV^{(*)}\,, \nonumber \\
&& qq\to qqV^{(*)} V^{(*)}\to qq\hsm,\quad \hsm\to\gamma\gamma,\,
\tau^+\tau^-, \,VV^{(*)}\,, \nonumber \\
&& gg, q\bar q\to t\bar t\hsm, \quad \hsm\to b\bar b,
\,\gamma\gamma, \,WW^{(*)}\,,\nonumber
\eeqa
where $V=W$ or $Z$.
The gluon-gluon fusion mechanism is the dominant Higgs production
mechanism at the LHC, yielding a total cross-section of about 30~pb
[15~pb] for $\mhsm=120$~GeV [$\mhsm=200$~GeV].
One also has
appreciable Higgs production via $VV$ electroweak gauge boson fusion,
with a total cross-section of about 6~pb
[3~pb] for the Higgs masses quoted above.
The electroweak gauge boson fusion mechanism can be separated from the
gluon fusion process by employing a forward jet tag
and central jet vetoing techniques.
Note that for $2\mw\lsim\mhsm\lsim 2\mz$, the Higgs branching ratio
to $ZZ^*$ is quite suppressed with respect to $WW$ (since one of the $Z$
bosons is off-shell).  Hence, in this mass window, $\hsm\to
W^+W^-\to \ell^+\nu\ell^-\bar\nu$ is the main Higgs discovery 
channel~\cite{dreiner}, as exhibited in \fig{fig:smhiggsathadron}(b). 

The cross-section for $t\bar t\hsm$ production can be significant for
Higgs masses in the intermediate mass range~\cite{Beenakker},
0.8~pb [0.2~pb] at $\mhsm=120$~GeV
[$\mhsm=200$~GeV], although this cross-section falls faster with Higgs
mass as compared to the gluon and gauge boson fusion mechanisms.
Finally, we note that the preferred channel at the Tevatron,
$q\bar q\to W\hsm\to \ell\nu b\bar b$, is not a discovery mode at the
LHC due in part to the larger background cross-sections at
$\sqrt{s}=14$~TeV.  Nevertheless, with 300~fb$^{-1}$ of data, it may be
possible to observe a SM Higgs signal (with $S/\sqrt{B}\geq 5$)
if $\mhsm\lsim 125$~GeV~\cite{dmull}.

\begin{figure}[t!]
\begin{center}
\resizebox{.85\textwidth}{!}{
\includegraphics*[angle=90]{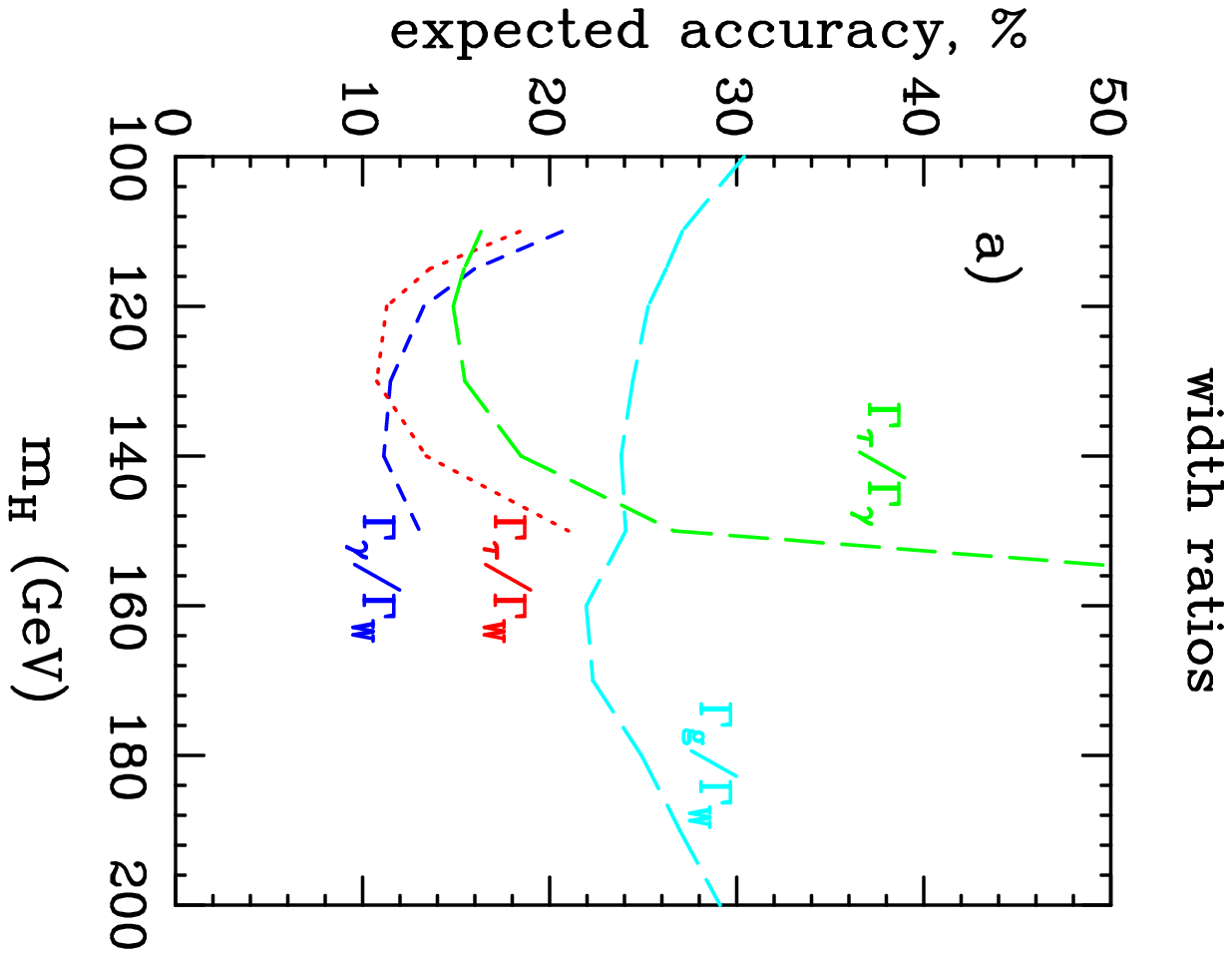}
\hspace*{3mm}
\includegraphics*[angle=90]{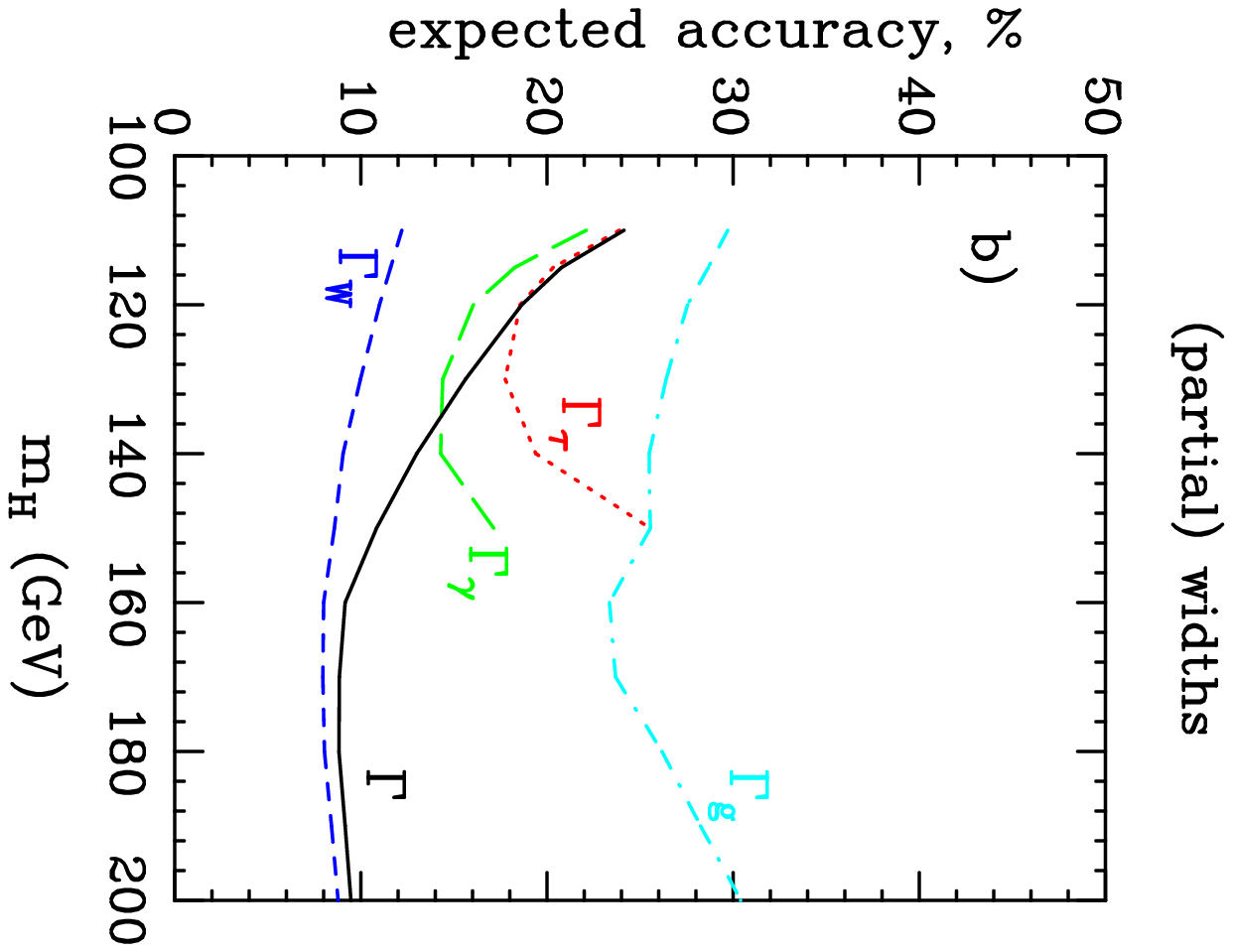}
}
\end{center}
\capt{\label{fig:LHCwidths}
Relative accuracy expected at the LHC with
200~fb${^{-1}}$ of data for (a)~various ratios of Higgs boson partial
widths and (b)~the indirect determination of partial and total
widths.  Expectations for width
ratios assume $W$, $Z$ universality;
indirect width measurements also assume $b$, $\tau$ universality and
a small branching ratio for unobserved modes.  Taken from the parton-level
analysis of \protect\Ref{zeppenfeld}.}
\end{figure}

The measurements of Higgs decay branching ratios can
be used to infer the
ratios of the Higgs couplings, and provide
an important first step in clarifying the
nature of the Higgs boson~\cite{zeppenfeld,PlehnRainwaterZeppenfeld,mumu}.  
These can be extracted
from a variety of Higgs signals that are observable over a limited
range of Higgs masses.
In the mass range $110~{\rm GeV}\lsim\mhsm
\lsim 150$~GeV, the Higgs boson can be
detected [with 100 fb$^{-1}$ of data]
in the $\gamma\gamma$ and the $\tau^+\tau^-$ channels
indicated above.  (The $\mu^+\mu^-$ channel was considered in
\Ref{mumu}.  With 300 fb$^{-1}$ of data, a $3\sigma$ excess above
background may be possible 
for $110~{\rm GeV}\lsim\mhsm\lsim 140$~GeV.)
For $\mhsm\gsim 130$~GeV, the Higgs boson can also be detected
in gluon-gluon fusion through its decay to $WW^{(*)}$,
with both final gauge bosons decaying leptonically~\cite{rainzepp},
and to $ZZ^{(*)}$
in the four-lepton decay mode~\cite{LHCreps,branson}.
There is additional
sensitivity to Higgs production via $VV$ fusion followed by its decay
to $WW^{(*)}$ for $\mhsm\gsim 120$~GeV.
These data can be used to extract
the ratios of the Higgs partial widths to gluon pairs,
photon pairs, $\tau^+\tau^-$,
and $W^+W^-$~\cite{zeppenfeld,PlehnRainwaterZeppenfeld}.
The expected accuracies in Higgs width
ratios, partial widths, and the total Higgs width
are exhibited in \fig{fig:LHCwidths}.
These results are obtained under the assumption that the
partial Higgs widths to $W^+W^-$ and $ZZ$ are fixed by electroweak
gauge invariance, and the ratio of the
partial Higgs widths to $b\bar b$ and $\tau^+\tau^-$ are fixed by
the universality of Higgs couplings to down-type fermions.
One can then extract the total Higgs width under the assumption
that all other unobserved modes, in the Standard Model and beyond,
possess small branching ratios of order 1\%.
Finally, we note that
the specific Lorentz structure predicted for the $\hsm W^+W^-$
coupling by the Higgs
mechanism can be tested in angular correlations between the spectator
jets in $WW$ fusion of the Higgs boson at the
LHC~\cite{PlehnRainwaterZeppenfeld}.

With an integrated luminosity of 100 fb$^{-1}$ per
experiment, the relative accuracy expected at the LHC for
various ratios of Higgs partial widths $\Gamma_i$
(or equivalently, ratios of Higgs squared-couplings)
range from 10\% to 30\%, as shown in \fig{fig:LHCwidths}. 
The ratio $\Gamma_\tau/\Gamma_W$ measures the coupling of
down-type fermions relative to the  Higgs couplings to gauge bosons.
To the extent that the one-loop $\hsm\gamma\gamma$ amplitude is
dominated by the $W$-loop, the partial width ratio
$\Gamma_\tau/\Gamma_\gamma$ probes the same relationship.  In
contrast, under the usual assumption that the one-loop $\hsm gg$ amplitude
is dominated by the top-quark loop, the ratio $\Gamma_g/\Gamma_W$
probes the coupling of up-type fermions relative to the $\hsm WW$
coupling.  In addition, one can measure the $\hsm t\bar t$ coupling directly 
by making use of the $t\bar t\hsm$ production mode at the LHC.
Recent studies suggest that for an integrated luminosity of
100~fb$^{-1}$, this signal is viable for the
$\hsm\to b\bar b$~\cite{ATLASstudy,CMSstudy} and
$\hsm\to \tau^+\tau^-$~\cite{breina} decay modes
if $\mhsm\lsim 130$--140~GeV, and for the $\hsm\to WW^{(*)}$ decay
mode for $130\lsim\mhsm\lsim 200$~GeV~\cite{mrw}.
In this way, one expects to be able to measure
the Higgs-top quark Yukawa coupling with a relative accuracy in the
range of 10--20\%.  

Finally, the measurement of
the triple Higgs self-coupling with an accuracy of order 25\%
can be obtained, in a very limited Higgs mass window, with
3000~fb$^{-1}$ of data~\cite{baur} 
(which requires at least a factor-of-ten
luminosity upgrade of the LHC~\cite{lhcupgrade}).

\subsection{Standard Model Higgs Boson Searches at the LC}
\label{sec:23}

The next generation of high energy $e^+e^-$ linear colliders
is expected to operate
at energies from 300 GeV up to about 1 TeV (JLC, NLC, TESLA),
henceforth referred to as the LC~\cite{jlc,nlc,tesla}.
The possibility of a multi-TeV linear collider
operating in an energy range of 3--5 TeV (CLIC) is also under
study~\cite{clic}.
With the expected high luminosities up to 1~ab$^{-1}$,
accumulated within a few years in a clean experimental environment,
these colliders are ideal instruments for reconstructing the
mechanism of electroweak symmetry breaking in a comprehensive and
conclusive form.

\begin{figure}[t!]
\begin{center}
\resizebox{\textwidth}{!}{
\scalebox{1.0}[1.15]{
\includegraphics*[width=5cm]{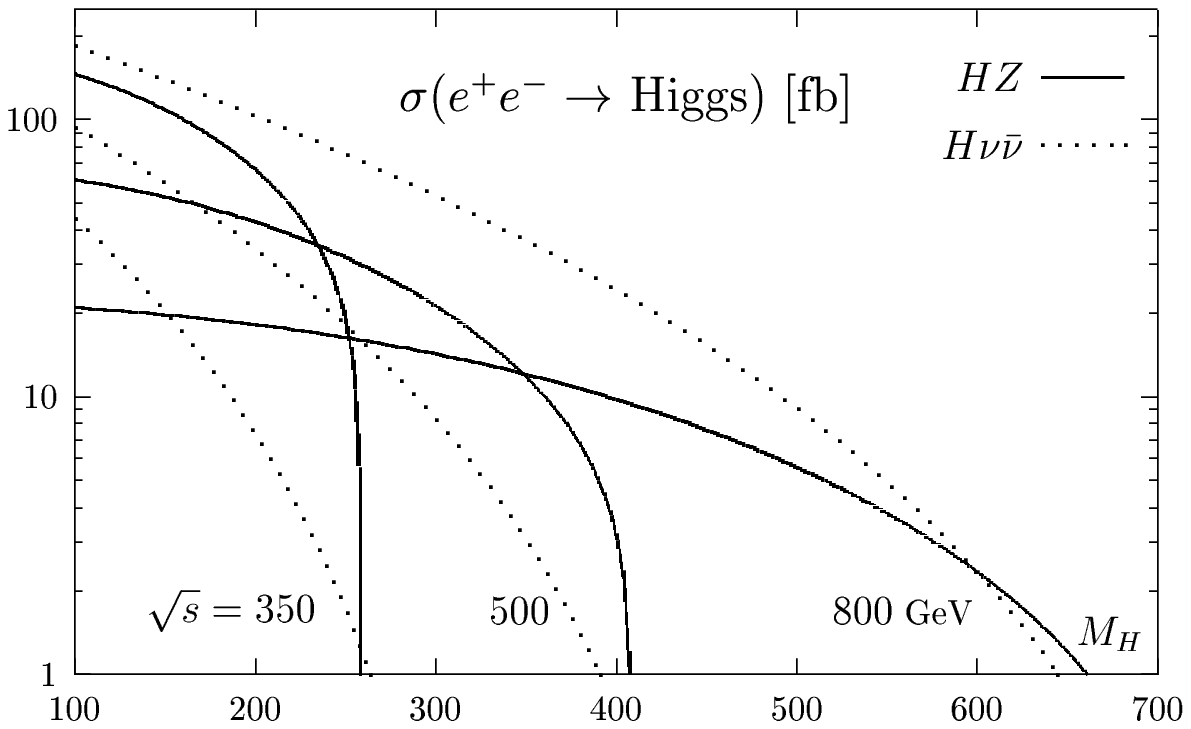}}
\hspace*{3mm}
\includegraphics*[width=5cm]{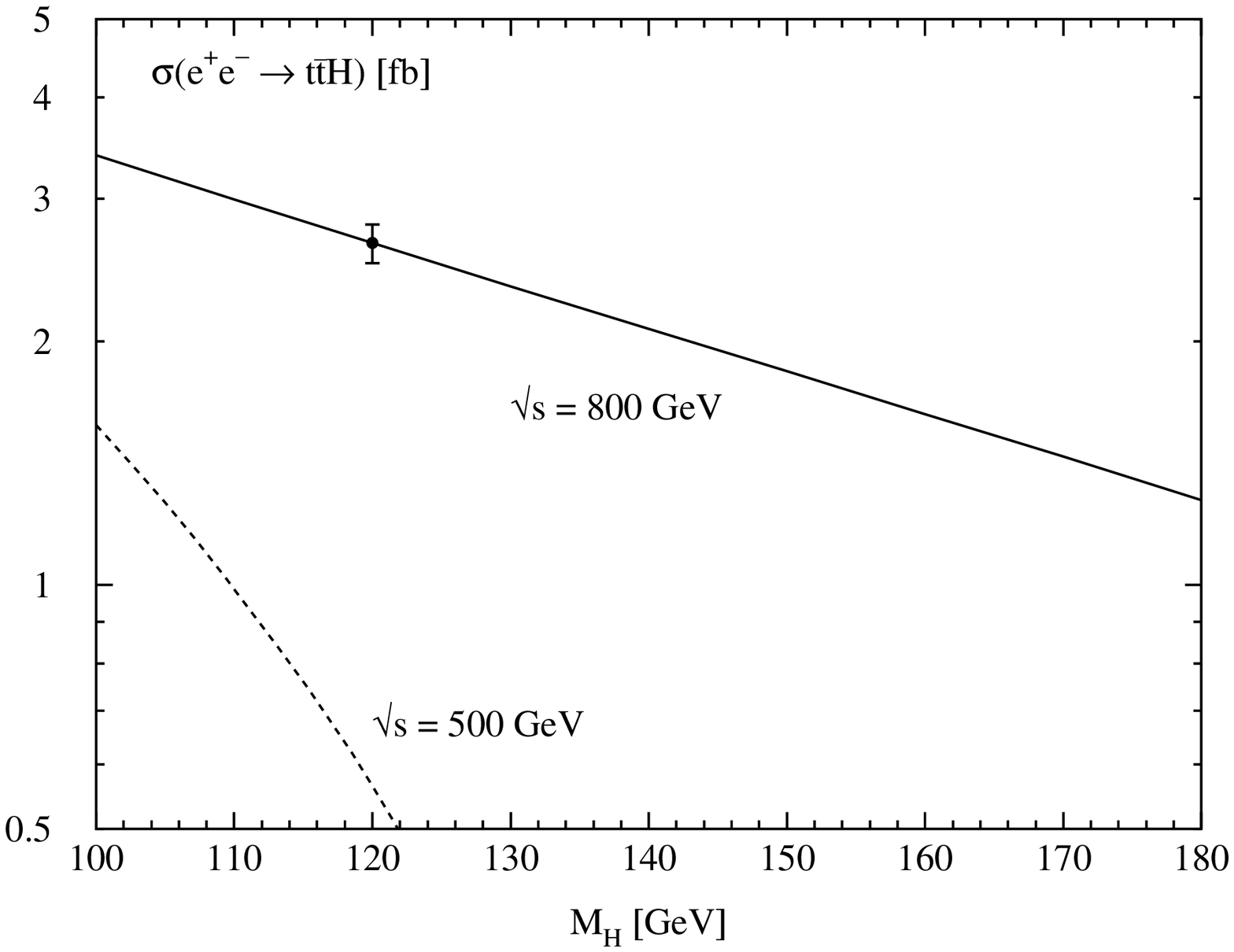}
}
\end{center}
\capt{\label{LCxsecs}
(a) The Higgs-strahlung and WW fusion production cross-sections
as functions of $\mhsm$ for $\sqrt{s}$ = 350 GeV, 500 GeV  and 800 GeV.
(b)The cross-section for $e^+e^-\to t \bar t \hsm$, including NLO QCD
corrections~\protect\cite{tth1}, as a function of  
$\mhsm$ for $\sqrt{s}$ = 500 GeV
and 800 GeV with the expected experimental accuracy for
$\mhsm=120$~GeV  shown by the dot with
error bar for an integrated luminosity of 1000 fb$^{-1}$.
Taken from \protect\Ref{Tesla-TDR}.}
\end{figure}

Weakly-coupled electroweak symmetry breaking dynamics
involving an elementary scalar Higgs field
can be established experimentally
in three steps.  First,
the Higgs boson must be observed clearly and unambiguously, and its
basic properties---mass, width, spin and C and P quantum numbers---must be
determined.  Second, the
couplings of the Higgs boson to the $W^\pm$ and $Z$
bosons and to leptons and quarks must be measured.  Demonstrating that
these couplings scale with the mass of the corresponding
particle would provide critical support for the Higgs mechanism
based on scalar dynamics as the agent responsible
for generating the masses of the fundamental particles.  Finally,
the Higgs potential must be reconstructed by measuring the self-coupling
of the Higgs field. The specific form of the potential shifts the ground
state to a non-zero value, thereby providing the mechanism for
electroweak symmetry breaking based on the self-interactions of scalar
fields.
Essential elements of this program can be realized at a high-luminosity
$e^+e^-$ linear collider, and
high-precision analyses of the Higgs boson are possible in
these machines~\cite{Tesla-TDR,LC-Orange,JLC-TDR}.

The main production mechanisms of the SM Higgs boson at the LC
are the Higgs-strahlung process~\cite{strahlung,thacker},
$e^+e^- \to Z\hsm$,
and the $WW$ fusion process~\cite{fusion} 
$e^+ e^- \to \bar\nu_e \nu_e W^* W^*\to \bar{\nu}_e \nu _e \hsm$.
With an accumulated luminosity of 500 fb$^{-1}$, about $10^5$ Higgs
bosons can be produced by Higgs-strahlung
in the theoretically
preferred intermediate mass range below 200 GeV. As $\sqrt{s}$ is increased,
the cross-section for the Higgs-strahlung process decreases as $s^{-1}$ and
is dominant at low energies, while the cross-section for the $WW$
fusion process grows as $\ln(s/\mhsm^2)$ 
and dominates at high energies (see \Ref{kniehlfusion}
for a convenient form for the corresponding cross-sections),
as shown in 
\fig{LCxsecs}(a).\footnote{The cross-sections shown in
\fig{LCxsecs}(a) are based on a tree-level computation.  The 
complete one-loop electroweak corrections for $e^+e^-\to Z\hsm$ have been
computed~\cite{eezhrad,kniehlfusion},
whereas only the leading one-loop corrections to the vector boson fusion
process are known~\cite{Kniehlee}.}
The $ZZ$ fusion mechanism, $e^+e^- \rightarrow e^+e^-Z^*Z^* \to
e^+e^- \hsm$,
also contributes to Higgs production, with a cross-section suppressed
with respect to that of $WW$ fusion
by a factor of $16 \cos^4\theta_W/[1+(1-4\sin^2\theta_W)^2]\simeq 9.4$.
The cross-sections for the Higgs-strahlung and the $WW$ fusion processes
are shown in \fig{LCxsecs}(a) for three values of $\sqrt{s}$.
The Higgs-strahlung process, $e^+e^-\to Z \hsm$,
with $Z \rightarrow \ell^+ \ell^-$, offers a very distinctive
signature. For $\sqrt{s} =350$ and 500 GeV and an integrated
luminosity of 500~fb$^{-1}$, this ensures the
observation of the SM Higgs
boson up to the production kinematical limit independently of its
decay \cite{Tesla-TDR}.
At $\sqrt{s} = 500$ GeV, the Higgs-strahlung and the
$WW$ fusion processes have approximately the same cross-sections,
${\cal O}$(50 fb), for 100 GeV  $\lsim \mhsm \lsim$ 200 GeV.
At $\sqrt{s}=800$~GeV with  500~fb$^{-1}$ of data,
the analysis of \Ref{heavyHiggs}
suggests that a Higgs
boson with mass up to about 650 GeV will be observable at the LC.
Finally, the process $e^+e^- \rightarrow t \bar t \hsm$~\cite{ttH}
yields a distinctive signature consisting
of two $W$ bosons and four $b$-quark jets, and can be observed at the
LC given sufficient energy and luminosity if the Higgs mass is not too
large.  The QCD-corrected cross-sections for this process~\cite{tth1}
for $\sqrt{s}=500$~GeV and 800~GeV are shown in \fig{LCxsecs}(b).

\begin{figure}[t!]
\begin{center}
\resizebox{\textwidth}{!}{
\includegraphics[width=2.923in,height=2.7in]{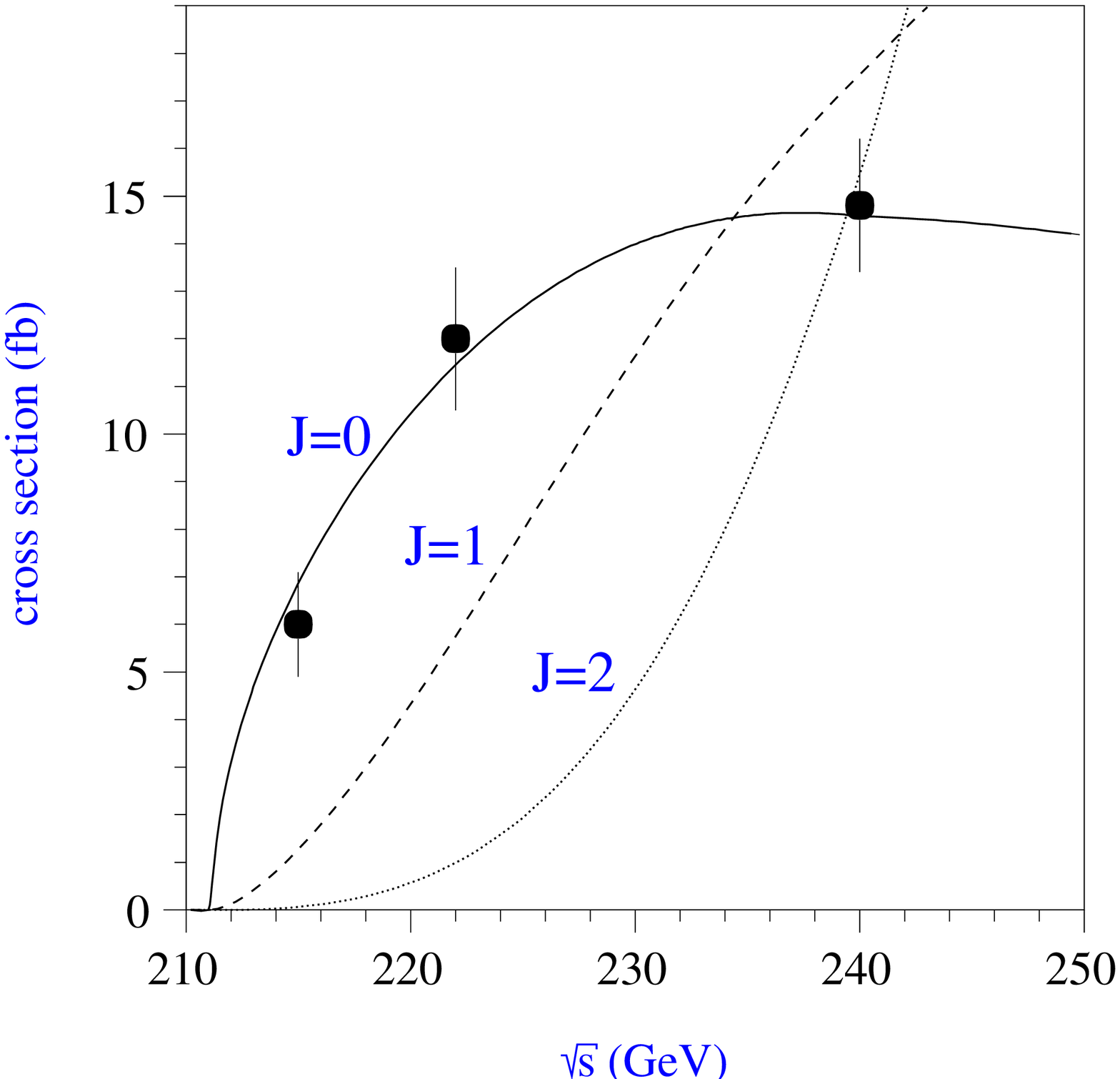}
\hspace*{10.7mm}\vspace*{-0.8mm}
\includegraphics[width=2.864in,height=2.67in]{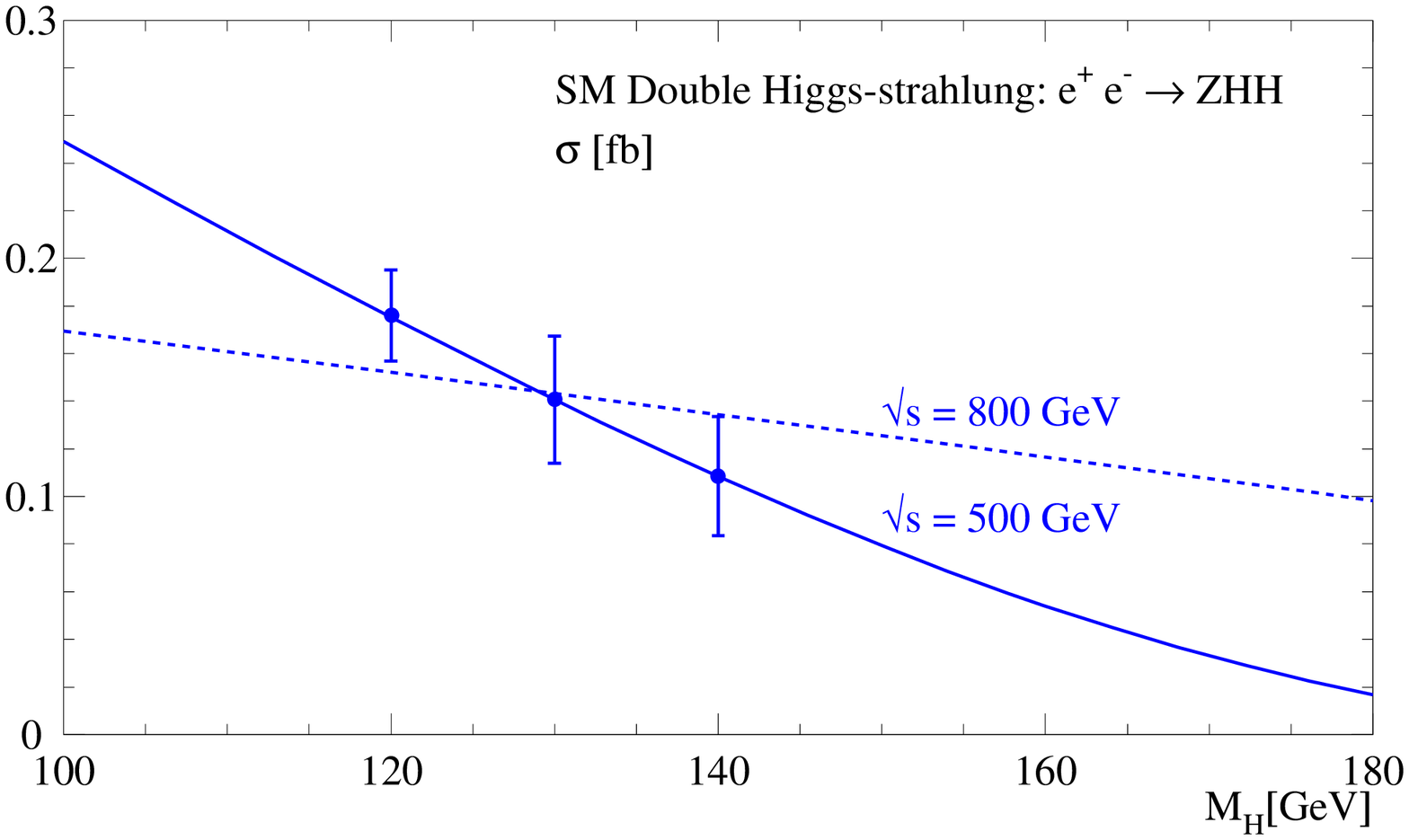}
}
\end{center}
\capt{\label{fig:LChiggsproperties} (a)~Simulated measurement of the
 $e^+e^-\to Z\hsm$ cross-section  for $\mhsm = 120$~GeV  with
 20~fb$^{-1}$ per point at three center of mass energies compared to the
predictions for spin-0 (solid line) and typical examples of spin-1
(dashed line) and spin-2 (dotted line)~\protect\cite{hspin}.
(b)~Cross-section for the
double Higgs-strahlung process $e^+e^-\to Z\hsm\hsm$ at
$\sqrt{s}=500$~GeV (solid line) and 800~GeV (dashed
line)~\protect\cite{n36a}.  The data points show the 
accuracy for 1~ab$^{-1}$.}        
\vspace{-0.5in}
\end{figure}

The phenomenological profile of the Higgs boson can be determined by
precision measurements.  For example, consider the case of
$\mhsm=120$~GeV at the LC with $\sqrt{s}=350$~GeV and 500~fb$^{-1}$ of
data~\cite{Tesla-TDR}.  The spin and parity of the Higgs boson can be
determined unambiguously from the steep onset of the excitation curve in
Higgs-strahlung near the threshold
(see \fig{fig:LChiggsproperties}(a)~\cite{hspin})
and the angular correlations in this process~\cite{angcorr}.
By measuring final state angular distributions and various 
angular and polarization asymmetries,
one can check whether the Higgs boson is a state of definite CP,
or whether it exhibits CP-violating
behavior in its production and/or decays~\cite{cpasymmetries}.
The Higgs mass can be measured to an accuracy of 40 MeV
by reconstructing the Higgs boson in $Z\hsm$ production and combining
the results from the various final state channels.
The Higgs width
can be inferred in a model-independent way, with an accuracy of about
6\%, by combining
the partial width to $W^+W^-$, accessible in the vector boson fusion
process, with the $WW^*$ decay branching ratio.  Similar results (with
precisions within a factor of two of those quoted above) are obtained
for larger Higgs masses in the intermediate mass regime.
The $\hsm ZZ$, $\hsm Z\gamma$ and $\hsm t\bar t$ couplings can be determined 
(with some sensitivity to possible anomalous couplings, if present) by the
optimal-observable method~\cite{optimalobserve},
which makes optimal use of the polarized
angular distributions and asymmetries 
for Higgs-strahlung and $\hsm t\bar t$ production.

\begin{table}[b!]
        \begin{center} \begin{tabular}{|c||c|c|}
\hline
        Higgs coupling  & $\delta {\rm BR}/\rm{BR}$ & $\delta g/g$ \\
\hline\hline
        $hWW$     &      5.1\%     &  1.2\% \\
        $hZZ$     &        ---     &  1.2\% \\ \hline
        $htt$     &      ---     &  2.2\% \\
        $hbb$     &      2.4\%     &  2.1\% \\
        $hcc$     &      8.3\%     & 3.1\% \\
        $h \tau\tau$ &      5.0\%     &  3.2\% \\
        $h \mu\mu$   &      $\sim 30\%$  &  $\sim 15\%$
\\ \hline
        $h gg$       &      5.5\% & \\
        $h \gamma\gamma$ &   16\% & \\ 
        $hhh$    &   --- & $\sim 20\%$ \\ \hline
        \end{tabular} \end{center}
\capt{ \label{tab:BRmeas} Expected fractional uncertainties
for measurements of Higgs 
branching ratios [BR($h\to X\overline X$)]
and couplings [$g_{hXX}$], for various choices of final state
$X\overline X$,
assuming $m_h=120$~GeV at the LC.
In all but four cases, the results shown are based on  500~fb$^{-1}$ of data
at $\sqrt{s}=500$~GeV~\protect\cite{BattagliaDesch}. 
Results for $h\gamma\gamma$~\protect\cite{boosphotons},
$h t\bar t$~\protect\cite{BattagliaDesch}, 
$h \mu\mu$~\protect\cite{mumulc} and $hhh$~\protect\cite{n36a,bby}
are based on 1~ab$^{-1}$ of data
at $\sqrt{s}=500$~GeV (for $\gamma\gamma$ and $hh$) and $\sqrt{s}=800$~GeV
(for $tt$ and $\mu\mu$), respectively.} 
\end{table}

Higgs decay branching ratios can be measured very precisely for
$\mhsm\lsim 150$~GeV~\cite{BattagliaDesch,boosphotons,mumulc,Brau}.  When
such measurements are combined with measurements of Higgs production
cross-sections, the absolute values of the Higgs couplings to the
$W^\pm$ and $Z$ gauge bosons and the Yukawa couplings to leptons and
quarks can be determined to a few percent in a model-independent way.
In addition, the Higgs-top quark Yukawa coupling can be inferred from
the cross-section for Higgs emission off $t\bar t$
pairs~\cite{tth1,tth2}.  As an example, Table~\ref{tab:BRmeas}
exhibits the anticipated fractional uncertainties in the measurements
of Higgs branching ratios for $\mhsm=120$~GeV at the
LC.\footnote{Here, BR($\hsm\to gg$) is assumed to be roughly equal to
the Higgs branching ratio into light hadrons ({\it i.e.}, excluding hadrons
that contain $c$ and $b$ quarks).}
Using this
data, a program {\texttt{HFITTER}} was developed in
\Ref{BattagliaDesch} to perform a Standard Model global fit based on
the measurements of the $Z\hsm$, $\nu\bar\nu\hsm$ and $t\bar t\hsm$
cross-sections and the
Higgs branching ratios listed in Table~\ref{tab:BRmeas}.  The output
of the program is a set of Higgs couplings along with their fractional
uncertainties (which are also exhibited in Table~\ref{tab:BRmeas}).
These results should be considered representative of what can
eventually be achieved at the LC.  For example, a comprehensive analysis
of LC Higgs data will have to take into account
the relevant electroweak loop corrections
(which are not presently included in {\texttt{HFITTER}}).
One should also note that theoretical uncertainties for the predicted Higgs
couplings have not been taken into account in this analysis.
The theoretical uncertainty in $g_{\hsm c\bar c}$ is
the most significant among the channels listed in
Table~\ref{tab:BRmeas}, due to the uncertainties in
$c$ quark-mass and in $\alpha_s$ (which governs the
running of the quark masses from the quark mass to the Higgs mass).
\Rref{chlm} estimates a theoretical fractional uncertainty in
$g_{hc\bar c}$ of about 12\%, significantly greater than the
experimental uncertainty listed in Table~\ref{tab:BRmeas}.
In contrast, the theoretical fractional uncertainty in
$g_{hb\bar b}$ is about 1.8\% due to the uncertainty of the $b$-quark
mass.  Although this is less than the anticipated
experimental uncertainty, it should not be neglected in
the determination of the overall $g_{\hsm b\bar b}$ uncertainty.
The theoretical uncertainties in the other channels listed above
are not significant compared to the quoted experimental uncertainties.

The measurement of the Higgs self-couplings is a very
ambitious task that requires the highest luminosities possible at the
LC, which possess unique capabilities for addressing this
question. The trilinear Higgs self-coupling can be measured in double
Higgs-strahlung, in which a virtual Higgs boson splits into two real
Higgs particles in the final state~\cite{trilinear}.  A simulation
based on 1~ab$^{-1}$ of data is exhibited in
\fig{fig:LChiggsproperties}(b)~\cite{n36a}.  In this way, for
$\mhsm=120$~GeV, the cubic term of the scalar potential can be
established at the LC with a precision of about 20\%~\cite{n36a,bby}.
Such a measurement is a prerequisite for determining the form of the
Higgs potential that is responsible for spontaneous electroweak
symmetry breaking generated by scalar sector dynamics.
Finally, the total SM Higgs width can be obtained indirectly by
using $\Gamma_{\rm tot}=\Gamma_{\hsm WW}/
{\rm BR}(\hsm\to WW^*)$.  The partial width is proportional to
$g^2_{\hsm WW}$, so the fractional uncertainty in $\Gamma_{\rm tot}$
can be obtained from the results of Table~\ref{tab:BRmeas}.
For $\mhsm\lsim 150$~GeV, an accuracy in the range of 5--10\%
can be achieved for the total Higgs width~\cite{desch}.

If the SM Higgs mass is above 150~GeV, then the precision
determination of Higgs couplings will have to be reconsidered.
The Higgs branching ratios into $c\bar c$,
$gg$ and $\tau^+\tau^-$ are now too small to be accurately measured.
Due to the growing importance of the $WW$ and $ZZ$ modes, one can
perform a precision measurement of the Higgs
branching ratios to $WW$ and $ZZ$, while the precision of the $b\bar b$
branching ratio is significantly reduced as $\mhsm$ increases to 200~GeV
and beyond~\cite{derwent}.  Moreover, due to the
rapid decline of $t\bar t\hsm$ production cross-section
with increasing $\mhsm$,
the Higgs-top quark Yukawa coupling cannot be extracted until
$\mhsm\gsim 2m_t$, at which point
the $t\bar t\hsm$ coupling can be obtained by observing Higgs
bosons produced by vector boson fusion which subsequently
decay to $t\bar t$.  The analysis of Ref.~\cite{topHiggs} finds that
at the LC with $\sqrt{s}=800$~GeV and 1~ab$^{-1}$ of data,
the $t\bar t\hsm$ Yukawa coupling can be determined with an accuracy
of about 10\% for a Higgs mass in the range 350---500~GeV.
The total Higgs width can be obtained directly from measuring the Higgs
boson line-shape if $\mhsm\gsim 200$~GeV.

The $e^+e^-$ linear collider with center-of-mass energy $\sqrt{s}$
can also be designed to operate in a
$\gamma\gamma$ collision mode.
This is achieved by using Compton
backscattered photons in the scattering of intense laser
photons on the initial polarized $e^\pm$ beams~\cite{n36b,boos}.
The resulting
$\gamma\gamma$ center of mass energy is peaked for proper
choices of machine parameters at about $0.8\sqrt{s}$.  The luminosity
achievable as a function of the photon beam energy depends strongly on
the machine parameters (in particular, the choice of laser
polarizations).  The $\gamma\gamma$ collider provides additional
opportunities
for Higgs physics~\cite{boos,hggpheno,higgsgamgam,asner,velasco}.
The Higgs boson can be produced as an
$s$-channel resonance in $\gamma\gamma$ collisions,
and one can perform
independent measurements of various Higgs couplings.
For example, the product
$\Gamma(\hsm\to\gamma\gamma){\rm BR}(\hsm\to b\bar b)$ can be measured
with a
statistical accuracy of about $2$---$10\%$ for 120~GeV$\lsim\mhsm\lsim
160$~GeV
with about 50~fb$^{-1}$ of data~\cite{higgsgamgam,asner,velasco}.
In order to reach such a precision, it is critical to control the
overwhelming two-jet background (with efficient $b$-tagging) and
overcome the irreducible $\gamma\gamma\to b\bar b$ background by
optimal use of the polarization of the photon beams and by judicious
kinematic cuts.  Knowledge of the QCD corrections to signal and
background processes is essential for this
task~\cite{higgsgamgam,Melles:1999xd}.

Using values for BR$(\hsm\to b\bar b)$ and
BR$(\hsm\to\gamma\gamma)$ measured at the $e^+e^-$ linear collider,
one can obtain a value for the total Higgs width with
an error dominated by the expected error in BR$(\hsm\to\gamma\gamma)$.
For heavier Higgs bosons, $\mhsm\gsim 200$~GeV, the total Higgs
width can in principle be measured {\it directly}
by tuning the collider to scan across the Higgs resonance.
One can also use the polarization of
the photon beams to measure various asymmetries in Higgs production
and decay, which are sensitive to the CP quantum numbers of the Higgs
boson~\cite{asner}.

\section{The Higgs Bosons of Low-Energy Supersymmetry}
\label{sec:3}

Electroweak symmetry breaking dynamics driven by a weakly-coupled
elementary scalar sector requires a mechanism for the stability of
the electroweak symmetry breaking scale with respect to the
Planck scale~\cite{susynatural}.
Supersymmetry-breaking effects, whose origins may lie at energy scales
much larger than 1~TeV, can induce a radiative breaking of the
electroweak symmetry due to the effects of the large
Higgs-top quark Yukawa coupling~\cite{radewsb}.
In this way, the origin of the
electroweak symmetry breaking scale is intimately tied to the
mechanism of supersymmetry breaking.  Thus, supersymmetry provides an
explanation for the stability of the hierarchy of scales, provided
that supersymmetry-breaking masses in the low-energy effective
electroweak theory are of $\mathcal{O}(1~{\rm TeV})$ or
less~\cite{susynatural}.

A fundamental theory of supersymmetry-breaking is presently unknown.
Nevertheless, one
can parameterize the low-energy theory in terms of the most general
set of soft-supersymmetry-breaking terms~\cite{susybreaking}.
The simplest realistic model of
low-energy supersymmetry is a minimal supersymmetric extension of the
Standard Model (MSSM), which employs the minimal supersymmetric particle
spectrum.  However, even in this minimal model with
the most general set of soft-supersymmetry-breaking terms, more than
100 new supersymmetric parameters are introduced~\cite{mssm124}.
Fortunately, most
of these parameters have no impact on Higgs phenomenology.
Thus, we will focus primarily on the Higgs sector of the MSSM and
identify the parameters that govern the main properties of
the Higgs bosons.

\subsection{The Tree-Level Higgs Sector of the MSSM}
\label{sec:31}

Both hypercharge $Y=-1$ and $Y=+1$ complex Higgs doublets are
required in any Higgs sector of
an anomaly-free supersymmetric extension of the Standard Model.
The supersymmetric structure of the theory also requires (at least) two
Higgs doublets to generate mass for both ``up''-type and ``down''-type
quarks (and charged leptons) \cite{susyhiggs}.
Thus, the MSSM
contains the particle spectrum of a two-Higgs-doublet extension of the
Standard Model and the corresponding supersymmetric partners.

The two-doublet Higgs sector \cite{hhgchap4} contains eight scalar
degrees of freedom:
one complex $Y=-1$ doublet, {\boldmath $\Phi_d$}$=(\Phi_d^0,\Phi_d^-)$
and one complex $Y=+1$ doublet, {\boldmath
$\Phi_u$}$=(\Phi_u^+,\Phi_u^0)$.  The notation reflects the
form of the MSSM Higgs sector coupling to fermions: $\Phi_d^0$
[$\Phi_u^0$] couples exclusively to down-type [up-type] fermion pairs.
When the Higgs potential is minimized, the neutral components of the
Higgs fields acquire vacuum expectation values:\footnote{The
phases of the Higgs fields can be chosen such that the vacuum
expectation values are real and positive.  That is, the tree-level MSSM
Higgs sector conserves CP, which implies that the neutral Higgs mass
eigenstates possess definite CP quantum numbers.}
\beq
\langle {\mathbold{\Phi_d}} \rangle={1\over\sqrt{2}} \left(
\begin{array}{c} v_d\\ 0\end{array}\right), \qquad \langle
{\mathbold{\Phi_u}}\rangle=
{1\over\sqrt{2}}\left(\begin{array}{c}0\\ v_u
\end{array}\right)\,,\label{potmin}
\eeq
where the normalization has been chosen such that
$v^2\equiv v_d^2+v_u^2={4\mw^2/ g^2}=(246~{\rm GeV})^2$.
Spontaneous electroweak symmetry breaking results in
three Goldstone bosons, which
are absorbed and become the longitudinal components of
the $W^\pm$ and $Z$.  The remaining five physical Higgs particles
consist of a charged Higgs pair
\beq \label{hpmstate}
\hpm=\Phi_d^\pm\sinb+ \Phi_u^\pm\cosb\,,
\eeq
one CP-odd scalar
\beq \label{hastate}
\ha= \sqrt{2}\left({\rm Im\,}\Phi_d^0\sinb+{\rm Im\,}\Phi_u^0\cosb
\right)\,,
\eeq
and two CP-even scalars:
\beqa
\hl &=& -(\sqrt{2}\,{\rm Re\,}\Phi_d^0-v_d)\sin\alpha+
(\sqrt{2}\,{\rm Re\,}\Phi_u^0-v_u)\cos\alpha\,,\nonumber\\
\hh &=& (\sqrt{2}\,{\rm Re\,}\Phi_d^0-v_d)\cos\alpha+
(\sqrt{2}\,{\rm Re\,}\Phi_u^0-v_u)\sin\alpha\,,
\label{scalareigenstates}
\eeqa
(with $\mhl\leq \mhh$).
The angle $\alpha$ arises when the CP-even Higgs
squared-mass matrix (in the $\Phi_d^0$---$\Phi_u^0$ basis) is
diagonalized to obtain the physical CP-even Higgs states (explicit
formulae will be given below).

The supersymmetric structure of the theory imposes constraints on the
Higgs sector of the model.  For example, the Higgs
self-interactions are not independent parameters; they can be expressed
in terms of the electroweak gauge coupling constants.  As a result,
all Higgs sector parameters at tree-level
are determined by two free parameters: the ratio of the two
neutral Higgs field vacuum expectation values,
\beq \label{tanbdef}
\tanb\equiv {v_u\over v_d}\,,
\eeq
and one Higgs mass, conveniently chosen to be $\mha$. In particular,
\beq
\mhpm^2 =\mha^2+\mw^2\,,
\label{susymhpm}
\eeq
and the CP-even Higgs bosons $\hl$ and $\hh$ are eigenstates of the
following squared-mass matrix
\beq
{\cal M}_0^2 =    \left(
\matrix{\mha^2 \sin^2\beta + m^2_Z \cos^2\beta&
           -(\mha^2+m^2_Z)\sin\beta\cos\beta \cr
  -(\mha^2+m^2_Z)\sin\beta\cos\beta&
  \mha^2\cos^2\beta+ m^2_Z \sin^2\beta }
   \right)\,.\label{kv}
\eeq
The eigenvalues of ${\cal M}_0^2$ are
the squared-masses of the two CP-even Higgs scalars
\beq
  m^2_{H,h} = \half \left( \mha^2 + m^2_Z \pm
                  \sqrt{(\mha^2+m^2_Z)^2 - 4m^2_Z \mha^2 \cos^2 2\beta}
                  \; \right)\,,\label{kviii}
\eeq
and $\alpha$ is the angle that diagonalizes the CP-even Higgs
squared-mass matrix.
From the above results, one obtains:
\beq
\cos^2(\beta-\alpha)={\mhl^2(\mz^2-\mhl^2)\over
\mha^2(\mhh^2-\mhl^2)}\,.
\label{cbmasq}
\eeq
In the convention where $\tanb$ is positive ({\it i.e.},
$0\leq\beta\leq\pi/2$), the angle $\alpha$ lies in the range
$-\pi/2\leq\alpha\leq 0$.

An important consequence of \eq{kviii} is that there is an upper bound
to the mass of the light CP-even Higgs boson, $\hl$.  One finds that:
\beq \label{kx}
\mhl\leq\mz |\cos 2\beta|\leq\mz\,.
\eeq
This is in marked contrast to the Standard Model, in which the theory
does not constrain the value of $\mhsm$ at tree-level.  The origin of
this difference is easy to ascertain.  In the \SM, $\mhsm^2=\half\lambda v^2$
is proportional to the Higgs self-coupling $\lambda$, which is a free
parameter.  On the other hand, all Higgs self-coupling
parameters of the MSSM are related to the squares of the electroweak
gauge couplings.

Note that the Higgs mass inequality [\eq{kx}] is saturated in the limit
of large $\mha$.  In the limit of $\mha\gg\mz$, the expressions for the
Higgs masses and mixing angle simplify and one finds
\beqa
\mhl^2 &\simeq & \ \mz^2\cos^2 2\beta\,,\label{largema1} \\[3pt]
\mhh^2 &\simeq & \ \mha^2+\mz^2\sin^2 2\beta\,,\label{largema2}\\[3pt]
\mhpm^2& = & \ \mha^2+\mw^2\,,\label{largema3}\\[3pt]
\cos^2(\beta-\alpha)&\simeq\ & {\mz^4\sin^2 4\beta\over 4\mha^4}\,.
\label{largema4}
\eeqa
Two consequences are immediately apparent.
First, $\mha\simeq\mhh
\simeq\mhpm$, up to corrections of ${\cal O}(\mz^2/\mha)$.  Second,
$\cos(\beta-\alpha)=0$ up to corrections of ${\cal O}(\mz^2/\mha^2)$.
This limit is known as the {\it decoupling} limit \cite{decoupling}
because when $\mha$ is
large, there exists an effective low-energy theory below the scale
of $\mha$ in which the effective Higgs sector consists only of one
CP-even Higgs boson, $\hl$.  As we shall demonstrate below, the
tree-level couplings of $\hl$ are precisely those of the Standard
Model Higgs boson when $\cos(\beta-\alpha)=0$.
From \eq{largema4}, one can also derive:
\begin{equation}  \label{cotalf}
\cot\alpha = -\tan\beta - \frac{2\mz^2}{\mha^2} \tan\beta \cos
2\beta + {\cal O}\left(\frac{\mz^4}{\mha^4}\right)\,.
\end{equation}
This result will prove useful in evaluating the CP-even Higgs boson
couplings to fermion pairs in the decoupling limit.

The phenomenology of the Higgs sector depends in detail on
the various couplings of the Higgs bosons to gauge bosons, Higgs
bosons and fermions.  The couplings of the two CP-even Higgs bosons
to $W$ and $Z$ pairs
are given in terms of the angles $\alpha$ and $\beta$ by
\beq
g\ls{\hl VV}= g\ls{V} m\ls{V}\sinbma \,,\qquad\qquad
           g\ls{\hh VV}= g\ls{V} m\ls{V}\cosbma\,,\label{vvcoup}
\eeq
where
$g\ls V\equiv 2m_V/v$ for $V=W$ or $Z$.
There are no tree-level couplings of $\ha$ or $\hpm$ to $VV$.
The couplings of $V$ to two neutral Higgs bosons
(which must have opposite CP-quantum numbers) are given by
$g_{\phi\ha Z}(p_\phi-p_\ha)$, where $\phi=\hl$ or $\hh$ and the
momenta $p_\phi$ and $p_\ha$ point into the vertex, and
\beq
g\ls{\hl\ha Z}={g\cosbma\over 2\cos\theta_W}\,,\qquad\qquad\quad
           g\ls{\hh\ha Z}={-g\sinbma\over 2\cos\theta_W}\,.
           \label{hvcoup}
\eeq
From the expressions above, we see
that the following sum rules must hold separately for $V=W$ and~$Z$:
\beqa
g_{\hh V V}^2 + g_{\hl V V}^2 &=&
                      g\ls{V}^2m\ls{V}^2\,,\label{vvsumrule} \\[3pt]
g_{\hl\ha Z}^2+g_{\hh\ha Z}^2&=&
             {g^2\over 4\cos^2\theta_W}\,,  \\[3pt]
 g^2_{\phi ZZ} + 4m^2_Z g^2_{\phi \ha Z}& =& {g^2m^2_Z\over
        \cos^2\theta_W}\,,\qquad \phi=\hl, \hh \,.\label{hxi}
\eeqa
Similar considerations also hold for
the coupling of $\hl$ and $\hh$ to $W^\pm H^\mp$.
Four-point couplings of vector bosons and Higgs bosons
can be found in \Ref{hhg}.  The properties of the three-point and
four-point Higgs boson-vector boson couplings are conveniently summarized
by listing the couplings that are proportional
to either $\sin(\beta-\alpha)$ or $\cos(\beta-\alpha)$, and the couplings
that are independent of $\alpha$ and $\beta$~\cite{hhg}:
\beq
\renewcommand{\arraycolsep}{1cm}
\let\us=\underline
\begin{array}{lll}
\us{\cos(\beta-\alpha)}&  \us{\sin(\beta-\alpha)} &
\us{\hbox{\rm{angle-independent}}} \\ [3pt]
\noalign{\vskip3pt}
       \hh W^+W^-&        \hl W^+W^- &  \qquad\longdash   \\
       \hh ZZ&            \hl ZZ  & \qquad\longdash  \\
       Z\ha\hl&          Z\ha\hh  & ZH^+H^-\,,\,\,\gamma H^+H^-\\
       W^\pm H^\mp\hl&  W^\pm H^\mp\hh & W^\pm H^\mp\ha \\
       ZW^\pm H^\mp\hl&  ZW^\pm H^\mp\hh & ZW^\pm H^\mp\ha \\
    \gamma W^\pm H^\mp\hl&  \gamma W^\pm H^\mp\hh & \gamma W^\pm H^\mp\ha \\
   \quad\longdash    &\quad\longdash  & VV\phi\phi\,,\,VV\ha\ha\,,\,VV H^+H^-
\end{array}
\label{littletable}
\eeq
where $\phi=\hl$ or $\hh$ and $VV=W^+W^-$, $ZZ$, $Z\gamma$ or
$\gamma\gamma$.
Note in particular that {\it all} vertices
in the theory that contain at least
one vector boson and {\it exactly one} non-minimal Higgs boson state
($\hh$, $\ha$ or $\hpm$) are proportional to $\cos(\beta-\alpha)$.
This can be understood as a consequence of unitarity sum rules which
must be satisfied by the tree-level amplitudes of the
theory \cite{unitarity,thacker,weldon,wudka}.

In the MSSM, the tree-level Higgs couplings to fermions obey the following
property: $\Phi_d^0$ couples exclusively to down-type fermion pairs and
$\Phi_u^0$ couples exclusively to up-type fermion pairs. This pattern
of Higgs-fermion couplings defines the
Type-II two-Higgs-doublet model \cite{wise,hhg}.
The gauge-invariant Type-II Yukawa interactions (using 3rd family
notation) are given by:
\begin{equation} \label{typetwo}
-{\cal L}_{\rm Yukawa}= h_t\left[\bar t P_L t \Phi^0_u-\bar t P_L b
\Phi^+_u\right] + h_b\left[\bar b P_L b \Phi^0_d-\bar b P_L t
\Phi^-_d\right] + {\rm h.c.}\,,
\end{equation}
where $P_L\equiv\half(1-\gamma_5)$ is the left-handed projection
operator.  [Note that ($\overline\Psi_1 P_L \Psi_2)^\dagger=
\overline\Psi_2 P_R \Psi_1$, where $P_R\equiv\half(1+\gamma_5)$.]
Fermion masses are generated when the neutral Higgs components acquire
vacuum expectation values.  Inserting \eq{potmin} into \eq{typetwo}
yields a relation between the quark masses and the Yukawa couplings:
\beq
h_b = {\sqrt{2}\,m_b\over v_d}={\sqrt{2}\, m_b\over
v\cos\beta}\,,\qquad\qquad
h_t = {\sqrt{2}\,m_t\over v_u}={\sqrt{2}\, m_t\over v\sin\beta}\,.
\label{hdef}
\eeq
Similarly, one can define the Yukawa coupling of the Higgs boson to
$\tau$-leptons (the latter is a down-type fermion).  The couplings
of the physical Higgs bosons to the third generation fermions is
obtained from \eq{typetwo} by using \eqs{hpmstate}{scalareigenstates}.
In particular, the couplings of the neutral Higgs bosons to $f\bar f$
relative to the Standard Model
value, $gm_f/2\mw$, are given by
\begin{eqaligntwo}
\hl b\bar b \;\;\; ({\rm or}~ \hl \tau^+ \tau^-):&~~~ -
{\sin\alpha\over\cos\beta}=\sin(\beta-\alpha)
-\tan\beta\cos(\beta-\alpha)\,, \label{qqcouplingsa}\\[3pt]
\hl t\bar t:&~~~ \phm{\cos\alpha\over\sin\beta}=\sin(\beta-\alpha)
+\cot\beta\cos(\beta-\alpha)\,, \label{qqcouplingsb}\\[3pt]
\hh b\bar b \;\;\; ({\rm or}~ \hh \tau^+ \tau^-):&~~~
\phm{\cos\alpha\over\cos\beta}=
\cos(\beta-\alpha)
+\tan\beta\sin(\beta-\alpha)\,, \label{qqcouplingsc}\\[3pt]
\hh t\bar t:&~~~ \phm{\sin\alpha\over\sin\beta}=\cos(\beta-\alpha)
-\cot\beta\sin(\beta-\alpha)\,, \label{qqcouplingsd}\\[3pt]
\ha b \bar b \;\;\; ({\rm or}~ \ha \tau^+
\tau^-):&~~~\phm\gamma_5\,{\tan\beta}\,, \label{qqcouplingse}\\[3pt]
\ha t \bar t:&~~~\phm\gamma_5\,{\cot\beta}\,, \label{qqcouplingsf}
\end{eqaligntwo}
(the $\gamma_5$ indicates a pseudoscalar coupling), and the
charged Higgs boson couplings to fermion pairs,
with all particles pointing into the vertex, are given by
\begin{eqaligntwo}
g_{H^- t\bar b}= &~~~ {g\over{\sqrt{2}\mw}}\
\Bigl[m_t\cot\beta\,P_R+m_b\tan\beta\,P_L\Bigr]\,,
\label{hpmqq}\\[3pt]
g_{H^- \tau^+ \nu}= &~~~ {g\over{\sqrt{2}\mw}}\
\Bigl[m_{\tau}\tan\beta\,P_L\Bigr]\,.\label{hpmll}
\end{eqaligntwo}

\begin{figure}[t!]
\begin{center}
\resizebox{\textwidth}{!}{
\includegraphics[width=5cm]{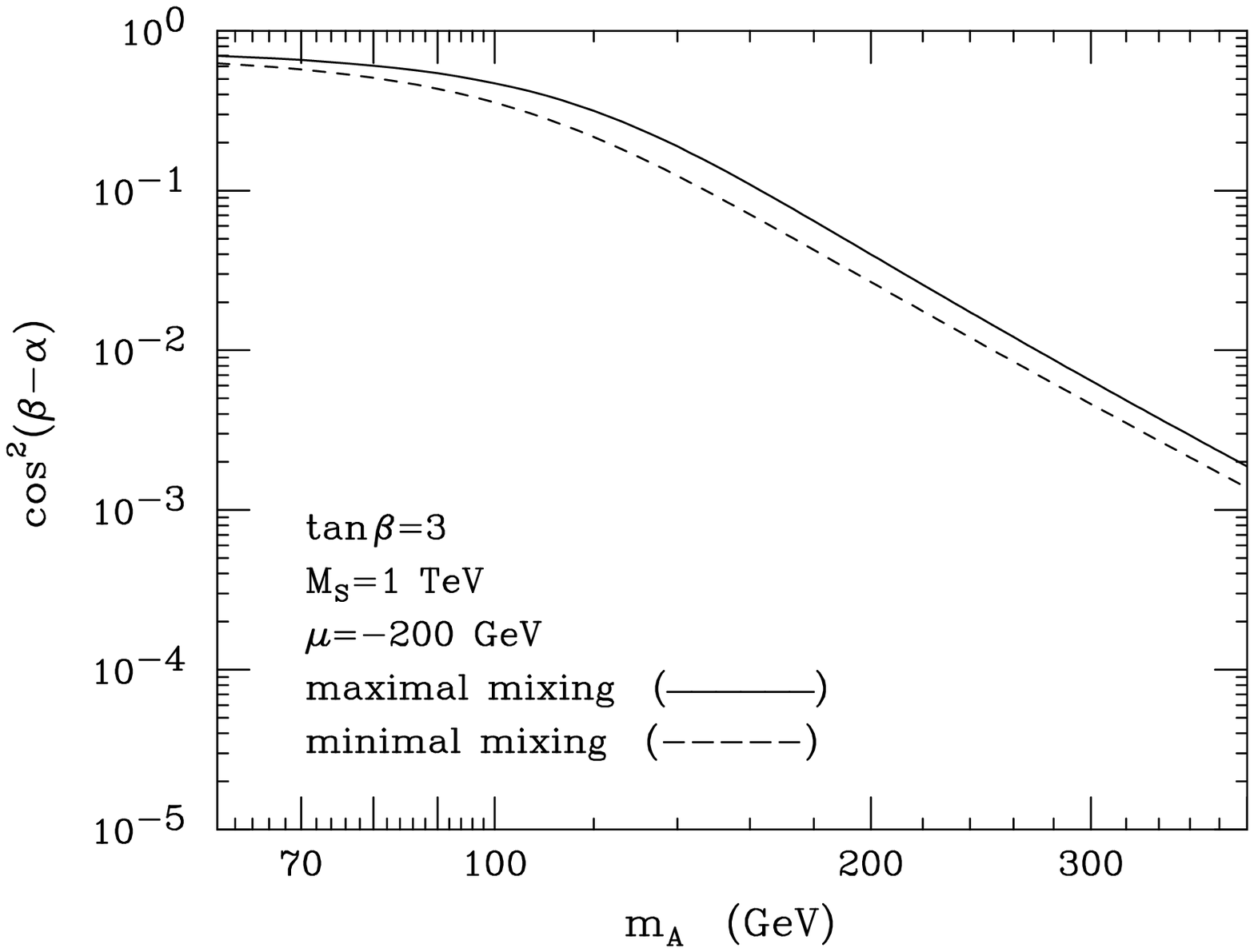}
\hfill
\includegraphics[width=5cm]{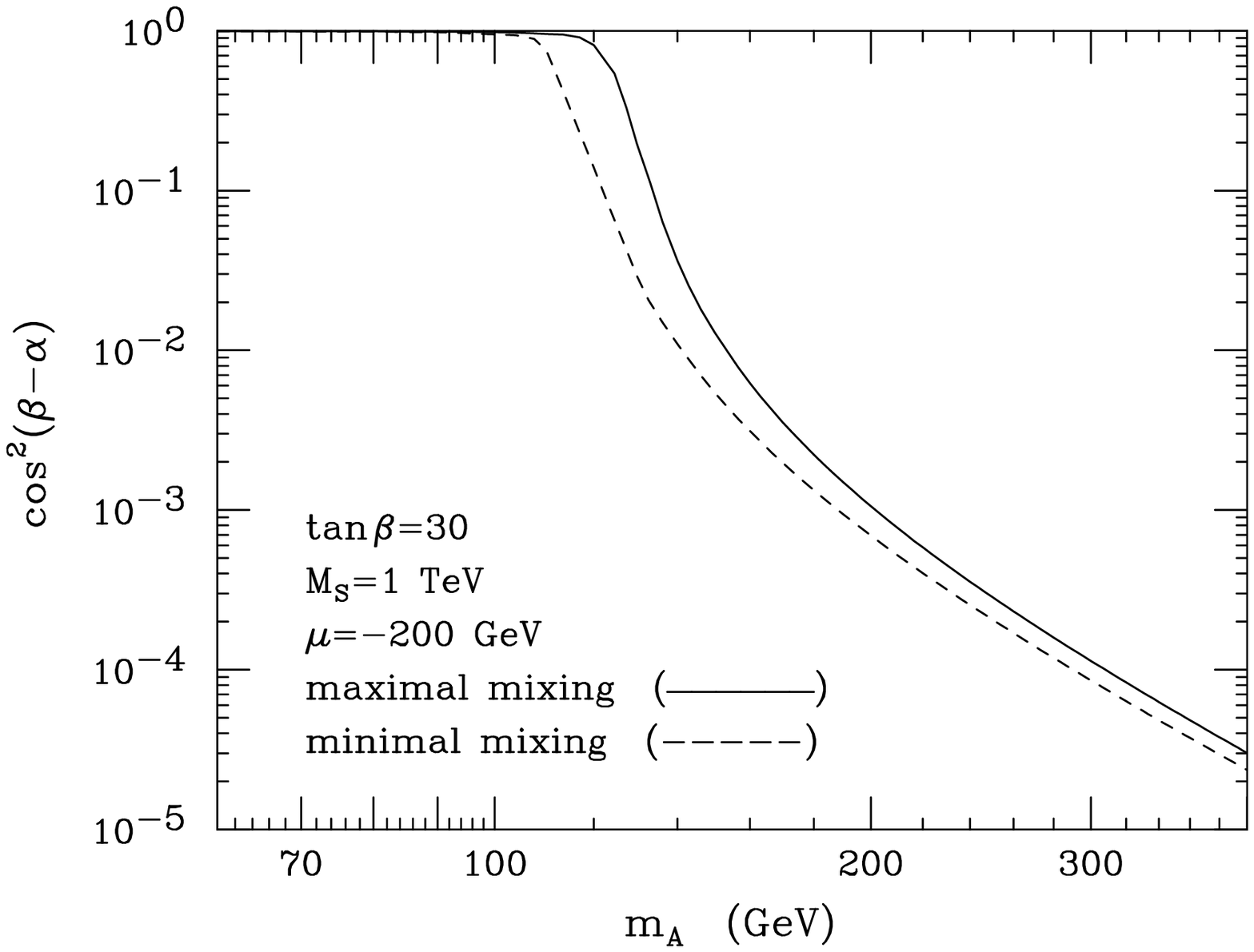}
}
\end{center}
\capt{\label{cosgraph} The value of $\cos^2(\beta-\alpha)$
is shown as a function of
$\mha$ for two choices of $\tan\beta = 3$ and $\tan\beta = 30$.
When radiative-corrections are included, one can define an approximate
loop-corrected angle $\alpha$ as a function of $\mha$, $\tan\beta$ and
the MSSM parameters.  In the figures above, we
have incorporated radiative corrections, assuming that
$\MSUSY\equiv M_Q=M_U=M_D=1$~TeV.  In addition,
two extreme cases for the squark mixing parameters
are shown (see \Secs{sec:32}{sec:33} for further discussion of the
radiative corrections and their dependence on the supersymmetric
parameters). The decoupling effect expected from
\protect\eq{largema4}, in which $\cos^2(\beta-\alpha)\propto \mz^4/\mha^4$
for $\mha\gg m_Z$,
continues to hold even when radiative corrections are included.}
\end{figure}

We next examine the behavior of the Higgs couplings at large
$\tan\beta$.  This limit is of particular interest since
at large $\tan\beta$, some of the Higgs couplings to down-type
fermions can be significantly enhanced.\footnote{In models of low-energy
supersymmetry, there is some theoretical prejudice that suggests that
$1 < \tanb \lsim m_t/m_b$, with the fermion running masses evaluated
at the electroweak scale.
For example, $\tanb\lsim 1$
[$\tanb > \mt/\mb$] is disfavored
since in this case, the Higgs--top-quark [Higgs--bottom-quark]
Yukawa coupling blows up at an
energy scale significantly below the Planck scale.}
Consider two large $\tan\beta$ regions of interest:
\textit{(i)}~If $\mha\gg\mz$,
then the decoupling limit is reached, in which $|\cos(\beta-\alpha)|\ll
1$ and $\mhh\simeq\mha$.
From \eqs{largema4}{qqcouplingsf}, it follows that the $b\bar b\hh$
and $b\bar b\ha$ couplings have equal strength and are significantly
enhanced (by a factor of
$\tanb$) relative to the $b\bar b\hsm$ coupling, whereas the $VV\hh$
coupling is negligibly 
small. In contrast, the values of the $VV\hl$ and $b\bar b\hl$ couplings
are equal to the corresponding couplings of the \SM\ Higgs boson.
To show that the value of the $b\bar b\hl$ coupling
[\eq{qqcouplingsa}]
reduces to that of $b\bar b\hsm$ in the decoupling limit,
note that \eq{largema4} implies that $|\tan\beta\cos(\beta-\alpha)|\ll 1$
when $\mha\gg \mz$ even when $\tan\beta\gg 1$.  Indeed, $\hl$ is a
SM-like Higgs boson.
\textit{(ii)}~If $\mha\lsim \mz$ and $\tanb\gg 1$, then
$|\cos(\beta-\alpha)|\sim 1$ [see \fig{cosgraph}] and $\mhl\simeq\mha$.
In this case, the $b\bar b\hl$
and $b\bar b\ha$ couplings have equal strength and are significantly
enhanced (by a factor of
$\tanb$) relative to the $b\bar b\hsm$ coupling, while the $VV\hl$
coupling is negligibly small.  Using \eq{vvsumrule} it follows that
the $VV\hh$ coupling is equal in strength to the $VV\hsm$ coupling.
In this case, it is conventional to refer to $\hh$ as a SM-like Higgs
boson.  However, this nomenclature is somewhat inaccurate, since
the value of the $b\bar b\hh$ coupling can differ
from the corresponding $b\bar b\hsm$ coupling when $\tanb\gg 1$
[since in case \textit{(ii)}, where $|\sin(\beta-\alpha)|\ll 1$,
the product $\tan\beta\sin(\beta-\alpha)$ need not
be particularly small]. Note that in both cases
\textit{(i)} and \textit {ii)} above,
only two of the three neutral Higgs bosons have enhanced
couplings to $b\bar b$.

The decoupling limit of $\mha\gg\mz$ is effective for all values of
$\tan\beta$.  It is easy to check that the pattern of all Higgs
couplings displayed in \eqs{vvcoup}{qqcouplingsf}
respect the decoupling limit.  That is, in the limit where $\mha\gg\mz$,
$\cos(\beta-\alpha)={\cal O}(\mz^2/\mha^2)$, which means that
the $\hl$ couplings to
Standard Model particles approach values corresponding precisely
to the couplings of the SM Higgs boson.
There is a significant
region of MSSM Higgs sector parameter space in which the decoupling
limit applies, because
$\cos(\beta-\alpha)$ approaches zero quite rapidly once
$\mha$ is larger than about 200~GeV, as shown in \fig{cosgraph}.
As a result, over a significant region of the MSSM parameter space, the
search for the lightest CP-even Higgs boson
of the MSSM is equivalent to the search for the \SM\
Higgs boson.  This result is more general; in
many theories of non-minimal Higgs sectors, there is a significant
portion of the parameter space that approximates the decoupling limit.
Consequently, simulations of the \SM\ Higgs signal are also relevant for
exploring the more general Higgs sector.

\subsection{Radiatively-Corrected MSSM Higgs Masses}
\label{sec:32}

The discussion of \Sec{sec:31} was based on a tree-level analysis of the
Higgs sector.  However, radiative corrections can have a significant
impact on the predicted values of Higgs masses and couplings.
The radiative corrections involve both loops of Standard Model particles
and loops of supersymmetric partners.
The dominant effects arise from
loops involving the third generation quarks and squarks
and are proportional to the corresponding Yukawa couplings.
Thus, we first review the parameters that control the masses
and mixing of the third-generation squarks.  (We shall neglect
intergenerational mixing effects, which have little impact on the
discussion that follows.)

For each left-handed and right-handed quark of fixed flavor, $q$,
there is a corresponding supersymmetric partner
$\widetilde q_L$ and $\widetilde q_R$, respectively.  These are the
so-called interaction eigenstates, which mix according to the squark
squared-mass matrix. The mixing angle that diagonalizes the squark mass
matrix will be denoted by $\theta_{\widetilde q}$.   The
squark mass eigenstates, denoted by $\widetilde q_1$ and $\widetilde q_2$,
are obtained by diagonalizing the following $2\times 2$ matrix
\begin{equation}
 \left(  
\matrix{M_{Q}^2+m_f^2+D_L & m_f X_f \crr
   m_f X_f & M_{R}^2+m_f^2+D_R}
\right)  \,,
\label{stopmatrix}
\end{equation}
where $D_L\equiv (T_{3f}-e_f\sin^2\theta_W)\mz^2\cos2\beta$ and
$D_R\equiv e_f\sin^2\theta_W\mz^2\cos2\beta$.  In addition, $f=t$,
$M_R\equiv M_U$, $e_t=2/3$ and $T_{3f}=1/2$ for the top-squark
squared-mass matrix, and $f=b$, $M_R\equiv M_D$, $e_b=-1/3$ and
$T_{3f}=-1/2$ for the bottom-squark squared-mass matrix. The
squark mixing parameters are given by
\beq
 X_t  \equiv A_t-\mu\cot\beta\,,\qquad\qquad
 X_b  \equiv A_b-\mu\tan\beta\,.
\eeq
Thus, the top-squark and bottom-squark masses and mixing angles
depend on the supersymmetric Higgsino mass parameter $\mu$ and
the soft-supersymmetry-breaking parameters: $M_Q$, $M_U$, $M_D$,
$A_t$ and $A_b$.  For simplicity, we shall initially assume
that $A_t$,
$A_b$ and $\mu$ are real parameters.  That is,
we neglect
possible CP-violating effects that can enter the MSSM Higgs sector
via radiative corrections.  The impact on new MSSM sources of CP-violation
on the Higgs sector will be addressed in \Sec{sec:323}.

\subsubsection{Radiatively-corrected Higgs masses in the CP-conserving
MSSM}
\label{sec:321}

\begin{figure}[t!]
  \begin{center}
\resizebox{\textwidth}{!}{
\includegraphics[width=5cm]{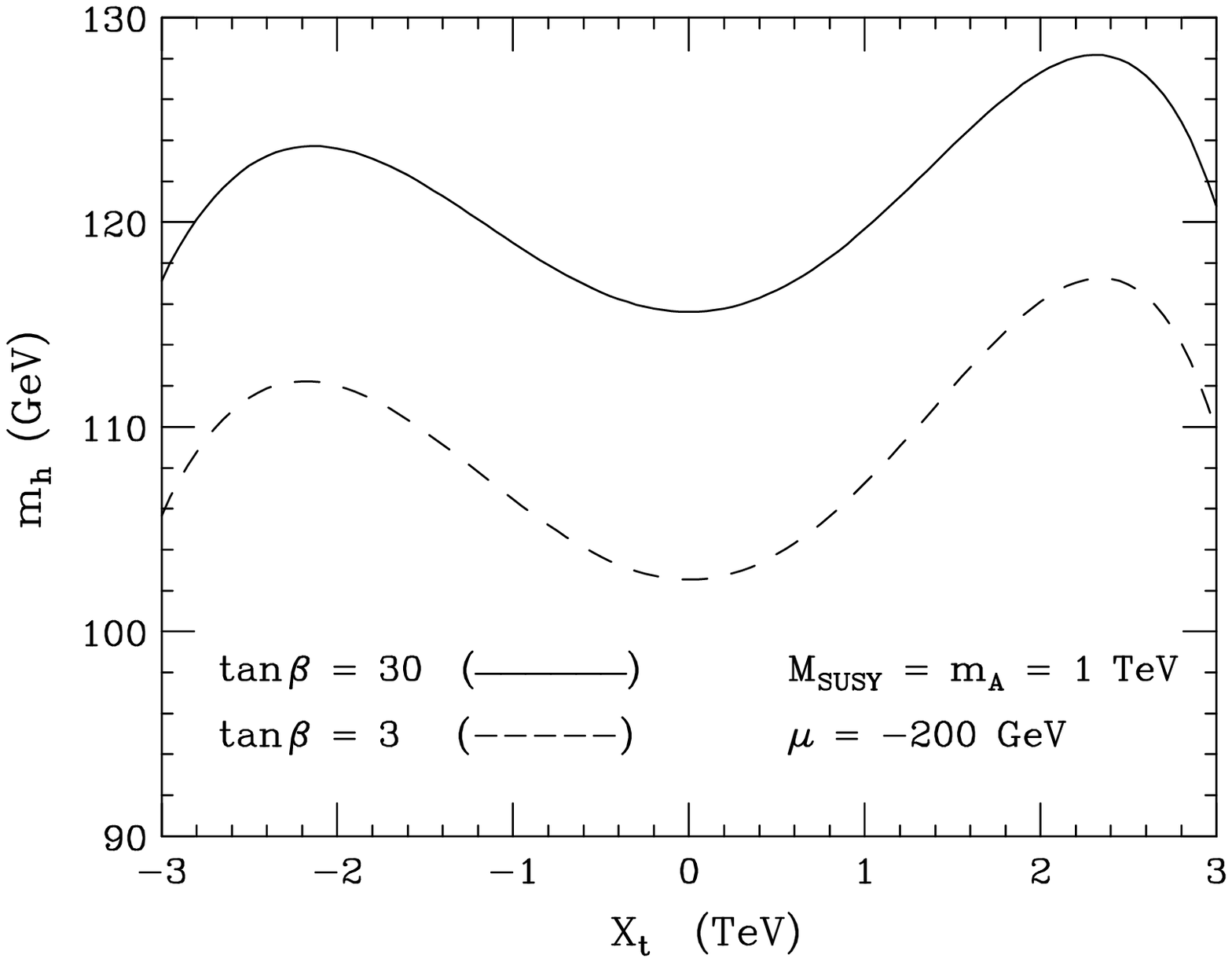}
\hfill
\includegraphics[width=5cm]{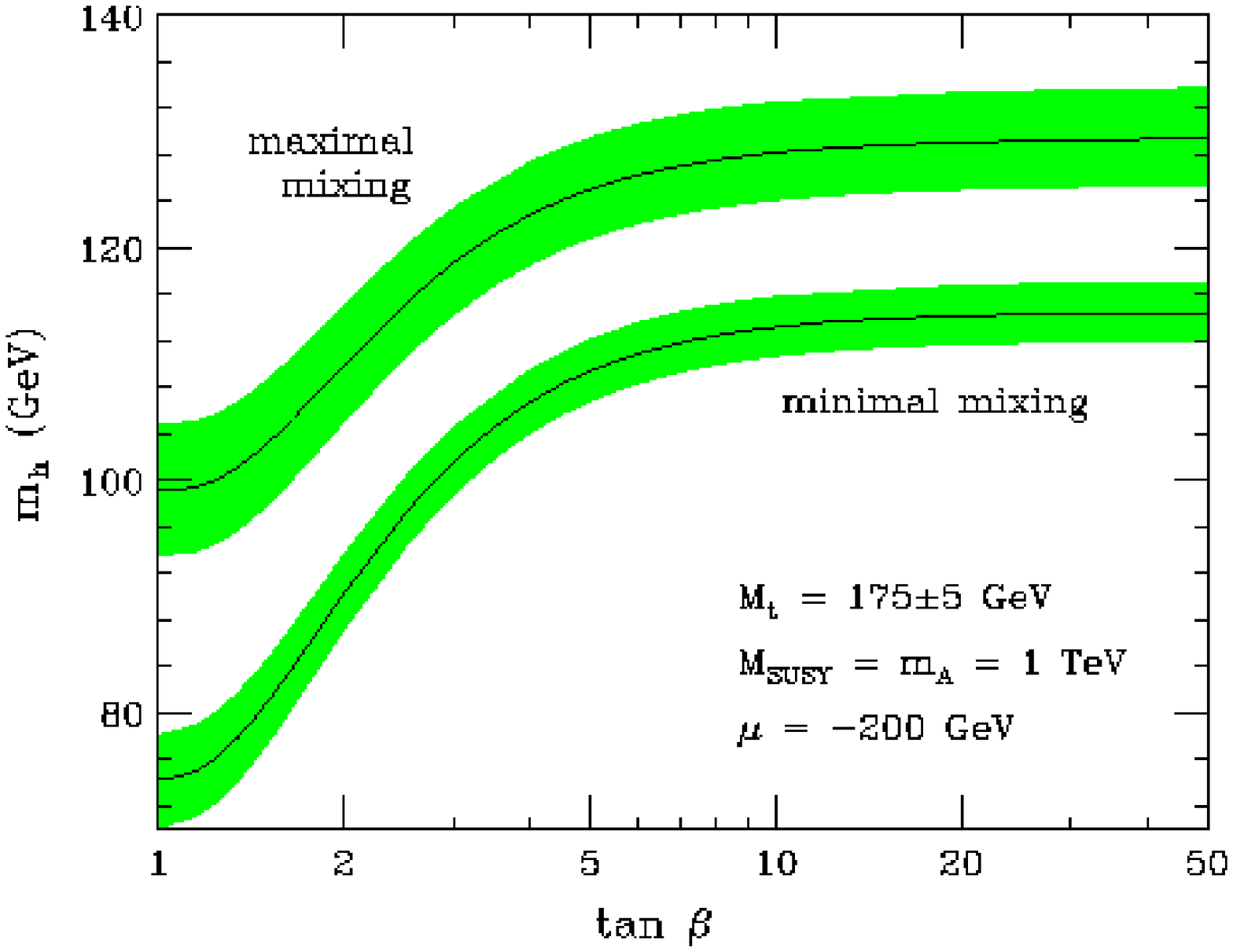}
}
\end{center}
\capt{\label{mhxt} The radiatively corrected light
CP-even Higgs mass is plotted (a) as a function of $X_t$, where
$X_t\equiv A_t-\mu\cot\beta$, for $M_t=174.3$~GeV and two choices
of $\tanb=3$ and 30, and (b) as a function of $\tanb$, for the
maximal mixing [upper band] and minimal mixing [lower band]
benchmark cases. In (b), the central value of the shaded bands
corresponds to $M_t=175$~GeV, while the upper [lower] edge of the
bands correspond to increasing [decreasing] $M_t$ by 5~GeV. In
both (a) and (b), $\mha=1$~TeV and the diagonal soft squark
squared-masses are assumed to be degenerate: $\MSUSY\equiv
M_Q=M_U=M_D=1$~TeV. }
\end{figure}

The radiative corrections to the Higgs squared-masses have been
computed by a number of techniques, and using a variety of
approximations such as the effective potential at 
one-loop~\cite{early-veff,veff,berz} and 
two-loops~\cite{zhang,espizhang,EZ2,bdsz} ,
and diagrammatic methods
\cite{hhprl,1-loop,completeoneloop,deltamb2,hempfhoang,weiglein}.
Complete one-loop diagrammatic computations of the MSSM Higgs
masses have been presented by a number of groups
\cite{completeoneloop,deltamb2}; and partial two-loop diagrammatic
results are also known~\cite{hempfhoang,weiglein}. These include
the ${\cal O}(m_t^2 h_t^2 \alpha_s)$ contributions to the neutral
CP-even Higgs boson squared-masses in the on-shell
scheme~\cite{weiglein}.  Finally, renormalization group methods
(to be discussed further below) provide a powerful technique for
identifying many of the most important contributions to the
radiatively corrected Higgs masses~\cite{rge,llog,carena,hhh}.
Typical results for the radiatively corrected value of $\mhl$ as a
function of the relevant supersymmetric parameters are shown in
\fig{mhxt}.

One of the most striking effects of the radiative corrections to the
MSSM Higgs sector is the modification of the upper bound of the
light CP-even Higgs mass, as first noted in \Refs{early-veff}{hhprl}.
Consider the region of parameter space
where $\tanb$ is large and $\mha\gg\mz$.  In this limit, the
{\it tree-level} prediction for $\mhl$ corresponds to its theoretical
upper bound, $\mhl=\mz$.  Including radiative corrections, the theoretical
upper bound is increased.  The dominant effect arises from an incomplete
cancellation\footnote{In
certain regions of parameter space (corresponding to large $\tanb$ and
large values of $\mu$), the
incomplete cancellation of the bottom-quark and bottom-squark loops can
be as important as the corresponding top sector contributions.  For
simplicity, we ignore this contribution in \eq{deltamh}.} 
of the top-quark and top-squark loops (these
effects cancel in the exact supersymmetric limit).
The qualitative behavior of the radiative corrections can be most easily
seen in the large top squark mass limit, where in addition, the
splitting of the two diagonal entries and the off-diagonal entry
of the top-squark squared-mass matrix are both small in comparison to
the average of the two top-squark squared-masses:
\beq \label{mstwo}
\msusyy\equiv\half(\mstopa^2+\mstopb^2)\,.
\eeq
In this case, the upper bound on the lightest CP-even Higgs
mass is approximately given by
\beq \label{deltamh}
\mhl^2\lsim \mz^2+{3g^2\mt^4\over
8\pi^2\mw^2}\left[\ln\left({M_S^2\over\mt^2}\right)+{X_t^2\over M_S^2}
\left(1-{X_t^2\over 12M_S^2}\right)\right]\,.
\eeq

The more complete treatments of the radiative corrections
cited above show that
\eq{deltamh} somewhat overestimates the true upper bound of $\mhl$.
Nevertheless, \eq{deltamh} correctly reflects some noteworthy features
of the more precise result.  First, the increase of the
light CP-even Higgs mass bound beyond $\mz$ can be significant.  This is
a consequence of the $m_t^4$ enhancement of the one-loop radiative
correction.
Second, the dependence of the light Higgs mass on the top-squark mixing
parameter $X_t$ implies that (for a given value of
$\msusy$) the upper bound of the light Higgs mass
initially increases with $X_t$ and reaches its {\it maximal} value
for $X_t=\sqrt{6}\msusy$.  This point is referred to as the {\it maximal
mixing} case (whereas $X_t=0$ corresponds to the {\it
minimal mixing} case).  In a more complete computation that includes
both two-loop logarithmic and non-logarithmic corrections, the $X_t$
values corresponding to maximal and minimal mixing are
shifted and exhibit an asymmetry under $X_t\to -X_t$ as shown in
\fig{mhxt}.
In the numerical
analysis presented in this and subsequent figures in this
section, we assume for simplicity that the third generation
diagonal soft-supersymmetry-breaking
squark squared-masses are degenerate: $\MSUSY\equiv
M_Q=M_U=M_D$, which defines the parameter $\MSUSY$.\footnote{We also
assume that $\MSUSY\gg\mt$, in which case it follows that
$M_S^2\simeq\MSUSY^2$ up to corrections of ${\cal O}(\mt^2/\MSUSY^2)$.}

\begin{figure}[t!]
\begin{center}
\resizebox{\textwidth}{!}{
\includegraphics[width=5cm]{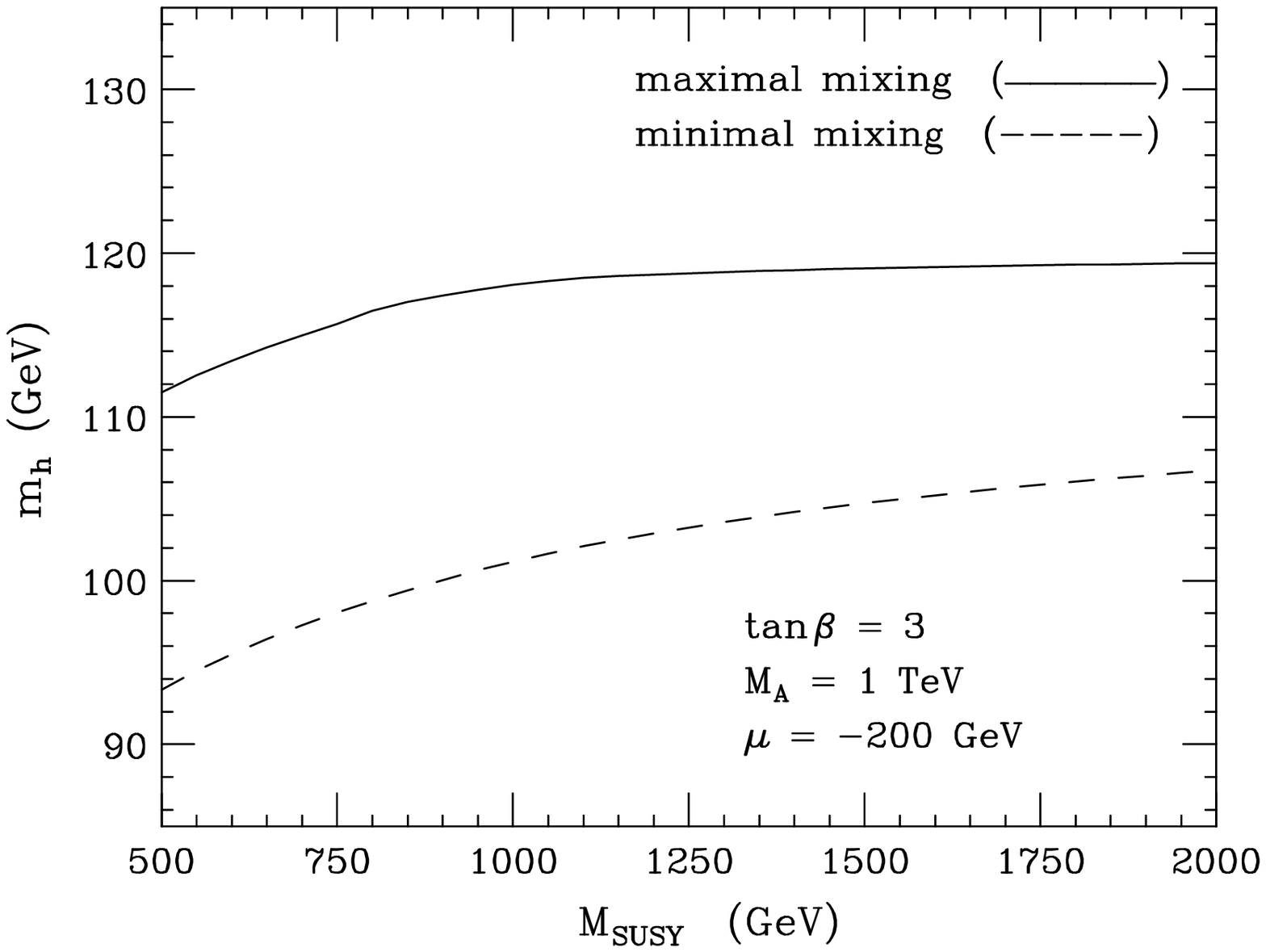}
\hfill
\includegraphics[width=5cm]{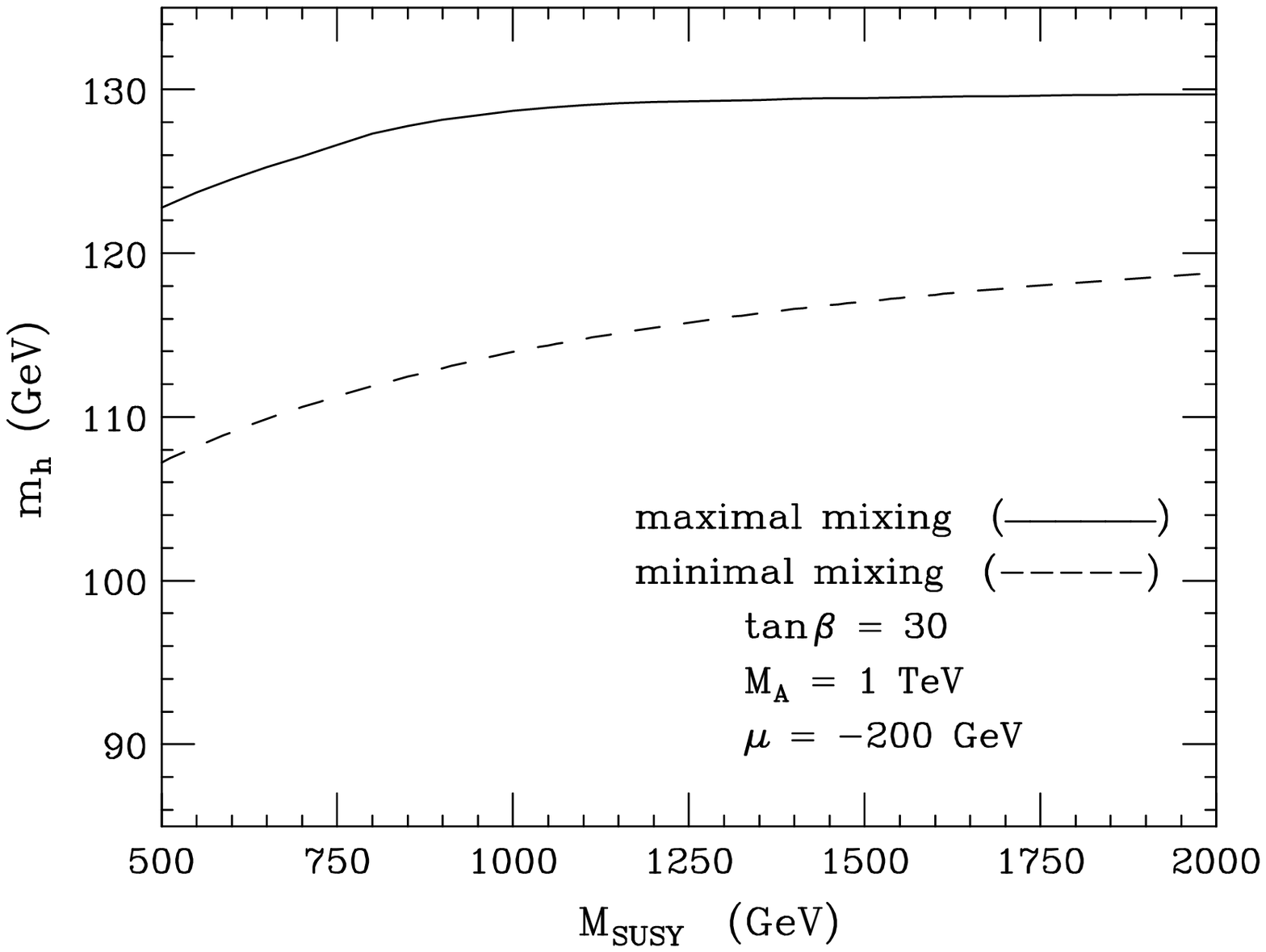}
}
\end{center}
\vskip1pc
  \capt{\label{mhmsusy} The radiatively corrected light CP-even
Higgs mass is plotted as a function of $\MSUSY\equiv M_Q=M_U=M_D$,
for $M_t=174.3$~GeV, $\mha= 1$~TeV and two choices of $\tanb=3$
and $\tanb=30$.  Maximal mixing and minimal mixing are defined
according to the value of $X_t$ that yields the maximal and
minimal Higgs mass as shown in \protect\fig{mhxt}(a). }
\end{figure}

Third, note the logarithmic sensitivity to the top-squark masses.
Naturalness arguments that underlie low-energy supersymmetry imply
that the supersymmetric particle masses should not be larger than
a few TeV.  Still, the precise upper bound on the light Higgs mass
depends on the specific choice for the upper limit of the
top-squark masses.  The dependence of the light Higgs mass
obtained by the more complete computation as a function of
$\MSUSY$ is shown in \fig{mhmsusy}.\footnote{The flattening of the
curves in \fig{mhmsusy}
as a function of $\MSUSY$ in the maximal mixing scenario is
due to the squark-mixing contributions at two-loops
which partially cancel the contributions that grow logarithmically
with $\MSUSY$.}

As noted above, the largest contribution to the one-loop radiative
corrections is enhanced by a factor of $m_t^4$ and grows
logarithmically with the top squark mass.  Thus, higher order
radiative corrections can be non-negligible for large top squark
masses, in which case the large logarithms must be resummed.
Renormalization group (RG) techniques for resumming the leading
logarithms have been developed by a number of authors
\cite{rge,llog,carena}. The computation of the
RG-improved one-loop corrections requires numerical integration of
a coupled set of RG equations~\cite{llog}. Although this procedure
has been carried out in the literature, the analysis is unwieldy
and not easily amenable to large-scale Monte-Carlo studies. It
turns out that over most of the parameter range, it is sufficient
to include the leading and sub-leading logarithms at two-loop
order. (Some additional non-logarithmic terms, which cannot be
ascertained by the renormalization group method, must also be
included \cite{chhhww}.) Compact analytic expressions have been
obtained for the dominant one and two-loop contributions to the
matrix elements of the radiatively-corrected CP-even Higgs
squared-mass matrix:
\beq \label{calmatrix}
{\cal M}^2\equiv
\left( \matrix{{\cal M}_{11}^2 &  {\cal M}_{12}^2 \crr {\cal
M}_{12}^2 &  {\cal M}_{22}^2 } \right) ={\cal M}_0^2+\delta {\cal
M}^2\,,
\eeq
where the tree-level contribution ${\cal M}_0^2$ was
given in \eq{kv} and $\delta {\cal M}^2$ is the contribution from
the radiative corrections.  The dominant corrections to
$\mathcal{M}^2$, coming from the one-loop top and bottom quark and
top and bottom squark contributions plus the two-loop leading
logarithmic contributions, are given to $\mathcal{O}(h_t^4,
h_b^4)$ by~\cite{carena,hhh,CMW1}
\beqa
        \delta \mathcal{M}^2_{11} &\simeq&
        - \bar \mu^2 x_t^2 \frac{h_t^4 v^2}{32 \pi^2}
        s^2_{\beta}
        \left[ 1 + c_{11}
        \ln \left( \frac{M_S^2}{m_t^2} \right) \right]
        - \bar \mu^2 a_b^2 \frac{h_b^4 v^2}{32 \pi^2}
        s^2_{\beta}
        \left[ 1 + c_{12}
        \ln \left( \frac{M_S^2}{m_t^2} \right) \right], 
\label{eq:M11approx}\\[6pt]
        \delta \mathcal{M}^2_{12} &\simeq&
        - \bar \mu x_t \frac{h_t^4 v^2}{32 \pi^2} (6 - x_t a_t)
        s^2_{\beta}\left[ 1 + c_{31}
        \ln \left( \frac{M_S^2}{m_t^2} \right) \right]
        + \bar \mu^3 a_b \frac{h_b^4 v^2}{32 \pi^2}
        s^2_{\beta}
        \left[ 1 + c_{32}
        \ln \left( \frac{M_S^2}{m_t^2} \right) \right]\,,
\label{eq:M12approx}\\[6pt]
        \delta \mathcal{M}^2_{22} &\simeq&
        \frac{3 h_t^4 v^2}{8 \pi^2} s^2_{\beta}
        \ln \left( \frac{M_S^2}{m_t^2} \right)
        \left[ 1 + \half c_{21}
        \ln \left( \frac{M_S^2}{m_t^2} \right) \right] \nonumber \\[5pt]
        &&
        + \frac{h_t^4 v^2}{32 \pi^2} s^2_{\beta} x_t a_t (12 - x_t
        a_t)
        \left[ 1 + c_{21}
        \ln \left( \frac{M_S^2}{m_t^2} \right) \right]
        - \bar \mu^4 \frac{h_b^4 v^2}{32 \pi^2} s^2_{\beta}
        \left[ 1 + c_{22}
        \ln \left( \frac{M_S^2}{m_t^2} \right) \right]\,,
\label{eq:M22approx}
\eeqa
where $s_{\beta} \equiv \sin\beta$, $c_{\beta} \equiv
\cos\beta$, and the coefficients $c_{ij}$ are:
\begin{equation}
c_{ij}\equiv {t_{ij}h_t^2+b_{ij}h_b^2-32g_3^2\over 32\pi^2}\,,
\end{equation}
$(t_{11},t_{12},t_{21},t_{22},t_{31},t_{32})=(12,-4,6,-10,9,-7)$
and $(b_{11},b_{12},b_{21},b_{22},b_{31},b_{32})=(-4,12,2,18,-1,15)$.
Above, $h_t$ and $h_b$ are the top and bottom quark Yukawa couplings
[see \eqs{yuklag}{tyukmassrel}],
$g_3$ is the strong QCD coupling,
$v = 246$ GeV is the SM Higgs
vacuum expectation value, and $M_S^2 = \half(M^2_{\tilde t_1} +
M^2_{\tilde t_2})$ is the average squared top squark
mass.\footnote{\Eqs{eq:M11approx}{eq:M22approx} 
have been derived under the assumption that
$|M^2_{\tilde t_1} - M^2_{\tilde t_2}|/(M^2_{\tilde t_1} + M^2_{\tilde t_2})
\ll 1$.  The approximate forms of \eqs{eq:M11approx}{eq:M22approx} 
are sufficient
to provide insight on the dependence of the radiatively-corrected
Higgs masses and couplings on the MSSM parameters, although our
numerical work is based on more exact forms for these expressions.}
The $\delta\mathcal{M}_{ij}^2$ also depend on the MSSM parameters $A_t$,
$A_b$ and $\mu$ that enter the off-diagonal top-squark and
bottom-squark squared-mass matrices.
We employ the following notation:
$\bar \mu \equiv \mu / M_S$,
$a_t\equiv  A_t / M_S$, $a_b \equiv A_b / M_S$ and
$x_t \equiv X_t / M_S$, where $X_t \equiv A_t - \mu \cot\beta$.
Diagonalizing the CP-even Higgs squared-mass matrix
yields radiatively-corrected values for
$\mhl^2$, $\mhh^2$ and the mixing angle $\alpha$.\footnote{Although
${\cal M}_{12}^2$ is {\it negative}
at tree level (implying that $-\pi/2\leq\alpha\leq 0$), it is possible
that radiative corrections flip the sign of ${\cal M}_{12}^2$.
Thus, the range of the radiatively corrected angle
$\alpha$ can be taken to be $-\pi/2\leq\alpha\leq\pi/2$.}
The end result is a
prediction for the Higgs mass in terms of running parameters in the
$\overline{\rm MS}$ scheme.  It is
a simple matter to relate these parameters to the corresponding
on-shell parameters used in the diagrammatic
calculations~\cite{espizhang,chhhww}.

Additional non-logarithmic two-loop contributions, which can
generate a non-negligible shift in the Higgs mass (of a few GeV),
must also be included.\footnote{An improved procedure for
computing the radiatively-corrected neutral Higgs mass matrix and
the charged Higgs mass in a self-consistent way (including
possible CP-violating effects), which incorporates one-loop
supersymmetric threshold corrections to the Higgs--top-quark and
Higgs--bottom-quark Yukawa couplings, can be found in
\Ref{cpcarlos2}.} A compact analytical expression that
incorporates these effects at ${\cal O}(m_t^2 h_t^2 \alpha_s)$ was
given in \Ref{compact} (with further refinements provided by
\Ref{arbitrarystops} to take into account the possibility of
arbitrary top-squark splitting), 
and the corresponding corrections proportional
to $h_b^2\alpha_s$ can be found in \Ref{bdsz}.
An important source of such contributions are the one-loop
supersymmetric threshold corrections to the relation between the
Higgs--top-quark and Higgs--bottom-quark Yukawa couplings and the
corresponding quark masses [\eqns{byukmassrel}{tyukmassrel}].
These generate a non-logarithmic two-loop shift of the radiatively
corrected Higgs mass proportional to the corresponding squark
mixing parameters. One consequence of these contributions
\cite{chhhww} is the asymmetry in the predicted value of $\mhl$
under $X_t\to -X_t$ as noted in \fig{mhxt}(a). Recently, the
computation of $\mhl$ has been further refined by the inclusion of
genuine two-loop corrections of ${\cal O}(m_t^2 h_t^4)$~\cite{EZ2}, 
and estimates of the two-loop corrections proportional to 
$h_b^2 h_t^2$ and $h_b^4$~\cite{bdsz} 
(which can be numerically relevant for values of $\tanb\gsim m_t/m_b$). 
These non-logarithmic corrections, which depend on the third generation
squark mixing parameters, can slightly increase the value of the
radiatively-corrected Higgs mass. 

\begin{figure}[t!]
\begin{center}
\includegraphics[height=7cm]{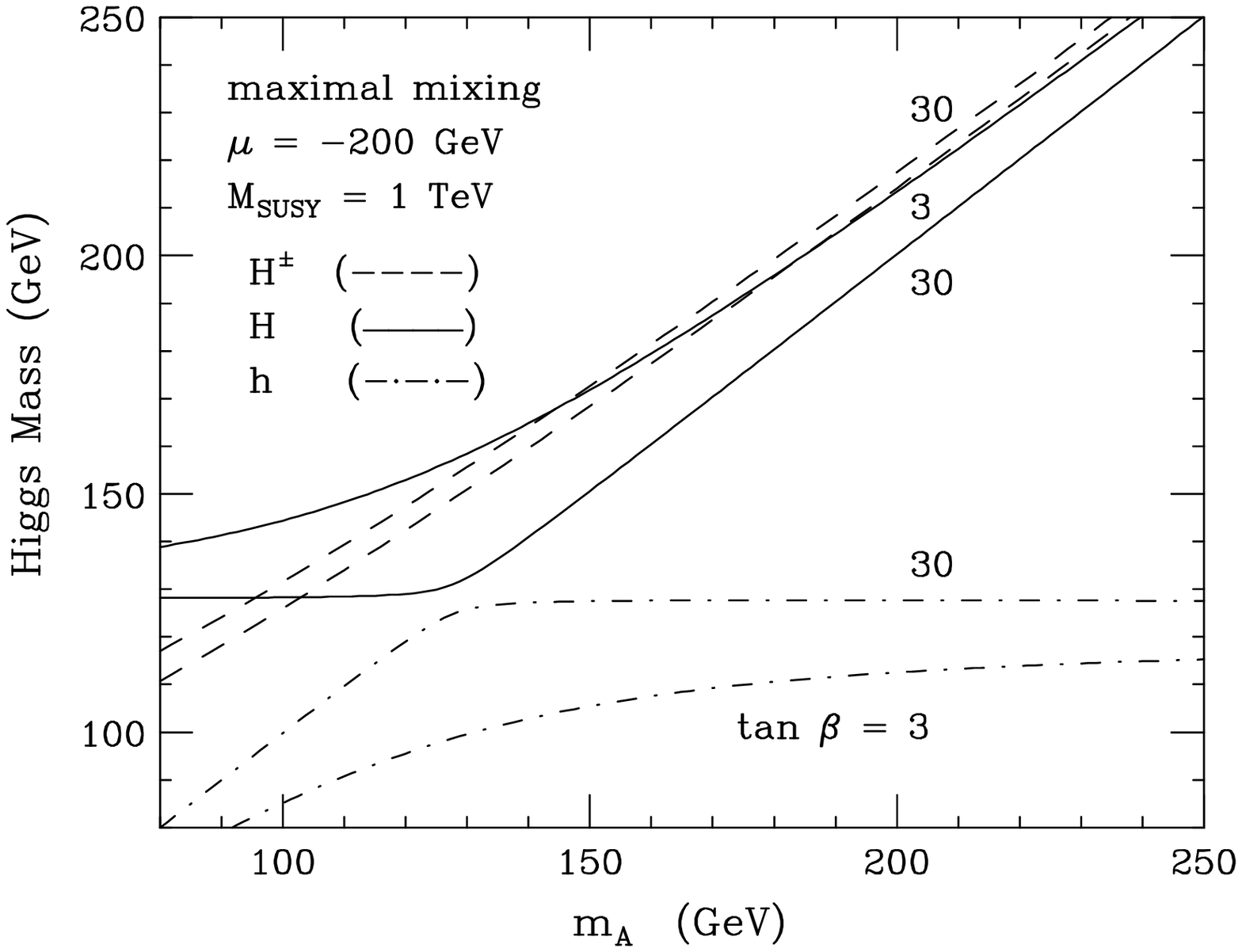}
\end{center}
  \capt{\label{massvsma} Lightest CP-even Higgs mass ($\mhl$),
heaviest CP-even Higgs mass ($\mhh$) and charged Higgs mass
($\mhpm$) as a function of $\mha$ for two choices of $\tan\beta=3$
and $\tan\beta=30$.  Here, we have taken $M_t=174.3$~GeV, and we
have assumed that the diagonal soft squark squared-masses are
degenerate: $\MSUSY\equiv M_Q=M_U=M_D=1$~TeV.  In addition, we
choose the other supersymmetric parameters corresponding to the
maximal mixing scenario. The slight increase in the charged Higgs
mass as $\tan\beta$ is increased from 3 to 30 is a consequence of
the radiative corrections.}
\end{figure}

The numerical results displayed in
figs.~\ref{cosgraph}--\ref{massvsma} are based on the calculations
of \Refs{carena}{hhh}, with improvements as described in
\Refs{weiglein}{chhhww}. The supersymmetric parameters in the
maximal and minimal mixing cases have been chosen according to the
first two benchmark scenarios of \Ref{benchmark}. Of particular
interest is the upper bound for the lightest CP-even Higgs mass
($\mhl$).  At fixed $\tan\beta$, the maximal value of $\mhl$ is
reached for $\mha\gg\mz$ (see \fig{massvsma}). Taking $\mha$
large, \fig{mhxt}(b) illustrates that the maximal value of the
lightest CP-even Higgs mass bound is realized at large $\tanb$ in
the case of maximal mixing. For each value of $\tan\beta$, we
denote the maximum value of $\mhl$ by $\mhmax(\tan\beta)$ [this
value also depends on the third-generation squark mixing
parameters]. Allowing for the uncertainty in the measured value of
$\mt$ and the uncertainty inherent in the theoretical analysis,
one finds for $\MSUSY\lsim 2$~TeV that $\mhl\leq
\mhmax\equiv\mhmax(\tan\beta\gg 1)$, where 
\beqa
\label{mhmaxvalue} \mhmax&\simeq & 122~{\rm GeV}, \quad
\mbox{if top-squark mixing is minimal,} \nonumber \\
\mhmax&\simeq & 135~{\rm GeV}, \quad
\mbox{if top-squark mixing is maximal.}
\eeqa
In practice, parameters leading to maximal
mixing are not expected in typical models of
supersymmetry breaking. Thus, in general, the  upper bound
on the lightest Higgs boson mass is expected to be
somewhere between the two extreme limits quoted above.
Cross-checks among various programs \cite{higgsprogs}
and rough estimates of higher order corrections not yet computed
suggest that the results for Higgs masses should be accurate to within
about 2 to 3 GeV over the parameter ranges displayed
in figs.~\ref{mhxt}--\ref{massvsma}.

In \fig{massvsma}, we exhibit the masses of
the CP-even neutral and the charged Higgs masses as a function of
$\mha$.  The squared-masses of the lighter and heavier neutral CP-even
Higgs are related by\footnote{At tree level, \eq{mhmaxtb} is a 
consequence of the sum rule: $\sum_n m_{\phi_n}^2 g_{\phi_n ZZ}^2
=\nicefrac{4}{3} m_Z^4 g_{GGGG}$, first derived in \Ref{weldon}, where the
$\phi_n$ are neutral Higgs bosons of a multi-Higgs-doublet
model and $g_{GGGG}$ is the quartic coupling of neutral Goldstone
bosons.  (In the MSSM, $g_{GGGG}=\nicefrac{3}{4}g^2\cos^2 2\beta/
\cos^2\theta_W$~\cite{hhg}.)
A general discussion of related tree-level Higgs mass sum rules and
bounds can be found in \Ref{comelli}.
In fact, \eq{mhmaxtb} is more general and applies to
the radiatively-corrected MSSM Higgs sector~\cite{bound,CMW2} 
in the approximation where the
renormalized $\alpha$ is determined as discussed in \Sec{sec:331}.}
\beq \label{mhmaxtb}
\mhh^2\cosbmaii+\mhl^2\sinbmaii=[\mhmax(\tan\beta)]^2\,.
\eeq
It is interesting to consider the behavior of the
CP-even Higgs masses in the large $\tan \beta$ regime.
For large values of $\tanb$ and for $\mha/\tan\beta \ll
\mhmax(\tan\beta)$, the off-diagonal elements of
the Higgs squared-mass matrix ${\cal M}^2$
become small compared to the diagonal elements
$|{\cal M}_{12}^2| \ll    {\cal M}_{11}^2 + {\cal M}_{22}^2 $;
${\cal M}_{12}^4 \ll    {\cal M}_{11}^2 {\cal M}_{22}^2 $.
Hence the two CP-even Higgs squared-masses
are approximately given by the diagonal elements of ${\cal M}^2$.
As above, we employ the notation where
$\mhmax$ refers to the asymptotic
value of $m_h$ at large $\tanb$ and $\mha$ (the actual numerical
value of $\mhmax$ depends primarily on the assumed values of
the third generation squark mass and mixing parameters).
If $\mha>\mhmax$, then $\mhl\simeq\mhmax$ and
$\mhh\simeq\mha$, whereas if $\mha<\mhmax$, then
$\mhl\simeq\mha$ and $\mhh\simeq\mhmax$.
This behavior can be seen in \fig{massvsma}.

\subsubsection{MSSM Higgs mass limits after LEP}
\label{sec:322}

\begin{figure}[t!]
\begin{center}
\resizebox{\textwidth}{!}{
\includegraphics[width=5cm]{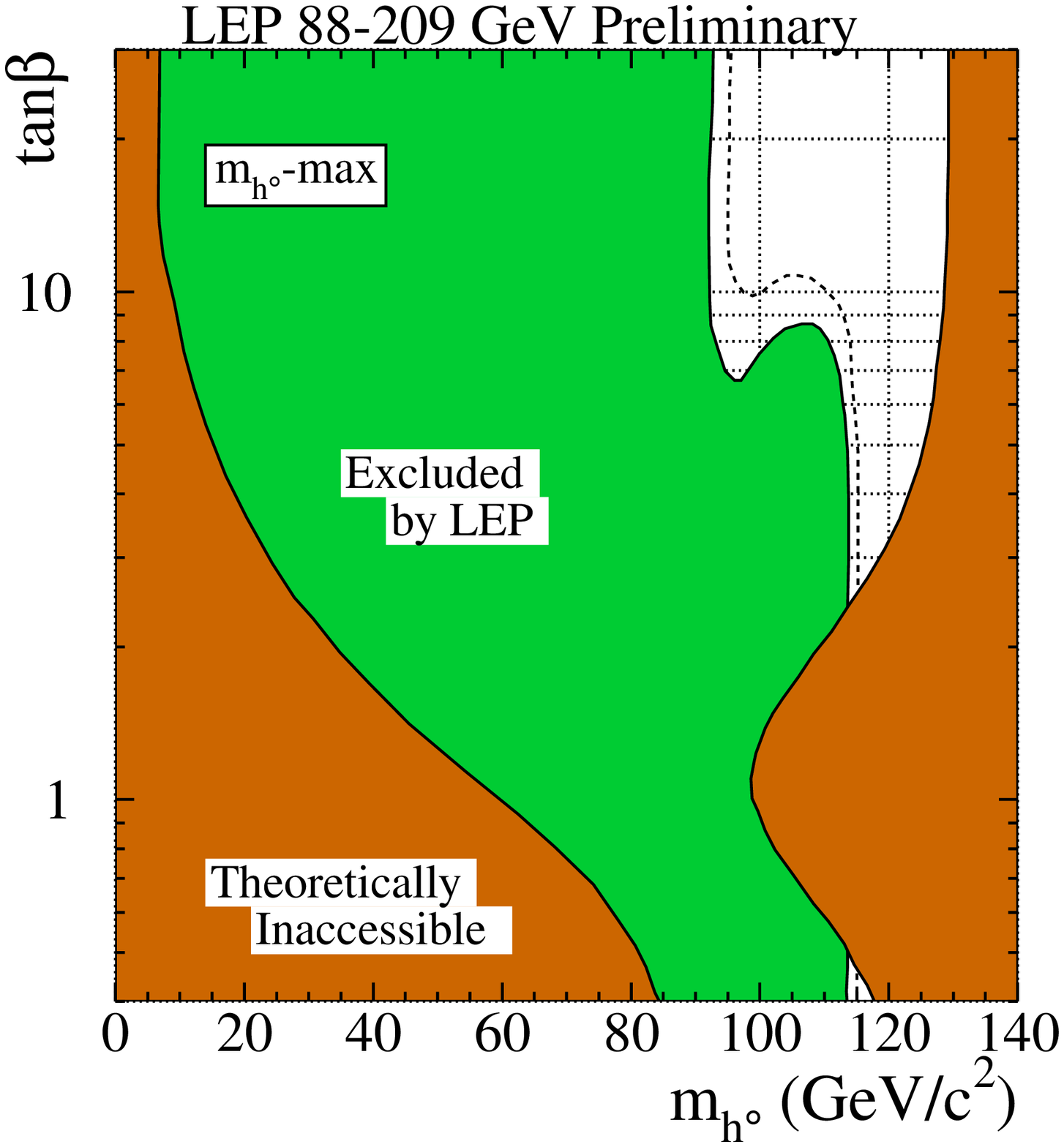}
\hfill
\includegraphics[width=5cm]{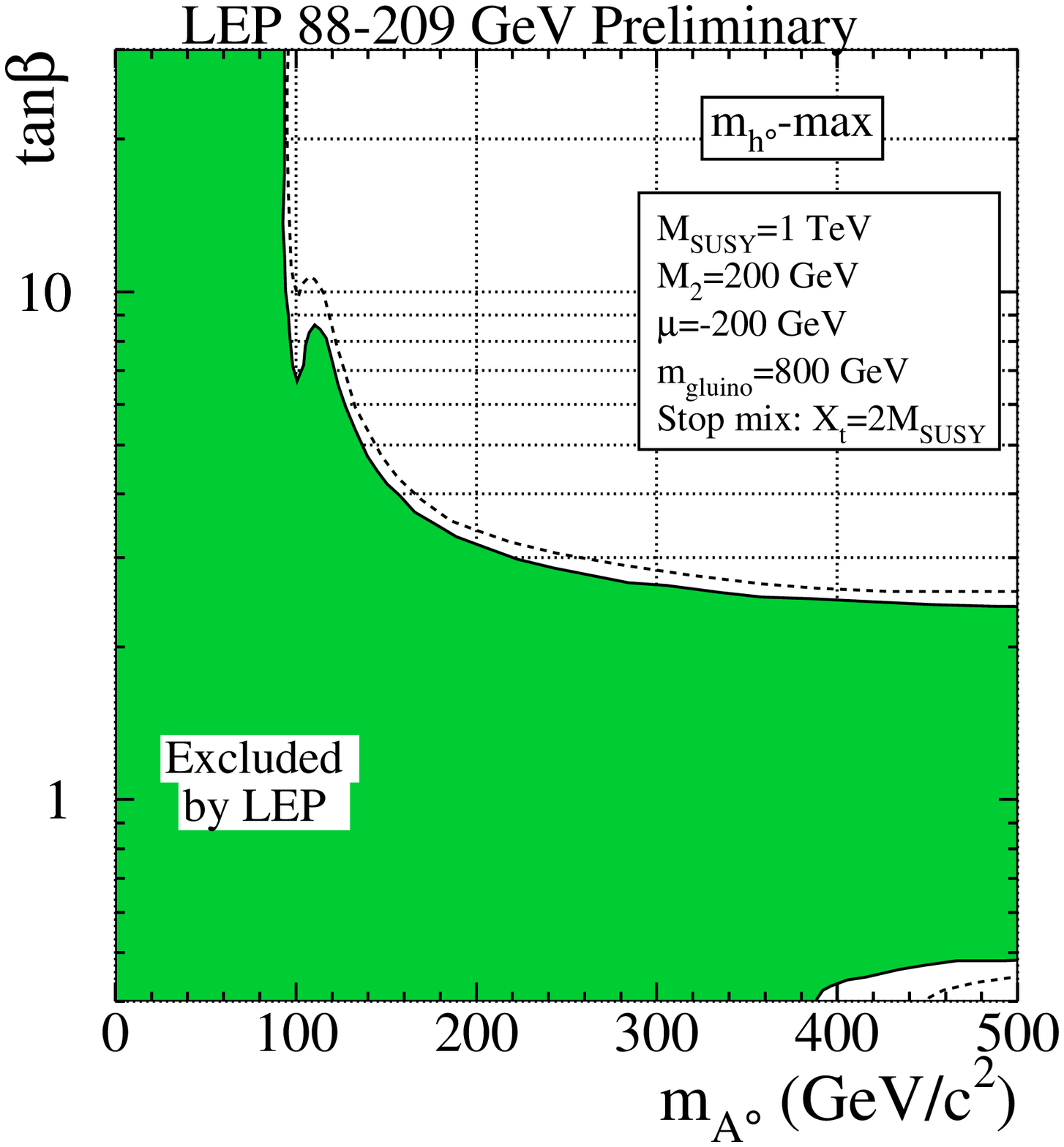}
}
\end{center}
\capt{\label{mh200} LEP2 contours of the 95\% CL exclusion
limits for MSSM Higgs sector parameters as a function of $\tan\beta$ and
(a) $\mhl$ and (b) $\mha$ (in GeV), taken from \protect\Ref{lepsusyhiggs}.
The contours shown have been obtained for MSSM Higgs parameters chosen
according to the maximal mixing benchmark of \protect\Ref{benchmark}.}
\end{figure}

No significant evidence for a Higgs signal has been detected at
LEP~\cite{lepsusyhiggs}.
As a result, one can obtain bounds on the possible MSSM Higgs
parameters.  These limits are often displayed in the $\mha$--$\tanb$
plane, although there is additional dependence on various MSSM
parameters that effect the radiative corrections to the Higgs masses
as discussed above.
In representative scans of the MSSM parameters, the LEP Higgs Working
Group~\cite{lepsusyhiggs}
finds that $\mhl>91.0$~GeV and $\mha>91.9$~GeV at 95\% CL.  These
limits actually correspond to the large $\tanb$ region in which $Z\hl$
production is suppressed, as shown in \fig{mh200}.
In this case, the quoted Higgs limits arise
as a result of the non-observation of $\hl\ha$ and $\hh\ha$ production.
As $\tanb$ is lowered, the limits on $\mhl$ and $\mha$ become more
stringent.  In this regime, the $\hl\ha$ production is suppressed
while the $Z\hl$ production rate approaches its SM value.  Thus, in this
case, the SM Higgs limit applies ($\mhl\gsim 114$~GeV) as shown in
\fig{mh200}(a).
The precise region of MSSM Higgs parameter space that is excluded
depends on the values of the
MSSM parameters that control the Higgs mass radiative corrections.
For example, a conservative exclusion
limit is obtained in the maximal mixing scenario, since
in this case the predicted value of $\mhl$ as a function of $\mha$ and
$\tanb$ is maximal (with respect to changes in the other MSSM parameters).
The excluded regions of the MSSM Higgs parameter space based on the
maximal mixing benchmark scenario of \Ref{benchmark}, are shown in
\fig{mh200}, and correspond to the exclusion of the range
$0.5<\tanb<2.4$ at the 95\%~CL.  However, the $\tan\beta$
exclusion region can still be significantly reduced (even to the point
of allowing all $\tan\beta$ values) by, {\it e.g.}, taking $\MSUSY=2$~TeV
and $m_t=180$~GeV (which still lies within the error bars of the 
experimentally measured value), and allowing for the theoretical
uncertainty in the prediction of $\mhmax$~\cite{Heinemeyer:1999zf}.

No evidence for the charged Higgs boson has yet been found.
The LEP Higgs Working Group quotes a limit of $\mhpm>78.6$~GeV at 
95\% CL~\cite{lepsusyhiggs2}, which holds for a more
general non-supersymmetric two-Higgs doublet model and assumes
only that the $H^+$ decays dominantly into $\tau^+\nu_\tau$ and/or
$c \bar s$.  Although the MSSM tree-level bound $\mhpm\geq\mw$ can be
relaxed somewhat by radiative corrections, the LEP bound quoted above
provides no useful additional constraints on the MSSM Higgs sector.

\subsubsection{Effect of explicit CP-violation on the
radiatively-corrected MSSM Higgs masses}
\label{sec:323}

In the Standard Model, CP-violation is due to the
existence of phases in the Yukawa couplings of the quarks to the
Higgs field, which results in one non-trivial phase in the CKM mixing
matrix. In the MSSM,
there are additional sources of CP-violation, due to phases in the
various supersymmetric mass parameters. In particular, the
gaugino mass parameters ($M_i$, $i=1,2,3$),
the Higgsino mass parameter, $\mu$, the
bilinear Higgs squared-mass parameter, $m_{12}^2$, and the trilinear couplings
of the squark and slepton fields ($\widetilde f$)
to the Higgs fields, $A_f$, may carry non-trivial
phases.  The existence of these CP phases can significantly affect the
MSSM Higgs sector through one-loop radiative
corrections~\cite{cprefs1,cprefs2,cpcarlos,cpcarlos2}.

Note that if one sets $\mu=M_i=A_f=m_{12}^2$, then the MSSM Lagrangian
possesses two independent global U(1) symmetries---a Peccei-Quinn (PQ)
symmetry and an R symmetry.  (The quantum numbers of the MSSM
fields with respect to U(1)$_{\rm PQ}$ and U(1)$_{\rm R}$
can be found in \Refs{thomas}{cpcarlos}.)
Consequently, in the MSSM with nonzero values for the above
parameters, there are two independent phase redefinitions of the
fields that can be used to remove two phases
from $\mu$, $M_i$, $A_f$ and $m_{12}^2$.  However, certain
combinations of these parameters remain invariant under such phase
redefinitions.  The simplest way to determine these combinations is to
treat the aforementioned parameters as spurions with quantum
numbers under the U(1)$_{\rm PQ}$ and U(1)$_{\rm R}$ symmetries
chosen such that the
full MSSM Lagrangian is invariant.  One can then easily check that the
phases of the parameter combinations,
${\rm arg}[\mu A_f (m_{12}^2)^*]$ and
${\rm arg}[\mu M_i (m_{12}^2)^*]$,
are indeed invariant under the U(1)$_{\rm PQ}$ and U(1)$_{\rm R}$
phase redefinitions of the MSSM fields~\cite{thomas,cpcarlos}.  
Therefore, if one
of these two quantities is different from zero (modulo $\pi$), one
should expect new CP-violating effects induced by
the production or exchange of supersymmetric particles.

We have already noted that the tree-level Higgs sector is
CP-conserving.  This is a consequence of the fact that
$m_{12}^2$ is
the only possible complex parameter that appears in the tree-level
Higgs potential.  Thus the phase of $m_{12}^2$  can be rotated away
by redefining the phases of the complex Higgs doublets appearing
in the Lagrangian. The same field redefinition implies
that one can choose
the vacuum expectation values of the two Higgs fields to be real and
positive.  However, at the
one loop-level, the Higgs potential acquires a dependence on
the parameters $\mu A_t$ and $\mu M_i$ through loops of third
generation squarks and weak gauginos, respectively, which induce non-trivial
CP-violating effects.
The most important of these
CP-violating effects is the generation of mixing between
the neutral CP-odd and CP-even Higgs boson states. Therefore, the
physical neutral Higgs bosons are no longer CP-eigenstates and the CP-odd
Higgs boson mass $m_A$ is no longer a physical parameter. The
charged Higgs mass is still physical and can be used as an input
for the computation of the neutral Higgs spectrum of the
theory~\cite{cpcarlos2}.
The Higgs mass spectrum can therefore be quite different from the CP-conserving
case.  For example, a large splitting between the masses of the
next-to-lightest and the heaviest neutral Higgs bosons is possible if
the charged Higgs boson is not too heavy.

For large values of the charged Higgs mass, the decoupling
limit applies, and the properties of the lightest neutral Higgs boson state
approach those of the SM Higgs boson.  That is, for $\mhpm\gg\mw$, the
lightest neutral Higgs boson is approximately a CP-even state, with
CP-violating couplings that are suppressed by terms of
$\mathcal{O}(\mw^2/\mhpm^2)$~\cite{ghk}.  In particular, the upper-bound
on the lightest neutral Higgs boson mass, which is reached in
the decoupling limit, takes the same value as in the CP-conserving
case~\cite{cpcarlos}.  Nevertheless, there still can be significant
mixing between the two heavier neutral mass eigenstates.
Quantitatively, the leading contribution to the
squared-mass terms that mix CP-even and CP-odd eigenstates,
$M^2_{SP}$ (in a convention where $m_{12}^2$ is real) is of order
\begin{equation}
M^2_{SP} \simeq \frac{3g^2 m_t^4|\mu A_t|}{64 \pi^2 m_W^2 M_S^2}
\sin\left({\rm arg}[\mu A_t]\right)\,.
\end{equation}
Under the reasonable assumption that $|\mu A_t| < 10 M_S^2$, it
is clear that the mixing effects between the lightest neutral Higgs boson
and the heavier Higgs states are small if the masses of the heavy
Higgs bosons are larger than $2m_t$.  In this limit, the two
heavier states are highly degenerate in mass, and the CP-violating
effects may still lead to non-trivial mixing of the two heavier
CP-eigenstates. For a detailed study of the Higgs mass spectrum and
parametric dependence of the Higgs mass radiative corrections, see
\Ref{cpcarlos2}.

\subsection{Radiatively-Corrected MSSM Higgs couplings}
\label{sec:33}

\subsubsection{Renormalization of \boldmath{$\cos(\beta-\alpha)$}}
\label{sec:331}

Radiative corrections also significantly modify the tree-level values
of the Higgs boson couplings to fermion pairs and to vector boson pairs.
As discussed in \Sec{sec:31}, 
the tree-level Higgs couplings depend crucially on
the value of $\cosbma$.  In first approximation,
when radiative corrections of the Higgs
squared-mass matrix are computed,
the diagonalizing angle $\alpha$ is shifted from its tree-level value.
Thus, one may compute a ``radiatively-corrected'' value for
$\cosbma$.  This provides one important source of the radiative
corrections of the Higgs couplings.   In \fig{cosgraph}, we show the
effect of radiative corrections on the
value of $\cosbma$ as a function of $\mha$  for different values of the
squark mixing parameters and $\tanb$.  One can then
simply insert the radiatively corrected value of $\alpha$ into
eqs.~(\ref{vvcoup}), (\ref{hvcoup}) and
(\ref{qqcouplingsa})--(\ref{qqcouplingsf})
to obtain radiatively-improved couplings of
Higgs bosons to vector bosons and to fermions.

The mixing angle $\alpha$ which diagonalizes the mass matrix in
\eq{calmatrix} can be expressed as:
\begin{equation}
        s_{\alpha} c_{\alpha} = \frac{\mathcal{M}^2_{12}}
        {\sqrt{({\rm Tr}\mathcal{M}^2)^2 - 4 \, {\rm det} \mathcal{M}^2}}\,,
\ \ \ \
        c^2_{\alpha} - s^2_{\alpha} =
        \frac{\mathcal{M}^2_{11} - \mathcal{M}^2_{22}}
        {\sqrt{({\rm Tr}\mathcal{M}^2)^2 - 4 \, {\rm det} \mathcal{M}^2}}\,,
        \label{eq:alpha}
\end{equation}
where $s_{\alpha} \equiv \sin\alpha$ and $c_{\alpha} \equiv \cos\alpha$.
Note that if $\mathcal{M}^2_{12} \to 0$, then either $\sin\alpha \to 0$
(if $\mathcal{M}_{11}^2 > \mathcal{M}_{22}^2$) or
$\cos\alpha \to 0$ (if $\mathcal{M}_{11}^2 < \mathcal{M}_{22}^2$).
At tree level, $\mathcal{M}^2_{12}$ is small for small
$\mha$ and/or large $\tan\beta$, but it cannot vanish.
However, radiative corrections to
$\mathcal{M}_{12}^2\equiv -(m_A^2+m_Z^2)s_\beta c_\beta+\delta
\mathcal{M}^2_{12}$ can be of the
same order as its tree level value
for small values of $\mha$ and large $\tan\beta$.  Hence,
it is possible for the one-loop contribution to
approximately cancel the tree-level result (with
two-loop corrections to $\mathcal{M}^2_{12}$
small compared to the corresponding one-loop result).
For moderate or large values of
$\tan\beta$, the vanishing of ${\cal M}_{12}^2$ [see
\eq{eq:M12approx}] leads
to the approximate numerical relation~\cite{CMW1}:
\begin{equation}
\left[{\mha^2\over \mz^2}
+ 1 \right] \simeq
{\mu x_t \tanb \over 100M_S} \left(
{2 a_t x_t} -
11 \right)
\left[ 1 -\frac{15}{16 \pi^2}  \ln\left({M_S^2\over\mt^2}\right) \right],
\label{suppression}
\end{equation}
where $h_t$, $\alpha_s$ and the weak gauge couplings have been replaced
by their approximate numerical values at the
the electroweak scale.
For low values of $\mha$ or large values of
the squark mixing parameters, a cancellation can easily
take place.

If ${\cal M}_{12}^2\simeq 0$ and $\tanb$ is large (values of
$\tan\beta \gsim 5$ are sufficient), the resulting pattern of Higgs
couplings is easy to understand.  In this limit,
${\cal M}^2_{11}\simeq\mha^2$ and ${\cal M}^2_{22}\simeq \mhmax$,
as noted at the end of \Sec{sec:321}.  Two cases must be treated
separately depending on the value of $\mha$.
First, if $\mha < \mhmax$, then
$\sin\alpha \simeq - 1$,
$\cos\alpha \simeq 0$ and $\sinb\simeq -\cos(\beta - \alpha) \simeq 1$.
In this case, the lighter CP-even Higgs boson
$\hl$ is roughly aligned along the $\Phi_d^0$ direction and
the heavier CP-even Higgs boson
$\hh$ is roughly aligned along the $\Phi_u^0$ direction [see
\eq{scalareigenstates}].  In particular, the coupling of $\hh$ to
$b\bar b$ and $\tau^+\tau^-$ is significantly diminished (since
down-type fermions couple to $\Phi_d^0$), while the $\hh VV$ couplings
[\eq{vvcoup}] are approximately equal to those of the Standard Model
[since $\cosbmaii \simeq 1$].
Consequently, the branching ratios of $\hh$ into
$gg$, $\gamma\gamma$, $c\bar c$, and $W^{+}W^{-}$ can be
greatly enhanced over \SM\ expectations~\cite{CMW1,CMW2,Wells,bdhty}.
Second, if $\mha \gg \mhmax$ then $\sin\alpha \simeq 0$ and
$\sinb\simeq\cosa\simeq \sin(\beta - \alpha) \simeq 1$ and
the previous considerations for $\hh$ apply now to $\hl$.

Although it is difficult to have an exact cancellation of the
off-diagonal element ${\cal M}_{12}^2$, in many regions of the MSSM parameter
space, a significant suppression of ${\cal M}_{12}^2$ may be present.
Generically, the leading radiative corrections
to $\mathcal{M}_{12}^2$ depend strongly on the sign of the product
$\mu X_t$ ($A_t \simeq X_t$ for large $\tan\beta$
and moderate $\mu$) and on the value of $|A_t|$.
For the same value of $X_t$, a change in the
sign of $\mu$ can lead to observable variations in the branching
ratio for the Higgs boson decay into bottom quarks. If $a_t^2\lsim 11/2$,
then the absolute value of $\mathcal{M}_{12}^2$ tends to be suppressed
[enhanced] for values of $\mu A_t < 0$  [$\mu A_t > 0$], which implies
a similar suppression [enhancement] for the coupling of bottom
quarks and $\tau$-leptons to the SM-like Higgs boson.
For larger values of $|a_t|$, the suppression [enhancement]
occurs for the opposite sign of $\mu A_t$.

\subsubsection{The decoupling limit revisited}
\label{sec:332}

Radiative corrections can also significantly affect the onset
of the decoupling limit.  Recall that at tree level [see \eq{largema4}],
$|\cos(\beta-\alpha)|\ll 1$ for $\mha\gg\mz$, in which case
the couplings of $\hl$ are
nearly identical to those of the SM Higgs boson.
Including the effects of $\delta\mathcal{M}^2$,
we use \eq{eq:alpha} to obtain
\beqa \label{eq:cosbma}
\cos(\beta-\alpha)&=&{(\mathcal{M}_{11}^2-\mathcal{M}_{22}^2)\sin 2\beta
-2\mathcal{M}_{12}^2\cos 2\beta\over
2(m_H^2-m_h^2)\sin(\beta-\alpha)} \nonumber \\[5pt]
&=&{m_Z^2\sin 4\beta+
({\delta\mathcal{M}}_{11}^2-{\delta\mathcal{M}}_{22}^2)\sin 2\beta
-2{\delta\mathcal{M}}_{12}^2\cos 2\beta\over
2(m_H^2-m_h^2)\sin(\beta-\alpha)}\,.
\eeqa
Since $\delta\mathcal{M}^2_{ij}\sim {\mathcal O}(m_Z^2)$,
and $m_H^2-m_h^2=m_A^2+\mathcal{O}(m_Z^2)$, one finds
\begin{equation} \label{cosbmadecoupling}
        \cos(\beta-\alpha)=c\left[{m_Z^2\sin 4\beta\over
        2m_A^2}+\mathcal{O}\left(m_Z^4\over m_A^4\right)\right]\,,
\end{equation}
in the limit of $\mha\gg\mz$, where
\begin{equation} \label{cdef}
        c\equiv 1+{{\delta\mathcal{M}}_{11}^2-{\delta\mathcal{M}}_{22}^2\over
        2m_Z^2\cos 2\beta}-{{\delta\mathcal{M}}_{12}^2\over m_Z^2\sin
        2\beta}\,.
\end{equation}
\Eq{cosbmadecoupling} exhibits the expected decoupling behavior
for $m_A\gg m_Z$.
However, \eq{eq:cosbma} illustrates another way in which
$\cos(\beta-\alpha)=0$ can be achieved---simply choose the
MSSM parameters (that govern the Higgs mass radiative
corrections) such that the numerator of \eq{eq:cosbma} vanishes.
That is,
\begin{equation}
        2 m_Z^2 \sin 2\beta =
        2\, \delta \mathcal{M}^2_{12}
        - \tan 2\beta
        \left(\delta \mathcal{M}^2_{11} - \delta \mathcal{M}^2_{22} \right)\,.
        \label{eq:tanbetadecoup}
\end{equation}
Note that \eq{eq:tanbetadecoup} is independent of the value of $m_A$.
For a typical choice of MSSM parameters,
\eq{eq:tanbetadecoup} yields a solution at large $\tan\beta$.  That
is, by
approximating $\tan 2\beta\simeq -\sin 2\beta \simeq -2/ \tan \beta$,
one can determine
the value of $\beta$ at which the decoupling occurs:
\begin{equation} \label{earlydecoupling}
\tan \beta\simeq \frac{2m_Z^2-
\delta\mathcal{M}_{11}^2+\delta \mathcal{M}_{22}^2}
{ \delta\mathcal{M}_{12}^2}\,.
\end{equation}
The explicit expressions for $\delta\mathcal{M}_{ij}^2$
quoted in \eq{eq:M12approx} confirm that the
assumption of $\tan\beta\gg 1$ used to derive this result is a consistent
approximation because $\delta \mathcal{M}^2_{12}$ is typically small.
We conclude that for the value of $\tan\beta$ specified
in \eq{earlydecoupling}, $\cos(\beta-\alpha)=0$ independently of
the value of $m_A$.  We shall refer to this phenomenon as
$\mha$-independent decoupling.
From \eq{eq:M12approx}, it follows that explicit solutions to
\eq{eq:tanbetadecoup} depend on ratios of MSSM parameters and
are thus insensitive to the overall supersymmetric mass scale,
modulo a mild logarithmic dependence on $M_S/m_t$.

\subsubsection{Corrections to tree-level Higgs-fermion Yukawa couplings}
\label{sec:333}

We have seen in \Sec{sec:331}
that Higgs couplings are modified at one loop due to
the renormalization of the CP-even Higgs mixing angle
$\alpha$.  Additional contributions from the
one-loop vertex corrections to tree-level Higgs couplings must also be
considered~\cite{hffsusy1l,hffsusyqcd,hffsusyprop,loganetal}.
These corrections are typically small and therefore
do not alter significantly the pattern of Higgs couplings.  However,
at large $\tan\beta$, the corrections to Higgs-fermion Yukawa
couplings can be enhanced, and thus require a careful analysis.

In the supersymmetric limit, bottom quarks only couple to $\Phi_d^0$
and top quarks only couple to $\Phi_u^0$.
However, supersymmetry is broken and a small coupling of
the bottom quark [top quark]
to $\Phi_u^0$ [$\Phi_d^0$] will be generated from the one-loop Yukawa
vertex corrections.
These results can be summarized by an effective
Lagrangian that describes the coupling of
the Higgs bosons to the third generation
quarks:\footnote{Due to weak isospin breaking, one should
allow for different radiatively induced couplings to charged and
neutral Higgs bosons.  For example, one should write
$\Delta h_b \bar b_R b_L \Phi^{0\ast}_u+\Delta 
\overline h_b \bar b_Rt_L \Phi^-_u$
in place of $\Delta h_b \bar b_R Q_L^k$%
{\boldmath{$\Phi_u$}}$^{\!\!\!\! k \ast}$, {\it etc.}
To the extent that weak isospin breaking effects are small in the loop
diagrams that generate $\Delta h_b$ and $\Delta \overline h_b$,
it follows that $\Delta h_b\approx\Delta \overline h_b$
(and similarly for the other radiatively generated coefficients), and
we may use \eq{yuklag} as written.\label{footispin}}
\begin{equation} \label{yuklag}
        -\mathcal{L}_{\rm eff} = \epsilon_{ij} \left[
        (h_b + \delta h_b) \bar b_R 
        {\mathbold{\Phi_d}}^{\!\!\!\! i}\, Q_L^j
        + (h_t + \delta h_t) \bar t_R Q_L^i
        {\mathbold{\Phi_u}}^{\!\!\!\! j} \right]
        + \Delta h_t \bar t_R Q_L^k {\mathbold{\Phi_d}^{\!\!\!\! k \ast}}
        + \Delta h_b \bar b_R Q_L^k {\mathbold{\Phi_u}^{\!\!\!\! k \ast}}
        + {\rm h.c.}\,,
\end{equation}
implying a modification of the tree-level relations between
$h_t$, $h_b$ and $m_t$, $m_b$ as
follows~\cite{deltamb,deltamb1,hffsusyqcd,deltamb2}:
\beqa
        m_b &=& \frac{h_b v}{\sqrt{2}} \cos\beta
        \left(1 + \frac{\delta h_b}{h_b}
        + \frac{\Delta h_b \tan\beta}{h_b} \right)
        \equiv\frac{h_b v}{\sqrt{2}} \cos\beta
        (1 + \Delta_b)\,, \label{byukmassrel} \\[5pt]
        m_t &=& \frac{h_t v}{\sqrt{2}} \sin\beta
        \left(1 + \frac{\delta h_t}{h_t} + \frac{\Delta
        h_t\cot\beta}{h_t} \right)
        \equiv\frac{h_t v}{\sqrt{2}} \sin\beta
        (1 + \Delta_t)\,. \label{tyukmassrel}
\eeqa
The dominant contributions to $\Delta_b$ are $\tan\beta$-enhanced,
with $\Delta_b\simeq (\Delta h_b/h_b)\tan\beta$; for
$\tan\beta\gg 1$, $\delta h_b/h_b$ provides a small correction to
$\Delta_b$.   In the same limit, $\Delta_t\simeq\delta h_t/h_t$, with
the additional contribution of $(\Delta h_t/h_t)\cot\beta$ providing a
small correction.\footnote{Because the one-loop
corrections $\delta h_b$, $\Delta h_b$, $\delta h_t$ and $\Delta h_t$
depend only on Yukawa and gauge couplings and the supersymmetric
particle masses, they
contain no hidden $\tan\beta$ enhancements \cite{Carena:2001uj}.}
Explicitly, one finds that for $\MSUSY\gg\mz$ (where $\MSUSY$
represents a typical supersymmetric mass that appears in the loops) and for
$\tan\beta \gg 1$~\cite{deltamb,deltamb1,deltamb2},
\beqa
        \Delta_b
        &\simeq&\left[
        \frac{2 \alpha_s}{3 \pi} \mu M_{\tilde g} \,
        I(M^2_{\tilde b_1}, M^2_{\tilde b_2}, M^2_{\tilde g})
        + \frac{h_t^2}{16 \pi^2} \mu A_t \,
        I(M^2_{\tilde t_1}, M^2_{\tilde t_2}, \mu^2)\right]\tan\beta\,,
        \label{eq:Deltab}\\[5pt]
        \Delta_t &\simeq&
        -\frac{2 \alpha_s}{3 \pi} A_t M_{\tilde g} I(M^2_{\tilde t_1},
        M^2_{\tilde t_2}, M^2_{\tilde g})
        - \frac{h_b^2}{16 \pi^2} \mu^2 I(M^2_{\tilde b_1},
        M^2_{\tilde b_2}, \mu^2)\,, \label{eq:Deltat}
\eeqa
where $\alpha_s\equiv g_3^2/4\pi$,
$M_{\tilde g}$ is the gluino mass,
$M_{\tilde b_{1,2}}$ are the
bottom squark masses, and smaller electroweak
corrections have been ignored.
The loop integral $I(a^2,b^2,c^2)$ is given by
\beq
I(a,b,c) = {a^2b^2\ln(a^2/b^2)+b^2c^2\ln(b^2/c^2)+c^2a^2\ln(c^2/a^2) \over
(a^2-b^2)(b^2-c^2)(a^2-c^2)}\,,
\eeq
and is of order $1/{\rm
max}(a^2,b^2,c^2)$ when at least one of its arguments is large
compared to $m_Z^2$.
Note that the Higgs coupling proportional to $\Delta h_b$ is a
manifestation of the broken supersymmetry in the low energy theory;
hence, $\Delta_b$ does not decouple
in the limit of large values of the supersymmetry breaking masses. Indeed,
if all supersymmetry breaking mass parameters (and $\mu$)
are scaled by a common factor, the correction
$\Delta_b$ remains constant.

Similarly to the case of the bottom quark, the relation between $m_\tau$ and
the Higgs--tau-lepton Yukawa coupling $h_\tau$ is modified:
\beq \label{yuktaumass}
m_\tau = {h_\tau v_d\over\sqrt{2}} (1+\Delta_\tau).
\eeq
The correction $\Delta_\tau$ contains a contribution from a
tau slepton--neutralino loop (depending on the two tau-slepton masses
$M_{\tilde \tau_1}$ and $M_{\tilde \tau_2}$ and the
mass parameter of the $\widetilde B$ component
of the neutralino, $M_{1}$) and a
tau sneutrino--chargino loop (depending on the tau sneutrino mass
$M_{\tilde \nu_\tau}$, the mass parameter of the $\widetilde W^\pm$
component of the chargino, $M_{2}$, and $\mu$).
It is given by \cite{deltamb1,deltamb2}:
\beq
\Delta_\tau = \left[{\alpha_1 \over 4\pi} M_1\mu\,
I(M_{\tilde\tau_1},
M_{\tilde\tau_2},M_1) - {\alpha_2 \over 4\pi} M_2\mu\,
I(M_{\tilde\nu_\tau},M_2,\mu)\right]\tan\beta\,,
\eeq
where $\alpha_2\equiv g^2/4\pi$ and $\alpha_1\equiv g^{\prime\,2}/4\pi$
are the electroweak gauge couplings.
Since corrections to $h_\tau$ are proportional to $\alpha_1$ and
$\alpha_2$, they are  expected to be smaller than
the corrections to $h_b$.

From \eq{yuklag} we can obtain the couplings of the physical Higgs
bosons to third generation fermions.  The resulting interaction
Lagrangian is of the form:
\beq \label{linthff}
{\cal L}_{\rm int} =  -\sum_{q=t,b,\tau} \left[g_{\hl q\bar q}\hl q \bar{q} +
 g_{\hh q\bar q}\hh q \bar{q}-
 i g_{\ha q\bar q} A \bar{q} \gamma_5 q\right] + \left[\bar b g_{H^- t\bar b}
t H^- + {\rm h.c.}\right]\,.
\eeq
Using \eqns{byukmassrel}{tyukmassrel}, one obtains:
\beqa
 g_{\hl b\bar b} &= & -{m_b\over v}{\sin\alpha \over \cos\beta}
\left[1+{1\over 1+\Delta_b}\left({\delta h_b\over h_b}-
\Delta_b\right)\left( 1 +\cot\alpha \cot\beta \right)\right]\,,
\label{hlbb}  \\[5pt]
 g_{\hh b\bar b} &= & {m_b\over v}{\cos\alpha \over \cos\beta}
\left[1+{1\over 1+\Delta_b}\left({\delta h_b\over h_b}-
\Delta_b\right)\left( 1 -\tan\alpha \cot\beta \right)\right]\,,
\label{hhbb} \\[5pt]
 g_{\ha b\bar b} &= & {m_b\over v}\tan\beta
\left[1+{1\over (1+\Delta_b)\sin^2\beta}\left({\delta h_b\over h_b}-
\Delta_b\right)\right]\,,
\label{habb} \\[5pt]
 g_{\hl t\bar t} & = & {m_t\over v}{\cos\alpha \over \sin\beta}
\left[1-{1\over 1+\Delta_t}{\Delta h_t\over h_t}
(\cot\beta+ \tan\alpha)\right]\,,
\label{hltt} \\[5pt]
 g_{\hh t\bar t} &= & {m_t\over v}{\sin\alpha \over \sin\beta}
\left[1-{1\over 1+\Delta_t}{\Delta h_t\over h_t}
(\cot\beta-\cot\alpha)\right]\,,
\label{hhtt} \\[5pt]
 g_{\ha t\bar t} &= & {m_t\over v}\cot\beta
\left[1-{1\over 1+\Delta_t}{\Delta h_t\over h_t}(\cot\beta+
\tan\beta)\right]\,,
\label{hatt}
\eeqa
and the $\tau$ couplings are obtained from the above equations
by replacing $m_b$, $\Delta_b$ and $\delta h_b$
with $m_{\tau}$, $\Delta_{\tau}$ and $\delta h_\tau$, respectively.
In addition, one must employ the renormalized value of $\alpha$ 
in the above formulae to incorporate the
radiative corrections discussed in section 3.3.1.
In writing out the Higgs-top quark couplings above, we found it
convenient to express the results in
terms of $\Delta_t$ and $\Delta h_t/h_t$, since $\Delta_t\simeq \delta
h_t/h_t$ and the corresponding contribution of $\Delta h_t/h_t$ is
$\tan\beta$ suppressed [\eq{tyukmassrel}].  Alternatively, 
\eqs{hlbb}{hatt} can be rewritten in a more symmetrical form by
using \eqns{byukmassrel}{tyukmassrel} 
to eliminate $\Delta_b$ and $\Delta_t$ from the numerators of the
corresponding expressions.

At large $\tan\beta$, terms involving
$\Delta_b\propto\tan\beta$ [\eq{eq:Deltab}]
provide the dominant corrections to the neutral Higgs couplings to
$b\bar b$.  The corrections proportional to $\delta h_b/h_b$ [see
\eqns{yuklag}{byukmassrel} and the discussion that follows]
are never $\tan\beta$-enhanced and are therefore numerically unimportant.
The sign of  $\Delta_b$
is governed by the sign of $M_{\tilde{g}}  \mu$,
since the  bottom-squark gluino loop  gives the dominant contribution
to \eq{eq:Deltab}. Thus, in a convention where $M_{\tilde{g}}>0$,
the radiatively corrected coupling  $g_{\ha b\bar b}$
is suppressed (enhanced) with respect to
its  tree level value for $\mu >0$ ($\mu<0$).
In contrast, the radiative corrections to
$g_{\hl b\bar b}$ and $g_{\hh b\bar b}$
have a more complicated
dependence on the supersymmetric parameters due to the dependence on
the CP-even mixing angle $\alpha$.
Since $\alpha$ and  $\Delta_b$  are governed by different
combinations of the supersymmetry breaking parameters, it is difficult
to exhibit in a simple way the behavior
of the radiatively corrected couplings of the CP-even Higgs bosons
to the bottom quarks as a function of the MSSM parameters.

One can check [using \eq{cotalf}]
that in the decoupling limit, $g_{\hl q\bar q}=
g_{\hsm q\bar q}=m_q/v$.
Away from the decoupling limit,
the Higgs couplings to bottom-type fermions can deviate significantly
from their tree-level values due to enhanced radiative corrections at
large $\tan\beta$ [where $\Delta_b\simeq\mathcal{O}(1)$].
In particular, because $\Delta_b\propto\tan\beta$, the leading
one-loop radiative correction to $g_{\hl b\bar b}$ is of
$\mathcal{O}(\mz^2\tanb/\mha^2)$, which formally decouples only when
$\mha^2\gg\mz^2\tanb$.  This behavior is called {\it delayed
decoupling} in \Ref{loganetal}.  In
addition, there are regions of MSSM parameter space in which there is
a strong suppression of the Higgs coupling to $b\bar b$ (or
$\tau^+\tau^-$) as compared to its tree-level value.  As a result,
there can be significant corrections to the tree-level relation
$g_{\hl b\bar b}/g_{\hl\tau^+\tau^-}=m_b/m_\tau$~\cite{Wells}.
In particular, in some parameter regimes,
the $\tau^+\tau^-$ decay mode can be the dominant $\hl$
decay channel, a result that would be fatal to certain Higgs search
strategies which assume that $\hl\to b\bar b$ is the dominant decay mode.

Assuming that weak isospin breaking effects in the loop corrections to
the charged Higgs fermion Yukawa couplings are small 
(see footnote\fnref{footispin}),
then $g_{H^-t\bar b}$ [defined in \eq{linthff}] is given by
\beqa
g_{H^-t\bar b} &\simeq & {\sqrt{2}\over v}\left\{m_t\cot\beta\left[1-
{1\over 1+\Delta_t}{\Delta h_t\over
h_t}(\cot\beta+\tan\beta)\right]P_R\right.\nonumber \\
&&\qquad\left.+m_b\tan\beta\left[1+
{1\over (1+\Delta_b)\sin^2\beta}\left({\delta h_b\over h_b}
-\Delta_b\right)\right]P_L\right\}\,,
\label{hmtb}
\eeqa
with a similar form for $g_{H^-\nu_\tau\tau^+}$ with the
replacements noted below \eq{hatt}.

\subsubsection{Effects of explicit CP-violation}
\label{sec:334}

In \Sec{sec:323}, we noted the possibility of mixing between the CP-even
and CP-odd eigenstates due to CP-violating effects that enter via the
one-loop radiative corrections.  In this case, the neutral scalar mass
eigenstates, denoted by $H_i$ ($i=1,2,3$),
are determined by diagonalizing a $3\times 3$ squared-mass
matrix.  Thus, one can no longer parameterize the various Higgs
couplings in terms of the CP-even Higgs mixing angle $\alpha$.
It is convenient to work in a convention where the two vacuum expectation
values are real and positive (by absorbing any potential phases into
the definition of the Higgs field) so that $\tan\beta=v_u/v_d$ as
before.  Then, \eqns{hastate}{scalareigenstates} are replaced by
\begin{equation} \label{cpvstates}
H_i= (\sqrt{2}\,{\rm Re\,}\Phi_d^0-v_d)O_{1i}+
(\sqrt{2}\,{\rm Re\,}\Phi_u^0-v_u)O_{2i}+
\sqrt{2}\left({\rm Im\,}\Phi_d^0\sinb+{\rm Im\,}\Phi_u^0\cosb\right)O_{3i}\,,
\end{equation}
where $O$ is a $3\times 3$ real orthogonal matrix.

\begin{figure}[t!]
\begin{center}
\vspace{-0.2in}
\resizebox{\textwidth}{!}{
\includegraphics*[height=2in]{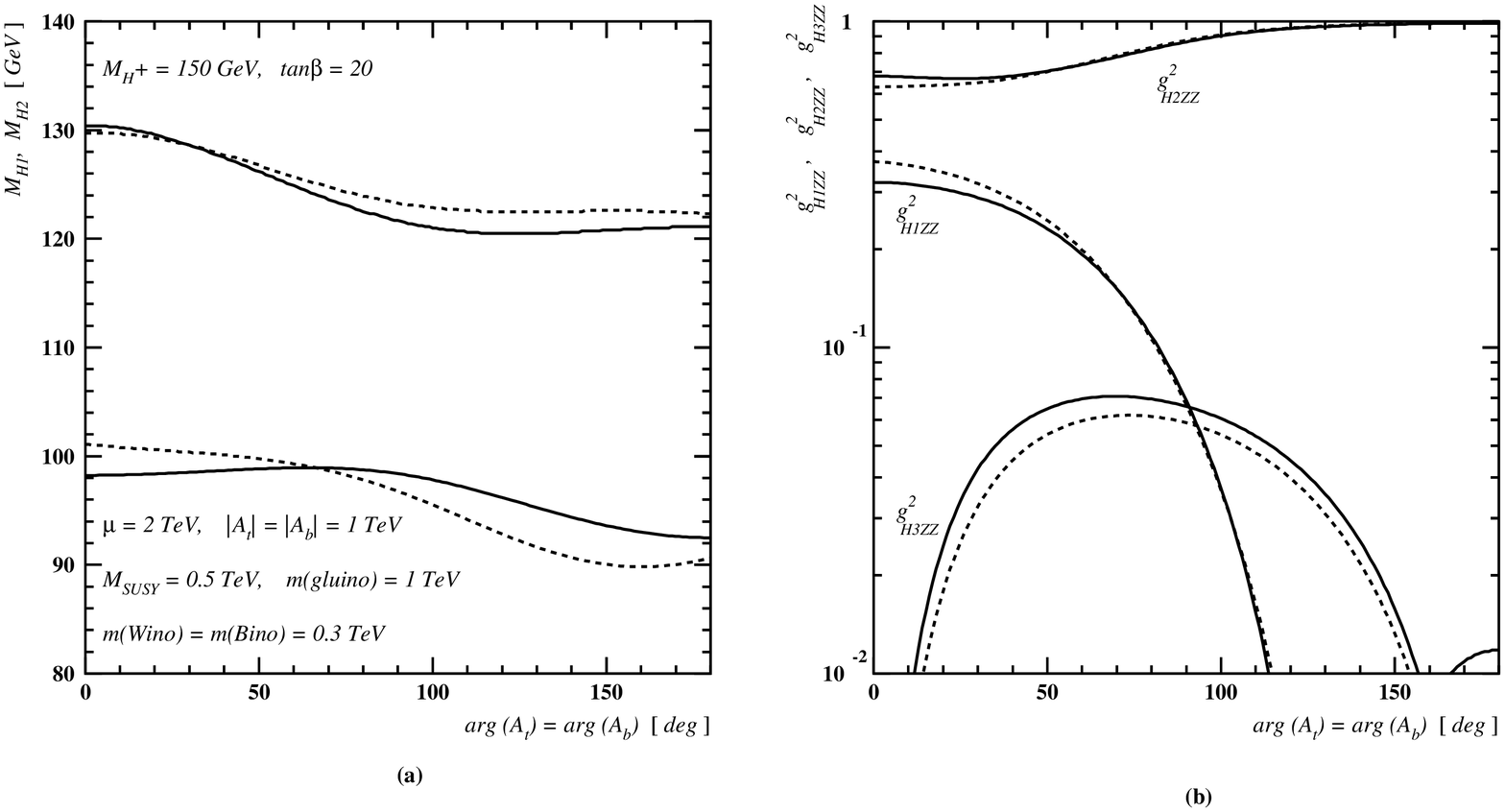}
}
\end{center}
\vspace{-0.4in}
  \capt{\label{cphtev} 
(a)~Lightest and next-to-lightest neutral Higgs masses and
(b)~relative couplings (normalized to the SM) of
the three neutral Higgs bosons to the $Z$ (or $W$)
as a function of the phase of $A_t$ for the indicated choices of the
MSSM parameters. Solid [dashed] lines are for
${\rm arg}(M_{\tilde{g}})$ = $0^\circ$ [$90^\circ$].  Taken from
\protect\Ref{cpcarlos2}.}
\end{figure}

In the CP-violating case,
vector boson pairs $VV$ ($V=W$ or $Z$) couple to all
three neutral Higgs mass eigenstates, $H_i$, with~\cite{cpcarlos}
\beq \label{cpvhvv}
g_{H_i VV}=O_{1i}\cos\beta+O_{2i}\sin\beta\,.
\eeq
\Fig{cphtev} shows the dependence of the
Higgs masses and the $H_i ZZ$
squared-couplings on the phase of $A_t$ for a
particular choice of MSSM parameters [as indicated in \fig{cphtev}(a)].
Clearly, these couplings can depend sensitively on the
phases of the complex supersymmetry-breaking parameters that
generate the mixing of the CP-even and CP-odd scalar eigenstates
through one-loop radiative effects.

The couplings of $V$ to a pair of neutral Higgs bosons are given by
$g_{H_i H_j Z}(p_{H_i}-p_{H_j})$, where the momenta $p_{H_i}$ and
$p_{H_j}$ point into the vertex, $g_{H_i H_j Z}$
is antisymmetric under the interchange of $H_i$ and $H_j$,
and~\cite{cpcarlos}
\beq \label{cpvhhz}
g_{H_i H_j Z}={g\over 2\cos\theta_W}\left[(O_{3i}O_{1j}-O_{1i}O_{1j})
\sin\beta-(O_{3i}O_{2j}-O_{2i}O_{3j})\cos\beta\right]\,.
\eeq
Using the orthogonality of $O$ (and ${\rm det}~O=1$), it is easy to
derive the relation~\cite{cpcarlos}:\footnote{One can easily check that
\eqns{vvcoup}{hvcoup} are recovered in the CP-conserving limit,
where $(H_1,H_2,H_3)=(\hl,\hh,\ha)$,
$O_{22}=-O_{11}=\sin\alpha$, $O_{12}=O_{21}=\cos\alpha$, $O_{33}=1$,
and all other elements $O_{ji}$ vanish.}
\beq \label{cpvcouprel}
g_{H_i H_j Z}={m_Z^2\over 2m_V^2}\,\epsilon_{ijk}\,g_{H_k VV}\,.
\eeq
The sum rules of \eqs{vvsumrule}{hxi} are then easily extended:
\beq
\sum_i g^2_{H_i VV}=g_V^2 m_V^2\,,\label{cpvsum1}
\eeq
which when combined with \eq{cpvcouprel} yields
\beqa
\sum_{i,j} g^2_{H_i H_j Z} &=&{g^2\over 2\cos^2\theta_W}\,,
\label{cpvsum2}
\eeqa
\beq
g_{H_i VV}\,g_{H_j VV}+{4m_V^4\over \mz^2}\sum_k g_{H_i H_k Z}\,g_{H_j
H_k Z}={m_V^4\over \mz^2\cos^2\theta_W}\delta_{ij}\,.\label{cpvsum3}
\eeq

The couplings of a neutral Higgs boson to $H^-W^+$ are
given by $g_{H_i H^-W^+}(p_{H_i}-p_{H^-})$, where the momenta
$p_{H^-}$ and $p_{H_i}$ point into the vertex, and~\cite{cpcarlos}
\beq
g_{H_i H^- W^+}=[g_{H_i H^+ W^-}]^* =\half
g\left[O_{1i}\sin\beta-O_{2i}\cos\beta-iO_{3i}\right]\,. 
\eeq

Another consequence of the CP-violating effects in the scalar sector
is that all neutral Higgs scalars can couple to both scalar and
pseudoscalar fermion bilinear densities ($\bar\psi \psi$ and
$\bar\psi\gamma\ls{5}\psi$, respectively).
The couplings of the mass eigenstate $H_i$ to fermions
depend on the loop-corrected fermion Yukawa
couplings, $h_{b,t}$, $\delta h_{b,t}$, $\Delta h_{b,t}$,
and on $\tan\beta$ and the $O_{ji}$.
It is convenient to adjust
the phases of the fields so that the quantities
$h_b+\delta h_b+\Delta h_b\tan\beta$ and $h_t+\delta h_t+\Delta h_t\cot\beta$
are both real and positive [{\it i.e.}, the physical fermion masses are still
given by \eqns{byukmassrel}{tyukmassrel}].
The resulting expressions are a
straightforward generalization of those presented above for the
CP-conserving case~\cite{cpcarlos2}:
\begin{equation}
  \label{Hff}
{\cal L}_{H\bar{f}f}\ =\ - \sum_{i=1}^3\, H_i\,
\bigg[\,\frac{m_b}{v}\, \bar{b}\,\Big( g^S_{H_ibb}\, +\,
ig^P_{H_ibb}\gamma_5 \Big)\, b\: +\: \frac{m_t}{v}
\, \bar{t}\,\Big( g^S_{H_i tt}\, +\,
ig^P_{H_itt}\gamma_5 \Big)\, t\, \bigg]\, ,
\end{equation}
with
\begin{eqnarray}
  \label{gSHbb}
g^S_{H_ibb} & =& \frac{1}{h_b\, +\, \delta h_b\, +\, \Delta h_b
\tan\beta }\ \bigg\{\,
 {\rm Re}(h_b + \delta h_b)\, \frac{O_{1i}}{\cos\beta}\:
+\: {\rm Re}(\Delta h_b)\, \frac{O_{2i}}{\cos\beta}
\nonumber\\
&&-\,  \Big[{\rm Im}(h_b + \delta h_b) \tan\beta
-  {\rm Im} (\Delta h_b)\Big]\, O_{3i}\, \bigg\}\, ,\\
  \label{gPHbb}
g^P_{H_ibb} & =&  \frac{1}{h_b\, +\, \delta h_b\, +\, \Delta h_b
\tan\beta }\ \bigg\{\, \Big[\, {\rm Re}\left(\Delta h_b\right)\, -\,
{\rm Re}(h_b + \delta h_b) \tan\beta\, \Big]\, O_{3i} \nonumber\\
&&-\, {\rm Im}(h_b + \delta h_b)\, \frac{O_{1i}}{\cos\beta}\:
-\:  {\rm Im}(\Delta h_b)\, \frac{O_{2i}}{\cos\beta}\  \bigg\}\ ,\\
  \label{gSHtt}
g^S_{H_itt} & = & \frac{1}{h_t\, +\, \delta h_t\, +\, \Delta h_t
\cot\beta }\ \bigg\{\,
 {\rm Re}(h_t + \delta h_t)\, \frac{O_{2i}}{\sin\beta}
+\: {\rm Re}(\Delta h_t)\, \frac{O_{1i}}{\sin\beta} \nonumber\\
&&-\, \Big[\, {\rm Im}(h_t + \delta h_t) \cot\beta\, -\,
{\rm Im}(\Delta h_t)\, \Big]\, O_{3i}\, \bigg\}\, ,\\
  \label{gPHuu}
g^P_{H_itt} & = & \frac{1}{h_t\, +\, \delta h_t\, +\, \Delta h_t
\cot\beta }\ \bigg\{\, \Big[\, {\rm Re}(\Delta h_t)\, -\,
{\rm Re}(h_t + \delta h_t) \cot\beta\, \Big]\, O_{3i} \nonumber\\
&&-\, {\rm Im}(h_t + \delta h_t)\, \frac{O_{2i}}{\sin\beta}\: -\:
{\rm Im}(\Delta h_t)\, \frac{O_{1i}}{\sin\beta}\, \bigg\}\, ,
\end{eqnarray}
where the Higgs scalar couplings are normalized with
respect to the corresponding SM values.  

For large values of the charged Higgs boson mass
and for heavy supersymmetric particles, the expressions of the lightest neutral
Higgs boson coupling to fermions reduce to those of the
(CP-conserving) SM Higgs
boson, as expected for the decoupling limit.  In contrast, the
two heavy neutral Higgs bosons are still admixtures of
CP-even and CP-odd eigenstates; hence, CP-violating effects
are still present in the  heavy neutral Higgs sector.
However, due to the high degeneracy in mass
of the heavy scalar sector (especially in the decoupling limit),
CP-violating effects may be difficult to observe without precision
measurements of the heavy neutral Higgs properties.

The couplings of the charged Higgs bosons to fermions are
of the form $\mathcal{L}_{\rm int}=\bar b\, g_{H^-t\bar b}\,tH^- +{\rm
h.c.}$, with 
\beq g_{H^-t\bar b}\simeq \left({\sqrt{2}\, m_b\over
v}\tan\beta-{\Delta h_b\over\cos\beta}\right) P_L + \left({\sqrt{2}\,
m_t\over v}\cot\beta-{(\Delta h_t)^*\over\sin\beta}\right) P_R \,.
\eeq
One can check that for real $\Delta h_b$ and $\Delta h_t$,
this result is equivalent to \eq{hmtb} [with the same caveats
noted in footnote\fnref{footispin}]. 
An explicit computation of the CP-violating
$H^-t\bar b$ vertex and its phenomenological implications can be
found in \Ref{cemk}.

\subsection{MSSM Higgs Boson Decay Modes}
\label{sec:34}

In the MSSM, we must consider the decay properties of three neutral
Higgs bosons and one charged Higgs pair.\footnote{Unless otherwise noted,
we shall neglect CP-violating effects
({\it e.g.}, by assuming that CP-violating effects induced by
radiative corrections are small).}
In the region of parameter space where $\mha\gg m_Z$
and the masses of supersymmetric particles are large, the
decoupling limit applies, and we find that the properties of $\hl$
are indistinguishable from the SM Higgs boson.
If supersymmetric particles are light, then
the decoupling limit does not strictly apply even in the limit of
$\mha\gg m_Z$.  In particular, the $\hl$ branching ratios are
modified, if the decays of $\hl$ into supersymmetric particles are
kinematically allowed.  In addition,
if light superpartners exist that can couple
to photons and/or gluons, then the one-loop $gg$ and $\gamma\gamma$
decay rates would also deviate from the corresponding Standard Model
Higgs decay rates due to the extra contribution of the
light superpartners appearing in the loops.
In both cases, the heavier Higgs states, $\hh$, $\ha$ and
$\hpm$, are roughly mass degenerate, and their decay branching ratios
depend crucially on $\tan\beta$ as shown below.

For values of $\mha\sim {\cal O}(\mz)$, all Higgs boson states lie
below 200~GeV in mass.  In this parameter
regime, there is a significant area of the parameter space in which
none of the neutral Higgs boson decay properties approximates that of
the SM Higgs boson.  For $\tan\beta\gg 1$, the
resulting Higgs phenomenology shows marked differences from that of
the SM Higgs boson~\cite{bdmv}.  In particular,
radiative corrections can significantly
modify the $b\bar b$ and/or the $\tau^+\tau^-$ decay rates with
respect to those of the SM Higgs boson, as noted in \Sec{sec:333}.
Additionally, the Higgs bosons can decay into new channels, either
containing lighter Higgs bosons or supersymmetric particles.
In the following,
the decays of the neutral Higgs bosons $\hl$, $\hh$ and
$\ha$ and the decays of charged Higgs
bosons are discussed with particular emphasis on differences
from Standard Model expectations.
In the following discussion, we exhibit results for $\tan\beta=3$ and 30
to illustrate
the difference between ``low'' and ``high'' $\tan\beta$.
The results shown below include the effects of the dominant
radiative corrections, which affect both
the masses and the couplings of the Higgs sector as described in
\Secs{sec:32}{sec:33}.

\begin{figure}[p]
\begin{center}
\resizebox{\textwidth}{!}{
\includegraphics*{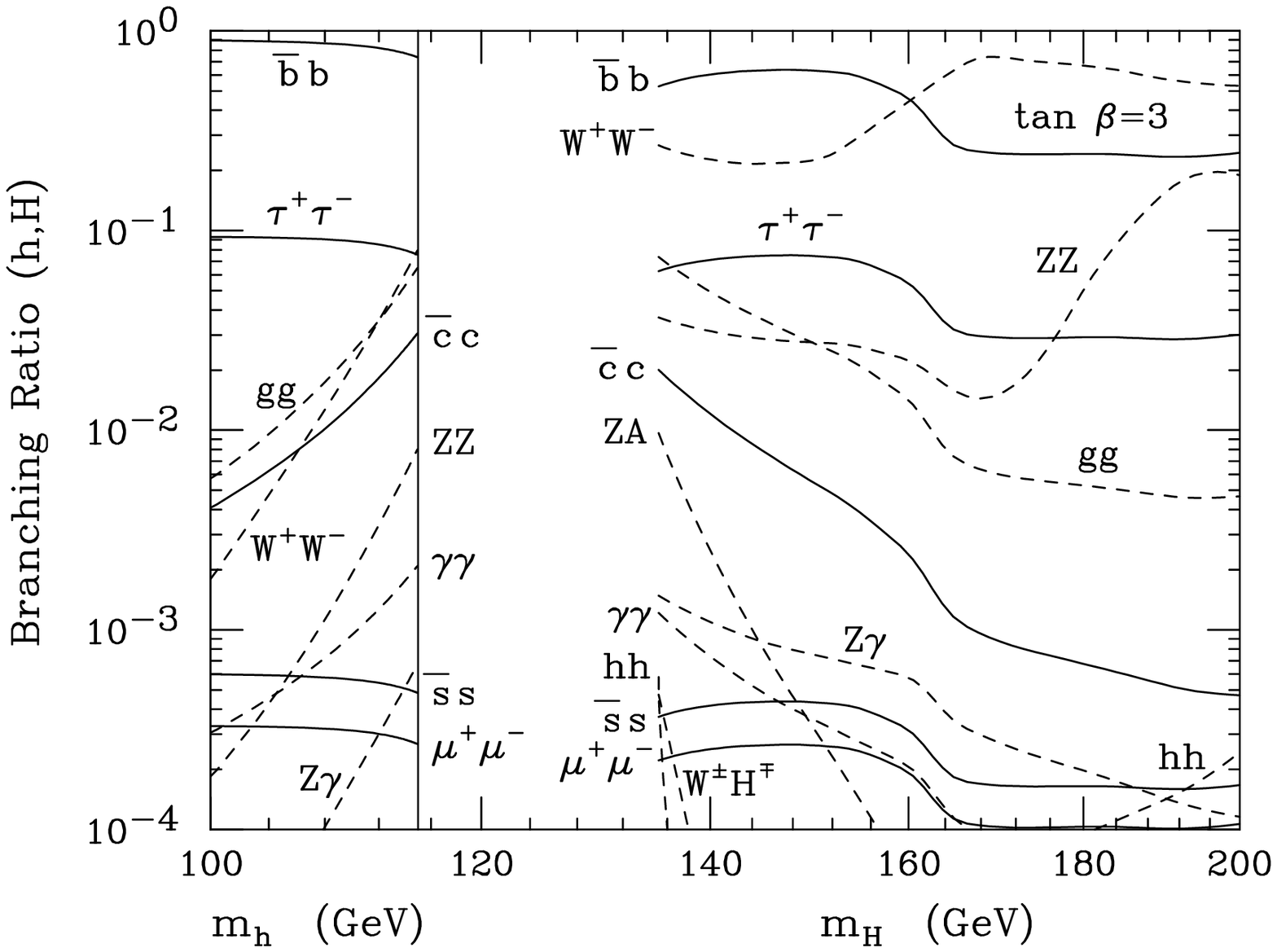}
\hfill
\includegraphics*{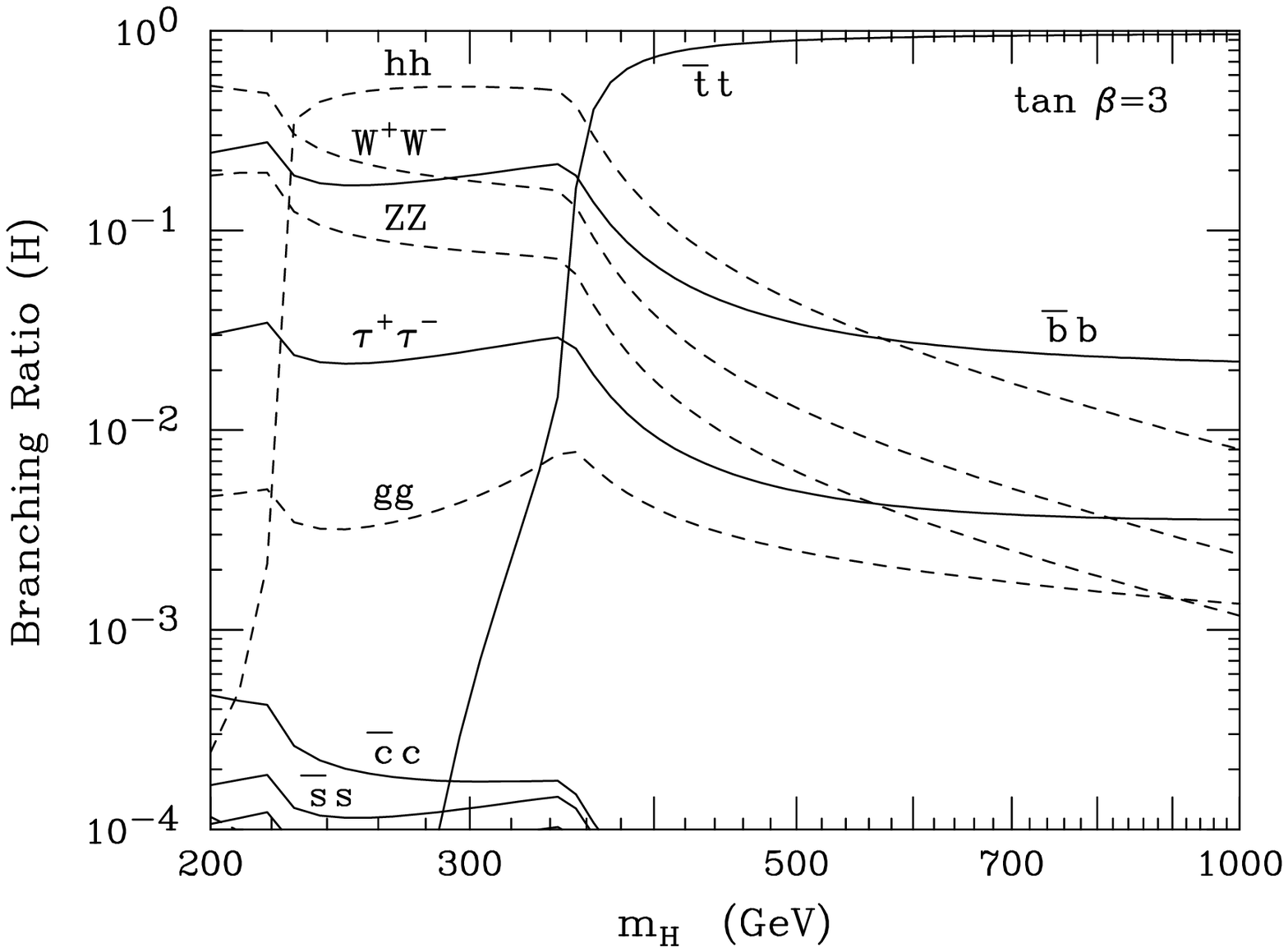}
}
\vskip2pc
\resizebox{\textwidth}{!}{
\includegraphics*{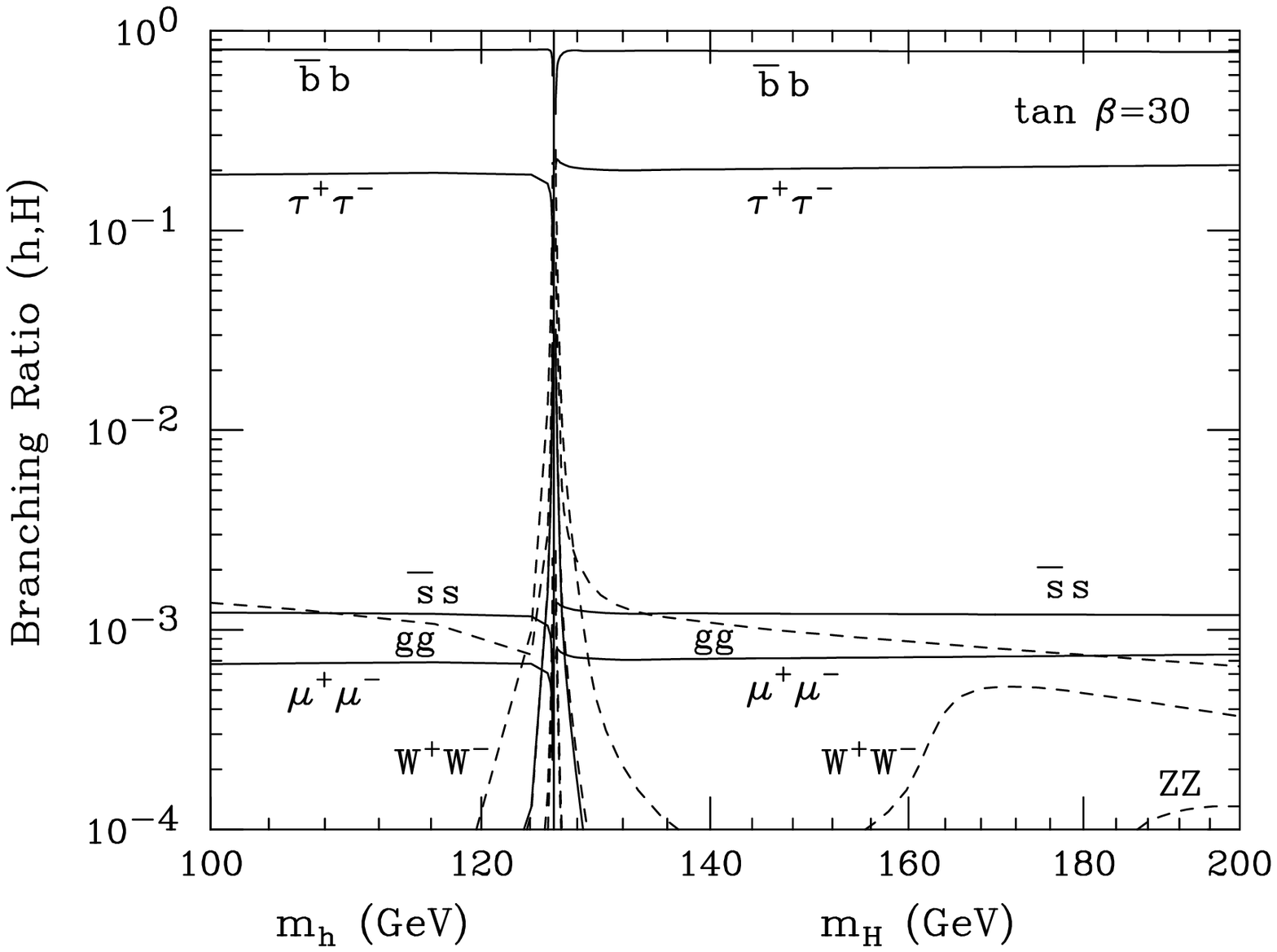}
\hfill
\includegraphics*{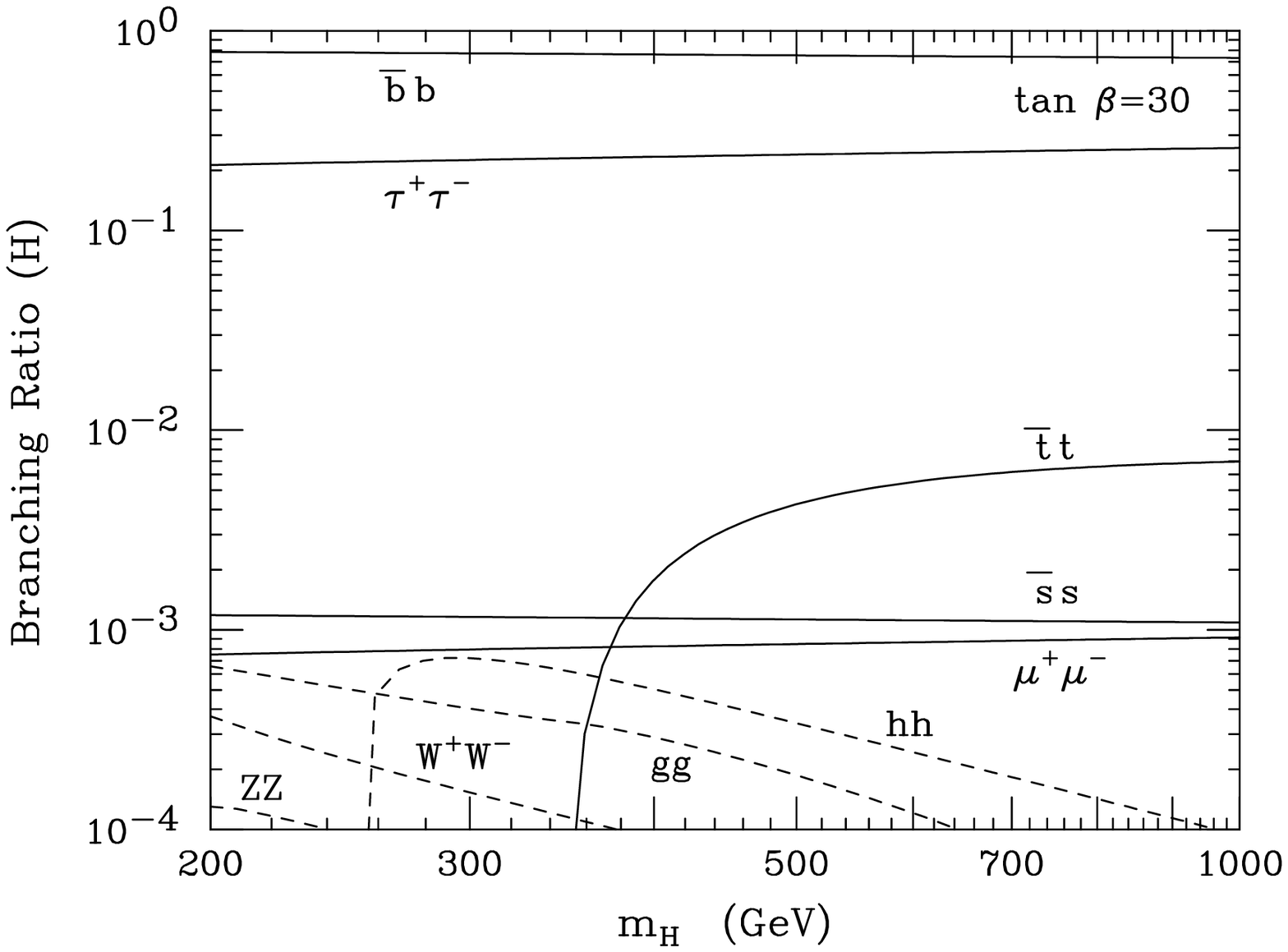}
}
\end{center}
\capt{\label{hlhhbr} Branching ratios of the MSSM Higgs bosons
$\hl$ and $\hh$, with $\tan\beta=3$ and $30$, respectively.
Final states labeled above include the possibility of one off-shell
final state particle below the corresponding two-particle decay threshold.
The above plots were made under the assumption
that the average top and bottom squark masses are 1~TeV and top-squark
mixing is maximal.
In this case, $\mhmax\simeq 115$~GeV
(125.9~GeV) for $\tan\beta=3$ (30),
corresponding to the limit of large $\mha$,
is indicated by the vertical line in
the two left-side plots.  The range of $\mhh$ shown corresponds to
varying $\mha$ between 90~GeV and 1~TeV, while $\mhl>100$~GeV
corresponds to $\mha>139$~GeV (104~GeV) for $\tan\beta=3$ (30).
Other supersymmetric parameters have been
chosen such that there are no supersymmetric particle decay modes in
the Higgs mass ranges shown above.}
\end{figure}

\begin{figure}[ht!]
\begin{center}
\resizebox{\textwidth}{!}{
\includegraphics*{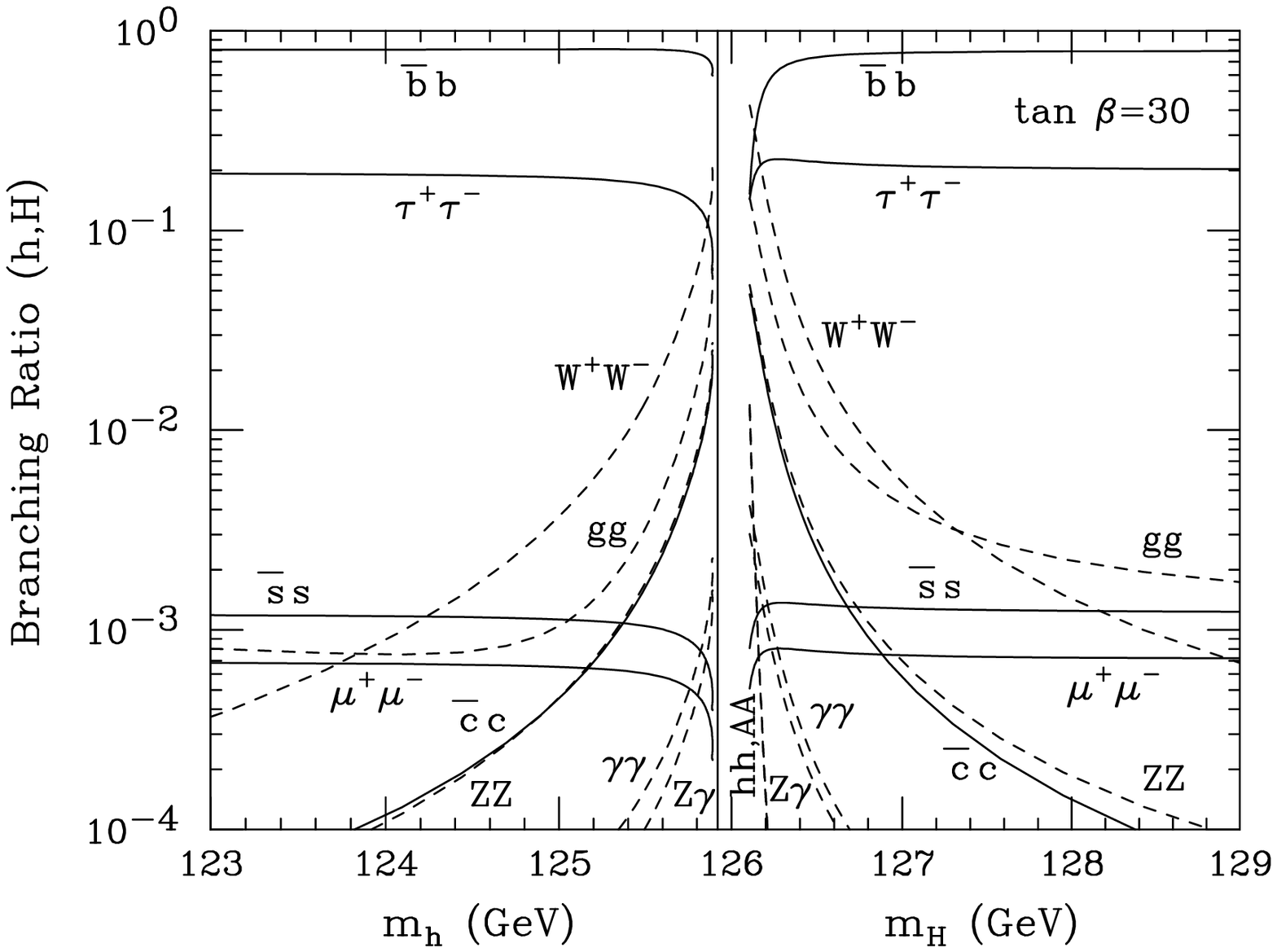}
\hfill
\includegraphics*{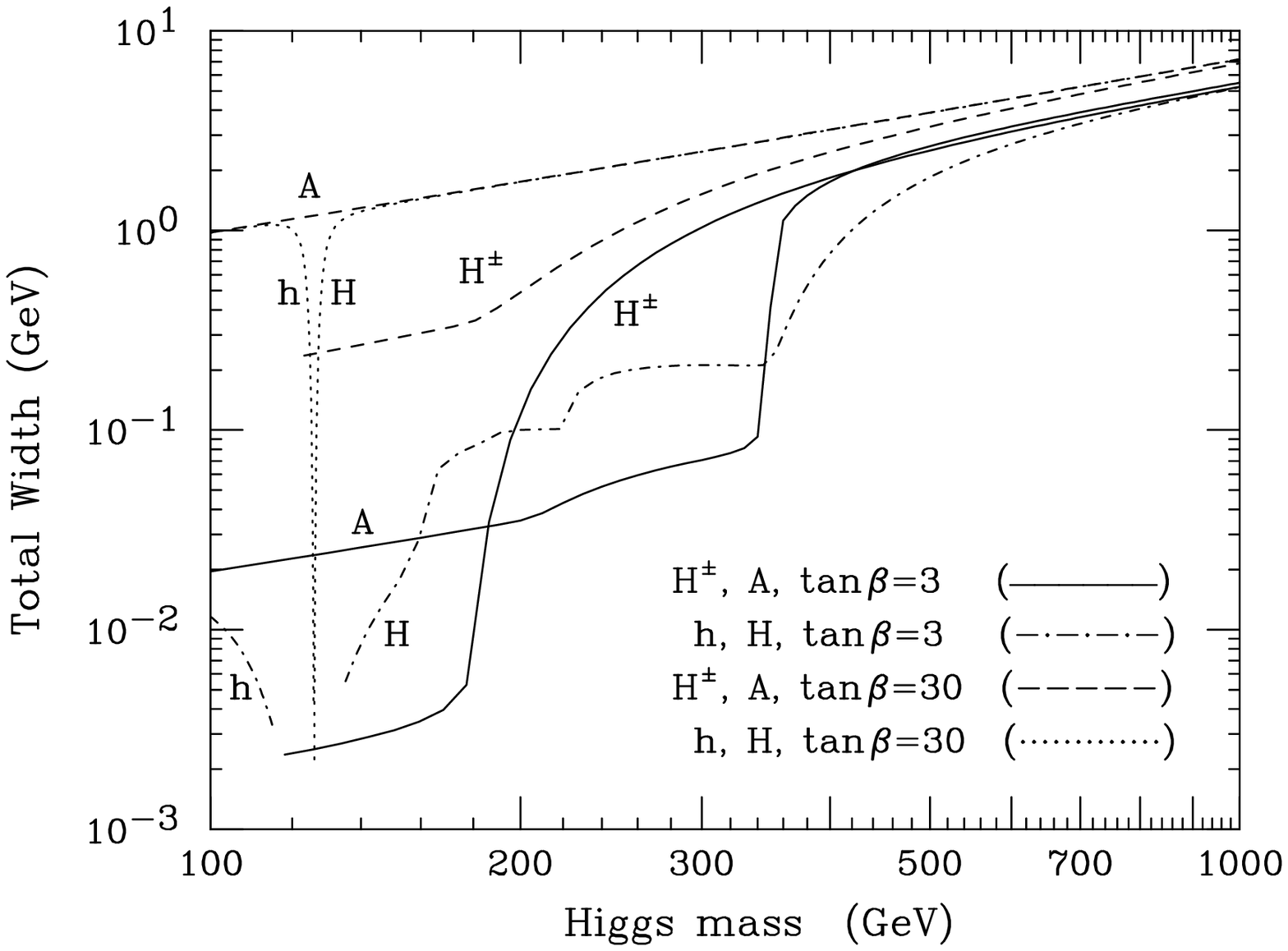}
}
\end{center}
\capt{\label{hlhhbra} 
(a) Branching ratios of the MSSM Higgs bosons
$\hl$ and $\hh$, with $\tan\beta=30$.  Here, we zoom in on the Higgs mass
regime within $\pm 3$~GeV of $\mhmax=125.9$~GeV
of \protect\fig{hlhhbr} in order to
get a clearer picture of the various decay modes.
The range of $\mhh$ shown corresponds to
$90~{\rm GeV}<\mha< 130$~GeV, whereas the range of
$\mhl$ shown corresponds to $128~{\rm GeV}<\mha< 1$~TeV.
(b) Total widths of the MSSM Higgs boson as a function of the
corresponding Higgs mass for $\tan\beta=3$ and 30, with the same parameter
assumptions employed in \protect\fig{hlhhbr}.}
\end{figure}

In order to display results for Higgs branching ratios, we must choose
a set of MSSM parameters.  We fix $\tan\beta$ (for two representative
choices) and vary $\mha$ from its LEP experimental lower bound of
90~GeV up to 1~TeV.
In addition, the gluino and
MSSM squark mass parameters have been chosen to be
$\MSUSY\equiv M_{\tilde g}=M_Q=M_U=M_D=1$~TeV, the squark
mixing parameter $X_t=A_t-\mu\cot\beta=2.4\MSUSY$, and
the gaugino mass matrix parameters, 
$\mu=M_2\simeq 2M_1=1$~TeV.  This differs somewhat from the
maximal mixing benchmark scenario of \Ref{benchmark}.  Nevertheless,
the value of $\mhl$ is still close to maximal (for fixed $\mha$ and
$\tan\beta$), so we will continue to loosely refer to
the above choice of MSSM parameters as
a maximal mixing scenario.  Our motivation
for choosing the gaugino mass parameters large is to avoid
possible supersymmetric decay modes for the Higgs bosons for Higgs
masses below 1~TeV.  We shall briefly comment on possible
supersymmetric decay modes at the end of this section.

\begin{figure}[p]
\begin{center}
\resizebox{\textwidth}{!}{
\includegraphics*{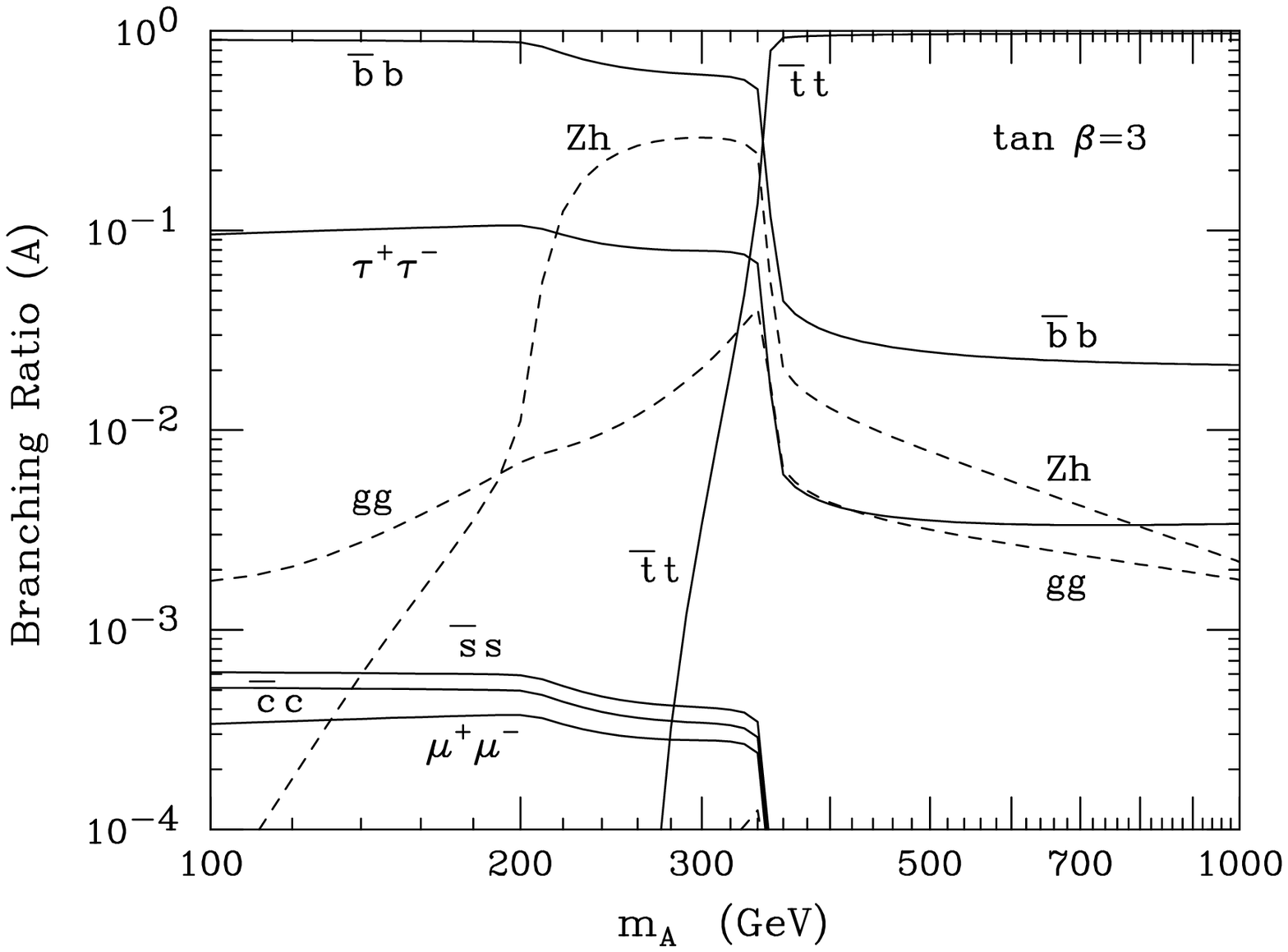}
\hfill
\includegraphics*{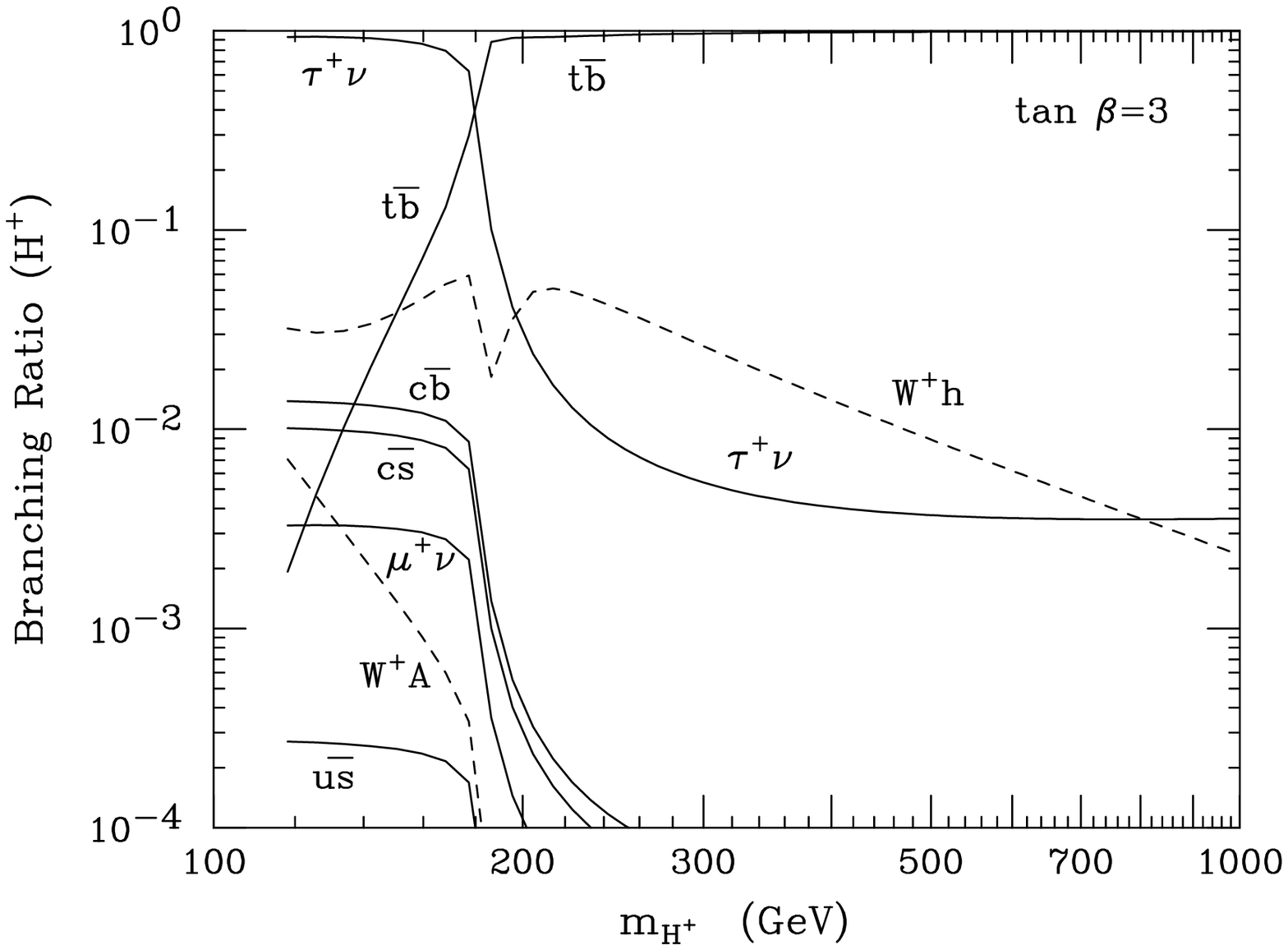}
}
\vskip2pc
\resizebox{\textwidth}{!}{
\includegraphics*{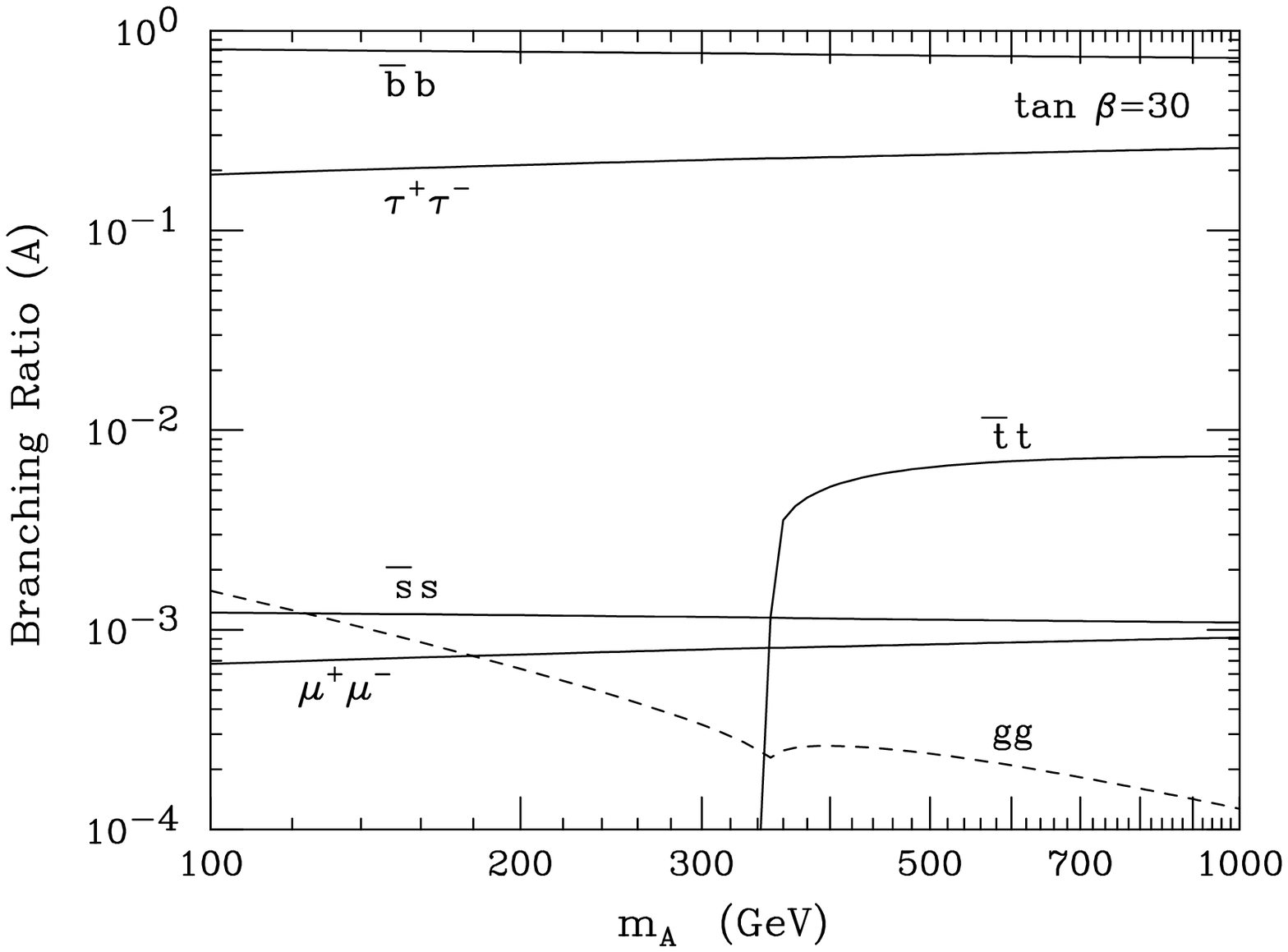}
\hfill
\includegraphics*{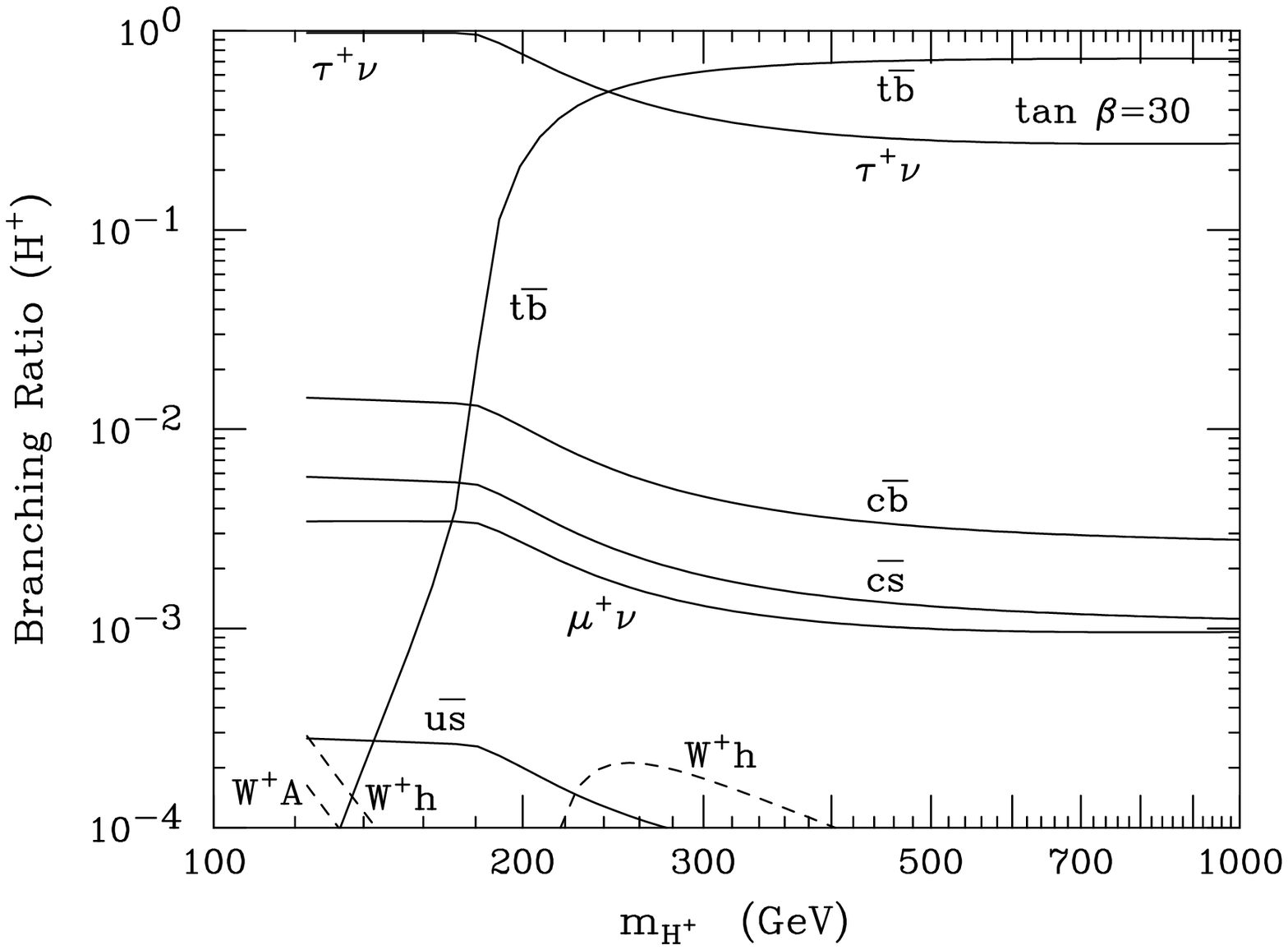}
}
\end{center}
\capt{\label{hahpbr} Branching ratios of the MSSM Higgs bosons
$\ha$ and $H^+$, with $\tan\beta=3$ and $30$, respectively.
Final states labeled above include the possibility of one off-shell
final state particle below the corresponding two-particle decay threshold.
The above plots were made under the assumption
that the average top and bottom squark masses are 1~TeV and top-squark
mixing is maximal.
The range of $\mhpm$ shown corresponds to
varying $\mha$ between 90~GeV and 1~TeV.
Other supersymmetric parameters have been
chosen such that there are no supersymmetric particle decay modes in
the Higgs mass ranges shown above.}
\end{figure}

The branching ratios for $\hl$ and $\hh$ as a function of their masses
are shown in \fig{hlhhbr}.  As $\mha$
varies from 90~GeV to 1~TeV, with the MSSM parameters as specified above,
$135~{\rm GeV}\lsim\mhh\lsim 1$~TeV when $\tan\beta=3$ and
$126.1~{\rm GeV}\lsim\mhh\lsim
1$~TeV when $\tan\beta=30$.
In contrast, most of the variation in $\mha$
occurs for values of $\mhl$ a few GeV below $\mhmax$.  Thus,
we also exhibit in \fig{hlhhbra}(a) the branching ratios for $\hl$ and
$\hh$ for $\tan\beta=30$ and Higgs mass values of $\mhmax\pm
3$~GeV.  This reveals a detailed pattern of branching ratios that is
not easily visible in \fig{hlhhbr}.
The branching ratios for
$\ha$ and $H^+$ as a function of their masses
are shown in \fig{hahpbr}.

The total Higgs decay widths as a function of the corresponding Higgs
mass are shown in \fig{hlhhbra}(b) for the two cases of $\tan\beta=3$
and 30 (and the other relevant MSSM parameters as described above).
Note that for large values of the Higgs mass, the corresponding widths
are considerably smaller than that of the SM Higgs boson.
This is due to the suppressed $\hh VV$ couplings at
large Higgs mass and to the
absence of tree-level $\ha VV$ and $H^+W^-Z$ couplings.  One can also
check that in the decoupling limit ($\mha\gg\mz$), the total width of
$\hl$ coincides with that of $\hsm$.  This is illustrated by
replotting the $\hl$ (and $\hh$) widths on the same plot as the $\hsm$
width [see \fig{fg:hwidth}(b)].  In particular, note that the dashed
and dot-dashed $\hl$ contours in \fig{fg:hwidth}(b) approach the $\hsm$
contour as $\mhl$ reaches its maximal value.
(which corresponds to the limit of large $\mha$ at fixed $\tanb$).
It is interesting to
note that in the opposite limit of small $\mha$ (especially at large
$\tan\beta$), $\cos(\beta-\alpha)\to 1$ and
it is $\hh$ that assumes many of the properties of
$\hsm$.  However, there can still be deviations in the $\hh
b\bar b$ coupling from the corresponding Standard Model value
at large $\tan\beta$, as noted below \eq{hpmll}.
This explains why the $\hh$ contours in \fig{fg:hwidth}(b)
do not quite coincide with the result of
the $\hsm$ contour as $\mhh$ approaches its lower
limit (with the discrepancy between the $\hh$ and $\hsm$ contours
more pronounced at large $\tan\beta$). 

The branching ratios and widths
in \figs{hlhhbr}{hahpbr} have been computed using
a modified version of the {\tt HDECAY} program~\cite{hdecay_mrenna}
that incorporates the
leading radiative corrections to the Higgs couplings discussed in
\Sec{sec:33}.\footnote{For the maximal mixing choice of MSSM
parameters used in \figs{hlhhbr}{hahpbr}, 
we find $\Delta_b\simeq 0.55$ for $\tanb=30$.
For large $\mha$ and $\tanb$, the partial widths of $\hh$, $\ha\to b\bar b$
and $H^+\to t\bar b$ (and likewise the corresponding total widths)
are suppressed by a factor of about $(1+\Delta_b)^2$ with
respect to the corresponding tree-level results.}  The
decay modes $\hl,\hh,\ha \to b\bar b$, $\tau^+\tau^-$ dominate
the neutral Higgs decay modes when $\tanb$ is large for all values of
the Higgs masses.  For small $\tanb$, these modes are significant
for neutral Higgs masses below $2m_t$ (although there are other
competing modes in this mass range), whereas
the $t\bar t$ decay mode dominates above the $t\bar t$
decay threshold.  In contrast to the SM Higgs boson, the vector
boson decay modes of $\hh$ are strongly suppressed at large $\mhh$ due
to the suppressed $\hh VV$ couplings in the decoupling limit.
For the charged Higgs boson, $H^+\to\tau^+\nu$ dominates below $t\bar
b$ threshold, while $H^+\to t\bar b$ dominates for large values of
$\mhpm$.  Note that final states labeled in \figns{hlhhbr}{hahpbr}
include the possibility of one off-shell
final state particle below the corresponding two-particle decay
threshold~\cite{offshell}.
For example, for $\mhpm< m_t+m_b$, the $t\bar b$ contour shown in
\fig{hahpbr} actually corresponds to an off-shell $t$ quark that
decays to $bW^+$.  That is, in this mass region, the $t\bar b$ contour
corresponds to the branching ratio for the three-body decay $H^+\to
W^+b\bar b$.  This decay mode can be especially significant at moderate
values of $\tan\beta$ due to the large Higgs-top Yukawa coupling.

As in the \SM\ case, the partial decay widths of the neutral Higgs
bosons into $b\bar b$ and $c\bar c$ are reduced by about 50--75\%
when QCD corrections are included ({\it e.g.}, by employing
running quark masses in the decay width formulae), whereas the QCD
corrections are less significant for Higgs decays into $t\bar
t$~\cite{DSZ}. The effects of the QCD radiative corrections on the
charged Higgs branching ratios~\cite{chhiggsdecay} are significant
in the region of $\tan\beta$ where the $c\bar s$ and
$\tau^+\nu_\tau$ decay modes are competitive or for large values
of $\tan \beta$ for the decay mode $H^+\to t\bar b$ (and for
$H^+\to W^+b\bar b$ below $t\bar b$
threshold~\cite{chhiggsdecay3}). Additional supersymmetric
radiative corrections discussed in \Sec{sec:33} can also
significantly affect the Higgs boson partial widths.  Some of
these corrections can be absorbed into the effective mixing angle
$\alpha$~\cite{hffsusyprop} as shown in \Sec{sec:331}. As a
consequence of this universal correction, the coupling of $\hl$ to
$b\bar b$ and $\tau^+\tau^-$ can be suppressed for small $\mha$
and large $\tan\beta$. For the $\hl\to b\bar b$ decay width, the
supersymmetric
corrections~\cite{CMW1,hffsusy1l,hffsusyqcd,hffsusyprop,loganetal}
proportional to the strong coupling constant $\alpha_s$ and the
Higgs-top quark Yukawa coupling $h_t$ can be significant for
large values of $\mu$ and $\tan\beta$. As shown in \Sec{sec:333},
this effect can be interpreted as a correction to the tree-level
relation between $m_b$ and $h_b$.

In addition to the decay modes of the neutral Higgs bosons into
fermion and gauge boson final states,
there exist new Higgs decay channels that involve scalars of
the extended Higgs sector and supersymmetric final states.
The unambiguous observation of these modes (as well as any decay mode
of a charged scalar) would clearly constitute
direct evidence of new physics beyond the Standard Model.
Higgs decays into
charginos, neutralinos and third-generation squarks and sleptons can become
important, once they are kinematically allowed \cite{13a}.
One interesting possibility is a significant branching
ratio for the decay of a neutral Higgs boson to the invisible mode
$\widetilde\chi_1^0 \widetilde\chi_1^0$ (where $\widetilde\chi_1^0$ is the
lightest supersymmetric particle).  In such a scenario, the discovery
of this neutral Higgs boson would be difficult
at a hadron collider~\cite{invisibly}.  In contrast, at lepton colliders,
methods exist for detecting an invisibly decaying
Higgs boson by observing a peak in the
missing mass recoiling against the produced $Z$.

\subsection{MSSM Higgs Boson Production at Hadron Colliders}
\label{sec:35}

\subsubsection{Cross-sections at hadron colliders}
\label{sec:351}

The production mechanisms for the SM Higgs boson at hadron
colliders can also be relevant for the production of the MSSM neutral
Higgs bosons.  However, we must take into account the possibility of
enhanced or suppressed couplings (with respect to those of the
Standard Model).  For example, the $HVV$ couplings are very suppressed
in the decoupling limit, and tree-level $AVV$ couplings are completely
absent.  On the other hand, at large $\tan\beta$, typically two of the
three neutral Higgs couplings to bottom-type quarks are enhanced.
These effects can significantly modify the neutral Higgs production
cross-sections.   New production mechanisms must be considered for
charged Higgs production.

\begin{figure}[p]
\begin{center}
\resizebox{\textwidth}{!}{
\includegraphics[width=5cm,angle=-90]{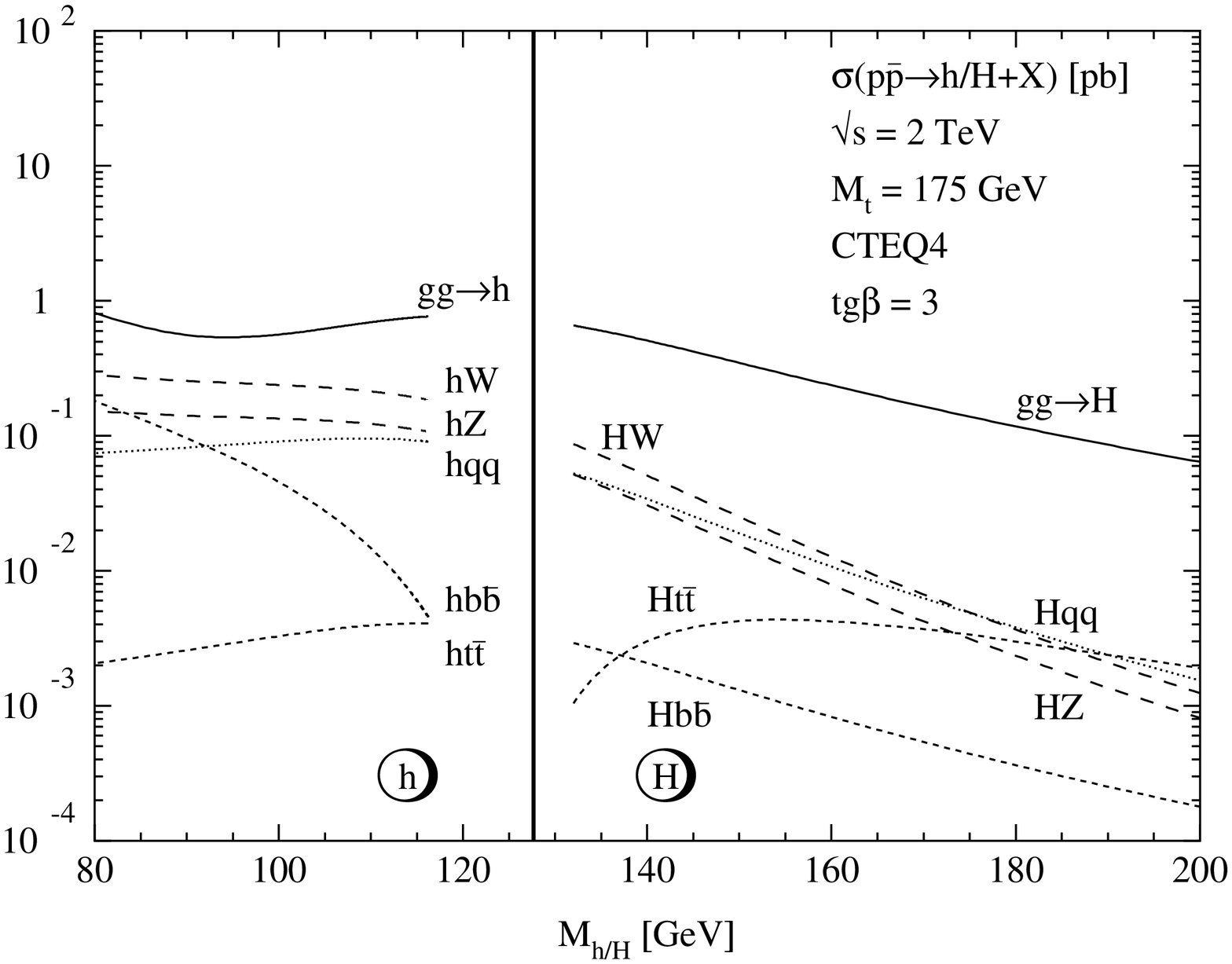}
\hfill
\includegraphics[width=5cm,angle=-90]{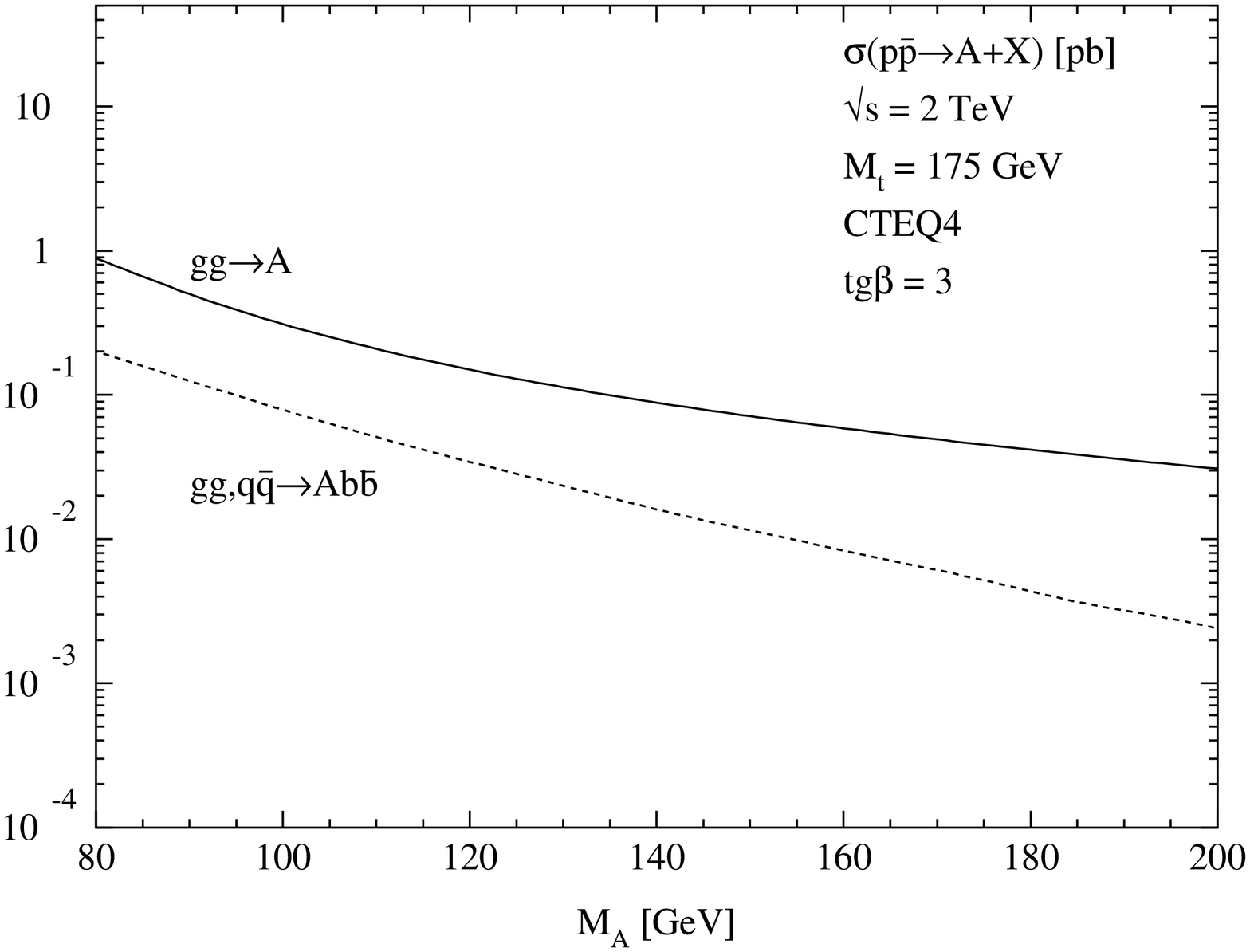}
}
\resizebox{\textwidth}{!}{
\includegraphics[width=5cm,angle=-90]{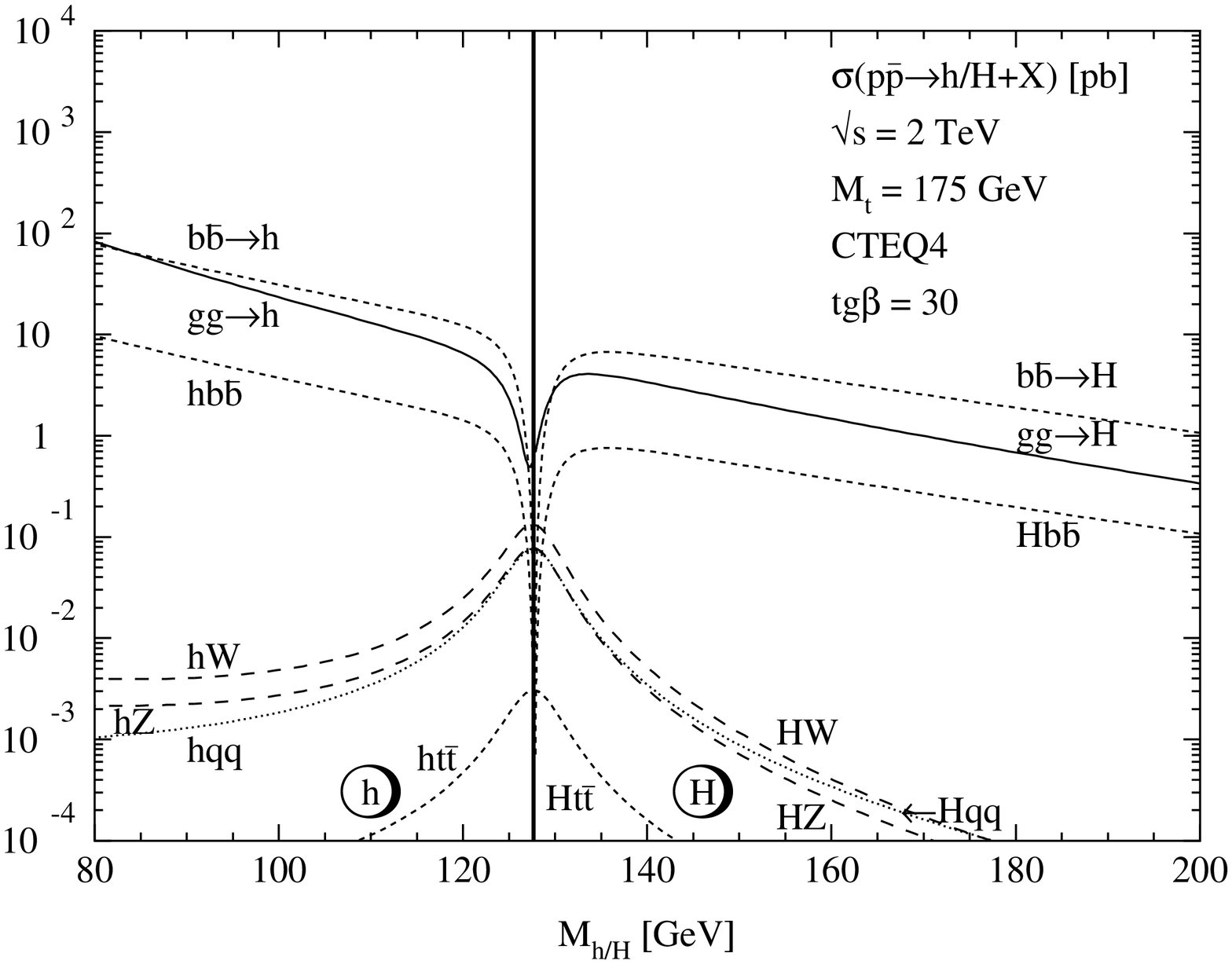}
\hfill
\includegraphics[width=5cm,angle=-90]{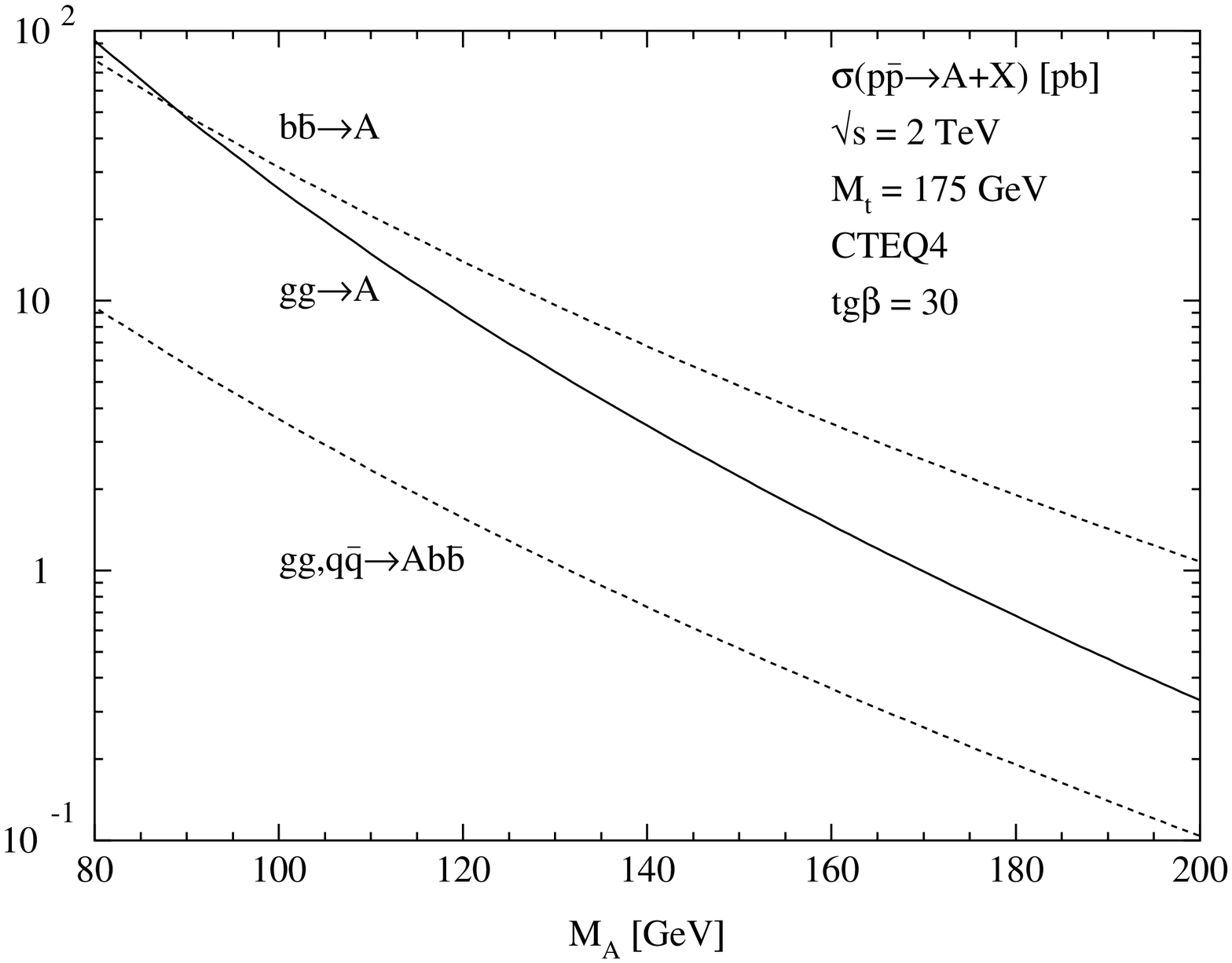}
}
\end{center}
\vskip1pc
\capt{\label{tevmssmxsec} Neutral MSSM Higgs production
cross-sections
at the Tevatron [$\sqrt{s}=2$ TeV] for gluon fusion $gg\to \phi$,
vector-boson fusion $qq\to qqV^*V^* \to qqh$,
$qqH$, vector-boson bremsstrahlung $q\bar q\to V^* \to \hl V/\hh V$ and
the associated
production $gg,q\bar q \to \phi b\bar b/ \phi t\bar t$ including all known
QCD corrections, where $\phi=\hl$, $\hh$ or $\ha$~\protect\cite{9,hxsec}.
As in \protect\fig{fg:4}, in the vector boson fusion process,
$qq$ refers to both $ud$ and $q\bar q$ scattering.
The four panels exhibited above show
(a) $h,H$ production for $\tanb=3$, (b) $\ha$ production for $\tanb=3$,
(c) $\hl$, $\hh$ production for $\tanb=30$,  (d) $A$ production for
$\tanb=30$.}
\end{figure}

\begin{figure}[p]
\begin{center}
\resizebox{\textwidth}{!}{
\includegraphics{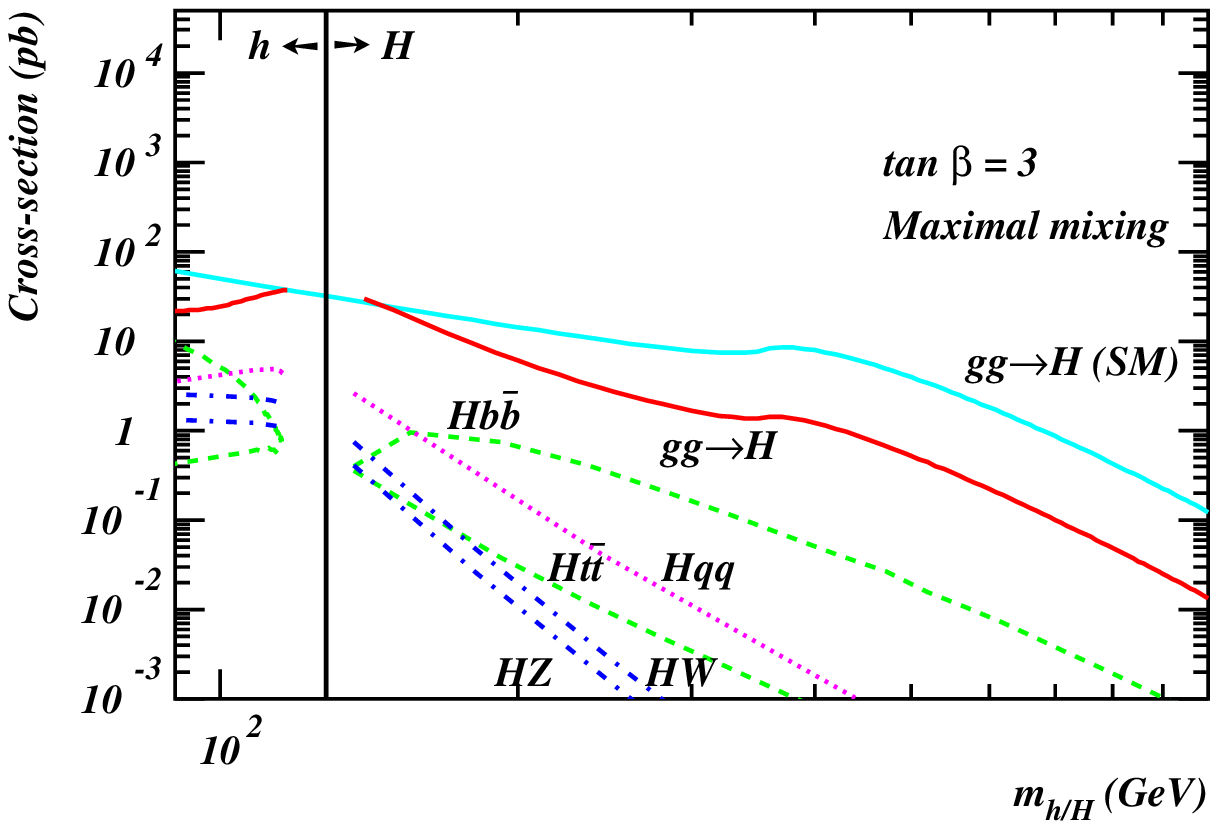}
\hfill
\includegraphics{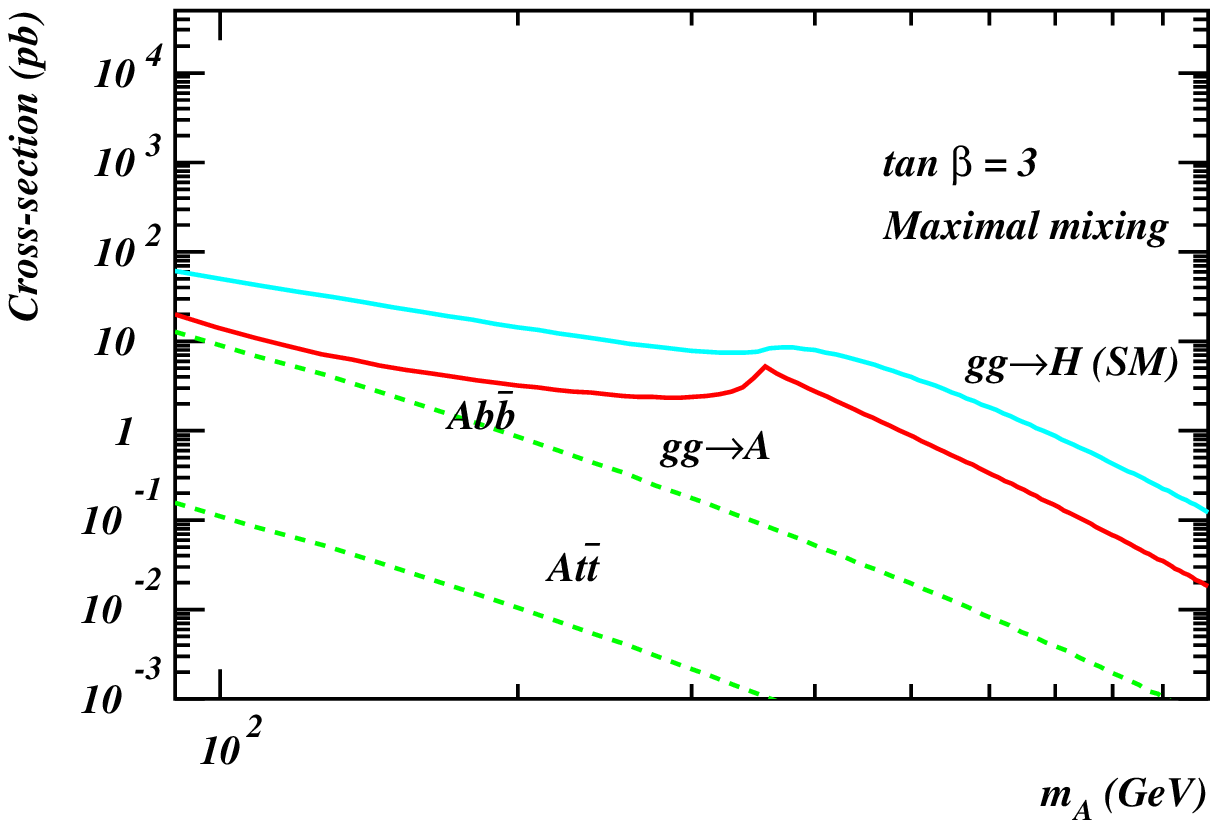}
}
\resizebox{\textwidth}{!}{
\includegraphics{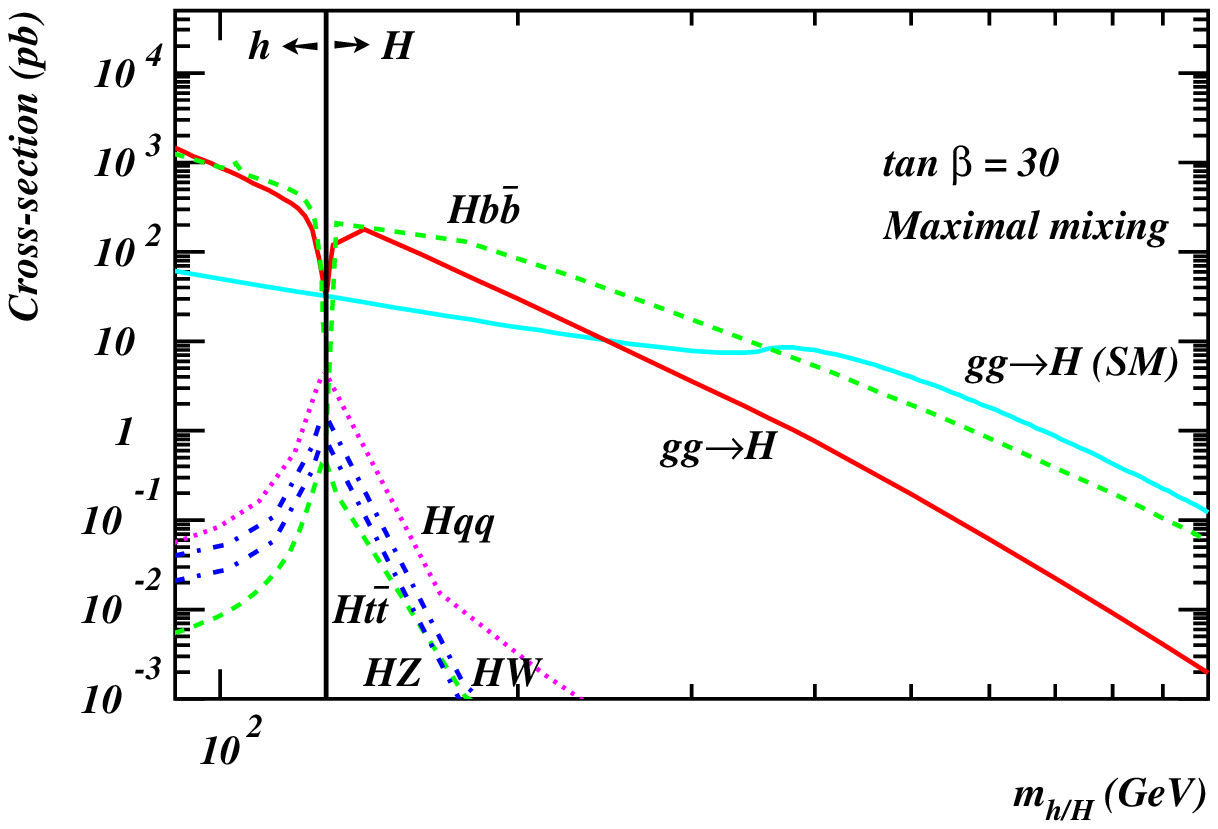}
\hfill
\includegraphics{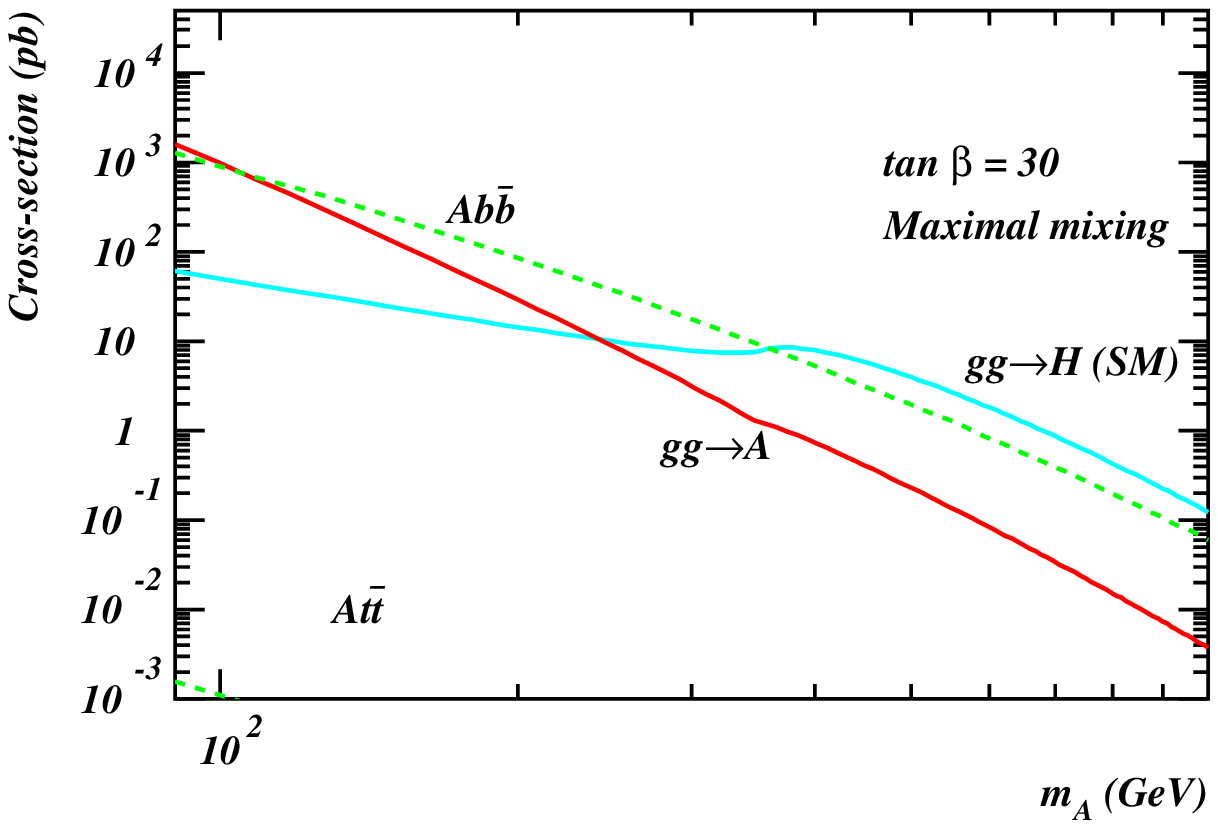}
}
\end{center}
\vskip1pc
\capt{\label{lhcmssmxsec} Neutral MSSM Higgs production
cross-sections
at the LHC [$\sqrt{s}=14$ TeV] for gluon fusion $gg\to \phi$,
vector-boson fusion $qq\to qqV^*V^* \to qqh$,
$qqH$, vector-boson bremsstrahlung $q\bar q\to V^* \to \hl V/\hh V$ and
the associated
production $gg,q\bar q \to \phi b\bar b/ \phi t\bar t$ including all known
QCD corrections, where $\phi=\hl$, $\hh$ or $\ha$~\protect\cite{hxsec,xseca}.
The four panels exhibited above show the cross-section in pb {\it vs.}
the Higgs mass, ranging from 90~GeV to 1~TeV, for
(a) $h,H$ production for $\tanb=3$, (b) $\ha$ production for $\tanb=3$,
(c) $\hl$, $\hh$ production for $\tanb=30$,  (d) $A$ production for
$\tanb=30$.  For comparison, the cross-section for gluon-gluon fusion
to a SM Higgs boson is also shown.}
\end{figure}

As in the case of Higgs branching ratios,
the predicted cross-sections are sensitive to the MSSM Higgs
parameters.  Again, we consider two representative values of
$\tan\beta$: a low value of $\tan\beta=3$ and a high value of
$\tan\beta=30$.  We then vary $\mha$, evaluate the
other Higgs masses, and compute each Higgs cross-section
as a function of the corresponding Higgs mass.  The Higgs
masses and cross-sections depend on other MSSM parameters through
radiative corrections.  
As in \Sec{sec:34}, we work in a maximal squark mixing scenario in
which the value of $\mhl$ for a fixed choice of $\tan\beta$ and $\mha$
is maximal.  In addition, because the squark masses
are assumed to be heavy (of order 1~TeV), potential supersymmetric
contributions to the one-loop Higgs-gluon-gluon
vertex (due to squark loops) are suppressed.
Cross-sections for neutral MSSM Higgs production at the Tevatron and
the LHC are shown in \figns{tevmssmxsec}{lhcmssmxsec} respectively.
The dominant Higgs production mechanism over much of the MSSM parameter
space is gluon-gluon fusion, which is
mediated by heavy top and bottom quark triangle loops and the
corresponding supersymmetric partners \cite{gunhab2,dds,djospi}.
The gluon-gluon fusion results shown in
\figns{tevmssmxsec}{lhcmssmxsec}
include NLO QCD corrections~\cite{spiramssm}.\footnote{A recent
computation of the NNLO QCD corrections to $\ha$ production via gluon
fusion exhibits a 20--30\% increase over the corresponding NLO
cross-section~\cite{ggannlo}.}  

The cross-sections for the production of
the neutral CP-even Higgs bosons ($\phi=\hl$ or $\hh$)
via gauge boson fusion $V^*V^*\to\phi$ ($V=W$ or $Z$)~\cite{32} and via
the process $q\bar q\to V^*\to V \phi$~\cite{23},
including first-order QCD corrections,\footnote{The supersymmetric-QCD
corrections due to the
exchange of virtual squarks and gluinos are
known to be small~\cite{djospi}.}  are also exhibited
in \figns{tevmssmxsec}{lhcmssmxsec}.
Recall that the CP-even scalar
$\phi$ has SM-like couplings to the vector bosons in two cases:
(i)~in the decoupling regime for the lightest Higgs boson, where
$\phi=\hl$ and (ii)~for large $\tanb$ and low $\mha$, where
$\phi=\hh$.  In either case, the SM-like Higgs scalar,
$\phi$, has a mass less than or about equal to 130~GeV, and the corresponding
cross-sections for $V^*V^*\to\phi$ and $q\bar q\to V^*\to V\phi$
are phenomenologically relevant.  The other (non-SM-like) CP-even scalar
has suppressed couplings to $VV$, and the corresponding cross-sections
are generally too small to be observed.

Higgs boson radiation off bottom quarks becomes important for large $\tanb$
in the MSSM, where
the Higgs coupling to bottom-type fermions is enhanced.
Thus, the
theoretical predictions, including full NLO computations, are
crucial for realistic simulations of the MSSM Higgs signals in these
channels.\footnote{As mentioned in the corresponding discussion for
$q\bar q$, $gg\to b\bar b\hsm$ [see \Sec{sec:221}], one also needs to
evaluate $gb \to b\phi$ and $b\bar b\to \phi$ with suitable subtraction
of the logarithms due to quasi-on-shell quark exchange (to avoid double
counting) in order to obtain the total inclusive cross-section for
$\phi$ production.  For example, the $\tan\beta$ enhancement
can lead to copious $s$-channel production of Higgs bosons via $b$-quark
fusion \cite{DSSW,bhy}.}
Moreover, as discussed in \Sec{sec:333}, vertex
corrections to the $b \bar b \phi$ coupling
play a very important role in enhancing or suppressing (depending on
the MSSM parameters) these production
cross-sections at large $\tanb$ \cite{CMW1,CMW2,bdhty}.

\begin{figure}[t!]
\begin{center}
\resizebox{\textwidth}{!}{
\scalebox{1.0}[0.75]{
\includegraphics*[height=10cm]{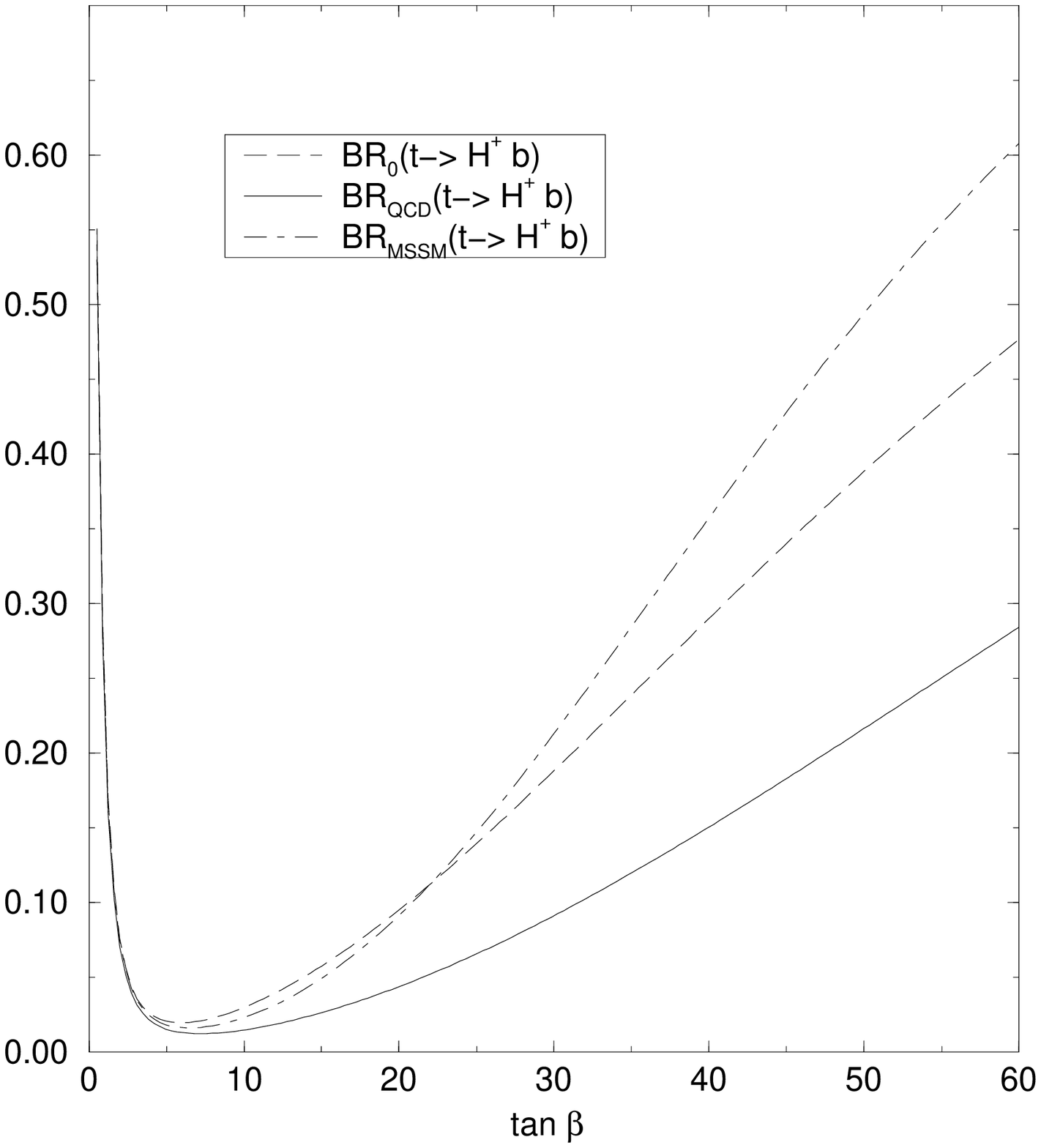}}
\hspace*{3mm}
\includegraphics*[height=10cm,angle=90]{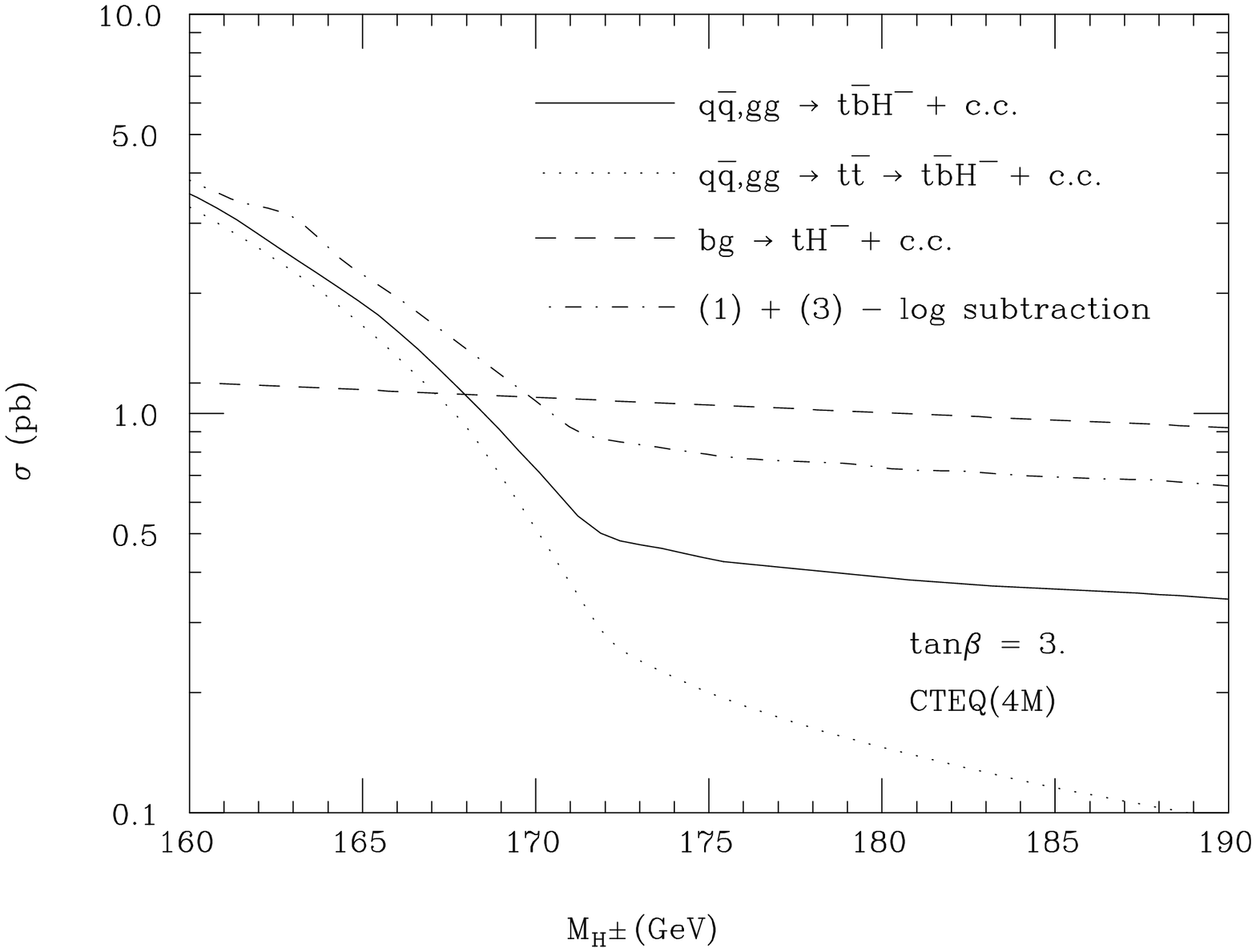}
}
\end{center}
\capt{\label{toptohiggs} 
(a) Branching ratio for $t\to bH^+$ in the MSSM as a function of
$\tan\beta$ for $\mhpm=120$~GeV.
The three curves shown are the results of a computation
that (i) is at tree-level; (ii) includes one-loop QCD corrections; and
(iii) incorporates both one-loop QCD, electroweak
and the effects of MSSM particle
exchange (taken from \protect\Ref{solaetal}); the parameters
chosen in (iii) correspond to a rather light supersymmetric
spectrum: $M_{\tilde g}=300$~GeV, $M_{\tilde t_1}=100$~GeV,
$M_{\tilde b_1}=150$~GeV, $A_t=A_b=300$~GeV, $M_2=150$~GeV, and
$M_{\tilde u}= M_{\tilde\nu}=200$~GeV.
Curves (i) and (ii) are also applicable to a Model-II
two-Higgs doublet model without supersymmetry.  (b) The
charged Higgs production cross-section at the LHC near the threshold
for $t\to bH^+$ for $\tan\beta=3$ (taken from \protect\Ref{morettietal}).}
\end{figure}

We now turn to charged Higgs production.
If $\mhpm<m_t-m_b$, then $H^\pm$ can be
produced in the decay of the top quark via $t\to bH^+$ (and $\bar t\to
\bar b H^-$) \cite{TDECAY}.  The $t \to bH^+$ decay
mode can be competitive with the dominant \SM\ decay mode, $t\to
bW^+$, depending on the value of $\tan\beta$, as shown in
\fig{toptohiggs}(a) for $\mhpm=120$~GeV.
This figure, taken from \Ref{solaetal} illustrates the effects of including
one-loop radiative corrections.  The curved labeled BR$_{\rm QCD}$,
which incorporates the one-loop QCD corrections
(first computed in \Ref{toptochhiggsdecay}), is applicable to a more
general (non-supersymmetric)
Type-II two-Higgs doublet model [based on the tree-level Higgs-fermion
couplings of \eqs{qqcouplingsa}{hpmll}].
Note that the supersymmetric corrections
can be particularly significant at large $\tan\beta$ due to the effect
of $\Delta_b$ [see \eq{hmtb}]
depending on the choice of MSSM parameters.
A full one-loop calculation of $\Gamma (t\rightarrow
H^{+}\,b)$ in the MSSM including
all sources of large Yukawa couplings can be found in
\Refs{susytoptochhiggsdecay}{solaetal}.
A treatment including
resummation of the leading QCD quantum effects and the dominant
contributions from loop effects arising from supersymmetric particle
exchange can be found in ref.~\cite{chhiggstotop2}.

For $\mhpm<m_t-m_b$, the
total cross-section for charged Higgs production (in the narrow-width
approximation) is then given by:\footnote{Note that if one evaluates
$\sigma(p\bar p\to H^+\bar t b+X)$ in the region of $\mhpm<m_t-m_b$,
one obtains the {\it single} charged Higgs
inclusive cross-section,
$\sigma(p\bar p\to H^+ +X)={\rm BR}(t\to bH^+)\sigma(p\bar p
\to t\bar t+X)$,
rather than full charged Higgs inclusive cross-section
of \eq{sighplus}.  The latter is not quite a factor of two
larger than the former since $X$ can contain a charged Higgs boson; one
must subtract off $[{\rm BR}(t\to bH^+)]^2\sigma(p\bar p\to t\bar t+X)$
to avoid double-counting.}
\begin{equation} \label{sighplus}
\sigma(p\bar p\to \hpm+X)=\left(1-[{\rm BR}(t\to bW^+)]^2\right)
\sigma(p\bar p\to t\bar t+X)\,.
\end{equation}
With $\sigma(p \bar p \rightarrow t \bar t) \simeq 7$~pb at $\sqrt{s} = 2$
TeV and $\sigma(pp \rightarrow t \bar t) \simeq 1$ nb at $\sqrt{s} =
14$~TeV~\cite{ttbarxsec},
roughly 1400 $t\bar t$ pairs per detector will be produced per year
in Run 2a of the Tevatron (assuming a yearly luminosity of
2~fb$^{-1}$), while about $10^7$--$10^8$ $t\bar t$ pairs will be
produced at the LHC (assuming a yearly luminosity of
10--100~fb$^{-1}$).
Folding in the top quark branching ratio,
it is a simple matter to compute the inclusive charged
Higgs cross-section.  For values of $\mhpm$ near $m_t$, the width
effects are important.  In addition, the full $2\to 3$ processes $p\bar p\to
H^+\bar t b+X$ and $p\bar p \to H^-t \bar b+X$ must be considered.
In this case \eq{sighplus} no longer provides an accurate estimate of
the charged Higgs cross-section~\cite{Guchait:2001pi},
as illustrated in \fig{toptohiggs}(b) (taken from \Ref{morettietal}).
The results of \fig{toptohiggs}(a) imply that for $\mhpm<m_t-m_b$, the
discovery of the charged Higgs boson at the Tevatron and/or LHC
(given sufficient luminosity)
is possible if $\tanb\gg 1$ or $\tanb\lsim 1$ (the latter is
theoretically disfavored).  The precise bound on $\tanb$
(as a function of $\mhpm$) depends
somewhat on the details of the other MSSM Higgs parameters.

If $\mhpm>m_t-m_b$, then charged Higgs boson production occurs
mainly through radiation off a third generation quark.
Single charged Higgs associated
production proceeds via the $2\to 3$ partonic processes
$gg,\, q\bar q\to t\bar b H^-$ (and the charge conjugate final state).
As in the case of $b\bar b\hsm$ production,
large logarithms $\ln(\mhpm^2/m^2_b)$ arise
for $\mhpm\gg m_b$ due to quasi-on-shell
$t$-channel quark exchanges, which can be resummed by absorbing them
into the $b$-quark parton densities.  Thus, the proper procedure for
computing the charged Higgs production cross-section is to add the
cross-sections for $gb\to tH^-$ and $gg\to t\bar b H^-$
and subtract out the large logarithms accordingly
from the calculation of the $2\to 3$ process \cite{soper,OT}.
This procedure avoids double-counting of the large logarithms at
${\cal O}(\alpha_s)$, and correctly resums the leading logs to all
orders.  In particular,
the contribution to the total cross-section coming from the kinematical
region of the gluon-initiated $2\to 3$ process in which one of the
two gluons splits into a pair of $b$-quarks (one of which is
collinear with the initial proton or antiproton), is incorporated into
the $b$-quark parton density.  A cruder calculation would
omit the contribution of the $2\to 2$ process and simply include
the results of the unsubtracted $2\to 3$ process.  The latter procedure
would miss the resummed leading logs that are incorporated into the
$b$-quark density.  However, the numerical difference between the two
procedures is significant only for $\mhpm\gg m_t$.

\begin{figure}[t!]
\begin{center}
\resizebox{\textwidth}{!}{
\includegraphics*{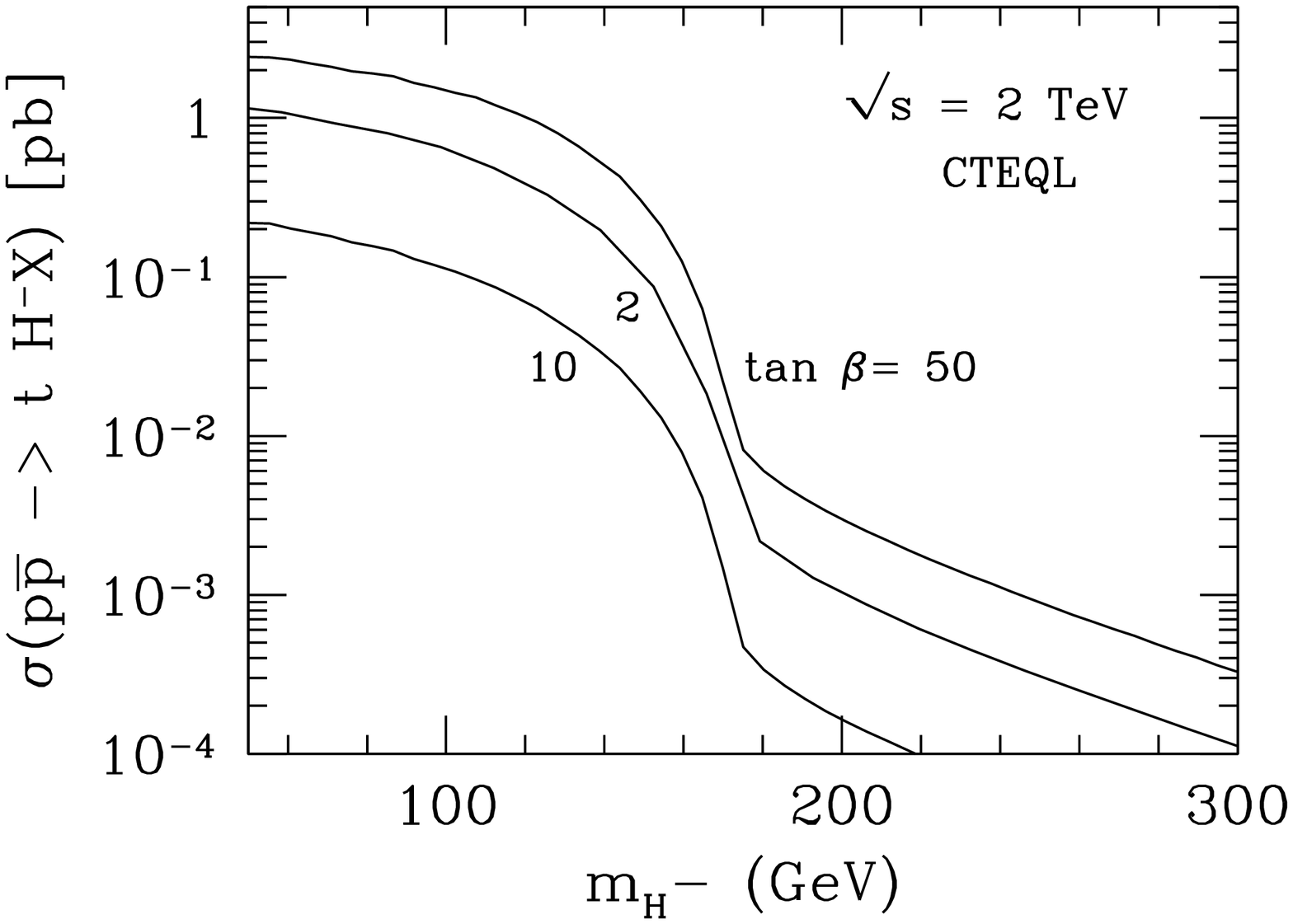}
\hspace*{3mm}
\includegraphics*{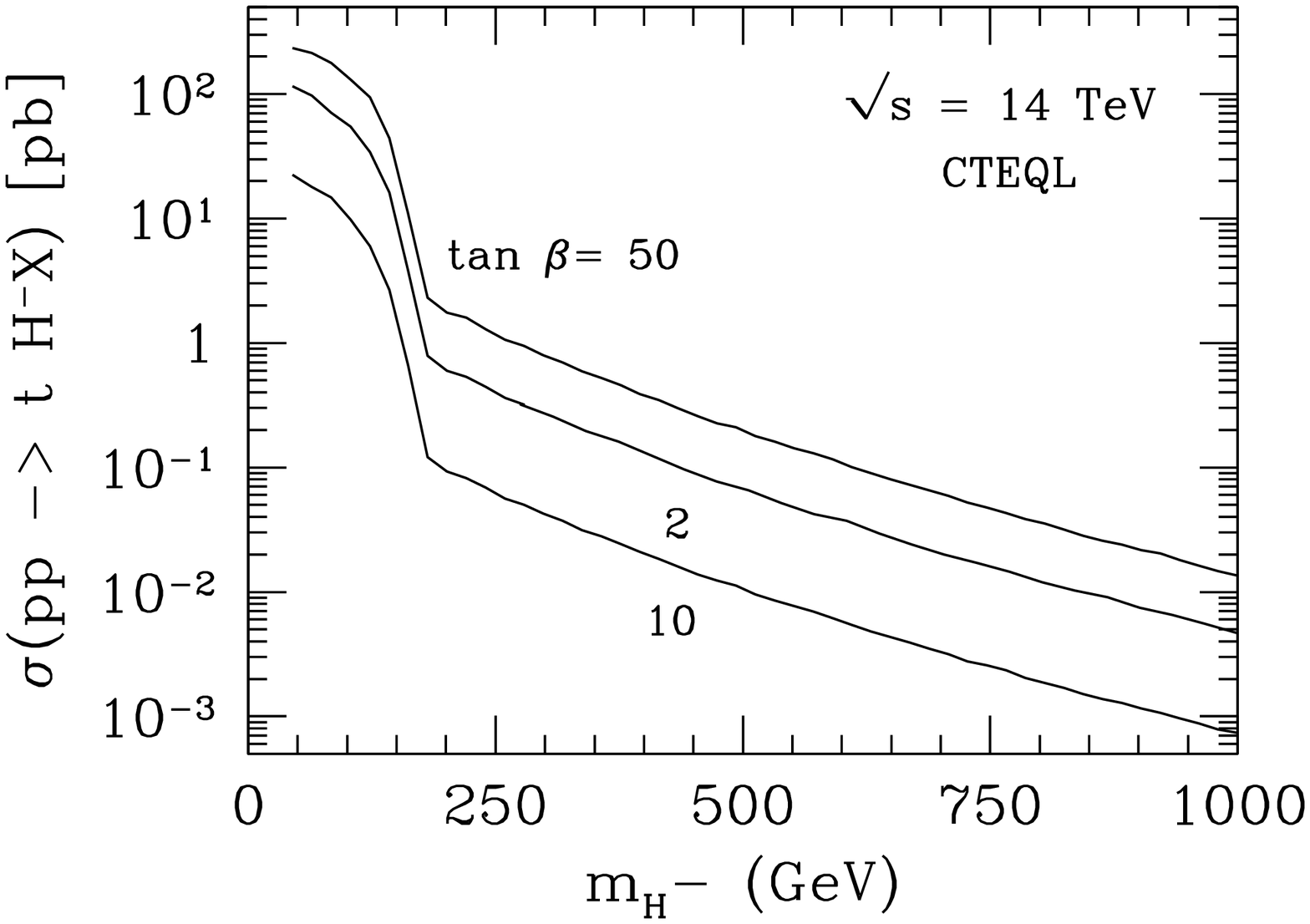}
}
\end{center}
\vskip1pc
\capt{\label{TeVsum}
The leading-order production cross-sections for charged Higgs production
at (a) the Tevatron ($p\bar p\to t\bar b H^- +X$)
and (b) the LHC ($\pp\to t\bar b H^- +X$) are shown
as a function of $\mhpm$ for
three values of $\tan\beta = 2$, 10 and 50.
The cross-sections are obtained by adding the
contribution of the $2 \to 2$ processes, $g b \to t H^-$, to those of
the $2 \to 3$ processes, $ g g \to t \bar{b} H^-$ and
$ q \bar{q} \to t \bar{b} H^-$ (suitably subtracted to avoid double
counting).  Renormalization and factorization scales have been both
set to $m_t+\mhpm$.
These results are taken from \protect\Ref{francesca}.}
\end{figure}

The single inclusive charged Higgs cross-sections at the Tevatron and LHC
are exhibited in \fig{TeVsum} as a function of the charged Higgs mass, for
$\tan\beta=2$, 10 and 50.  Note that the cross-sections shown include
the region of charged Higgs mass below $\mhpm=m_t-m_b$ corresponding
to the case discussed above where the charged Higgs cross-section is
dominated by $t\bar t$ production followed by $t\to bH^-$.
These results are based on the calculations of
\Ref{francesca} and include the contributions of
the $2\to 2$ process and suitably subtracted $2\to 3$ process as described
above.  Similar results have also been obtained in \Ref{bggs}.  The
impact of the leading electroweak and MSSM radiative corrections has
been studied in \Ref{chhiggstotop}.  In
addition, the NLO QCD corrections to the $2\to 2$ process
$gb\to H^+ t$ have recently been evaluated~\cite{zhu}.  These
corrections typically increase the tree-level cross-section by
a factor of 1.3 to 1.6, depending on the
value of the charged Higgs mass and $\tan\beta$, with some
additional dependence
on the choice of renormalization and factorization scales.

\begin{figure}[t!]
\begin{center}
\resizebox{\textwidth}{!}{
\includegraphics*{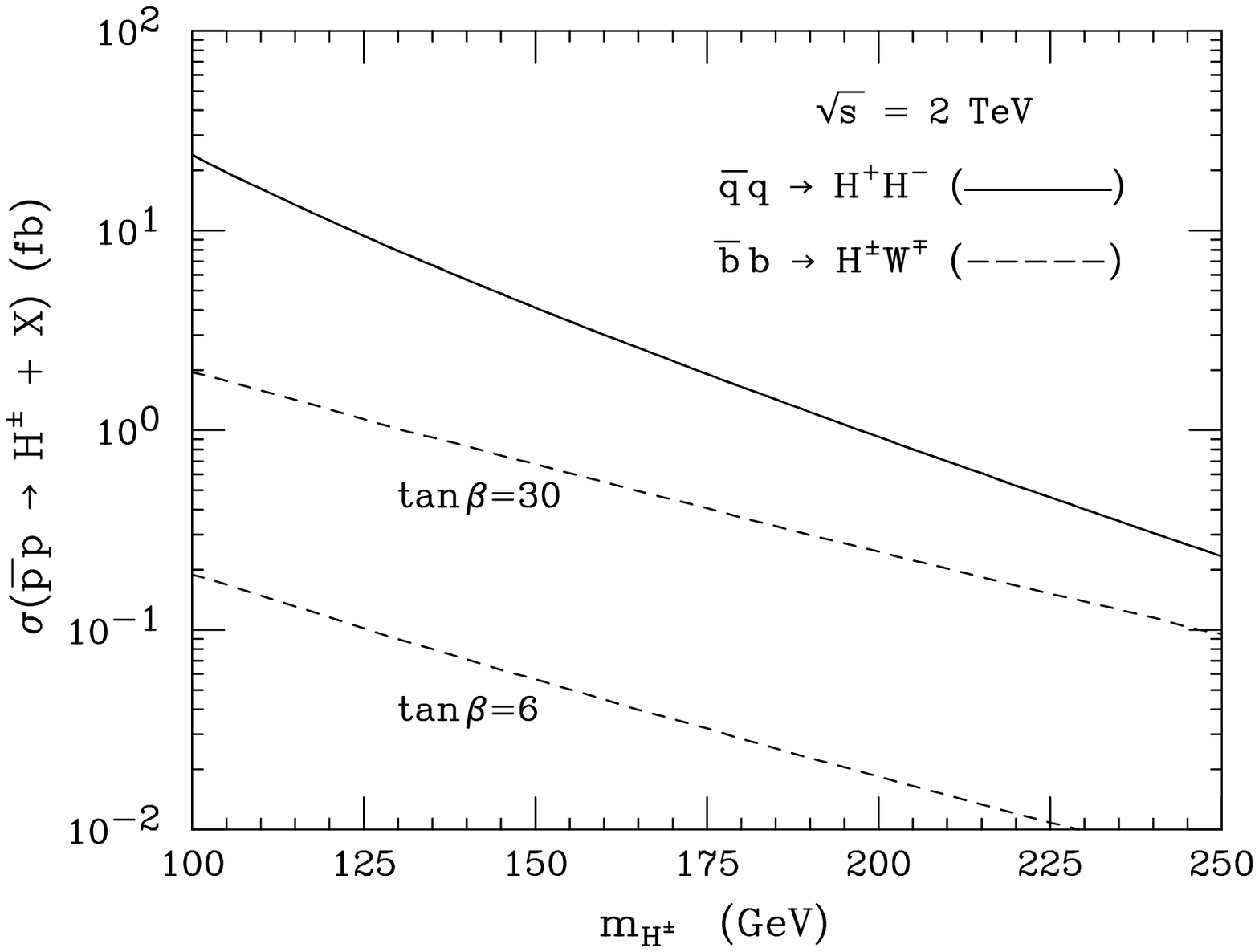}
\hspace*{3mm}
\includegraphics*{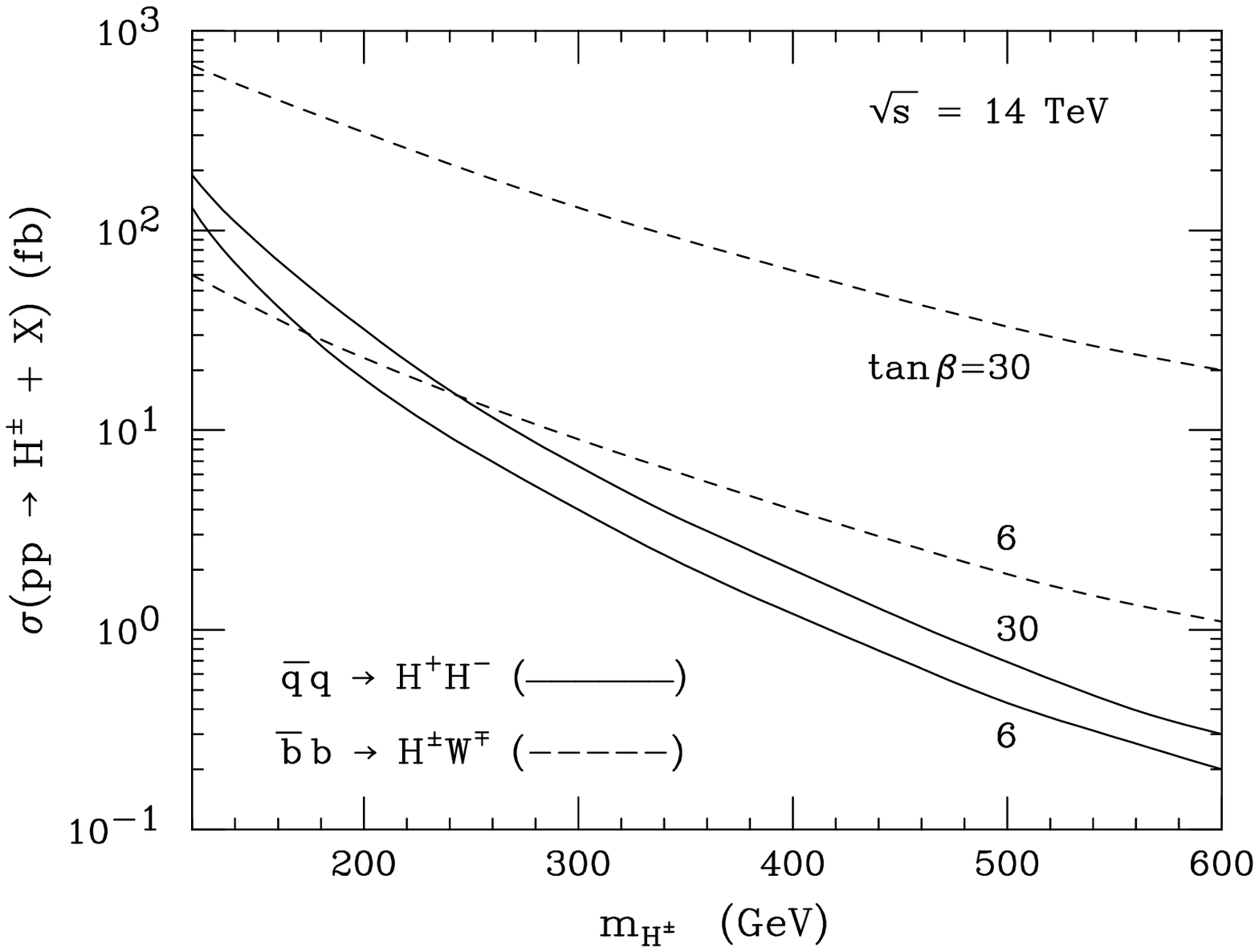}
}
\end{center}
\capt{\label{kniehl} 
Total cross-section (in fb) for inclusive production of 
(i)~$H^+ H^-+X$~\protect\cite{kniehl1} (solid line) and
(ii)~$\hpm W^\mp +X$~\protect\cite{kniehl2} (dashed lines)
as a function of $\mhpm$ for $\tan\beta=6$ and 30.
Curves for (a) $p\bar p\to H^+ H^-+X$ at the Tevatron and
(b) $pp\to H^+ H^-+X$ at the LHC are exhibited.  Note that the
dependence of process (i) on $\tan\beta$ is negligible
at the Tevatron, while there is some $\tan\beta$ dependence at the LHC
due to the enhancement of $b\bar b\to H^+H^-$ at large $\tan\beta$.
The contribution of $b\bar b$ annihilation to process (ii) dominates
over the $gg$ fusion scattering mechanism.
$M_t=174.3$~GeV and a fixed $b$-quark pole mass
of $M_b=4.7$~GeV are used to fix the Higgs--fermion Yukawa coupling.
The leading-order CTEQ5L parton distribution functions are used.}
\end{figure}

Associated production of a charged Higgs boson and a $W^\pm$ can occur
via $b\bar b$ annihilation and $gg$-fusion \cite{kniehl2}.  The
contribution of
$b\bar b$ annihilation to $\sigma(p\bar p\to \hpm W^\mp+X)$
[$\sigma(pp\to \hpm W^\mp+X)$] at the Tevatron [LHC] (both charge
combinations are included)
are shown as function of the charged Higgs mass for $\tan\beta =
6$ and 30 in \fig{kniehl}.
The loop-induced $gg$ fusion contribution is significantly suppressed
relative to the tree-level $b\bar b$ annihilation
if $\tan\beta\gsim 6$, independently of the value of $\mhpm$~\cite{gluehw}.

Charged Higgs bosons can also be produced in pairs
via Drell-Yan $q\bar q$
annihilation.  The dominant contribution, which  arises from $u\bar u$
and $d\bar d$ annihilation into a virtual photon or $Z$, is
independent of $\tan\beta$.  Some $\tan\beta$ dependence enters through
$b\bar b$ annihilation via $t$-channel top-quark exchange, although this
effect is more than one order of magnitude suppressed relative to the
dominant contribution at the Tevatron.  The $b\bar b$ annihilation is
more significant at the LHC (at large $\tan\beta$ where the $H^-t\bar
b$ coupling is enhanced).
The tree-level results for $\sigma(p\bar p\to H^+
H^- +X)$ at the Tevatron and $\sigma(pp\to H^+H^- +X)$ at the LHC
are shown in \fig{kniehl}.  These results are obtained~\cite{kniehl1} with
the Higgs--fermion Yukawa coupling based on
a fixed $b$-quark pole mass of $M_b=4.7$~GeV.  The 
contribution of the loop-induced $gg\to
H^+H^-$ is typically less important than that of $q\bar q$
annihilation~\cite{gluehphm1,gluehphm2}, and 
is not included in \fig{kniehl}.  However, the $gg$ fusion
contribution can become significant at large $\tanb$, 
with further enhancements
in some regions of MSSM parameter space in which the squarks 
(which appear in the loop) are light and strongly mixed~\cite{gluehphm2}.
Nevertheless, the inclusive
$H^+H^-$ cross-section lies below the cross-section for single charged
Higgs associated production ({\it c.f.} \figns{TeVsum}{kniehl}).

Finally, one can compute the cross-sections for double neutral Higgs
production at hadron colliders.  These include the inclusive
production of $\hl\hl$, $\hl\hh$, $\hh\hh$, $\hl\ha$, $\hh\ha$ and
$\ha\ha$.  Cross-sections can be found in
refs.~\cite{ggdoublehiggs,doublehiggs,doublehiggstwo} and
\cite{doublehiggs1,doublehiggs2} (QCD corrections to these
cross-sections are evaluated in \Ref{doublehiggs}).
In general, the rates for these processes are considerably smaller than
for the corresponding single Higgs production rates.  However, in
certain regions of supersymmetric parameter space, squark loops can
enhance the cross-section for pair production of two CP-even Higgs
bosons by as much as two orders of magnitude~\cite{doublehiggs1}.
In some cases, observation
of double Higgs production provides some information on three-Higgs
couplings.  For example, for low to moderate values of $\tan\beta$,
gluon fusion to a virtual Higgs boson, which splits into $\hl\hl$, is
dominant over $b\bar b\to\hl\hl$.   Thus, the overall rate for
$pp\to\hl\hl+X$ would provide a measure of the $\hl\hl\hl$ vertex.

Additional sources for Higgs boson production can arise from the decay
of supersymmetric particles into final states containing one or more
Higgs boson in the decay chain~\cite{scottkon}.  These processes
depend in detail on the details of the supersymmetric particle
spectrum and their couplings.   For example, the production of $\hl$
in supersymmetric particle decay followed by
the decay $\hl\to b\bar{b}$ can yield a signal above background
at LHC~\cite{susytohiggs}.
Processes of this type provide additional
channels for possible Higgs discovery and precision study, and deserve
further analysis.

\subsubsection{Benchmarks for Higgs searches}
\label{sec:352}

In the search for the MSSM Higgs bosons, 
one must first search for the lightest Higgs scalar,
which is expected (in almost all cases) to be the neutral
CP-even scalar, $\hl$.  In the decoupling region of the MSSM Higgs
parameter space (where $\mha\gg\mz$), the
search techniques already outlined for $\hsm$ are relevant for $\hl$,
since the properties of $\hl$ approximately coincide with
those of the SM Higgs boson.  The $\hl$
discovery reach can be mapped out as a region of $\mha$--$\tan\beta$
parameter space, since these two parameters (along with the
MSSM parameters that determine the size of the radiative
corrections) fix the value of $\mhl$.  Next, it is critical to identify
deviations of the properties of $\hl$ from those of $\hsm$.  Positive
evidence for such a deviation would signal the existence of additional
scalar states of the non-minimal Higgs sector.  The difficulty of this
step depends on how close the
model is to the decoupling limit.  
After the discovery of $\hl$, the Higgs boson search will focus on the
non-minimal Higgs states of the model.

For values of
$\mha\sim\mz$, all the Higgs bosons of the MSSM are of a similar order
of magnitude, and the properties of $\hl$ will no longer resemble those
of $\hsm$.  In principle, one can then discover multiple scalar
states in one experiment.  Since the two CP-even scalars share the
coupling to vector boson pairs [\eq{vvsumrule}], one may identify the
CP-even scalar whose squared-coupling to $VV$ is larger than
$0.5 g_{\hsm VV}^2$.
The Tevatron and LHC production cross-sections of this
scalar (compared to that of $\hsm$) are reduced by no more than 50\% (by
assumption), while the Higgs branching ratio into $b\bar b$ is similar
to that of $\hsm$ over most of the MSSM parameter space.  Thus, the
Tevatron and LHC SM Higgs search results
also apply here modulo minor
modifications (which account for the somewhat suppressed production
cross-section and the effects of supersymmetric corrections to the
third generation  Yukawa couplings).

The general MSSM parameter space involves many {\it a priori} unknown
parameters.
In practice, only a small subset of these parameters govern the
properties of the Higgs sector.  Nevertheless, a
full scan of this reduced subset is still a formidable task.
However, a detailed study of a few appropriately chosen points
of the parameter space can help determine the ultimate MSSM Higgs
discovery reach of the Tevatron and LHC.
It is convenient to choose a set of benchmark MSSM parameters that
govern the Higgs radiative corrections~\cite{benchmark,benchmark2}.
These include the supersymmetric Higgs mass parameter
$\mu$, the third generation squark mixing parameters, $A_t$ and $A_b$,
the gluino mass $M_{\tilde g}$, the diagonal
soft-supersymmetry-breaking third generation squark squared-masses
(which we take for simplicity to be degenerate and equal to $\MSUSY$),
and the top quark mass (which is held fixed at $m_t=174.3$~GeV). 
The {\it maximal mixing} benchmark scenario, is
defined as the one in which the squark mixing parameters are such that
they maximize the value of the lightest CP-even Higgs boson mass for
fixed values of $\mha$, $\tan \beta$ and $\MSUSY$.  Here, we choose
$X_t\equiv A_t-\mu\cot\beta\simeq \sqrt{6}$, $A_b=A_t$,
$M_2 = -\mu = 200$~GeV and $M_{\tilde{g}}=M_{\rm SUSY}= 1$ TeV 
[corresponding to $\mhmax=129$~GeV].

The maximal mixing scenario poses a challenge for Higgs searches,
since the predicted Higgs mass takes on its maximal value for a given set of
MSSM parameters.  However, different regions of the MSSM Higgs
parameter space pose new challenges.
For example, regions of parameter space exist
in which the CP-even neutral Higgs boson with
SM-like couplings to the $W,Z$ and $t$  has suppressed couplings to
$b\bar b$. The benchmark scenario denoted by
``suppressed $V \phi \to V b\bar b$ production''
is an example of this behavior.  In this case, we take
$\mu=-A_t=1.5$~TeV, $A_b=0$, $M_2 =200$~GeV and
$M_{\tilde{g}}=M_{\rm SUSY}= 1$ TeV
[corresponding to $\mhmax=120$~GeV].
The regions of strongly
suppressed ${\rm BR}(\phi\to b \bar b)$ correspond to a suppressed
$\hh b\bar b$ coupling at lower $\mha$ and a suppressed
$\hl b\bar b$ coupling at larger $\mha$.  In particular,
the suppression for large $\tan\beta$ extends to
relatively large values of $\mha\sim 300$~GeV,
indicating a delay in the onset of
the decoupling limit.
Moreover, in the suppressed $V\phi\to Vb\bar b$ benchmark
scenario, {\it all} the Higgs couplings to $b\bar b$ are generally
suppressed, since $0<\Delta_b\ll 1$ and $\sin 2\alpha\simeq 0$.
From the analytic formulae, it can be deduced that $\mu A_t<0$ and
large values of $|A_t|$, $|\mu|$ and $\tan\beta$ are needed.

The coverage in the $m_A$--$\tan\beta$ plane
by different Higgs production and decay channels can
vary significantly, depending on the choice of MSSM parameters.
In the last example in which the CP-even Higgs boson with the larger
coupling to the $W$ and $Z$  has a strongly suppressed coupling to
bottom quarks, the Higgs searches at the Tevatron will become
more problematical, while the LHC search for Higgs production followed
by its decay into photons becomes more favorable~\cite{CMW2}.
At the same time the LHC Higgs discovery reach via
vector bosons fusion to Higgs production followed by its decay
into $\tau^+\tau^-$ pairs can be significant~\cite{tautau}.

\subsubsection{MSSM Higgs Boson searches at the Tevatron}
\label{sec:353}

We first consider the Tevatron MSSM Higgs search.
Specifically, we make use of the Tevatron $\hsm$ search techniques,
where $\hsm$ is replaced by either $\hl$ or $\hh$.  If $\tanb\gg 1$, a
new search mode becomes viable, due to the possibility of enhanced
couplings of the neutral Higgs boson states to 
$b\bar b$~\cite{tevbbbb,bdhty}.  Thus, we
also consider the possibility of the $b\bar b\phi\to b\bar b b\bar b$
signature, where $\phi=\hl$, $\hh$, and/or $\ha$.
If $\tan\beta$ is large, two of the neutral Higgs boson states,
$\phi=\ha$ and $\hl$ [$\hh$] are
produced with enhanced rates if $\mha\lsim\mz$ [$\mha\gg\mz$],
as noted below \eq{hpmll}.
We may combine the results for the various channels to provide summary plots
of the MSSM Higgs discovery reach of the upgraded Tevatron
collider.  We consider here the results based on a generic MSSM
analysis~\cite{tevreport}; see \Ref{bht} for a similar analysis in the
context of a variety of models of supersymmetry breaking.  In the
latter case, the results obtained will be somewhat more constraining
than the generic analysis, since the supersymmetry-breaking parameters
that control the radiative corrections to Higgs masses and
couplings are no longer arbitrary.

\begin{figure}[b!]
\begin{center}
\resizebox{0.98\textwidth}{!}{
\includegraphics[width=8cm]{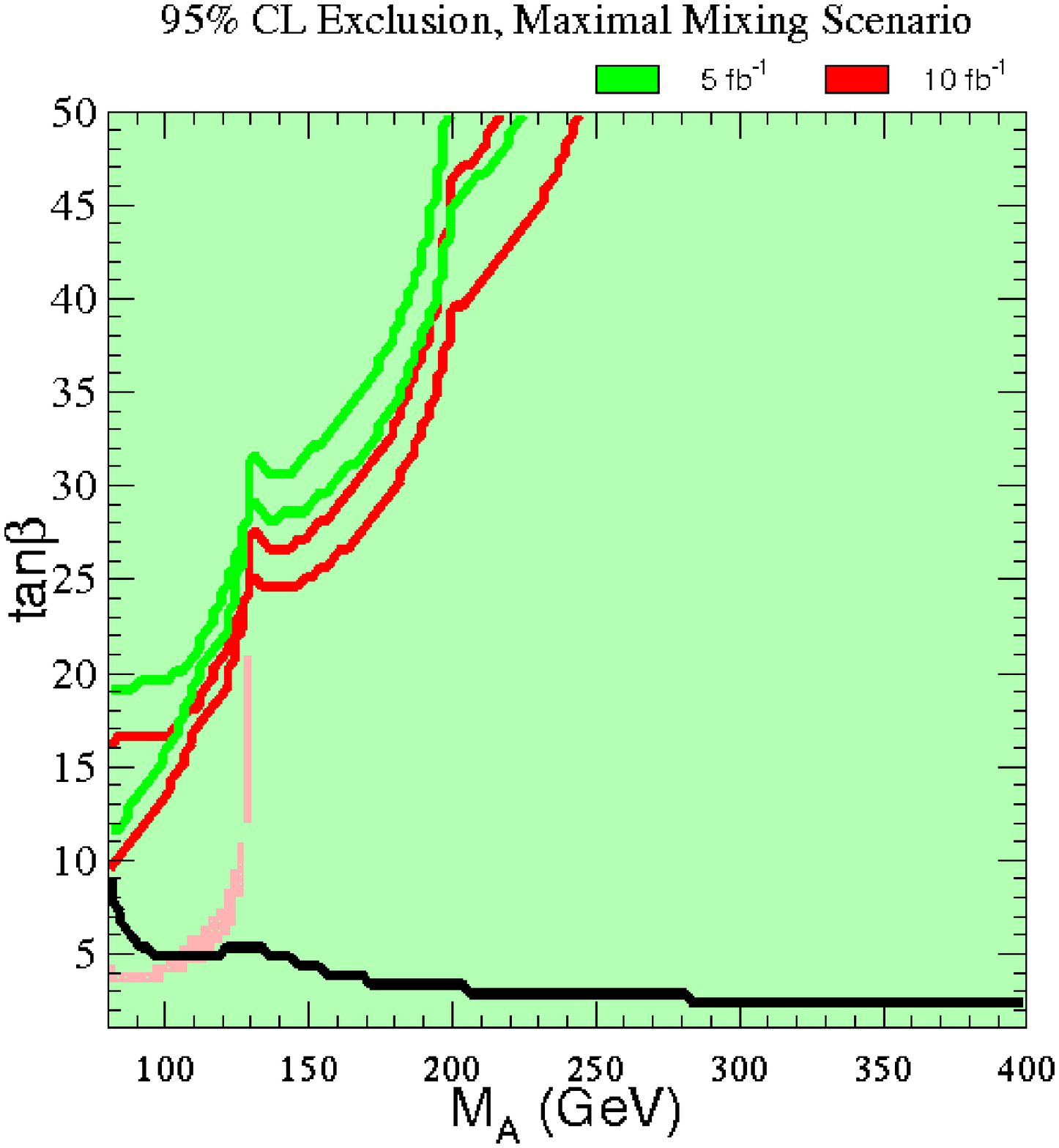}
\hskip1pc
\includegraphics[width=8cm]{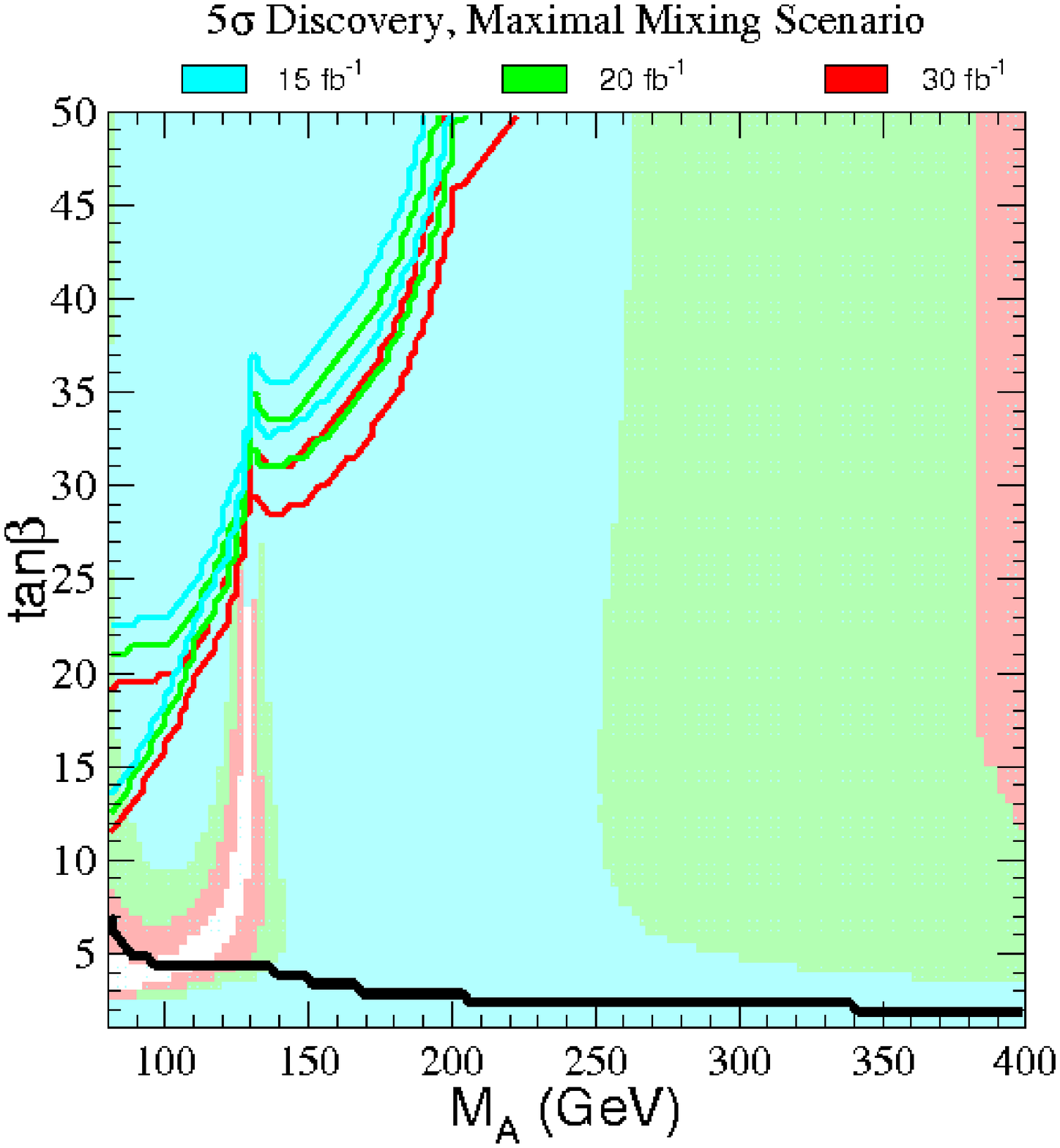}
}
\end{center}
\capt{\label{fullmhmax95} (a) 95$\%$ CL exclusion region
and (b) $5\sigma$ discovery region in the $\mha$--$\tan \beta$
plane, for the maximal mixing benchmark scenario (see \Sec{sec:352})
and two different search channels:
$q\bar q\to V\phi$ [$\phi=\hl$, $\hh$], $\phi\to b\bar b$
(shaded regions) and
$gg$, $q\bar q\to b\bar b\phi$ [$\phi=\hl$, $\hh$, $\ha$],
$\phi\to b\bar b$ (region in the upper left-hand corner bounded by the
solid lines; the two sets of lines correspond to CDF and D\O\ simulations).
The region below the solid black line near the bottom
of the plot is excluded by the absence of $e^+e^-\to Z\phi$
events at LEP.  Taken from \protect\Ref{tevreport}.
}
\end{figure}

\begin{figure}[t!]
\begin{center}
\resizebox{0.98\textwidth}{!}{
\includegraphics[width=8cm]{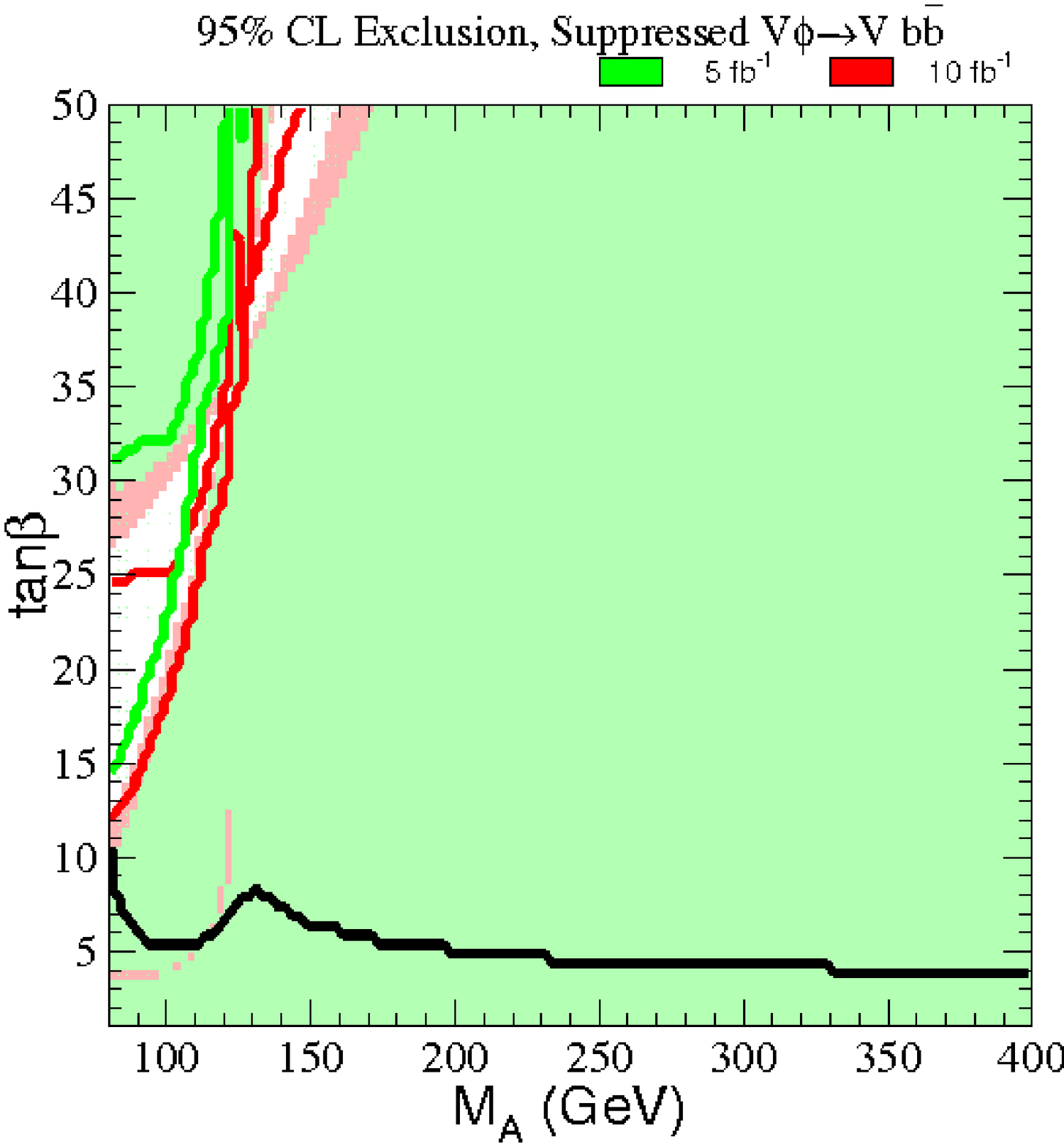}
\hskip1pc
\includegraphics[width=8cm]{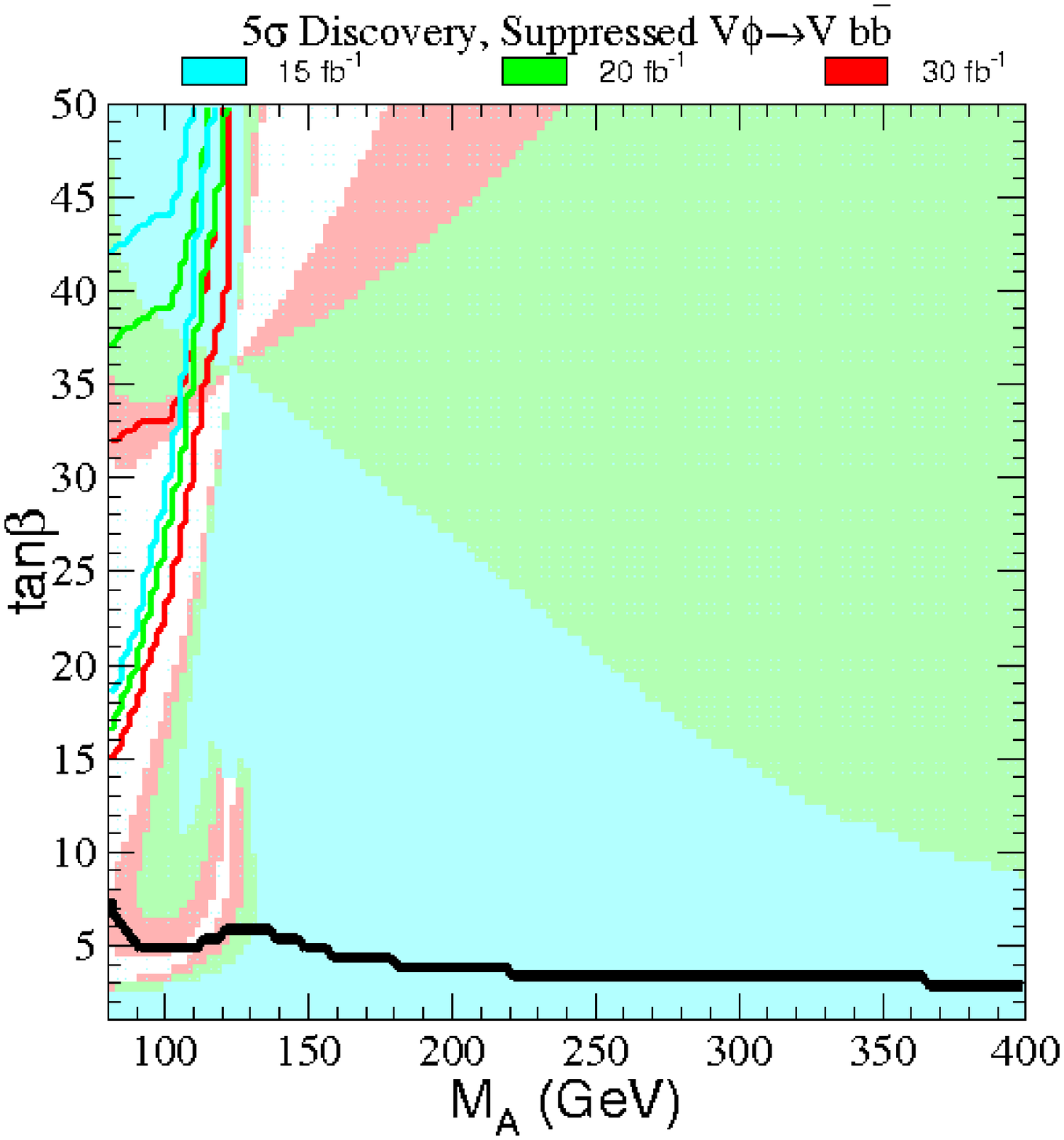}
}
\end{center}
\capt{\label{fullmix95} 
The same as \protect\fig{fullmhmax95} but for the suppressed
$V \phi \to V b\bar b$ production benchmark scenario
of \protect\Sec{sec:352}. Taken from \protect\Ref{tevreport}. 
}
\end{figure}

In \figns{fullmhmax95}{fullmix95},
we show the regions of 95$\%$ CL Higgs exclusion
and $5\sigma$ Higgs discovery on the $\mha$--$\tan \beta$
plane, for two representative MSSM parameter choices,
via the search of neutral Higgs bosons in the
channels: $q\bar q\to V\phi$ [$\phi=\hl$, $\hh$], $\phi\to b\bar b$
(shaded regions) and
$gg$, $q\bar q\to b\bar b\phi$ [$\phi=\hl$, $\hh$, $\ha$],
$\phi\to b\bar b$ (region in the upper left-hand corner bounded by
the solid lines), for different integrated
luminosities as indicated by the color coding.
The shaded regions presented in these figures reflect the results
of the SHW simulation of $q\bar q\to V\phi$
improved by neutral network techniques~\cite{tevreport}.
The two sets of lines (for a given color) bounding the regions accessible
by the $b\bar b\phi$ search correspond to the CDF and D\O\ simulations,
respectively.  The solid black line near the bottom
of each plot indicates the lower limit of $\tan\beta$ (as a function
of $\mha$) based on the absence of observed $e^+e^-\to Z\phi$
events at LEP~\cite{lepsusyhiggs}.
Note the importance of the complementarity between the
$q\bar q\to V\phi$ and $q\bar q\to b\bar b\phi$ channels
for improving the coverage of the MSSM parameter space in the
low $\mha$ region in \fig{fullmix95}(a).
The results of figs.~\ref{fullmhmax95}(a) and \ref{fullmix95}(a)
demonstrate that 5~fb$^{-1}$ of integrated
luminosity per experiment will be sufficient to cover
nearly all of the MSSM Higgs parameter space at 95\% CL
in the benchmark scenarios specified above.

To assure discovery of a CP-even Higgs boson at the 5$\sigma$ level,
the luminosity requirement becomes very important.
Figs.~\ref{fullmhmax95}(b) and \ref{fullmix95}(b)
show that
a total integrated luminosity of about 20~fb$^{-1}$ per experiment is
necessary in order to assure a significant, although not exhaustive,
coverage of the MSSM parameter space. In general, we observe that
the complementarity between the two channels,
$q\bar q\to V\phi$ and  $q\bar q\to b\bar b\phi$,
is less effective in assuring discovery of a Higgs boson as compared
with a 95\% CL Higgs exclusion.
This is due to
the much higher requirement of total integrated
luminosity combined with the existence of MSSM parameter regimes
which can independently suppress both Higgs production channels.
\Fig{fullmix95} exhibits one of the most difficult regions of
MSSM parameter space for Higgs searches at the Tevatron collider.
Nevertheless, even in this case,
a very high luminosity experiment can cover a significant fraction
of the available MSSM parameter space.

If explicit CP violation occurs
through nonzero phases of the supersymmetry breaking parameters, then
the three neutral Higgs bosons are a combination of CP-even and CP-odd
states and the phenomenology can become much more complicated.
In particular the couplings of the neutral Higgs bosons to the $W$ and $Z$
bosons are now shared by the three Higgs bosons and it may well be
that the lightest Higgs has such a weak coupling to the vector bosons
that it would have been missed at LEP and will be elusive at the
Tevatron.
\Fig{cphtev} shows an interesting example where the effects of CP
violation are such that for CP-violating phases of the parameter $A_t$
of about $90^\circ$,
the lightest Higgs boson cannot be detected at the Tevatron even though
its mass is below 100 GeV, but the second lightest Higgs has
SM-like couplings to the $W$ and $Z$ and thus can be detected
if sufficient luminosity is provided.

\subsubsection{MSSM Higgs Searches at the LHC}
\label{sec:354}

\begin{figure}[t!]
\begin{center}
\resizebox{\textwidth}{!}{
\scalebox{1.0}[1.25]{
\includegraphics*{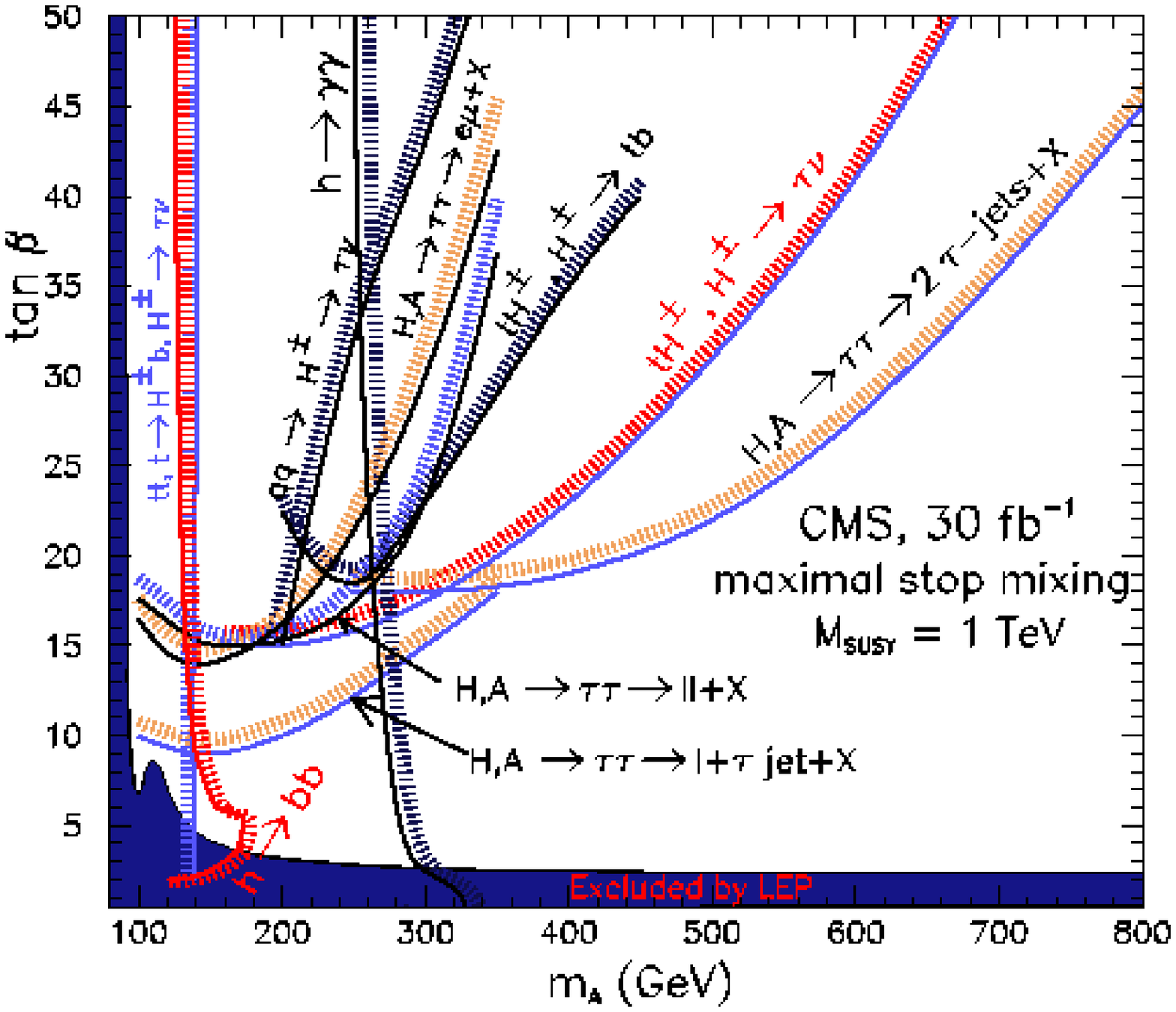}}
\hspace*{3mm}
\includegraphics*{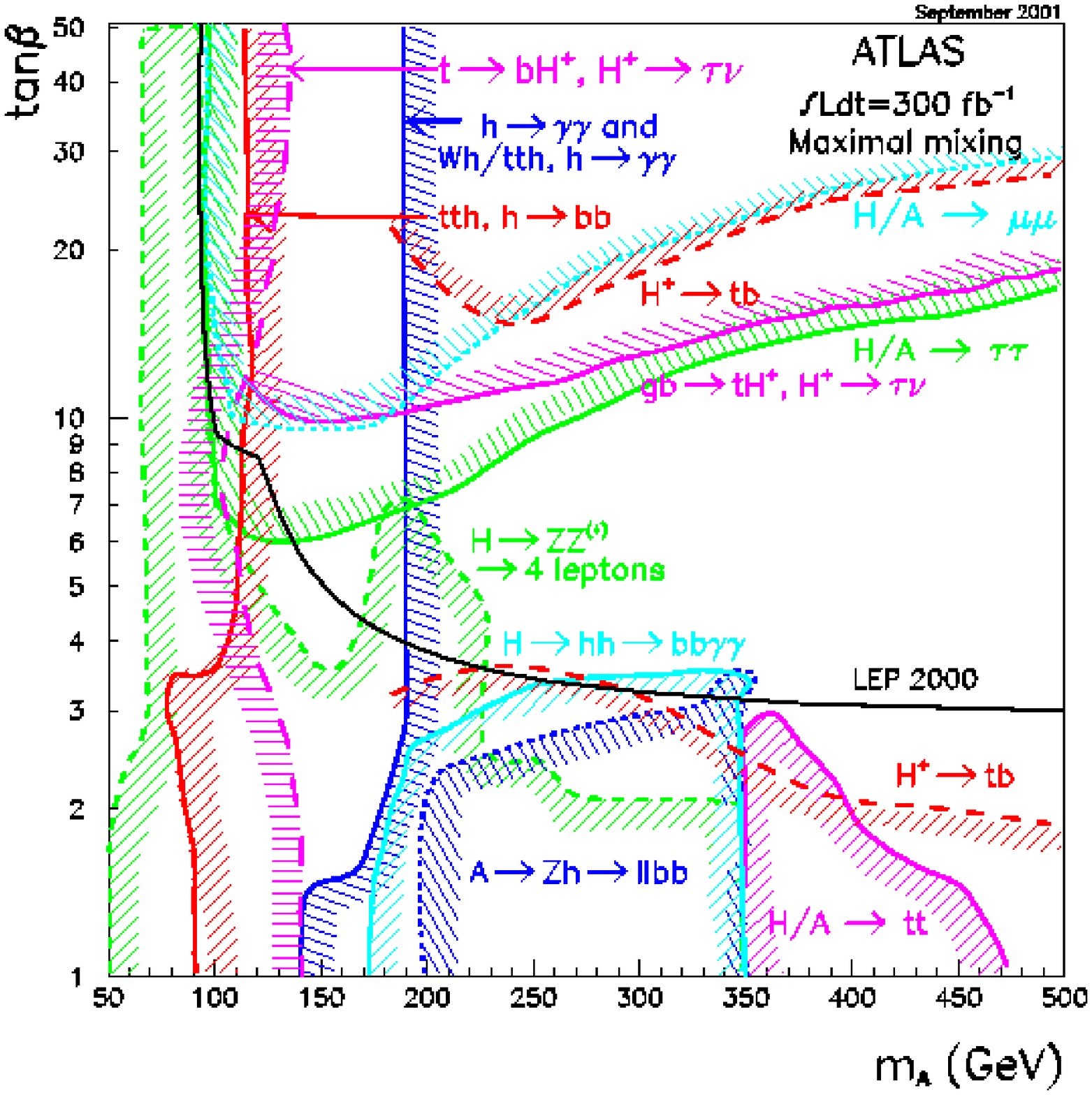}
}
\end{center}
\capt{\label{fig:susyhiggsathadron}
(a)~$5\sigma$ discovery contours for MSSM Higgs boson detection in
various channels in the $m_A$--$\tan\beta$ plane,
in the maximal mixing scenario, assuming
an integrated luminosity of $L=30~{\rm fb}^{-1}$
for the CMS detector~\protect\cite{cmsnote}. 
(b) As in (a), but for an integrated luminosity of
$L=300~{\rm fb}^{-1}$
for the ATLAS detector~\protect\cite{rwas}.}
\vspace{-0.2in}
\end{figure}

If no Higgs boson is discovered at the Tevatron, the LHC will cover the
remaining unexplored regions of the $\mha$--$\tanb$ plane, 
as shown in \figns{fig:susyhiggsathadron}{lhchiggses}.
That is, in the maximal mixing scenario (and probably in most regions
of MSSM Higgs parameter space),
at least one of the Higgs bosons 
is guaranteed to be discovered at either the Tevatron and/or the LHC.
A large fraction of the
parameter space can be covered by the search for
a neutral CP-even Higgs boson by employing the SM Higgs search
techniques, where the SM Higgs boson is replaced by $\hl$ or
$\hh$ with the appropriate re-scaling of the couplings.  Moreover,
in some regions of the parameter space, both $\hl$ and $\hh$ can be
simultaneously observed, and additional Higgs search
techniques can be employed to discover $\ha$, and/or $\hpm$
at the LHC.

\begin{figure}[t!]
  \begin{center}
\includegraphics[width=9.7cm]{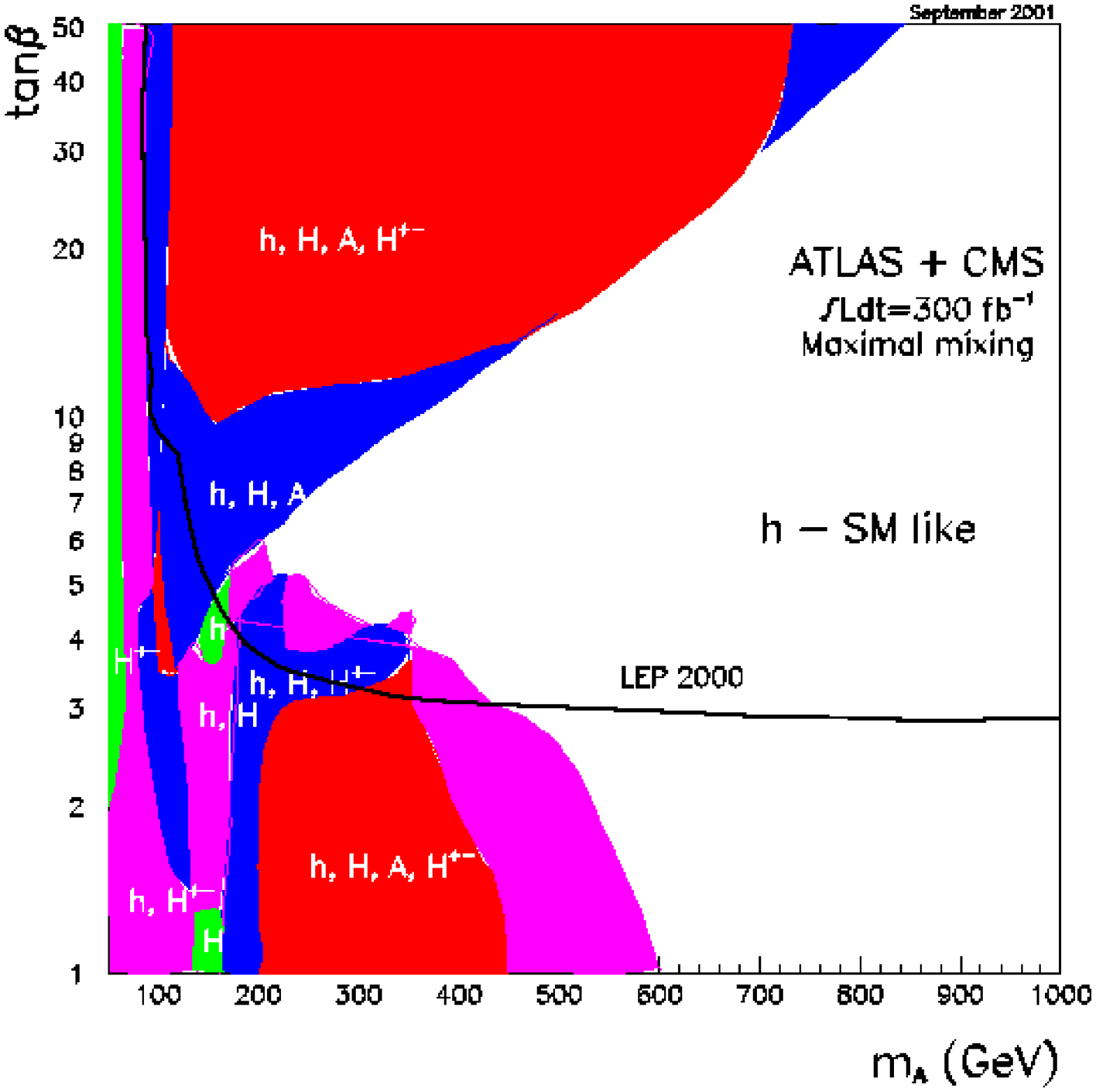}
  \end{center}
  \capt{\label{lhchiggses} 
Regions in the $\mha$--$\tan\beta$ plane in the maximal mixing scenario
in which up to four Higgs boson states
of the MSSM can be discovered at the LHC with
300~fb$^{-1}$ of data, based on a simulation that combines
data from the ATLAS and CMS detectors.  Taken from \protect\Ref{lhcupgrade}.
}
\vspace{-0.2in}
\end{figure}

\begin{figure}[t!]
\begin{center}
\resizebox{0.9\textwidth}{!}{
\includegraphics*{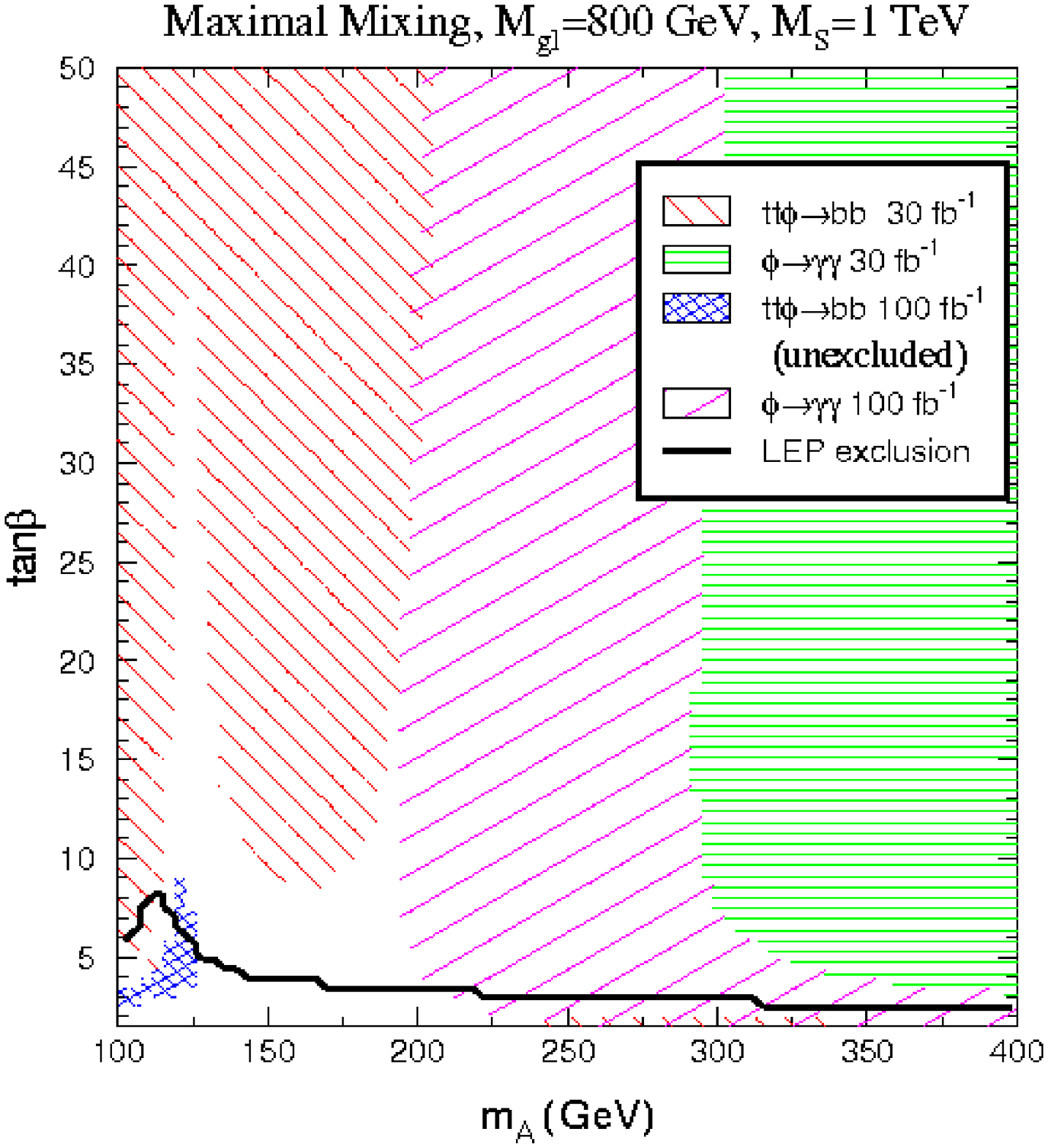}
\hspace*{3mm}
\includegraphics*{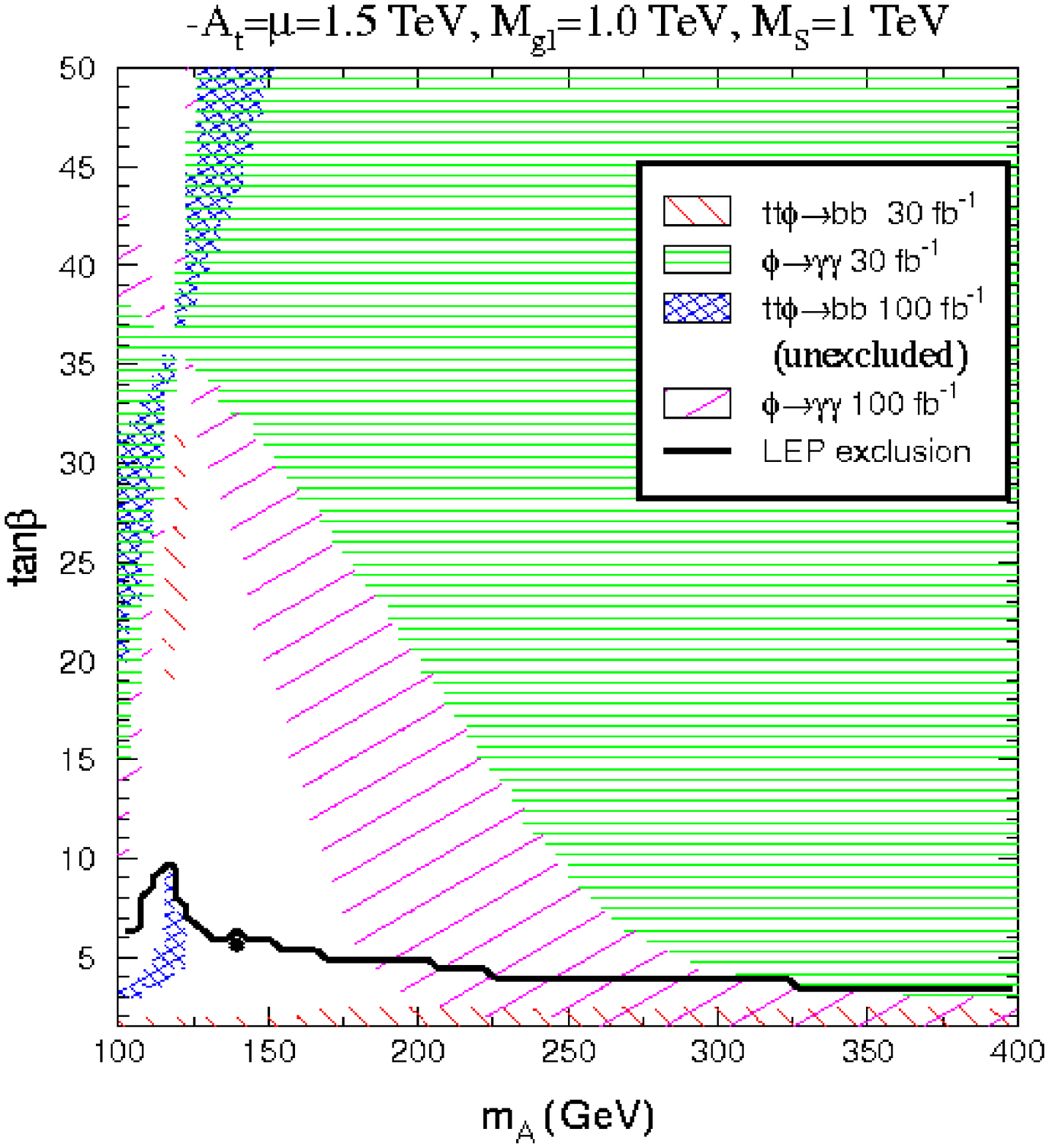}
}
\end{center}
\vspace{-0.15in}
\capt{\label{lhccomp} 
Complementarity between the LHC
searches for the decay modes $\phi\to b\bar b$ and $\phi\to\gamma\gamma$,
where $\phi=\hl$ or $\hh$
corresponds to the CP-even Higgs boson with the larger coupling to
$VV$.  Two different choices of MSSM parameters are
exhibited: (a)~the maximal mixing scenario and (b)~the
suppressed $Vb\to Vb\bar b$ production scenario.  In both cases, the
region corresponding to $5\sigma$ discovery of $\phi\to\gamma\gamma$
with 30~fb$^{-1}$ of data is shaded with parallel horizontal lines.  With
100~fb$^{-1}$ of data, these regions expand to include areas shaded
with diagonal parallel lines with positive slope.  The region corresponding
to $5\sigma$ discovery of $t\bar t\phi\to t\bar t b\bar b$ with
30~fb$^{-1}$ of data is shaded with diagonal parallel lines with
negative slope.  With 100~fb$^{-1}$ of data, these regions expand to
include the entire $\mha$--$\tan\beta$ plane {\it excluding} the
blue cross-hatched region.  The ``unexcluded'' region (where no discovery
of $t\bar t\phi\to t\bar t b\bar b$ is possible) occupies a small
region at low $\tanb$ and $\mha$ in both (a) and (b).  In addition, in
case (b), the excluded region also includes the two narrow wedge
regions at large $\tanb$ and low $\mha$. Taken from \protect\Ref{{cplhcyo}}.
}
\vspace{-0.2in}
\end{figure}

A CP-even Higgs boson, $\phi$, can be observed in a number of
different decay modes.  
If $\mha>\mhmax$, then
$\phi=\hl$ is SM-like (near the decoupling limit), whereas at
large $\tanb$ and $\mha<\mhmax$, $\phi=\hh$ is the SM-like Higgs
boson.\footnote{For $\mha<\mhmax$ and moderate $\tanb$ values, neither
CP-even Higgs boson is SM-like, although both Higgs masses lie below
about 150~GeV and will appear (albeit with reduced couplings to $VV$)
in the Higgs searches described above.}
It is possible to observe $\phi\to\gamma\gamma$
when $\phi$ is produced singly via $gg$ and $V^*V^*$ fusion, or when
produced in association with $W^\pm$ and/or $t\bar t$.  A second decay
mode, $\hl\to b\bar b$, can be observed in  $\phi t\bar t$ production.
Finally, it may be possible to observe $\phi\to\tau^+\tau^-$ when $\phi$
is produced via $V^*V^*$ fusion, where the forward jets are used to help
reduce backgrounds~\cite{tautau}.  Hence,
by using the complementarity of the various Higgs signatures
described above, one can discover
$\phi$ over nearly the entire MSSM parameter space, given
sufficient integrated luminosity.\footnote{One must
still demonstrate that it is possible at the LHC to 
discover the lightest CP-even Higgs boson, 
even if its branching ratios into $b\bar b$ 
and/or $\gamma\gamma$ are significantly suppressed
(either due to the effects of radiative corrections or due to the
existence of a significant branching fraction into invisible modes).
Such suppressions can occur in 
regions of the MSSM parameter space not yet considered by
the LHC Higgs search simulations.\label{footlhc}} 
In order to illustrate the complementarity of the $\gamma\gamma$ and
$b\bar b$ decay modes, we exhibit in \fig{lhccomp}
the regions of MSSM Higgs parameter space that can be covered for the
two benchmark scenarios of MSSM parameters described in \Sec{sec:352}.
The behavior illustrated in this figure can be understood by noting that
the $\phi b\bar b$ coupling can be significantly
suppressed (or enhanced), depending on the impact of the
radiative corrections discussed in \Sec{sec:33}.  As a result,
the branching ratio for $\phi\to \gamma\gamma$ is correspondingly
larger (or smaller), with obvious implications for the
$\phi\to b\bar b$ and $\phi\to\gamma\gamma$ searches.

We next focus on the potential for observing the heavier Higgs
states ($\hpm$, $\ha$ and $\hh$).  A number of recent
studies~\cite{branson,leshouches,cmsnote,nikitenko,morettiroy,assamagan} 
show that the following modes will
be effective in searching for the heavier MSSM Higgs bosons.
For the heavy neutral Higgs bosons, the most relevant decay
signatures are: $\ha,\hl\to\tau^+\tau^-$ (where the
$\tau$ is detected either via its leptonic or hadronic decay) and
$\ha$, $\hl\to\mu^+\mu^-$, which yield promising signals if $\tan\beta$
is large.  The $\tau^+\tau^-$ channel provides the largest discovery
reach in the heavy Higgs mass.  Other possible neutral Higgs decays:
$\ha$, $\hh\to t\bar{t}$; $\hh\to Z Z^* \to 4\ell$; $\hh\to\hl\hl$
and $\ha\to Z\hl$ are significant in regions of the parameter space
that are (nearly) ruled out by the LEP Higgs search.
For the charged Higgs boson, we must again consider
whether $\hpm$ can be produced in (on-shell) top-quark decays.
If this decay is forbidden, the
positively charged Higgs boson will be produced
primarily by $gb\to H^+t$
(see \Sec{sec:351}).  In either case, the observation of the charged
Higgs boson is possible if $\tanb\gg 1$ or
$\tanb\lsim\mathcal{O}(1)$~\cite{assamagan}.  For large $\tanb$, the decays
$\hp\to\tau^+\nu$ and $t\bar b$ (if kinematically allowed) provide the
most favorable signatures.  In particular, the $\tau\nu$ decay mode,
followed by the hadronic decay of the $\tau$ provides the largest discovery
reach for large $\mhpm$.  The ultimate charged Higgs
mass reach can depend significantly on the choice of MSSM parameters
that control the radiative corrections to the Higgs-bottom quark
Yukawa coupling~\cite {chhiggstotop2}
[see, {\it e.g.}, \eq{hmtb}].

Putting all of the above results together, it may be possible 
at the LHC to either {\it exclude} the entire $\mha$--$\tan\beta$ plane 
(thereby eliminating the MSSM Higgs sector as a viable model), or 
achieve a $5\sigma$ discovery of at least one of the MSSM Higgs bosons,
independently of the value of $\tanb$ and $\mha$.
For example, \Fig{fig:susyhiggsathadron} shows what
can be achieved by the CMS detector with 30 fb$^{-1}$~\cite{cmsnote} 
and by the ATLAS detector with 300 fb$^{-1}$~\cite{rwas}, 
assuming the maximal mixing scenario.
Note that over a significant fraction of the MSSM Higgs parameter space, at
least two Higgs bosons can be observed as shown in \fig{lhchiggses}.
Nevertheless, there is still a sizable wedge-shaped
region at moderate values of $\tan\beta$
opening up from about $\mha=200$~GeV to higher values in which
the heavier Higgs bosons cannot be discovered at the LHC.
In this parameter regime,
only the lightest CP-even Higgs boson can be discovered, and its
properties
are nearly indistinguishable from those of the SM Higgs
boson. Precision measurements of Higgs branching ratios and other
properties will be required in order to detect deviations from
SM Higgs predictions and
demonstrate the existence of a non-minimal Higgs sector.

Finally, we noted at the end of \Sec{sec:353} that CP-violating
effects in the Higgs sector can modify the usual
CP-conserving Higgs phenomenology.  As a result, the LHC
discovery reach of various Higgs channels discussed above may be
altered in a significant way.  It is therefore essential to make
complementary measurements in as many Higgs channels as possible in
order to cover the most general MSSM
parameter space~\cite{cplhcyo}.

\subsection{MSSM Higgs Boson Searches at the LC}
\label{sec:36}

The main production mechanisms for the MSSM Higgs
bosons are~\cite{MSSMprod}
\beqa
(i)~~&&e^+e^-\to Z\hl\,, Z\hh~~~\mbox{\rm via Higgs-strahlung}\,,\nonumber \\
(ii)~~&& e^+e^- \to \nu \bar{\nu}\hl\,,\nu\bar\nu\hh~~~{\rm
via}~W^+W^-~{\rm fusion}\,,\nonumber \\
(iii)~~&&e^+e^-\to\hl\ha\,, \hh\ha~~~{\rm via}~s\mbox{\rm-channel}~Z~{\rm
exchange}\,,\nonumber \\
(iv)~~&&e^+e^-\to H^+H^-~~~{\rm via}~s\mbox{\rm-channel}~\gamma\,,Z~{\rm
exchange}\,.
\eeqa
As in the SM Higgs search, process \textit{(i)} followed by
$Z\to\ell^+\ell^-$ allows the Higgs
boson recoiling against the $Z$ to be reconstructed, 
independently of the Higgs decay
channel.  Thus, $\hl$ [or $\hh$ if $\sin(\beta-\alpha)\ll 1]$ 
can be discovered at the LC even if it has a large branching
fraction into invisible modes.\footnote{The processes 
$e^+e^-\to e^+e^-\hl$, $e^+e^-\hh$ (via $ZZ$ fusion) also allow for
Higgs detection, independently of its decay channel, by reconstruction of
the Higgs boson recoiling against the final state $e^+e^-$ pair.
However, the $ZZ$ fusion rates are an order of magnitude smaller
than the corresponding $W^+W^-$ fusion rates [process \textit{(ii)}].}

\begin{figure}[t!]
\begin{center}
\resizebox{0.95\textwidth}{!}{
\includegraphics*{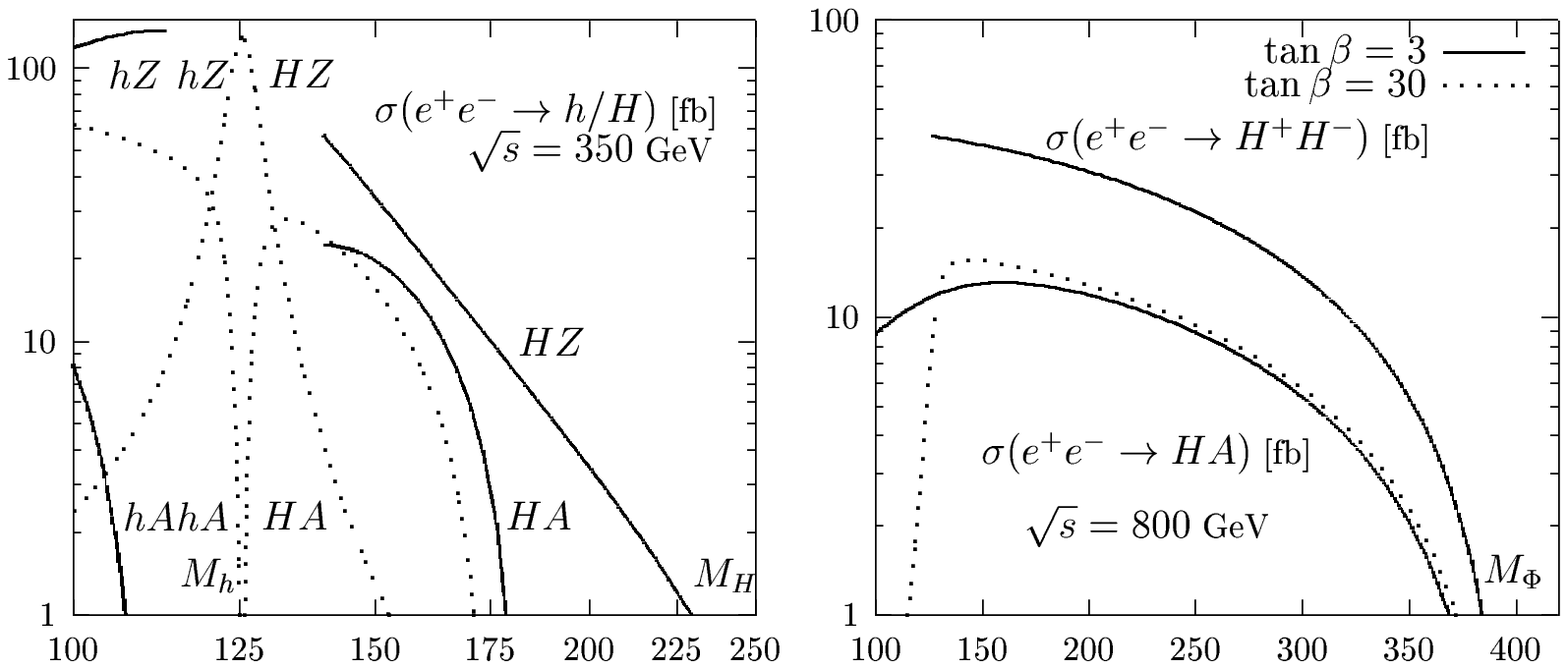}
}
\end{center}
\vspace{-0.15in}
\capt{\label{fig:mssmhiggsproduction} MSSM Higgs boson production
rates at the LC for two choices of $\tan\beta=3$ (solid) and 30
(dotted) for (a)~$\sqrt{s}=350$~GeV  as a function of the
mass of the produced CP-even neutral Higgs boson (either $\hl$ or
$\hh$); and for (b)~$\sqrt{s}=800$~GeV as a function of $\mhpm$ and
$\mha$, respectively.  Taken from \protect\Ref{abdel}.}
\vspace{-0.2in}
\end{figure}

Processes \textit{(i)} and \textit{(iii)} are complementary to
each other as a consequence of unitarity sum rules for
tree-level Higgs couplings~\cite{weldon,wudka}.  In particular, \eq{hxi}
implies that both $g^2_{\phi ZZ}$ and $g^2_{\phi
AZ}$ ($\phi=\hl$ or $\hh$)
cannot simultaneously vanish.  If $\mha\lsim\mhmax$, then all
the MSSM Higgs boson states have mass below 150~GeV, and
can be cleanly reconstructed at
the LC (with $\sqrt{s}\geq 350$~GeV) via the four production mechanisms
listed above~\cite{Tesla-TDR}.
On the other hand,
when $\mha\gsim 200$~GeV, one finds that $\mha\sim\mhh\sim\mhpm$
and $g_{HZZ}\sim g_{hAZ}\sim 0$,
and the couplings of $\hl$ are nearly identical to those of $\hsm$ as a
consequence of the decoupling limit.  Since $\mhl\lsim 135$~GeV, the LC
with a center-of-mass energy of 300 GeV is more than
sufficient to observe the $\hl$ [via processes
\textit{(i)} and \textit{(ii)}] and thus
cover the entire MSSM parameter space with certainty.  Moreover,
the cross-sections for $\hh Z$, $\hh\nu\bar\nu$ and $\hl\ha$ are
strongly suppressed [since $|\cos(\beta-\alpha)|\ll 1$].  The
cross-sections for $\hh\ha$ and $H^+ H^-$ production are unsuppressed if
kinematically allowed.\footnote{Due to the $p$-wave suppression at
threshold, the $\hh\ha$ and $H^+ H^-$ cross-sections fall off rapidly as
the corresponding Higgs masses approach $\sqrt{s}/2$.}
That is, the heavy Higgs bosons, $\hh$, $\ha$
and $\hpm$ can only be observed in pair production processes where both
Higgs states are heavy (and the minimum $\sqrt{s}$ required is somewhat
above $2\mha$).
These features are evident in \fig{fig:mssmhiggsproduction}, which
depicts cross-sections for Higgs-strahlung [process \textit{(i)}]
and associated Higgs pair production
[processes \textit{(iii)} and \textit{(iv)}] as a function of the
corresponding Higgs mass for two different choices of $\sqrt{s}$
and $\tanb$.  The cross-section for Higgs production via $W^+W^-$ fusion
[process \textit{(ii)}] is not shown.  The $\nu\bar\nu\phi$
production cross-section
is suppressed relative to the corresponding SM cross-section (shown in
\fig{LCxsecs}) by a factor of $\sin^2(\beta-\alpha)$
[$\cos^2(\beta-\alpha)$] for $\phi=\hl$ [$\phi=\hh$].

In addition to $H^+H^-$ production, there are a number of
mechanisms in which the charged Higgs boson is singly produced.
Charged Higgs bosons can be produced in top decays via
$t \rightarrow b + H^+$ if $\mhpm<m_t-m_b$, as discussed previously in
\Sec{sec:351}.  The  process $e^+e^- \to W^\pm H^\mp$, which
arises at one-loop~\cite{wpmhmp,logansu},
allows for the possibility
of producing a charged Higgs boson with $\mhpm>\sqrt{s}/2$,
when $H^+H^-$ production is kinematically forbidden.
With favorable MSSM parameters and moderate values of
$\tan\beta$, more than ten $W^{\pm}H^{\mp}$
events can be produced at the LC
for $\mhpm\lsim 350$ GeV with $\sqrt{s}=500$~GeV and 500~fb$^{-1}$ of
data, or for $\mhpm\lsim 600$~GeV with $\sqrt{s}=1$~TeV and
1~ab$^{-1}$~\cite{logansu}.
Other single charged Higgs production mechanisms
include $t\bar b H^-$/$\,\bar t bH^+$ production~\cite{ttH},
$\tau^+\nu H^-$/$\,\tau^-\bar\nu H^+$ production~\cite{moretti},
and a variety
of processes in which $H^\pm$ is produced in
association with a one or two other gauge and/or
Higgs bosons~\cite{kmo}.

\begin{figure}[t!]
\begin{center}
\resizebox{0.95\textwidth}{!}{
\includegraphics*[19,142][529,682]{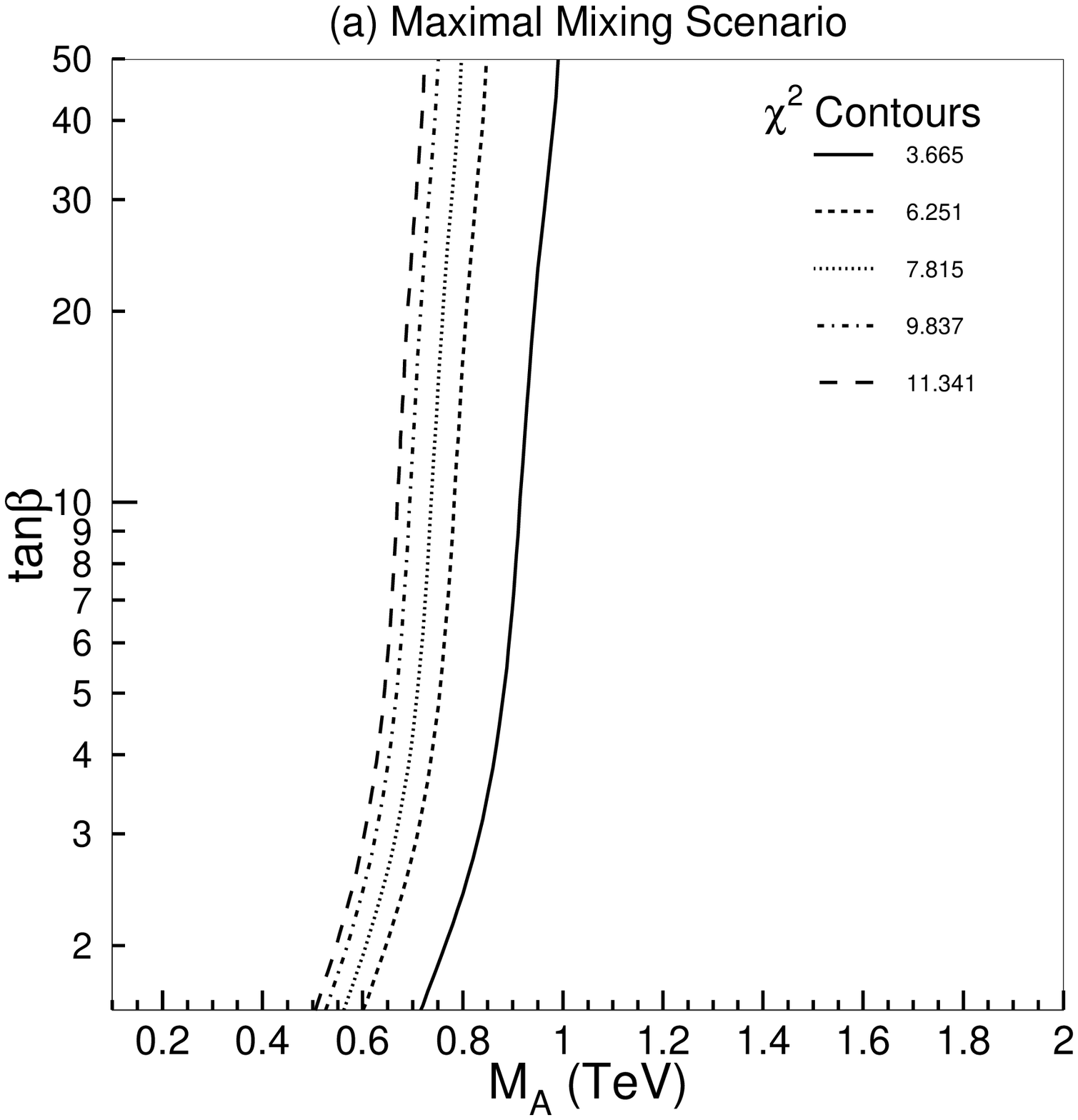}
\hspace*{3mm}
\includegraphics*[19,142][529,682]{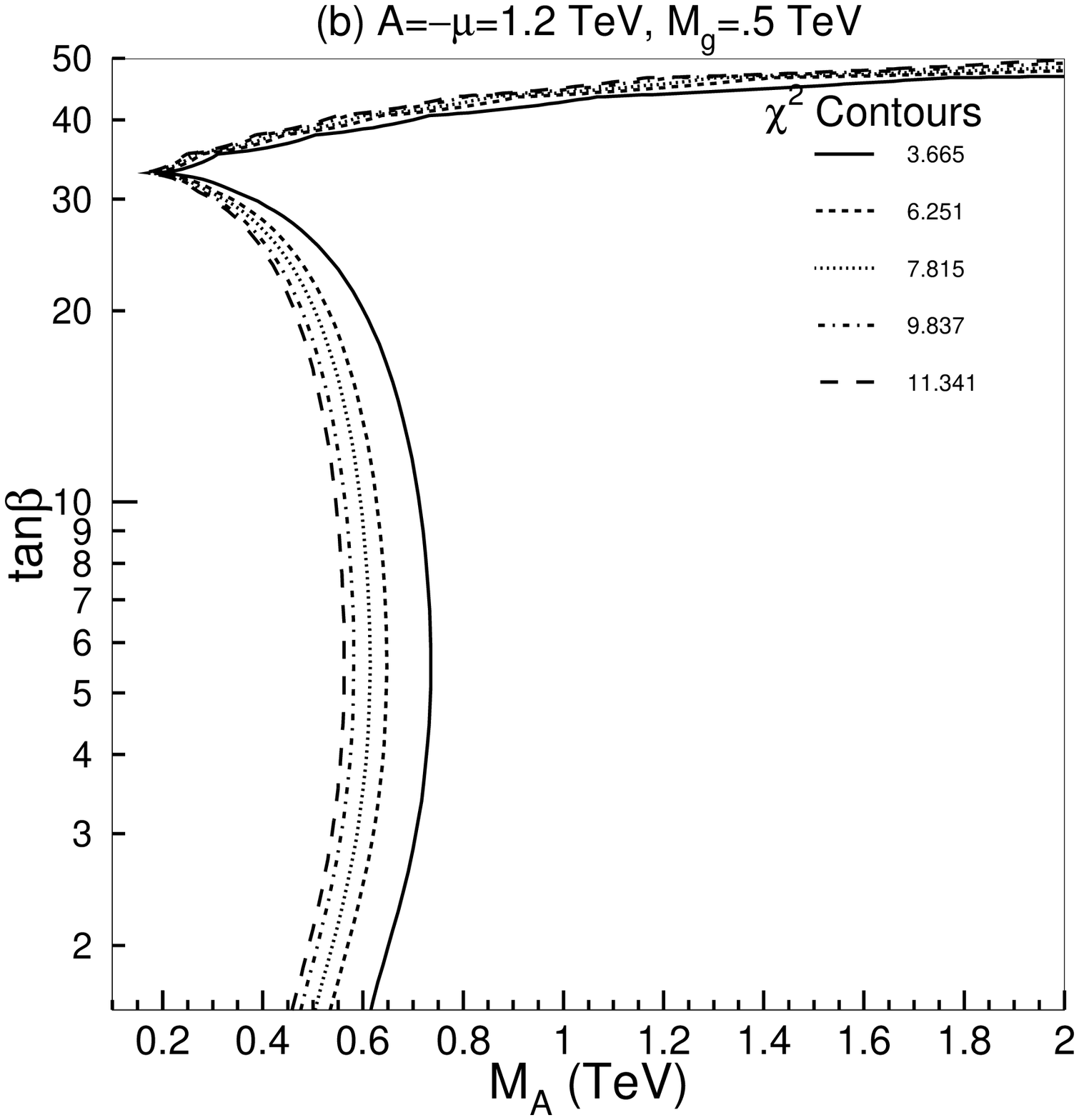}
}
\vspace{-0.15in}
\end{center}
\capt{\label{fig:chisquare}
 Contours of $\chi^2$ for Higgs
boson decay observables for (a) the maximal mixing scenario; and
(b) a choice of MSSM parameters for which the loop-corrected
$\hl b\bar b$ coupling is suppressed (relative to the corresponding
tree-level coupling) at large $\tanb$ and low $\mha$.
These results are based on Higgs partial width
measurements anticipated at the LC (shown in Table~1) with
$\sqrt{s}=500$~GeV and an integrated luminosity of 500~fb$^{-1}$.
The contours correspond to
68, 90, 95, 98 and 99\% confidence levels (right to left) for the 
observables $g^2_{hbb}$, $g^2_{h\tau\tau}$, and $g^2_{hgg}$. See
\protect\Ref{chlm} for additional details.}
\vspace{-0.2in}
\end{figure}

The heavier Higgs states could lie beyond the discovery reach
of the LC ($\sqrt{s} \leq 1$~TeV) and the LHC [{\it cf.} \fig{lhchiggses}].
In this case, the precision measurements of the $\hl$ decay
branching ratios and couplings achievable at the LC
are critical for distinguishing between $\hsm$ and $\hl$
of a non-minimal Higgs sector
with properties close to that of the SM Higgs boson.
To illustrate the challenge of probing the decoupling limit, suppose
that $m_A>\sqrt{s}/2$ so that only the light Higgs boson, $\hl$, can be
observed directly at the LC.  In this case,
the fractional deviation of the couplings of $\hl$ relative
to those of the SM Higgs boson scales as $m_Z^2/m_A^2$.  Thus,
if precision measurements reveal
a non-zero deviation, one could in principle derive a constraint 
({\it e.g.}, upper and lower bounds) on the
heavy Higgs masses of the model.  In the MSSM, the constraint is
sensitive to the MSSM parameters that control the radiative
corrections to the Higgs couplings.  This is illustrated in
\fig{fig:chisquare}, where the constraints on $\mha$ are derived
for two different sets of MSSM parameter 
choices~\cite{chlm}.  Here, a simulation of a
global fit of measured $hbb$, $h\tau\tau$ and $hgg$ couplings is made
(based on the results of Table 1)
and $\chi^2$ contours are plotted indicating the constraints in the
$m_A$--$\tan\beta$ plane, assuming a deviation from SM Higgs
boson couplings
is seen.\footnote{These results are similar to those
of \Ref{BattagliaDesch}, in which $\mhl=120$~GeV and an
integrated LC luminosity of 1000~fb$^{-1}$ were assumed.  The
upper bound on $\mha$ as a function of $\tanb$ was determined in which
68\%, 90\% and 95\% of the MSSM parameter space, respectively, yield
$\hl$ branching ratio predictions that differ from those of the SM at
the 95\% CL.  However, 
the MSSM parameter regimes in which the sensitivity to $\mha$ are 
weakest were not identified in \Ref{BattagliaDesch}.}   
In the maximal mixing scenario shown in \fig{fig:chisquare}(a),
the constraints on $\mha$ are significant and rather insensitive to the
value of $\tan\beta$. 
Similar results, in which deviations of
${\rm BR}(\hl\to b\bar b)/{\rm BR}(\hl\to\tau^+\tau^-)$ from the SM 
prediction yield limits on the allowed values of $\mha$ as a function of
$\tan\beta$, have been obtained in \Ref{siannah}.
However in some cases, as shown in \fig{fig:chisquare}(b),
a region of $\tan\beta$ may yield almost no constraint on $\mha$.
This is due to the phenomenon of $\mha$-independent decoupling noted
below \eq{earlydecoupling}, in which $\cos(\beta-\alpha)$ [which
controls the departure from the decoupling limit] vanishes at a
particular value of $\tanb$ independently of the value of $\mha$.
Thus, one cannot extract a fully model-independent upper bound on
the value of $\mha$ (beyond what can be deduced if no direct $\ha$
production is observed at the LC).
Of course, after supersymmetric particles are discovered,
information about the MSSM spectrum can be used to obtain a more
stringent bound on $\mha$, using the techniques described above.

The $e^+e^-$ linear collider running in the $\gamma\gamma$ collider mode
presents additional opportunities for the study of the MSSM Higgs
sector.
Resonance production $\gamma\gamma \to H$ and $A$
can be used to extend the reach in
Higgs masses beyond the limit set by $HA$ pair production in the
$e^+e^-$ mode~\cite{n41a,asner,velasco}.  Typically, one can probe the
heavy
Higgs masses out to $m_A\sim 0.8\sqrt{s}$ (where $\sqrt{s}$ is the
center of mass energy of the LC).  This expands the MSSM Higgs discovery
reach further into
regions of the $\mha$--$\tan\beta$ parameter space for which
the LHC is not sensitive in general
(the so-called ``blind wedge'' of large $\mha$ and
moderate values of $\tan\beta$ seen in \fig{lhchiggses}).

As noted above, at
least one Higgs boson must be observable at the LC in the MSSM.
In non-minimal supersymmetric models, additional Higgs bosons appear
in the spectrum, and the ``no-lose'' theorem of the MSSM must be
reconsidered.  For example, in the non-minimal supersymmetric
extension of the Standard Model (the so-called NMSSM where a Higgs
singlet is added to the model~\cite{nmssm}),
the lightest Higgs boson decouples
from the $Z$ boson if its wave function is dominated by the Higgs
singlet component.  However, in this case the second
lightest neutral CP-even
Higgs boson usually plays the role of $\hl$ of the 
MSSM.  That is, the mass of the second lightest 
Higgs boson is typically below 150 GeV with
significant couplings to $ZZ$, so that it can be produced by the
Higgs-strahlung process with an observable 
cross-section~\cite{okada,comelli}.
If the second lightest Higgs boson also decouples from the $Z$, then
the third lightest Higgs boson will play the role of $\hl$ of the MSSM
for which the observation is ensured, and so on.
Even in bizarre scenarios where all the neutral Higgs boson share
equally
in the coupling to $ZZ$ (with the sum of all squared couplings
constrained to equal the square of the $\hsm ZZ$
coupling~\cite{weldon,wudka}),
the ``no-lose'' theorem still applies---Higgs production at the LC
must be
observable~\cite{jack}.  In contrast, despite significant progress,
there is no complete guarantee
that at least one Higgs boson of the NMSSM must be discovered
at the LHC for all choices of the model parameters~\cite{nolose-lhc}.

One of the key parameters of the MSSM Higgs sector
is the value of the ratio of Higgs vacuum expectation values,
$\tan\beta$.  In addition to providing information about the structure
of the non-minimal Higgs sector, the measurement of this parameter
also provides an important check of supersymmetric structure, since
this parameter also enters the chargino, neutralino
and third generation squark mass matrices and couplings.
Thus, $\tan\beta$ can be measured independently using
supersymmetric processes and compared to the value obtained from
studying the Higgs sector.  Near the decoupling limit, the properties
of
$h$ are almost indistinguishable from those of $\hsm$, and thus
no information can be extracted on the value of $\tan\beta$.
However, the properties of
the heavier Higgs bosons are $\tan\beta$-dependent.
Far from the
decoupling limit, all Higgs bosons of the MSSM will be observable at
the LC and exhibit strong $\tan\beta$-dependence in their couplings.
Thus, to extract
a value of $\tan\beta$ from Higgs processes, one must observe the
effects of the heavier Higgs bosons of the MSSM at the LC.

The ultimate accuracy of the $\tan\beta$ measurement at the LC depends
on the value of $\tan\beta$.  In Ref.~\cite{gunionetal}, it is argued
that one must use a number of processes, including $b\bar b b\bar b$ final
states arising from $b\bar b H$, $b\bar b A$, and $HA$ production,
and $t\bar t b\bar b$ final states arising from $t\bar b H^+$,
$b\bar t H^-$ and $H^+ H^-$ production.  One subtlety that arises here
is that in certain processes, the determination of $\tan\beta$
may be sensitive to loop corrections that depend on the values of
other supersymmetric parameters.  One must settle on a consistent
definition of $\tan\beta$ when loop corrections are
included~\cite{tanbloopdef}.
A comprehensive analysis of the extraction of
$\tan\beta$ from collider data, which incorporates loop effects,
has not yet been given.

The study of the properties of the heavier MSSM Higgs bosons
(mass, width, branching ratios, quantum numbers, {\it etc.})
provides a number of additional challenges.
For example, in the absence of CP-violation,
the heavy CP-even and CP-odd Higgs bosons, $\hh$ and $\ha$, are
expected to be nearly mass-degenerate.  Their CP quantum numbers and
their separation can be investigated at the same time in the
$\gamma\gamma$ collider mode of the LC.  If the polarization states of
the two incoming linearly-polarized photons are parallel
[perpendicular] then only the
CP-even Higgs boson $\hh$
[CP-odd Higgs boson $\ha$] will be produced~\cite{gamgamcp}.
Thus, the determination of the Higgs boson CP quantum numbers
and the separation of the two
different states can be achieved.
In the case of a CP-violating Higgs sector,
the observation and measurement
of Higgs boson properties become much more challenging.  The
$\gamma\gamma$ collider
can provide new opportunities to test the nature of the couplings of the
Higgs neutral eigenstates (with indefinite CP quantum numbers)
to gauge bosons and fermions~\cite{velasco,gamgamcpv}.

Finally, once the heavy Higgs spectrum is revealed, one would like to
reconstruct the two-Higgs-doublet scalar
potential~\cite{mssmhiggspotential}.  This is not
likely to be accomplished at a first generation LC, although one can
make a start if the heavy Higgs masses are not too large.
To probe aspects of the Higgs potential one must observe multiple Higgs
production in order to extract the Higgs
self-couplings~\cite{trilinear,n36a,mssmhiggspotential,susytrilinear}.
Ultimately, such a program would require
an LC with very high energy and luminosity such as CLIC.

\section{Conclusions}
\label{sec:4}

The physical origin of electroweak symmetry breaking is not yet known.
In all theoretical approaches and models,
the dynamics of electroweak symmetry
breaking must be revealed at the TeV-scale or below.  This energy
scale will be thoroughly explored by hadron colliders, starting with
the Tevatron and followed later in this decade by the LHC.
Even though the various theoretical alternatives can only be confirmed
or ruled out by future collider experiments,
a straightforward interpretation of the electroweak precision data
suggests that electroweak symmetry breaking dynamics is
weakly-coupled, and a Higgs boson with mass between
100 and 200 GeV must exist.   With the
supersymmetric extension of the Standard Model, this interpretation
opens the route to grand unification of all the fundamental forces, with the
eventual incorporation of gravity in particle physics.

In this review we have summarized the theoretical
properties of the Standard Model Higgs boson and the Higgs bosons of
the MSSM, and surveyed the search strategies for discovering the Higgs
boson at hadron and lepton colliders.
We have assessed the Higgs boson discovery reach
of present and future
colliders, and described methods for measuring the various Higgs boson
properties (mass, width, CP quantum numbers, branching ratios and
coupling strengths).

The observation of a Higgs boson in the theoretically preferred mass range
below $200$~GeV may be possible at the
Tevatron, whereas experiments at the LHC can discover the SM Higgs
boson over the full Higgs mass range up to 1~TeV.  The Tevatron can
also extend the LEP search for Higgs bosons of the MSSM by either
discovering the lightest CP-even MSSM Higgs boson, $\hl$ (or in some
special cases discovering additional Higgs scalars of the model),
or by further constraining the MSSM Higgs parameter space.  The LHC is
sensitive to nearly the entire MSSM Higgs parameter space, in which
either $\hl$ alone can be discovered or multiple Higgs states can be
observed.  A program of Higgs measurements will be initiated at the
LHC to measure Higgs partial widths with an accuracy in the range of
10--30\%.

The discovery of the Higgs boson at the Tevatron and/or the LHC
is a crucial first step.  The measurement of Higgs properties at the
LHC will begin to test the dynamics of electroweak symmetry breaking.
However, a high-luminosity $e^+e^-$
linear collider, now under development, is
needed for a systematic program of precision Higgs measurements.
For example, depending on the value of the Higgs mass, branching
ratios and Higgs couplings can be determined in some cases at the
level of a few percent.
In this way, one can extract
the properties of the Higgs sector in a comprehensive way,
and establish (or refute) the existence of
scalar sector dynamics as the mechanism responsible
for generating the masses of the fundamental particles.

\section*{Acknowledgments}

We would like to thank Ulrich Baur, Sally Dawson, Jack Gunion, Heather Logan,
Carlos Wagner, Georg Weiglein, Scott Willenbrock and
Peter Zerwas for enlightening discussions.  We are especially grateful
to Steve Mrenna and Michael Spira for many useful suggestions and for
their dedicated work in providing
a number of plots shown in this review.  Finally, much of this
work would not have been possible without the collective wisdom gathered
from our colleagues of the 1998 Tevatron Higgs Working
Group and the 2001 Snowmass Electroweak Symmetry Breaking Working Group.

Fermilab is operated by Universities Research Association,
under contract no.~DE-AC02-76CH03000 with the U.S. Department of
Energy.  H.E.H. is supported in part by the U.S. Department of Energy
under grant no.~DE-FG03-92ER40689.

\end{document}